\documentclass[11pt,a4paper,twoside,openright,titlepage,fleqn,%
               headinclude,dottedtoc,footinclude,BCOR5mm,%
               numbers=noenddot,cleardoublepage=empty,%
               tablecaptionabove]{scrreprt}
               
\usepackage[italian,american,spanish,british]{babel}
\usepackage[utf8]{inputenc}
\usepackage[T1]{fontenc}
\usepackage{amsmath,amssymb,amsthm}
\usepackage{varioref}
\usepackage[titletoc]{appendix}

\usepackage[style=authoryear-comp,
hyperref,
date = year, 
dashed=false,
mincitenames = 1,
maxcitenames = 2, 
minbibnames = 3, 
uniquename=false,
uniquelist=false,
giveninits=true,
sortcites=false,
natbib]{biblatex}
\renewbibmacro{in:}{}
\AtEveryBibitem{\clearfield{eprint}\clearfield{title}\clearfield{month}\clearfield{url}}
\usepackage{xpatch}
\xpatchbibmacro{date+extradate}{%
  \printtext[parens]%
}{%
  \setunit{\addperiod\space}%
  \printtext%
}{}{}

\usepackage{chngpage}
\usepackage{calc}
\usepackage{listings}
\usepackage{graphicx}
\usepackage{subfig}
\usepackage{multicol}
\usepackage{makeidx}
\usepackage{fixltx2e}
\usepackage{relsize}
\usepackage{lipsum}
\usepackage{textgreek}
\usepackage{upgreek}
\usepackage[eulerchapternumbers,subfig,beramono,eulermath,pdfspacing,listings]{classicthesis}
\usepackage{arsclassica}

\newcommand{\myTitle}{Observatorio Virtual y aprendizaje automático para el estudio de objetos de baja masa en cartografiados espectroscópicos y fotométricos}
\newcommand{\myTitleEn}{Virtual Observatory and machine learning for the study of low-mass objects in photometric and spectroscopic surveys}

\let\orgtheindex\theindex
\let\orgendtheindex\endtheindex
\def\theindex{%
	\def\twocolumn{\begin{multicols}{2}}%
	\def\onecolumn{}%
	\clearpage
	\orgtheindex
}
\def\endtheindex{%
	\end{multicols}%
	\orgendtheindex
}

\makeindex

\definecolor{lightergray}{gray}{0.99}

\lstnewenvironment{code}%
{\setkeys{lst}{columns=fullflexible,keepspaces=true}%
\lstset{basicstyle=\small\ttfamily}%
}{}

\lstnewenvironment{sidebyside}{%
    \global\let\lst@intname\@empty 
    \setbox\z@=\hbox\bgroup 
    \setkeys{lst}{columns=fullflexible,%
    linewidth=0.45\linewidth,keepspaces=true,%
    breaklines=true,%
    breakindent=0pt,%
    boxpos=t,%
    basicstyle=\small\ttfamily
}%
    \lst@BeginAlsoWriteFile{\jobname.tmp}%
}{%
    \lst@EndWriteFile\egroup 
        \begin{center}%
            \begin{minipage}{0.45\linewidth}%
                \hbox to\linewidth{\box\z@\hss} 
            \end{minipage}%
            \qquad 
            \begin{minipage}{0.45\linewidth}%
            \setkeys{lst}{frame=none}%
                \leavevmode \catcode`\^^M=5\relax 
                \small\input{\jobname.tmp}%
            \end{minipage}%
        \end{center}%
}

\graphicspath{{Graphics/}}

\bibliography{Bibliography}

\defbibheading{bibliography}{%
\cleardoublepage
\manualmark
\phantomsection
\addcontentsline{toc}{chapter}{\tocEntry{\bibname}}
\chapter*{\bibname\markboth{\spacedlowsmallcaps{\bibname}}
{\spacedlowsmallcaps{\bibname}}}}     

  \DeclareCiteCommand{\citeyearpar}[\mkbibparens] 
  {\boolfalse{citetracker}%
   \boolfalse{pagetracker}%
   \usebibmacro{prenote}} 
  {\printtext[bibhyperref]{\printfield{year}}} 
  {\multicitedelim} 
  {\usebibmacro{postnote}} 

\makeatletter 
  \DeclareCiteCommand{\citetalias} 
  {\usebibmacro{prenote}} 
  {\usebibmacro{citeindex}%
   \bibhyperref{\@citealias{\thefield{entrykey}}}} 
  {\multicitedelim} 
  {\usebibmacro{postnote}} 
\makeatother 

\newcommand\tablefoot[1]{
  \par\vspace{1.5ex}
  \noindent
  \begin{minipage}{\linewidth}
    {\small\bfseries\small}~%
    \small
    \ignorespaces
    #1%
  \end{minipage}%
}

\defcitealias{bello2023}{Bello23}
\defcitealias{pass20}{Pass20}
\defcitealias{mar21}{Mar21}
\defcitealias{schw19}{Schw19}
\defcitealias{pass2019}{Pass19}

\usepackage[left=1.3in,right=1in,top=1in,bottom=1in]{geometry}

\counterwithin{figure}{chapter}
\counterwithin{table}{chapter}
\usepackage{pdflscape}
\usepackage[figuresright]{rotating}
\usepackage{longtable}
\usepackage{pdfpages}

\begin{document}
\pagenumbering{roman}
\pagestyle{plain}


\includepdf[pages={1}]{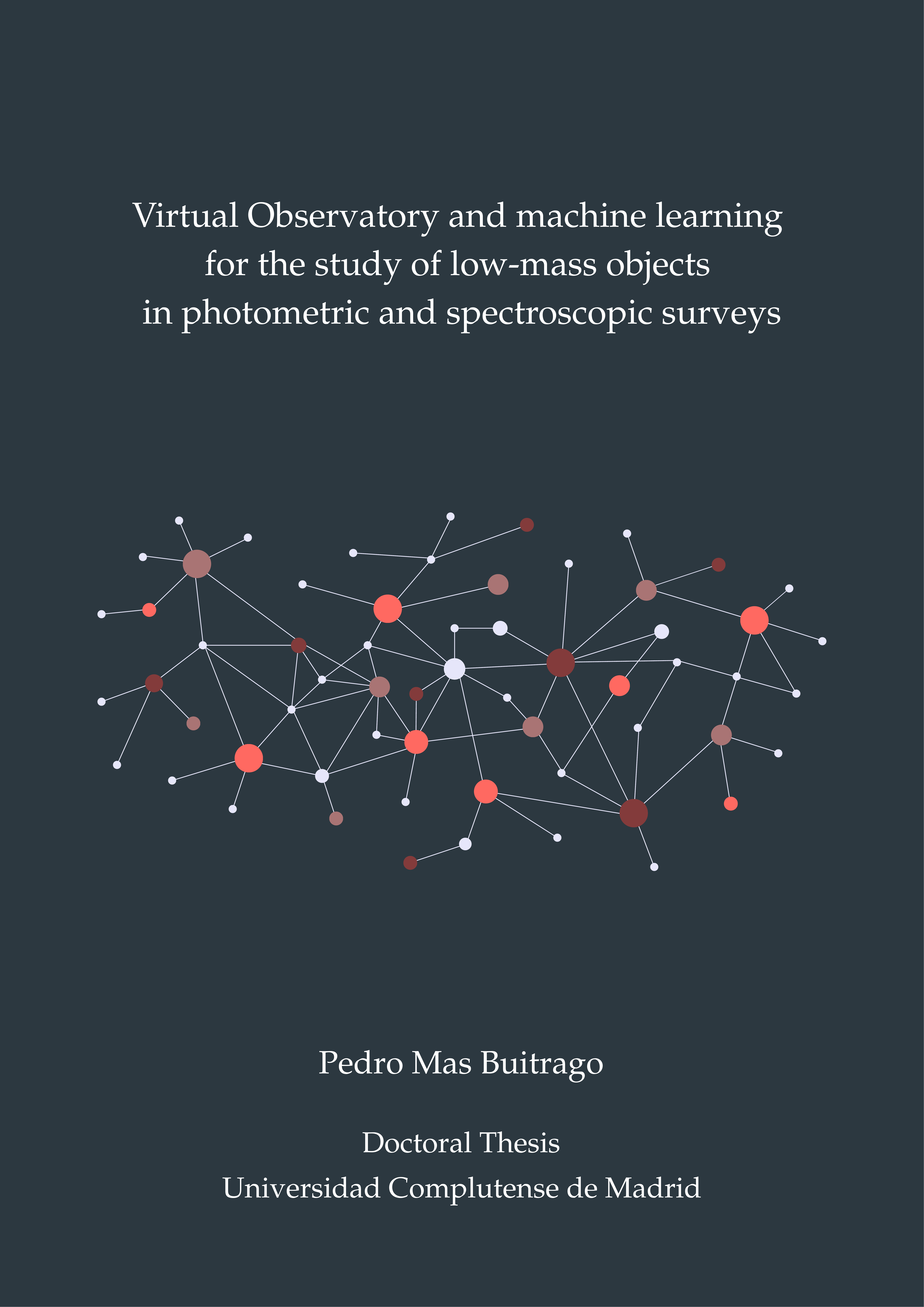}

\clearpage\hbox{}\thispagestyle{empty}\newpage
\clearpage

\begin{center}
\begin{LARGE}
{\textsc{Universidad Complutense de Madrid}}\\\vspace{.4cm}
\end{LARGE}

\begin{Large}
\textcolor{darkgray}{{Facultad de Ciencias F\'isicas}}\\
\end{Large}

\begin{Large}
\textcolor{darkgray}{{Doctorado en Astrofísica}}\\
\end{Large}

\begin{figure}[h!]
\vspace{0.8cm}
  \centering
    \includegraphics[width=.35 \textwidth]{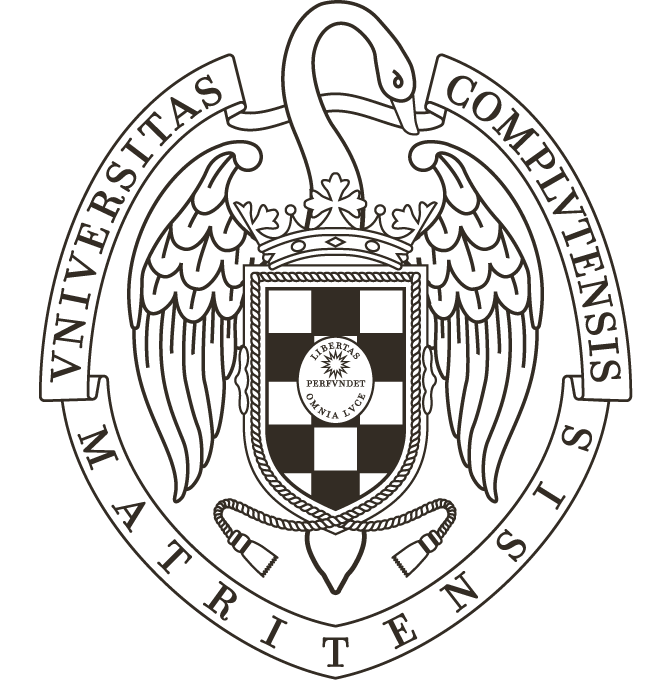}
\vspace{0.2cm}
\end{figure}

\begin{LARGE}
\textsc{Tesis doctoral}	
\end{LARGE}
\vspace{1 cm}

\begin{Large}
\textit{{\myTitle}}\\
\vspace{.6cm}
\textcolor{darkgray}{\textit{{\myTitleEn}}}\\\vspace{2.3 cm}
\end{Large}

\begin{large}
\textsc{Memoria para optar al grado de doctor\\ 
presentada por:}\\\vspace{0.5 cm}
\end{large}

\begin{Large}
	{Pedro Mas Buitrago}\\\vspace{1 cm}
\end{Large}

\begin{large}
\text{Supervisado por:} \\\vspace{0.4 cm}
Dr. Enrique Solano Márquez \\\vspace{0.4 cm}
Dra. Ana González Marcos\\\vspace{0.3 cm}
\end{large}

\begin{small}
2025 
\end{small}

\end{center}\vfill 

\clearpage\hbox{}\thispagestyle{empty}\newpage

\clearpage
\pdfbookmark{Agradecimientos}{Agradecimientos}

\begingroup
\let\clearpage\relax
\let\cleardoublepage\relax
\let\cleardoublepage\relax

\chapter*{Agradecimientos}

\textit{Suerte.} 

Dicen que es solo para quien la busca, pero yo la encontré en el instante en el que abrí los ojos por primera vez, en mi familia. Mamá, Papá, hermanito. Gracias por regalarme una vida repleta de felicidad y amor. Siempre habéis estado a mi lado iluminando el camino, protegiéndome, siendo el faro que me guía entre las turbulentas olas que navegas a medida que te vas haciendo un hueco en este mundo. Gracias por ser mi estrella. Gracias por ser mi hogar. Me habéis enseñado la importancia de ser justo, de tratar bien y respetar a los demás, de amar y cuidar a los tuyos por encima de todo. Lo pienso cada día: no hay tesoro más valioso que teneros a mi lado. Mi corazón es vuestro.

Marta, Adrián, Ana. Gracias por llenar de felicidad cada momento que paso con vosotros. No hay nada más bonito que ver una familia que irradia amor a raudales, que veros a los cuatro ser uno. A los niños de mis ojos: os deseo mucha suerte en esta vida en la que acabáis de aterrizar. Estoy seguro de que os depara un futuro precioso. Y lo estoy porque habéis tenido mi misma suerte, la de despertar en una familia que os quiere con locura.

Cuando me embarqué en esta nueva aventura, tuve la suerte de que me tocaron los mejores capitanes para liderarla. Gracias Enrique por el gran privilegio de compartir este camino contigo, por todas nuestras charlas en el despacho, por ser un ejemplo a seguir más allá del ámbito científico. Todo lo que he crecido gracias a ti durante estos años es incalculable, y siempre te estaré agradecido por ello. Espero seguir aprendiendo de ti muchos años. Gracias Ana por tu cercanía, por tu agradable y maravillosa forma de liderar. Siempre había querido adentrarme en este mundo apasionante de las redes neuronales, y no podría haber tenido una mejor mentora. La vida también me regaló un compañero con el que he compartido gran parte de este camino. Gracias Alberto por todas nuestras conversaciones locas, desde la Tribunona hasta el metaverso, pasando por el oso de cara corta. Todas esas horas que echamos juntos en el despacho y en el coche son de los mejor que me llevo de estos años. Llegaste como compañero de despacho y te quedas como un amigo de por vida. Y Diego, qué bonitos recuerdos me vienen cuando pienso en aquellos meses de locura que pasamos los tres juntos por aquí. Inmersos en una pandemia, nadie por las calles, y entre los tres conseguíamos que los días estuvieran llenos de risas. He tenido la suerte de tener muy cerca a personas increíbles que también se encuentran sumidas en esta caótica y ardua aventura. Muchas gracias por compartir este camino conmigo y mucha fuerza para los que estáis en la recta final. A los compañeros y compañeras del CAB, gracias por hacer del edificio D un segundo hogar. Gracias Margie por las innumerables gestiones y tu gran ayuda durante estos años. A todo el equipo del SVO por el increíble y cercano ambiente, y por ayudarme siempre que lo he necesitado. Al INTA, por financiar este proyecto a través de la beca PRE-OVE. Gracias al Cool Star Lab de la UCSD por el ambiente tan agradable que viví durante esos tres meses de ensueño.

La vida no es nada sin la gente con la que la compartes, y yo tengo la inmensa suerte de hacerlo con personas a las que admiro. Indy, Guille, Nacho, Alberto, David, Mario. Muchas gracias por estar siempre ahí, por ser un soporte cuando las cosas se ponen complicadas, por escucharme cada vez que necesito darle mil vueltas al mismo tema. Compartir hogar con tus mejores amigos es una experiencia maravillosa que poca gente puede experimentar a lo largo de su vida. Yo he tenido la suerte de disfrutarlo en dos ocasiones. Gracias Nacho e Indy por todos los momentos inolvidables en Bremen. Gracias a las ``Bremen Warriors'' por alegrarnos los días con nuestros gritos calle-terraza. Gracias Ester por las incontables llamadas en el coche, por escucharme siempre que lo he necesitado y quererme como un hermano, tenerte como hermana es un privilegio. A l$@$s Galáctic$@$s por tantas noches y viajes inolvidables. A Carpaso por reiniciar y refrescar mi cabeza cuando llegaba con todo nublado al partido. La vida es mucho más bonita con amigos y amigas como vosotros. Os quiero.

Y gracias a ti, Frerita, que has llegado a mi vida en la fase final de este viaje para llenarla de pasión. Pienso en estas líneas en un tren rumbo al norte del mundo, mientras acaricio tu pelo y duermes con la cabeza apoyada sobre mi pierna. Miro la nieve caer a través de la ventana. Los copos caen más despacio cuando te siento respirar, parece que bailan. Y pienso de nuevo en la enorme suerte que este mundo tenía reservada para mí. Y la felicidad invade mi cuerpo.

La única forma de devolverle al Universo toda esta suerte que me ha regalado es intentar comprender su naturaleza, adentrarme en sus entrañas y estudiar sus secretos. Esta tesis es esto. Es mi forma de intentar acercarnos un poco más al Universo.

\endgroup

\clearpage
\phantomsection
\pdfbookmark{\contentsname}{tableofcontents}
\setcounter{tocdepth}{2}
\begingroup 
    \let\clearpage\relax
    \let\cleardoublepage\relax

    \tableofcontents
\endgroup
\markboth{\spacedlowsmallcaps{\contentsname}}{\spacedlowsmallcaps{\contentsname}} 

\begingroup 
    \let\clearpage\relax
    \let\cleardoublepage\relax
\endgroup

\clearpage

\pdfbookmark{Abstract}{Abstract}
\begingroup
\let\clearpage\relax
\let\cleardoublepage\relax
\let\cleardoublepage\relax

\chapter*{Abstract}

Low-mass objects are ubiquitous in our Galaxy. Their low temperature provides them with complex atmospheres characterised by the presence of strong molecular absorption bands which, together with their faintness, have made their accurate characterisation a great challenge for astronomers over the last decades. M dwarfs account for 75\% of the census of stars within 10\,pc of the Sun, and their suitability as targets in the search for Earth-like planets has led many research groups to focus on the study of these objects, which is crucial for the understanding of the structure and kinematics of our Galaxy. Very low-mass stars and substellar objects with spectral types M7 or later, including the extended L, T, and Y spectral types, constitute the domain of ultracool dwarfs. The study of these objects, discovered definitively in 1995, is key for understanding the boundary between stellar and substellar objects and promises to experience a quantum leap thanks to the characteristics of new-generation surveys such as \textit{Euclid} or LSST.

Data analysis in the field of observational astronomy has undergone a paradigm shift during the last decades driven by an exponential growth in the volume and complexity of available data. In this revolution, the Virtual Observatory has become a cornerstone providing a system that fosters data access and interoperability between astronomical archives around the world. In response to this growth in data complexity, the astronomical community has increasingly adopted machine and deep learning techniques for the development of scalable, automated solutions capable of analysing huge amounts of data in an efficient way.

This thesis explores the discovery and characterisation of M dwarfs and ultracool dwarfs, always using a data-driven approach supported by Virtual Observatory technologies and protocols. We rely on a variety of machine and deep learning techniques to develop flexible methodologies aimed at advancing our understanding of M dwarfs and ultracool dwarfs in the coming years. In this context, we use J-PLUS multi-filter photometry to enrich the ultracool dwarf census by providing a characterised catalogue of candidates. In addition, we consolidate a novel deep transfer learning methodology to determine atmospheric stellar parameters of M dwarfs from high-resolution spectra, and provide new estimations for a sample of M dwarfs observed by the CARMENES survey. We demonstrate that this methodology can also be extended to the ultracool dwarf domain by adapting it to low-resolution spectroscopic data.

We expect that the work carried out in this thesis will lay the foundations for future advances in the low-mass domain. We make available to the astronomical community all the catalogues and methodologies developed throughout the thesis, in the hope that future researchers will find them valuable resources to advance the knowledge of these faint, cool, low-mass objects that populate our Universe.

\endgroup

\clearpage

\pdfbookmark{Resumen}{Resumen}
\begingroup
\let\clearpage\relax
\let\cleardoublepage\relax
\let\cleardoublepage\relax

\chapter*{Resumen}

Los objetos de baja masa son omnipresentes en nuestra Galaxia. Su baja temperatura les confiere atmósferas complejas dominadas por fuertes bandas moleculares de absorción que, junto con su baja luminosidad, han hecho de su caracterización precisa un gran reto para los astrónomos en las últimas décadas. Las enanas M representan el 75\% del censo de estrellas a menos de 10 pc del Sol, y su idoneidad como objetivos en la búsqueda de planetas similares a la Tierra ha llevado a muchos grupos de investigación a centrarse en el estudio de estos objetos, crucial para la comprensión de la estructura y cinemática de nuestra Galaxia. Las estrellas de muy baja masa y los objetos subestelares con tipos espectrales M7 o posteriores, incluidos los tipos espectrales extendidos L, T e Y, constituyen el dominio de las enanas ultrafrías. El estudio de estos objetos, descubiertos de manera definitiva en 1995, es de vital importancia para comprender la frontera entre los objetos estelares y subestelares y promete experimentar un gran impulso gracias a las características de varias misiones futuras.

El análisis de datos en el campo de la astronomía observacional ha experimentado un cambio de paradigma durante las últimas décadas, impulsado por un crecimiento exponencial en el volumen y la complejidad de los datos disponibles. En esta revolución, el Observatorio Virtual se ha convertido en una piedra angular proporcionando un sistema que permite el acceso a los datos y la interoperabilidad entre archivos astronómicos de todo el mundo. En respuesta a esta creciente complejidad de los datos, la comunidad astronómica ha adoptado cada vez más técnicas de aprendizaje automático y profundo para el desarrollo de soluciones escalables y automatizadas, capaces de analizar enormes cantidades de datos de manera eficiente.

Esta tesis explora el descubrimiento y la caracterización de enanas M y enanas ultrafrías, utilizando siempre un enfoque orientado a los datos y apoyado en tecnologías y protocolos del Observatorio Virtual. Utilizamos una variedad de técnicas de aprendizaje automático y profundo para desarrollar metodologías flexibles destinadas a avanzar en nuestra comprensión de las enanas M y las enanas ultrafrías en los próximos años. En este contexto, utilizamos fotometría multifiltro de J-PLUS para enriquecer el censo de enanas ultrafrías proporcionando un catálogo caracterizado de candidatas. Además, consolidamos una novedosa metodología de aprendizaje profundo por transferencia para determinar parámetros estelares atmosféricos de enanas M a partir de espectros de alta resolución, y proporcionamos nuevas estimaciones de estos parámetros para una muestra de enanas M observadas por CARMENES. Demostramos que esta metodología también puede extenderse al dominio de las enanas ultrafrías adaptándola a datos espectroscópicos de baja resolución.

Esperamos que el trabajo realizado en esta tesis siente las bases para futuros avances en el dominio de los objetos de baja masa. Ponemos a disposición de la comunidad astronómica todos los catálogos y metodologías desarrollados a lo largo de la tesis, con la esperanza de que futuros investigadores encuentren en ellos valiosos recursos para avanzar en el conocimiento de estos objetos débiles, fríos y de baja masa que pueblan nuestro Universo.

\endgroup

\pagestyle{scrheadings} 
\cleardoublepage


\pagenumbering{arabic}

\chapter{General Introduction} \label{chp:general_intro}


\section{M dwarfs and the substellar realm} \label{chp:mdwarfs_intro}

\subsection{M dwarfs$\dots$}

\begin{figure}
    \centering
	\includegraphics[width=.8\columnwidth]{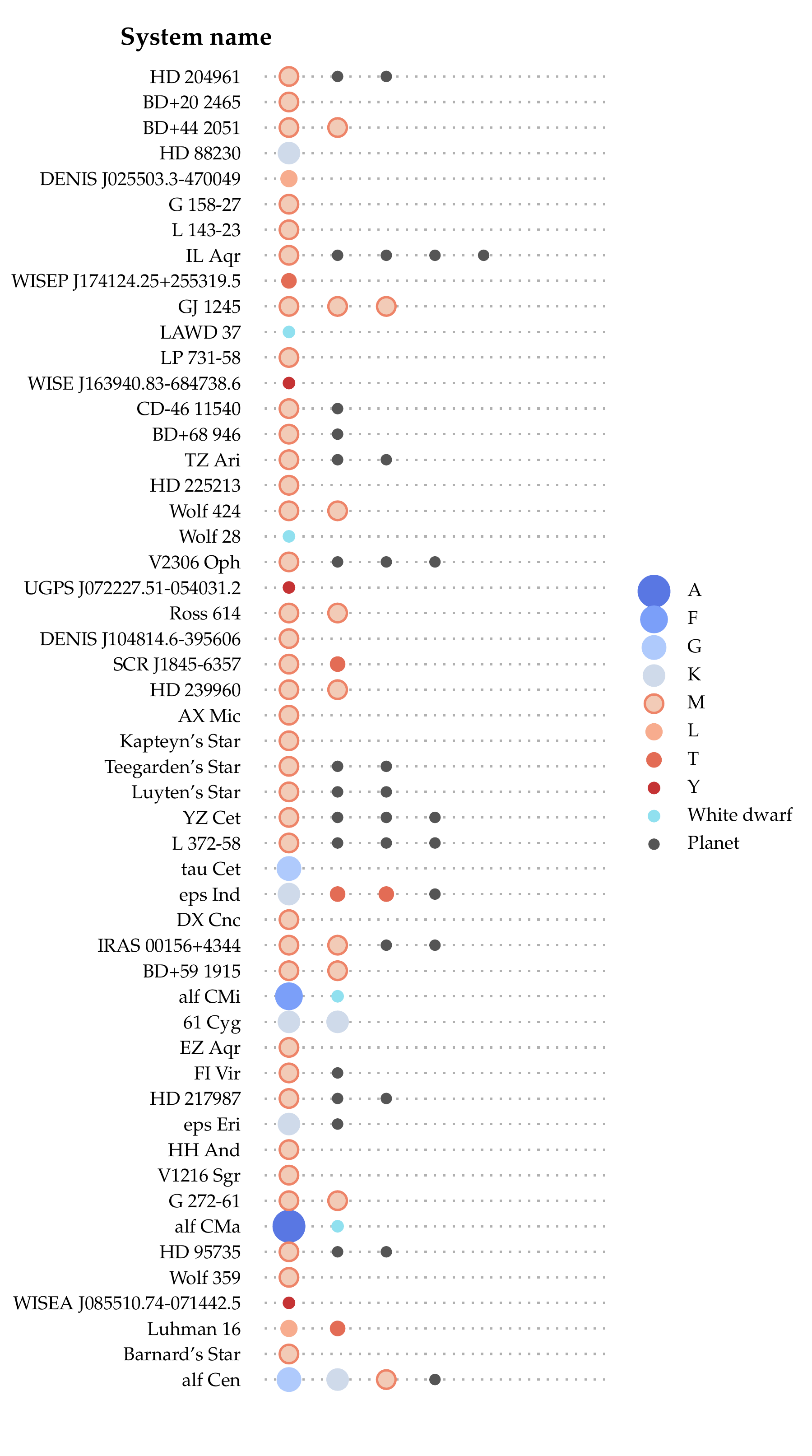}
    \caption{Sub-sample of the catalogue of stars, brown dwarfs, and exoplanets within 10\,pc from the Sun provided by \citet{reyle2021}. The Figure only shows a schematic representation of the systems within 5\,pc for visualisation purposes, as it is sufficient to illustrate the ubiquity of M dwarfs in our vicinity.}
    \label{fig:4pc}
\end{figure}

\begin{figure}
    \centering
	\includegraphics[width=.44\columnwidth]{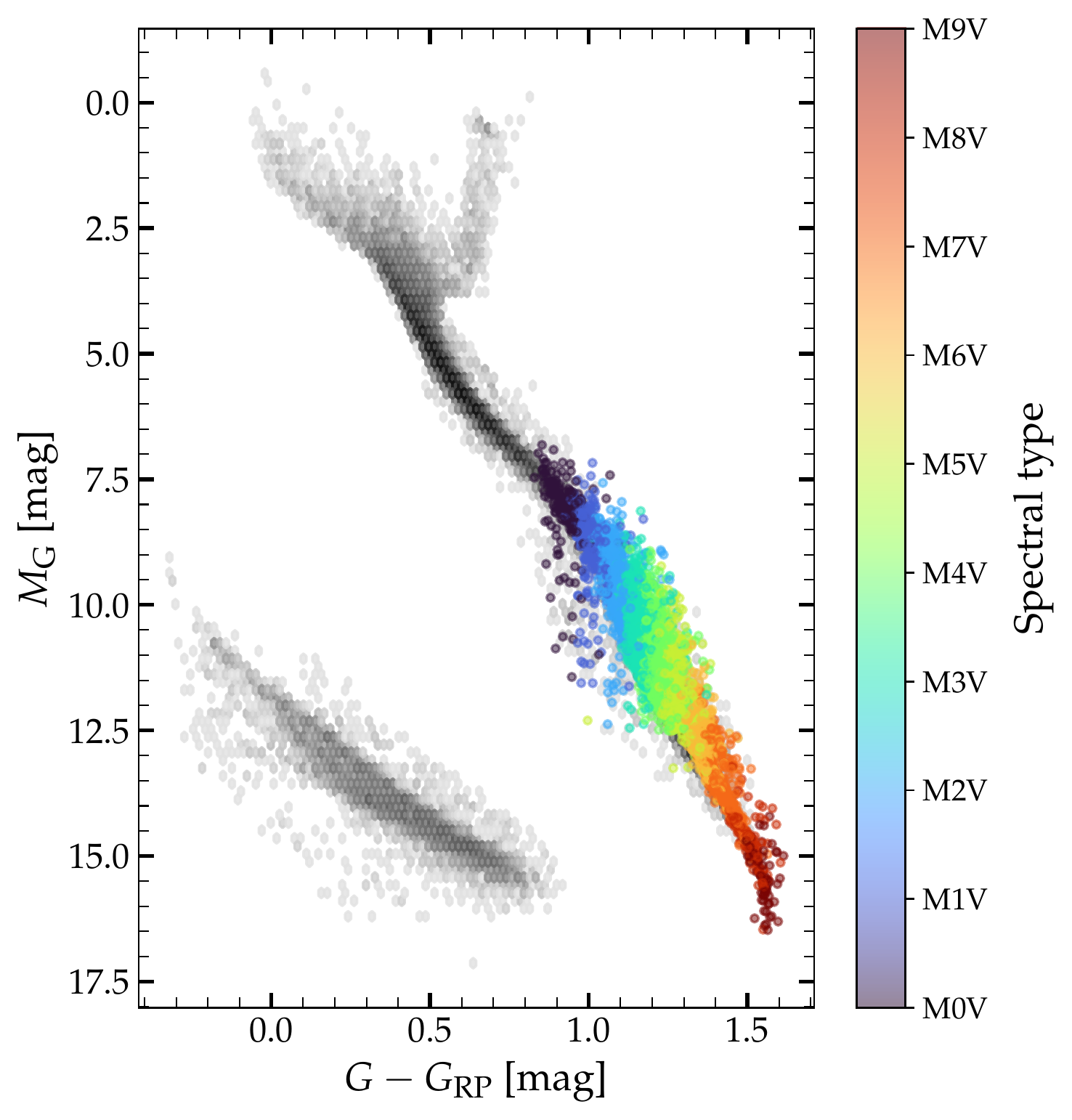}
	\includegraphics[width=.45\columnwidth]{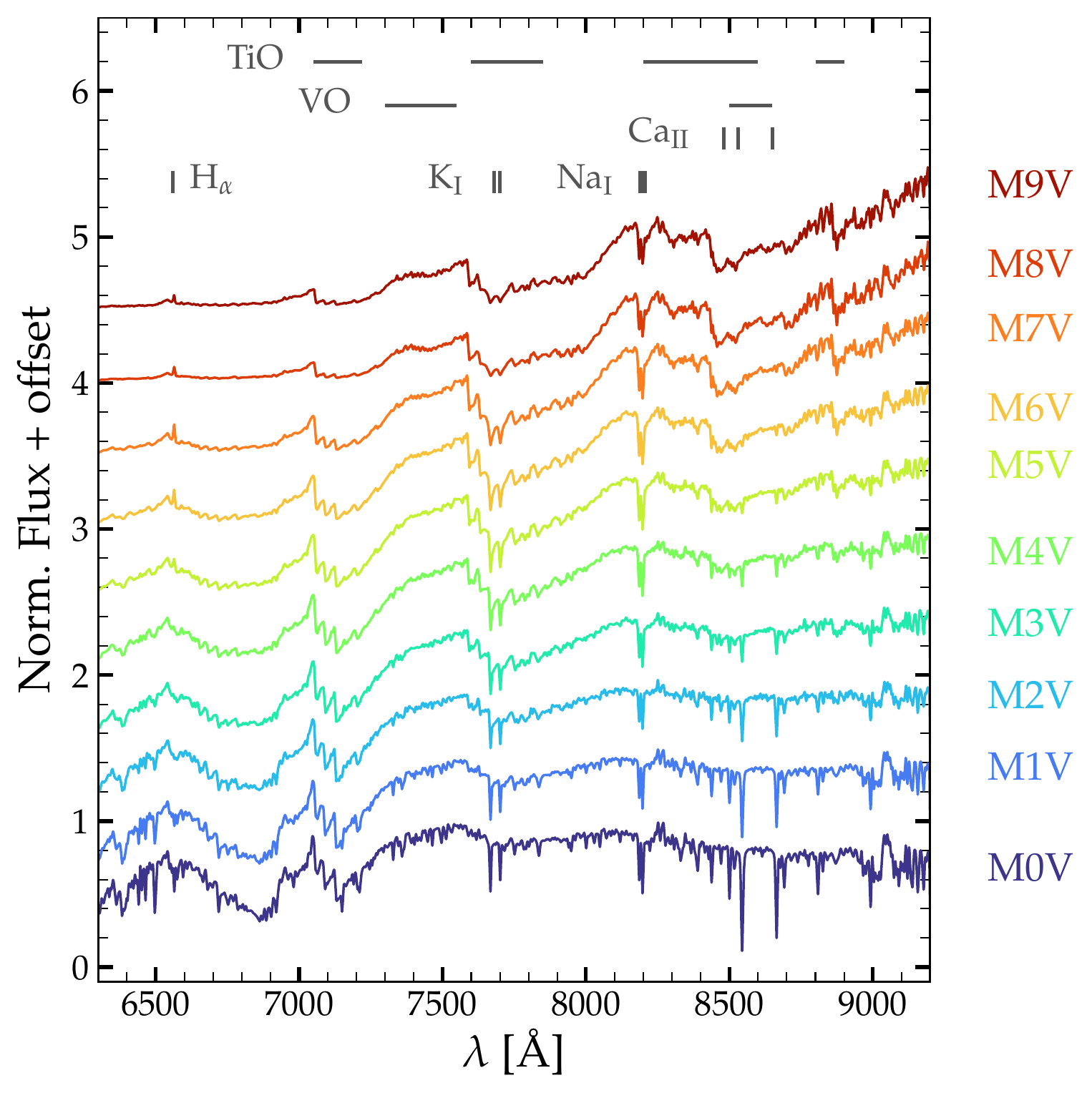}
    \caption{\textit{Left panel:} Sample of M dwarfs spectroscopically classified by \citet{west2011}, superimposed on a Hertzsprung-Russell diagram built with high-quality \textit{Gaia} DR3 \citep{gaiadr3} data. \textit{Right panel:} M dwarf spectral sequence constructed with optical template spectra from \citet{bochanski2007}. Key spectral features are highlighted.}
    \label{fig:hr_mdwarfs}
\end{figure}

For centuries, humans have gazed at the night sky, wondering what the bright objects up there might look like. Thanks to the technological advances in the last decades, which allow an ever more detailed exploration of our universe, we now know that most of our nearest neighbours were so faint that we could not see them with our naked eyes. Faint, cool, low-mass stars known as M dwarfs are by far the most common type of star in the Solar Neighbourhood \citep{henry1994,reid2004,bochanski2010,reyle2021,kirkpatrick2024}. As presented by \citet{reyle2021}, the ubiquity of M dwarfs is overwhelming in our vicinity (see Figure \ref{fig:4pc}), with three out of every four stars within 10\,pc being spectroscopically classified as M dwarfs \citep[see Table 2 in][]{henry2024}, often with planets orbiting around them. This abundance, together with their remarkable lifespan of tens of billions of years \citep{adams1997,laughlin1997}, makes them a fundamental piece in the study of Galactic structure and kinematics \citep{chabrier2003,chabrier2005,bochanski2007,caballero2008,ferguson2017,cortes2024}. Thus, an accurate characterisation of the M dwarf population, with masses ranging from $\sim0.6$\,$M_{\odot}$ to $\sim0.1$\,$M_{\odot}$ \citep{cifuentes_ucm} and located at the lower tail of the main sequence (see Figure \ref{fig:hr_mdwarfs}), is key to the understanding of our Galaxy.

One of the most active lines of research in stellar astrophysics at international level is the detection and characterisation of extrasolar planets. Along with projects dedicated to the search of terrestrial exoplanets in orbits up to the habitable zone of Sun-like stars, such as PLATO \citep{plato}, several programs have been established with the goal of identifying potentially habitable planets orbiting M dwarfs. Notable examples include the Transiting Exoplanet Survey Satellite \citep[TESS,][]{tess}, the Echelle Spectrograph for Rocky Exoplanet and Stable Spectroscopic Observations \citep[ESPRESSO,][]{pepe21} and its predecessor, the High-Accuracy Radial velocity Planet Searcher \citep[HARPS,][]{mayor2003,bonfils13}, or the Calar Alto high-Resolution search for M dwarfs with Exoearths with Near-infrared and optical Echelle Spectrographs \citep[CARMENES,][]{Quirrenbach16,Quirrenbach20}. The small size and low luminosity of M dwarfs, compared to Sun-like stars, make it easier to detect close-in terrestrial planets in their habitable zones \citep{zechmeister2019,Kossakowski2023,suarez2023,dreizler2024}. Moreover, M dwarfs have established themselves in recent years as very suitable targets in the search for Earth-like planets \citep{dressing2015,kopparapu2017,gillon2017,reiners2018,sabotta2021,nagel2023}, with  several studies confirming an elevated occurrence rate of Earth-like planets around M dwarfs \citep{gaidos2016,mulders2021,sabotta2021}. 

The precise determination of the physical parameters of planet-hosting stars is crucial to improve our understanding of planetary formation and evolution, which depends fundamentally on the thorough characterisation of their host stars \citep{souto2017,cifuentes2020}. However, well-established photometric and spectroscopic methods for determining the stellar atmospheric parameters of M dwarfs encounter several pitfalls, mainly due to the particular features of their cool atmospheres. The low temperatures, between $\sim2300$\,K and $\sim3900$\,K, of these atmospheres enable the formation of diatomic and triatomic molecules, with a spectral sequence characterised by the presence of strong molecular absorption bands, such as TiO and VO \citep{joy1947,keenan1952,boeshaar1976}, as shown in Figure \ref{fig:hr_mdwarfs}. Moreover, for late M-dwarfs (M5 or later), the outermost layers of the atmosphere are cool enough to form dust and clouds, which makes the modelling of these atmospheres and the consequent determination of their stellar parameters even more complex. This is further aggravated by the inherent faintness of M dwarfs, which makes it difficult to obtain high-S/N, high-resolution spectra, and their frequent manifestation of strong stellar activity. Despite these problems, numerous efforts have been devoted to estimating photospheric parameters in M dwarfs, including effective temperature, surface gravity, and metallicity. Several methods have proven successful in inferring these parameters, such as fitting synthetic spectra \citep{bayo2017,pass18,Rajpurohit2018,pass2019,schw19,Souto2020,mar21,Sarmento2021}, pseudo-equivalent widths \citep{Mann2013,Mann2014,Neves2014,Khata2020,almendros2022}, 
spectral indices \citep{RojasAyala2010,bayo2011,Rojas2012,Khata2020}, empirical calibrations \citep{casagrande08,Neves2012,rojasayala2014,Rodriguez2019}, interferometry \citep{Boyajian2012,Rabus2019}, and machine learning  \citep{Sarro2018,Antoniadis2020,pass20,Li2021,bello2023,masbuitrago2024,rains2024}. 

The main difference between M dwarfs and other stellar objects is that their stellar properties change significantly from early to late types. Especially, for spectral types $\sim M3-4$ and later \citep[masses below $0.35$\,$M_{\odot}$; ][]{chabrier1997}, M dwarfs become fully convective and experience a critical transition in their structure and behaviour. In this boundary, the radiative cores typical of earlier M dwarfs disappear and their interiors become fully convective, with energy transport dominated by convection throughout the stellar envelope \citep{delfosse1998,reiners2009}. As a result, a large fraction of low-mass stars, especially young, fast-rotating M dwarfs, are magnetically active, with a chromospheric activity often diagnosed by H$\alpha$ or Ca~{\sc ii} H and K line emission \citep{cincunegui2007,ibanezbustos2023}. After reaching the main sequence, low-mass stars slowly spin-down due to the loss of angular momentum by stellar winds, thus undergoing a decrease in their magnetic activity over time \citep{yang2017,davenport2019,raetz2020} that may also be dependent on stellar metallicity \citep{see24}. This abundant activity, combined with the proximity of the habitable zone of M dwarfs, makes exoplanets more exposed to energetic events related to stellar activity \citep{tilley2019,gunther2020,chen2021}, such as flares or coronal mass ejections, which are frequent in M dwarfs.

The faintness and low temperature of M dwarfs provide them with characteristics that push astronomers to the limit when it comes to accurately characterising them. But they are not the faintest. What do we find when we venture towards even lower masses? What separates our planet, the Earth, from the coolest stars?

\subsection{$\dots$and beyond} \label{sec:ucds_intro}

\begin{figure}
    \centering
	\includegraphics[width=.73\columnwidth]{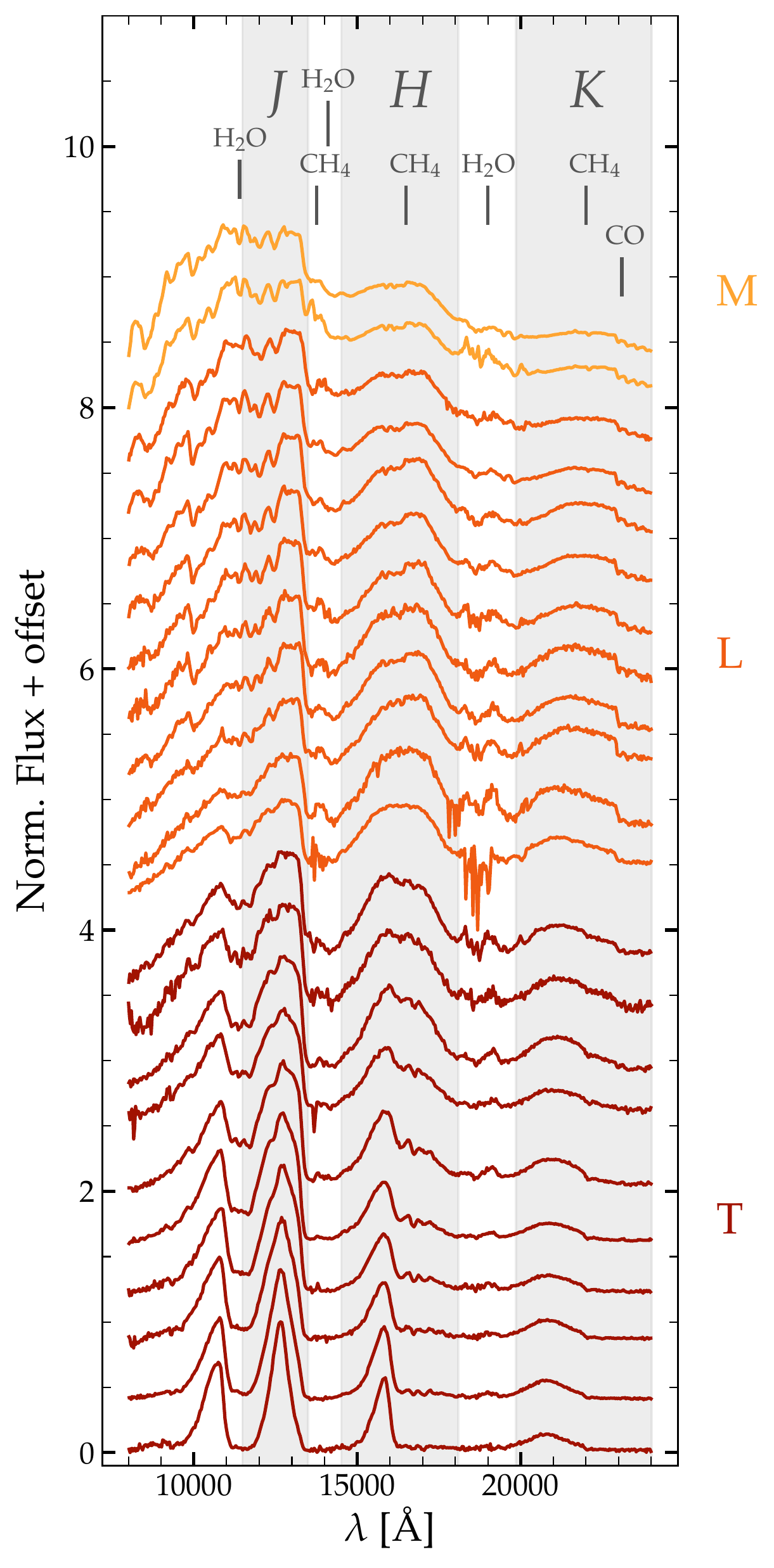}
    \caption{Near-infrared spectral sequence for ultracool dwarfs built using the standard spectra available in the SpeX Prism Library \citep{spex}. The relevant molecular absorption bands discussed in the text are highlighted.}
    \label{fig:comp_l6t6}
\end{figure}

``WHAT distinguishes a star from a planet? Could we call Jupiter a failed star?'' This is how Dr. Lorne Nelson began his article on page 102 of volume 377 of the journal Nature in September 1995. Twenty-seven pages below, \citet{rebolo1995} reported the discovery of an object, in the young Pleiades star cluster, located on the boundary between the stars and the giant planets. \citet{kumar1963b,kumar1963} and \citet{hayashi1963} had first postulated the existence of this substellar objects, termed as ``brown dwarfs'' in 1975 by Jill Tarter \citep{tarter2014}, unable to maintain stable hydrogen ($^1$H) fusion in their interior due to their low mass. This substellar boundary is established for $\sim0.072$\,$M_{\odot}$ ($\sim75$\,$M_{\mathrm{J}}$), depending on the models and the metallicity, beyond which the low mass makes objects unable to reach sufficient internal pressure and temperature to sustain thermonuclear processes of hydrogen-to-helium conversion. However, up to masses of $\sim13$\,$M_{\mathrm{J}}$ \citep{chabrier2000} these substellar objects are massive enough to sustain deuterium fusion in their interiors at some point in their evolution, and this limit is often used to define the boundary between brown dwarfs and giant exoplanets.

Decades after they were first proposed theoretically, 1995 marked a turning point in the exploration of the substellar realm, with the first solid discoveries of brown dwarfs and exoplanets. First, \citet{basri1995} presented evidence of lithium in PPl 15, identifying this object as a brown dwarf just below the substellar limit. This ``lithium test'', or detection of lithium in the atmosphere, was of paramount importance for the detection of the first brown dwarfs and was first proposed by \citet{Rebolo1992} to distinguish between very low-mass stars and brown dwarfs close to the substellar boundary. Unlike very low-mass stars, objects with masses below $\sim0.060$\,$M_{\odot}$ \citep[see Figure 2 in][]{chabrier2000} cannot reach the $^{7}Li$ burning temperature and preserve a significant amount of their original Li content, so the substellar nature of brown dwarfs can be confirmed by spectroscopic detection of the Li 670.8\,nm resonance line. This is the test that \citet{rebolo1996} used to finally confirm the brown dwarf nature of Teide\,1 and Calar\,3. In late 1995, at the same conference where the discovery of the first extrasolar planet was announced, the discovery of a cool brown dwarf \citep[GJ\,229B; ][]{nakajima1995,oppenheimer1995} was also reported.

\begin{figure}
    \centering
	\includegraphics[width=.5\columnwidth]{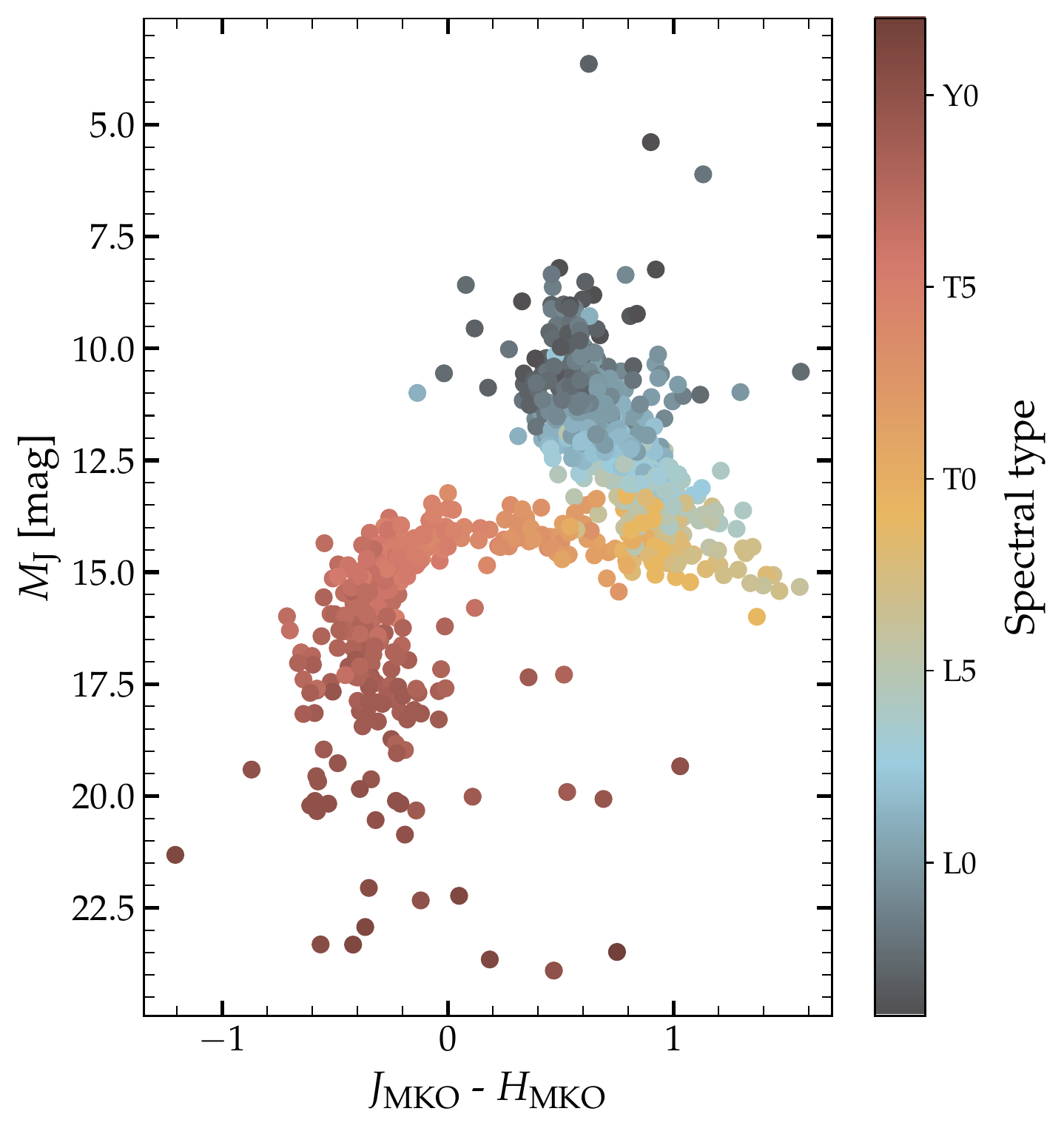}
	\includegraphics[width=.48\columnwidth]{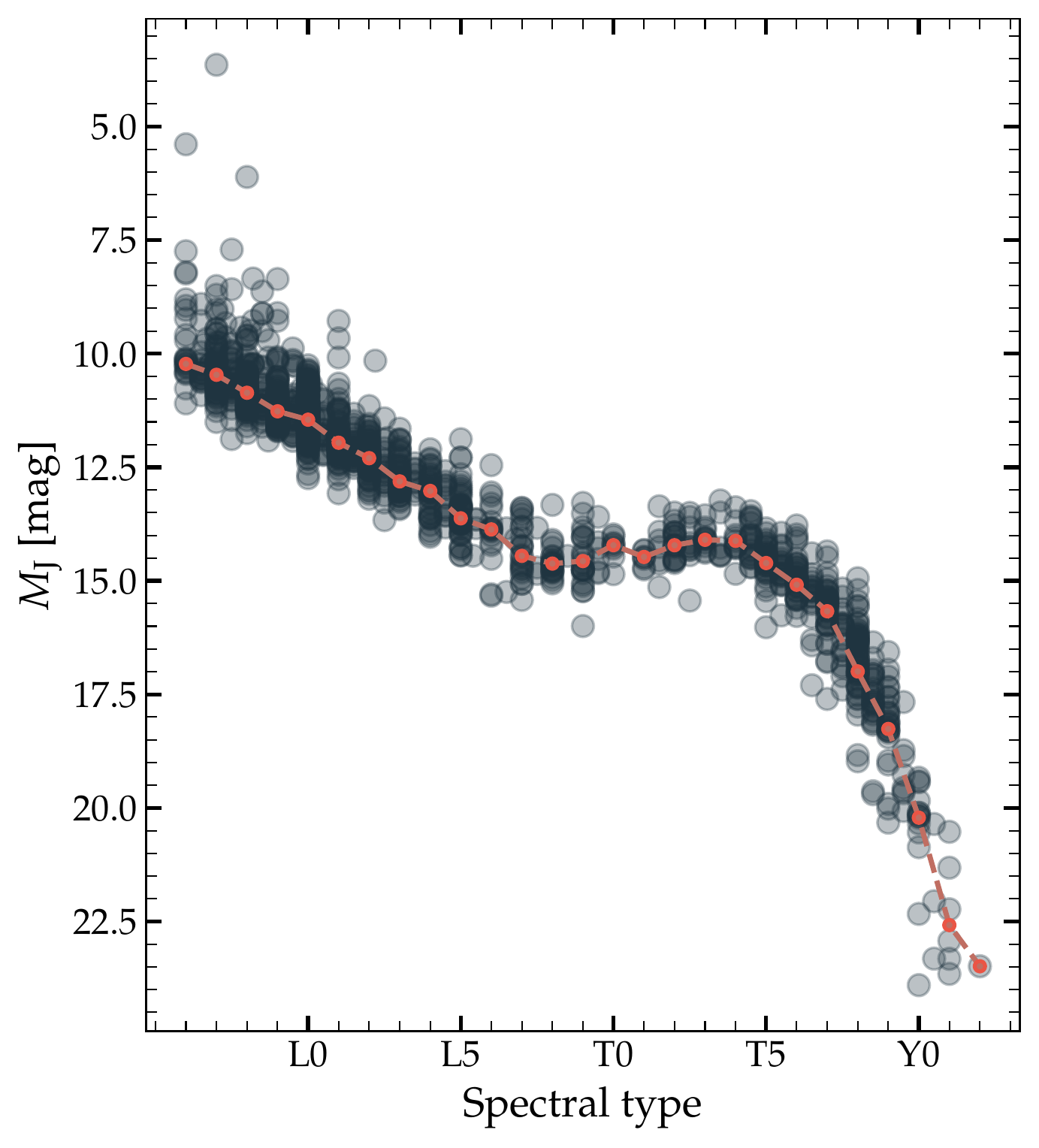}
    \caption{\textit{Left panel:} $M_{\mathrm{J}}$ vs. $\textit{J}-\textit{H}$ colour-magnitude diagram, with the dots colour-coded by spectral type, of our sample of ultracool dwarfs. \textit{Right panel:} Evolution of $M_{\mathrm{J}}$ with the spectral type for the same sample. The red dots indicate the median value for each of the spectral types. All values used in the figure have been taken from the UltracoolSheet catalogue.}
    \label{fig:ucds_diag}
\end{figure}

The ultracool dwarf domain covers very low-mass stars and substellar objects with spectral types M7 or later \citep{kirkpatrick1997}, including the extended L, T, and Y spectral types \citep{martin1997,martin1999b,kirkpatrick1999,burgasser2006,geballe2002,burningham2008,cushing2011}. With effective temperatures of $T_{\mathrm{eff}}\lesssim2800$\,K, the spectra of ultracool dwarfs are dominated by strong molecular absorption bands. The transition from late-M to L dwarfs (at about $2200$\,K) is characterised by the gradual disappearance of the TiO and VO oxide bands, the strengthening of H$_2$O and metal hydride (CrH, FeH, CaH) absorption bands, and a increasing steepness around the $6000-10000$\,$\AA$ interval \citep{kirkpatrick2000,reid2000,geballe2002}. Also, the neutral alkali metal absorption lines, especially Na~{\textsc{i}} and K~{\textsc{i}}, grow considerably by mid-L dwarfs in the optical. The beginning of the T dwarfs sequence, around $1300$\,K, is marked by the appearance of methane (CH$_4$) absorption in the near-infrared \textit{H} and \textit{K} bands, which strengthens along with H$_2$O absorption as the sequence evolves towards late-T spectral types. Due to the increasing depth of the CH$_4$ absorption bands, the flux in the \textit{H} and \textit{K} bands is reduced with respect to the \textit{J} band (see Figure \ref{fig:comp_l6t6}), and the near-infrared colours of T dwarfs become increasingly blue as compared to L dwarfs \citep{burgasser2002,geballe2002}. Finally, the transition to Y dwarfs, at about $500$\,K, is characterised by the presence of H$_2$O and ammonia (NH$_3$) photospheric clouds \citep{delorme2008,cushing2011}, in contrast to the CH$_4$ clouds typical of T dwarfs, and recent studies with the James Webb Space Telescope \citep[JWST,][]{jwst} have also found the presence of phosphine (PH$_3$) \citep{burgasser2024}.

The left panel in Figure \ref{fig:ucds_diag} shows a near-infrared $M_{\mathrm{J}}$ vs. $\textit{J}-\textit{H}$ colour-magnitude diagram of a clean sample of ultracool dwarfs with spectroscopic spectral classification. To obtain our sample, we queried the UltracoolSheet catalogue \citep{ucs} and applied selection criteria to retain only resolved ultracool dwarfs with reliable photometry and parallax, discarding objects with a photometric spectral type. The final sample contains $\sim1500$ ultracool dwarfs with a reliable spectroscopic classification of L0 or later. The colour-magnitude diagram shows how ultracool dwarfs, as they become fainter, evolve into redder $\textit{J}-\textit{H}$ colours until they reach the L/T transition. By spectral type L5, the photosphere is cool enough to allow the hydrogenation of CO to CH$_4$ \citep{noll2000,canty2015}, and CH$_4$ gradually becomes dominant over carbon monoxide (CO), typical of the photospheres of early- to mid-L objects. Throughout this transition to mid-T dwarfs, the absolute magnitude remains nearly constant while the $\textit{J}-\textit{H}$ colour grows bluer due to increased CH$_4$ absorption. Likewise, the right panel in Figure \ref{fig:ucds_diag} shows how the relation between $M_{\mathrm{J}}$ (and also effective temperature)  and spectral type is non-linear and exhibits a plateau in the L/T transition \citep{golimovsky2004,saumon2008,kirkpatrick2021}.

\begin{figure}
    \centering
	\includegraphics[width=.9\columnwidth]{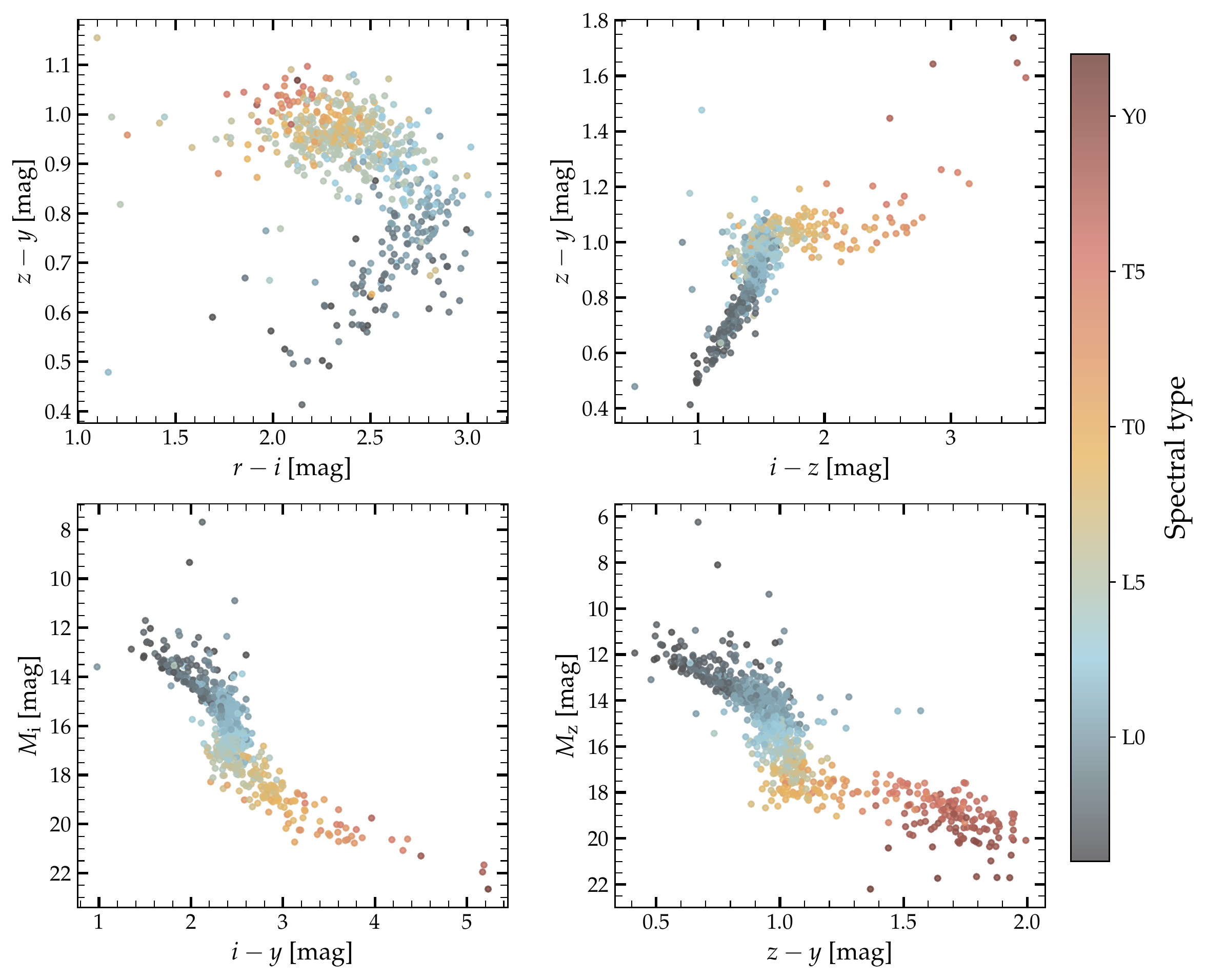}
    \caption{Optical colour-colour and colour-magnitude diagrams for our sample of ultracool dwarfs, using Pan-STARSS photometry. The dots are colour-coded by spectral type. All values used in the figure have been taken from the UltracoolSheet catalogue.}
    \label{fig:ucds_ccopt}
\end{figure}

\begin{figure}
    \centering
	\includegraphics[width=.9\columnwidth]{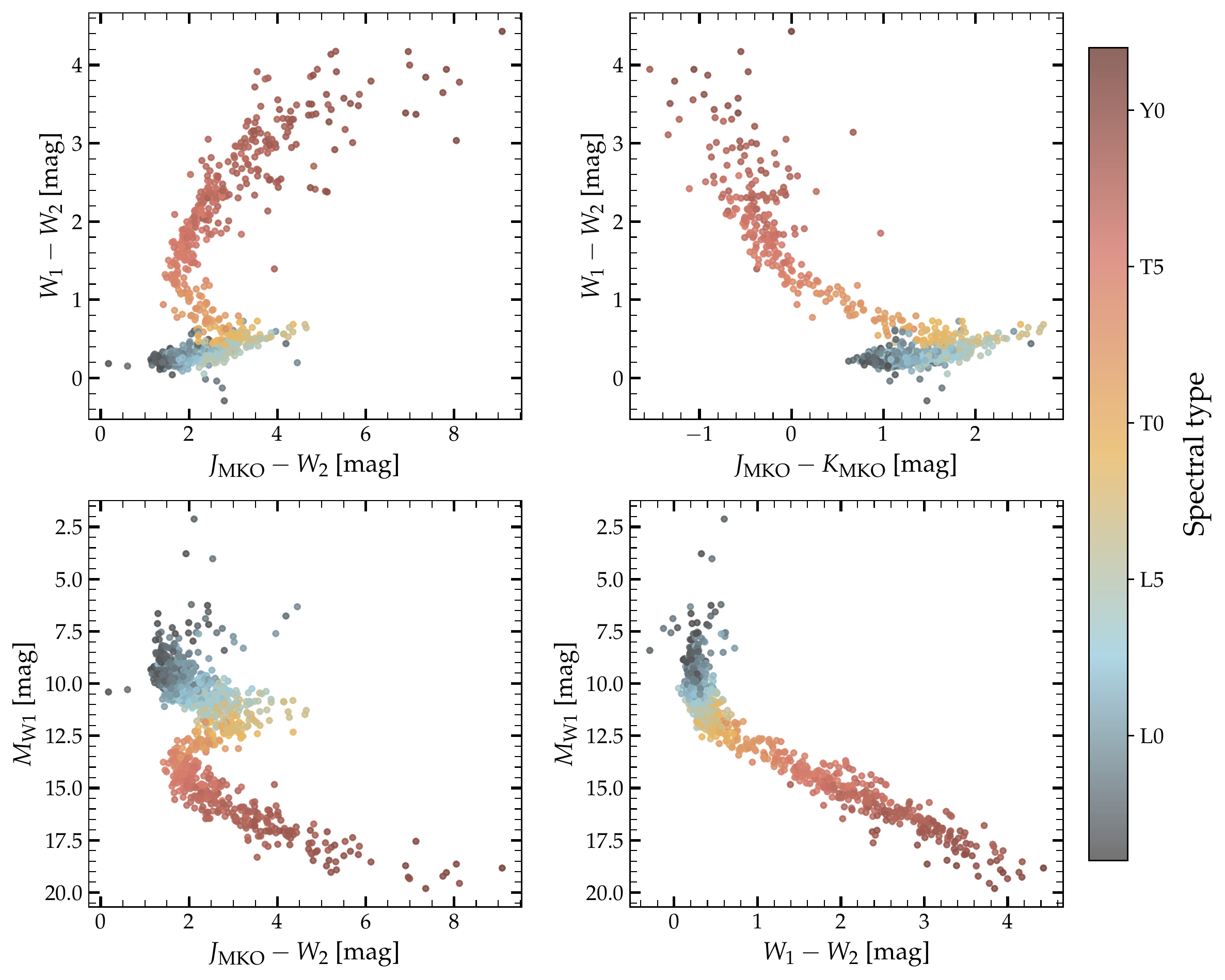}
    \caption{Near-infrared colour-colour and colour-magnitude diagrams for our sample of ultracool dwarfs, using MKO photometry available in the UltracoolSheet catalogue. The dots are colour-coded by the spectral type. All values used in the figure have been taken from the UltracoolSheet catalogue.}
    \label{fig:ucds_ccnir}
\end{figure}

Discoveries of ultracool dwarfs have primarily been driven by wide-field optical and infrared imaging surveys such as the Deep Near Infrared Survey of the Southern Sky \citep[DENIS;][]{denis}, the Sloan Digital Sky Survey \citep[SDSS;][]{sdss}, the Two-Micron All Sky Survey \citep[2MASS;][]{2mass}, the UKIRT Infrared Deep Sky Survey \citep[UKIDSS; ][]{ukidss}, the Wide-Field Infrared Sky Explorer \citep[WISE;][]{wise}, the Panoramic Survey Telescope and Rapid Response System \citep[Pan-STARRS;][]{panstarrs}, and the Javalambre Photometric Local Universe Survey \citep[J-PLUS;][]{Cenarro2019}. The \textit{Gaia} mission \citep{gaiadr3} has also contributed to the discovery of ultracool dwarfs in the whole sky. Despite all these efforts, the 20\,pc census of ultracool dwarfs is still incomplete \citep{kirkpatrick2024}, with a completeness volume of 15\,pc and 11\,pc for spectral types later than $\sim$T8.5 and $\sim$Y0, respectively. A consolidated approach for the identification of ultracool dwarfs in these surveys is the definition of a locus in colour-colour or colour-magnitude diagrams in the optical (see Figure \ref{fig:ucds_ccopt}) or the infrared (see Figure \ref{fig:ucds_ccnir}) using previously known objects \citep{skrzypek2016,smart2017,panstarrs1,Reyle2018,carnero2019,masbuitrago2022,sarro2023,euclid_ero}.

Spectral classification, and its calibration to temperature or luminosity classes, is of utmost importance for characterising the ultracool dwarf (and any) astronomical population. There are several ways of doing this, such as direct comparison of the observed spectra with anchored optical \citep{kirkpatrick1999} and infrared \citep{burgasser2006,kirkpatrick2010,cushing2011} spectral standards. Late-M and L dwarfs classification is tied to the red optical region of the spectrum, while the T dwarfs are often characterised from the near-infrared region due to the presence of strong H$_2$O and CH$_4$ bands. Another approach is the classification through the measurement of different spectroscopic indices, defined as flux ratios that measure the strength of specific absorption or pseudocontinuum features \citep{kirkpatrick1995,martin1997,Martin1999,burgasser2007}. Over the last decade, analysis toolkits such as \texttt{SPLAT} \citep{splat}\,\footnote{\url{https://github.com/aburgasser/splat}}  have incredibly facilitated this classification task for the astronomical community. Once the spectral classification is done, it can be converted to effective temperature or luminosity following predefined empirical relations \citep{pecaut2013,filippazzo2015,kirkpatrick2021}.

These methodologies have proven to be broadly consistent throughout the literature, but how do we mine the large astronomical archives that make them possible? Most of the workflows followed in astronomical research require the combination of multi-wavelength data from different surveys. And here the Virtual Observatory is king. Moreover, the advent of huge volumes of data that will be provided in the coming years by missions such as \textit{Euclid} \citep{euclid} demand the development of fully automated solutions for ultracool dwarf identification and characterisation. Will machines fill this gap? Will deep learning be a cornerstone in the future study of the stunning ultracool dwarfs domain?


\section{From stars to data: the Virtual Observatory} \label{chp:vo_intro}

Just as a song means nothing if no one listens to it, data gains purpose only through the lens of analysis. And in the late 1990s and early 2000s, astronomy faced a crucial challenge in this aspect. The rapid advancement of astronomical instrumentation in recent decades has led to an exponential increase in the volume of astronomical data and the complexity of their processing. Until 1990, astronomical data were collected mainly with ground-based telescopes, but the launch of the Hubble Space Telescope \citep{bahcall1986} would bring a revolution in digital astronomy, with an unprecedented volume of data. However, it would not be alone up there, as other space observatories, such as the Infrared Astronomical Satellite \citep{iras} and the International Ultraviolet Explorer \citep{boggess1978}, a pioneer in the development of astronomical archives, were already in operation. During the 1990s, the digitisation of photographic plates enabled the generation of catalogues such as the Guide Star Catalog \citep{lasker1990,lasker1996} and the USNO \citep{monet1998,monet2003}. Notable catalogues in the last years of the 1990s were the Tycho-2 \citep{tycho2} collected by the ESA Hipparcos satellite, with two-colour photometric data for 2.5 million stars, and other catalogues such as the ROSAT All-Sky Survey \citep{rosat} or the NRAO VLA Sky Survey \citep{nvss}. The first data releases from large astronomical surveys, such as 2MASS in 1999 and SDSS in 2003, further fuelled this data revolution, ultimately breaking the Big Data barrier in the 2010s with the launch of the \textit{Gaia} telescope \citep{gaiadr1}. \textit{Gaia} is the mission that pushed astronomy into the petabyte\,\footnote{$1$\,petabyte $ = 1\,048\,576$\,gigabytes} domain, and  has revolutionised observational astronomy by providing the largest, most precise map of the Milky Way. In short, all these efforts by the scientific community have meant that we are now living in the era of large astronomical catalogues such as Pan-STARRS1 \citep{panstarrs}, \textit{Gaia} DR3 \citep{gaiadr3}, SDSS DR12 \citep{sdssdr12} or UKIDSS DR9 \citep{ukidssdr9}, among many others\,\footnote{A comprehensive list of large catalogues can be found at: \url{https://vizier.cds.unistra.fr/vizier/welcome/vizierbrowse.gml?bigcat}}. But this is only the beginning, and this data tsunami will only get bigger and bigger with the next generation of observatories such as the Vera C. Rubin Observatory \citep{ivezic2019}, the Square Kilometre Array \citep{dewdney2009}, or the Nancy Grace Roman Space Telescope \citep{mosby2020}.

The ability to fully utilize these vast datasets poses a major challenge to the astronomical community, and the Virtual Observatory (VO) is the response to this revolution. Just as 1995 was the year of the ultracool dwarfs, 2000 marked a turning point for data mining in astronomy. Two conferences held during the summer of this year, ``Virtual Observatories of the Future'' in Pasadena and ``Mining the Sky'' in Garching, laid the foundations for what two years later would become the International VO Alliance, or IVOA\,\footnote{\url{https://ivoa.net/}}. At its core, the VO is an international initiative aimed at removing the barriers imposed by the geographical and structural fragmentation of astronomical archives, by developing data standards and protocols\,\footnote{\url{https://www.ivoa.net/documents/}} to enable this interoperability. The main goal of the VO is to create a unified and interoperable system that allows astronomers to efficiently discover, retrieve, and analyse astronomical observations, models, and simulations, from multiple archives around the world.

\begin{figure}
    \centering
	\includegraphics[width=.7\columnwidth]{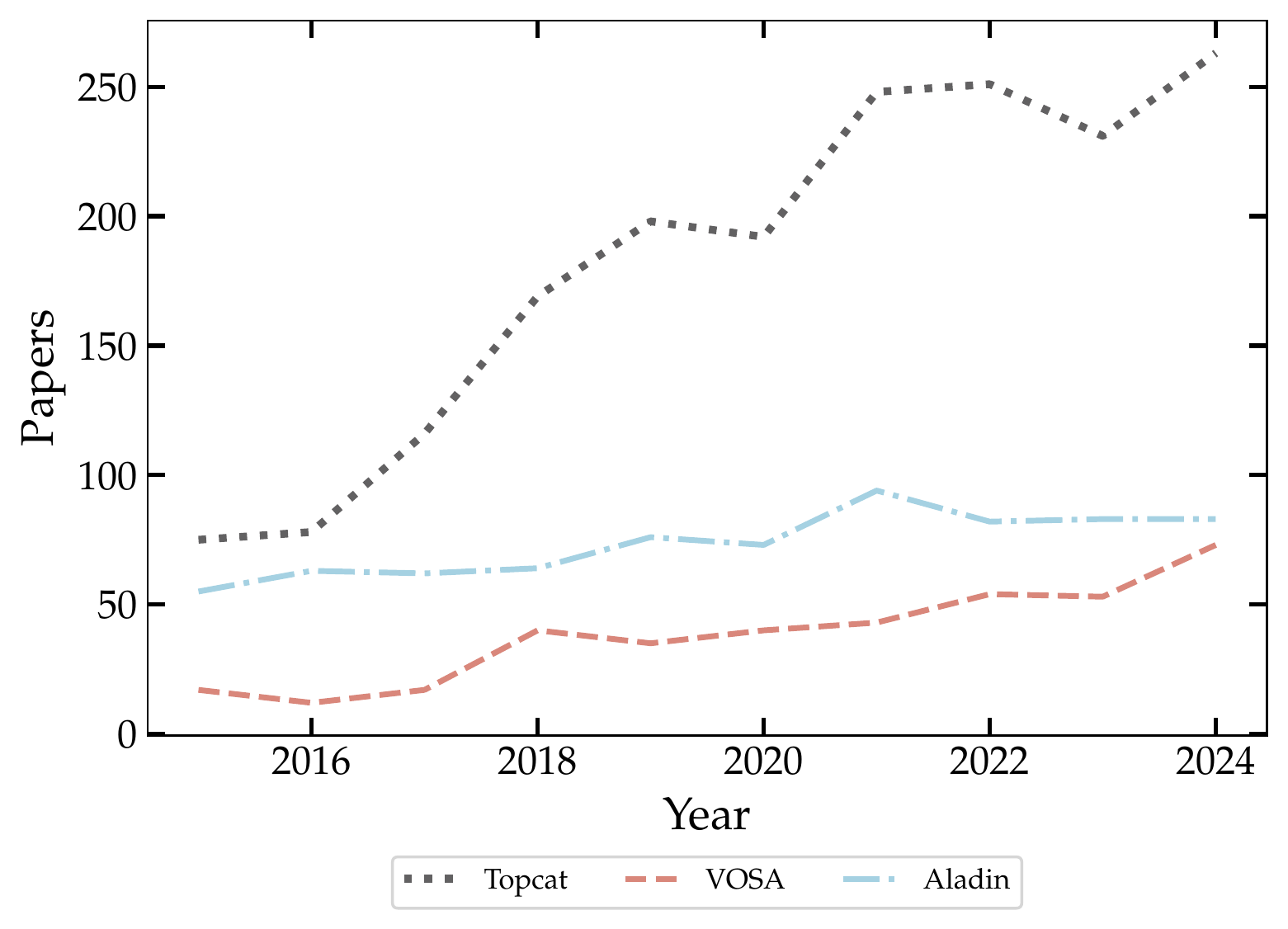}
    \caption{Evolution of the use of different VO tools over the last years, with the Y axis showing the number of referred papers in relevant astronomy journals. We note that the values should be taken as lower limits, as some of the works using the tools may not have been tracked. The data have been collected by the SVO team using the VOpubs tool: \url{https://sdc.cab.inta-csic.es/vopubs/}.}
    \label{fig:evo_VO}
\end{figure}

A key aspect of the VO is the development of data discovery and mining tools that benefit from this data standardisation and enable the access and analysis of multi-wavelength data. \texttt{TOPCAT} \citep{Taylor2005} is a tool that allows interactive manipulation and visualisation of tabular data, making it easier for the astronomer to access source catalogues, and to compare these catalogues with local data. Moreover, the \texttt{Aladin} interactive sky atlas \citep{aladin} is a service that provides simultaneous access to digitised sky images, astronomical catalogues and archives. Within IVOA, individual countries have developed national VO initiatives that further support and implement the VO framework. One particularly successful example is the Spanish VO (SVO)\,\footnote{\url{https://svo.cab.inta-csic.es/main/index.php}}, established in 2004 and coordinated by Dr. Enrique Solano Márquez at the Centro de Astrobiología. The SVO has played a pivotal role in developing and deploying VO services, such as the VO Sed Analyzer (VOSA)\,\footnote{\url{http://svo2.cab.inta-csic.es/theory/vosa/}} \citep{vosa}, a tool that fits observed photometry to different collections of theoretical models to estimate physical properties, such as the effective temperature or luminosity. In addition, VOSA offers a wide range of functionalities to the user, such as the possibility of querying several VO catalogues to enlarge the input data of the sources studied. Another of the flagship services of the SVO is the \textit{Carlos Rodrigo} Filter Profile Service (FPS)\,\footnote{\url{http://svo2.cab.inta-csic.es/theory/fps/}} \citep{fps}, which is widely used by the astronomical community. The FPS contains detailed information on more than ten thousand photometric filters, the largest public collection of its kind. In addition, the SVO is responsible for the management of important astronomical archives\,\footnote{\url{https://svo.cab.inta-csic.es/docs/index.php?pagename=Archives}}, notably the GTC and Calar Alto archives. 

In modern astronomy, most of the research studies require the use of multi-wavelength data that is often stored in separate archives, and with different formats or query mechanisms. Without standardisation and interoperability between the archives, conducting multi-wavelength or multi-messenger \citep{multimessenger} astronomy would require an arduous and time-consuming technical process that, thanks to the VO, is fast and transparent to the user. This, aided by the data mining and analysis tools provided by the VO, is what we know as VO-science. Figure \ref{fig:evo_VO} illustrates the significant adoption of VO-science by the astronomical community at an international level, with an increasing trend over the last few years in the use of the developed tools.

Among the plethora of astronomical archives available to the astronomical community, during this thesis we have made extensive use of two of them in particular. J-PLUS is a multi-filter survey conducted from the Observatorio Astrofísico de Javalambre \citep[OAJ;][]{OAJ} in Teruel, Spain, using the 0.83 m Javalambre Auxiliary Survey Telescope (JAST80). All data available in the J-PLUS archive\,\footnote{\url{https://archive.cefca.es/catalogues/jplus-dr3}} is accessible through VO protocols, such as ``Simple Image Access Protocol'', ``Simple Cone Search'', or ``Table Access Protocol''. Especially, we made use of the latter, which allows querying the archive using complex searches based on ADQL\,\footnote{\url{https://www.ivoa.net/documents/REC/ADQL/ADQL-20081030.pdf}}, which is an extension of the common SQL language to support astronomy-specific queries. The wide-field covered by J-PLUS (3\,192\,deg$^2$ in the last data release), combined with its unique system of 12 optical filters \citep{jpluscal} that allows an accurate estimation of stellar parameters such as the effective temperature, provide a suitable setting for the identification on ultracool dwarfs. The CARMENES instrument is installed at the 3.5\,m telescope at the Calar Alto Observatory, located in Almería, Spain, and stands as one of the leading instruments in the quest for searching for Earth-like planets within the habitable zones around M dwarfs using the radial velocity technique. It comprises two separate spectrographs: one for the visible (VIS) wavelength range (from 520 to 960\,nm) and the other for the near-infrared (NIR) range (from 960 to 1710\,nm), each offering high-spectral resolutions of R\,$\approx$\,94\,600 and 80\,500, respectively \citep{Quirrenbach20,reiners2018}. The high-S/N, high-resolution spectra provided by the CARMENES data archive\,\footnote{\url{http://carmenes.cab.inta-csic.es/gto/}}, which is part of the SVO, offer a unique opportunity to determine the photospheric stellar parameters of the observed M dwarfs.

The VO represents a transformative milestone in astronomical research. By breaking down barriers to data access and fostering interoperability between astronomical archives around the world, it has become a cornerstone of modern observational astronomy. As the era of exabyte-scale\,\footnote{$1$\,exabyte $ = 1\,073\,741\,824$\,gigabytes} archives approaches, the continued evolution of VO solutions and protocols will be essential to ensure that astronomy remains at the forefront of scientific discovery in the 21st century. In this sense, the development of science platforms that allow the user to bring the analysis to the data, and not the other way around, will be crucial for scientific analysis on massive amounts of data.  As such platforms are gaining prominence in recent years (e.g. ESA Datalabs\,\footnote{\url{https://datalabs.esa.int/}}), the VO is working on integrating its technologies and protocols into them. 

The revolution in data management and accessibility of the last decades did not come alone. The unprecedented scale and complexity of these datasets raised new challenges, as traditional approaches struggle to efficiently process, classify, and extract knowledge from them. This has led to the increasing adoption of artificial intelligence and machine learning, which provide scalable and automated solutions for data analysis, capable of analysing huge amounts of data in an efficient way. From detecting rare astronomical phenomena to refining stellar classifications, artificial intelligence was here to stay.


\section{The age of artificial intelligence} \label{chp:ml_intro}

``Can machines think?''. Six years passed from Turing's famous enquiry \citep{turing} until artificial intelligence was consolidated as a research field in the Dartmouth Summer Research Project on Artificial Intelligence conference in 1956, whose organiser, John McCarthy, coined the term ``artificial intelligence'' for the field \citep{McCarthy2006}. Already in 1943, \citet{mcculloh1943} had proposed the first computational model of a biological neuron. In the nearly 80 years since then, artificial intelligence has undergone a remarkable transformation, moving from theoretical explorations to real-world applications that have redefined entire fields, and we have even come to normalise coming across driverless taxis \citep{auto_driving} on the streets of San Francisco and having human conversations with large language models \citep{bubeck2023,deepseek}. Do machines think? Can machines be conscious? These questions has been at the centre of debates in recent years and depend heavily on how we define intelligence and consciousness. For a captivating discussion on this topic, we refer the reader to \citet{qin2025}. What is clear is that, nowadays, machines have the ability to accomplish very complex goals, and this can be of great help to us in building data-driven solutions that ensure that we do not miss out or delay scientific knowledge simply because we cannot cope with the vast amounts of data. 

\begin{figure}
    \centering
	\includegraphics[width=.7\columnwidth]{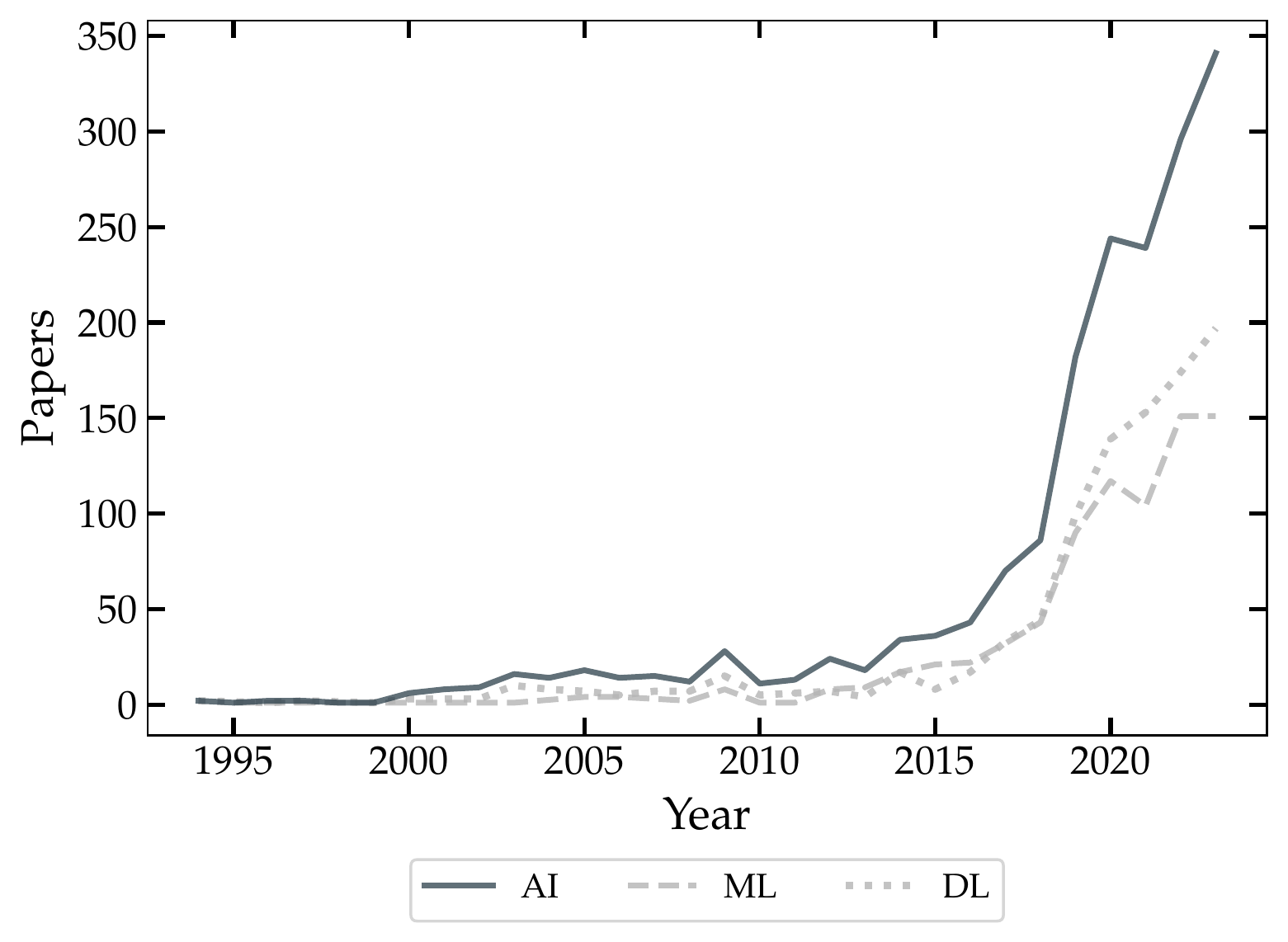}
    \caption{Diagram illustrating the increasing adoption of artificial intelligence by the astronomical community over the last years. The Y axis shows the number of referred papers inside arXiv:astro-ph. Papers were identified by searching for multiple keywords relevant to each of the categories in the abstracts. The raw data is publicly available at \url{https://www.kaggle.com/datasets/Cornell-University/arxiv}.}
    \label{fig:evo}
\end{figure}

Artificial intelligence, broadly defined as the field focused on the development of machines that mimic human intelligence to solve problems, is a domain that encompasses the well-known subfields of machine and deep learning. While machine learning refers to all systems that automatically learn from the data and make predictions without being explicitly programmed to do so, deep learning focuses on multi-layered neural networks that automatically extract features and create a hierarchical representation of the data. In traditional machine learning, an essential step is feature engineering, where domain experts manually select or design the most relevant features from raw data to improve model performance. For example, in the classification of stellar spectra, an astronomer might compute spectral indices such as the TiO and VO band strengths to distinguish between different M dwarf spectral types. These indices serve as handcrafted features that are then used by machine learning models like support vector machines or decision trees. In contrast, deep learning models, particularly convolutional neural networks, automatically extract relevant features from raw data without requiring manual input. For instance, instead of relying on predefined spectral indices, a convolutional neural network trained on stellar spectra is capable of learning patterns directly from the full spectrum, identifying subtle absorption lines and continuum variations that may be difficult to define explicitly. This automatic feature extraction can enhance classification accuracy and reveal new insights that might be overlooked with traditional methods.

Depending on the nature of the problem to be addressed, machine learning algorithms fall into different categories. Supervised algorithms such as support vector machines \citep{qu2003,huertas2008,kovacs2015,paschenko2018} or supervised decision trees and random forests \citep{carliles2010,moller2016,ishida2019,bluck2022},  are used to map a set of features to a target variable based on input-output pairs that are often based on domain expertise. On the other hand, unsupervised machine learning algorithms such as K-means \citep{balazs2996,sanchez2010,garcia2018}, hierarchical clustering \citep{hojnacki2007,baron2015,ma2018}, principal component analysis \citep{boroson1992,vandenberk2006,bailey2012}, or self-organising maps \citep{meusinger2012,armstrong2016,rahmani2018}, are used to learn complex relationships within an unlabelled dataset for data exploration and visualisation, dimensionality reduction, or outlier detection tasks. It is important to note that several algorithms, such as random forests or artificial neural networks, can be used in both a supervised and unsupervised setting. When a small set of labelled data is available, semi-supervised learning techniques allow leveraging unlabelled data to learn a structured representation of the data or create pseudo-labels \citep{richards2011,Slijepcevic2022}. Alternatively, self-supervised learning algorithms use large amounts of unlabelled data to supervise themselves, and have been wildly used in representation learning \citep[introduced in astronomy by ][]{serra1993}, where algorithms extract meaningful compressed representations (embeddings) of complex high-dimensional data,  during recent years \citep{Yang2015,hayat2021,sarmiento2021,masbuitrago2024}. Reinforcement learning is an active branch of machine learning that optimises control tasks by interacting with a dynamic environment, evaluating outcomes and refining the actions of the system based on long-term rewards, and holds great promise as an approach for adaptive optics in astronomy \citep{Nousiainen2021,nousiainen2022,gutierrez2024}. 

\begin{figure}
    \centering
	\includegraphics[width=.75\columnwidth]{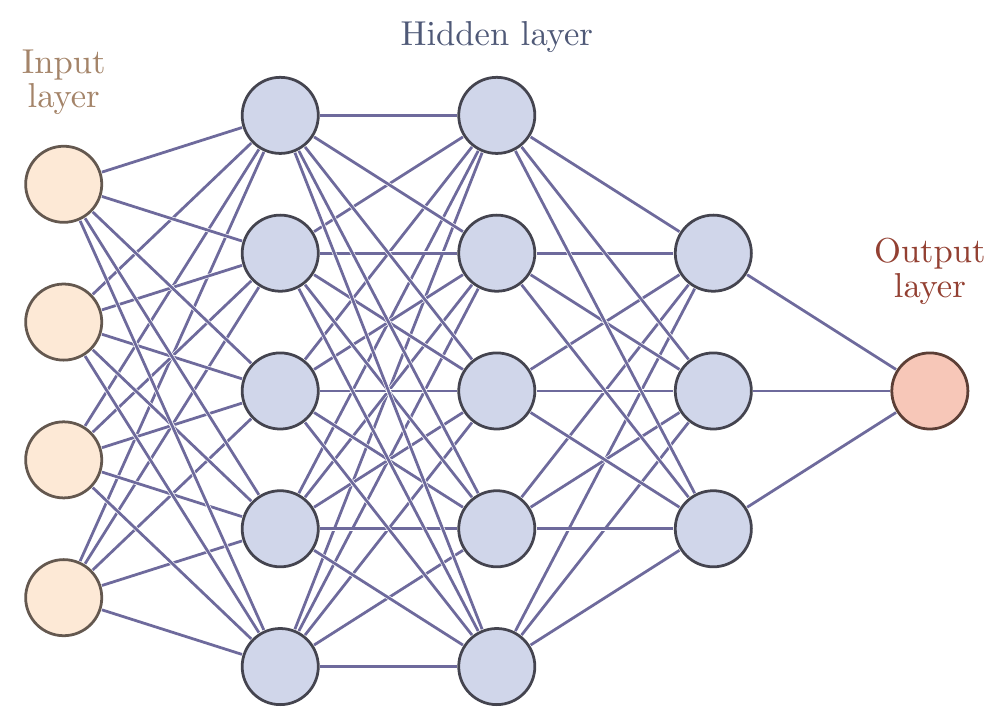}

    \caption{Schematic representation of a multilayer perceptron architecture. Source: \url{https://tikz.net/neural_networks/}.}
    \label{fig:ann}
\end{figure}

The last few years have witnessed an explosion in the number of deep learning methodologies (see Figure \ref{fig:evo}), driven by major advances in the field since the refinement of training techniques for deep neural networks \citep{bengio2006,hinton2006a,hinton2006b} and the popularisation of convolutional neural networks \citep{alexnet2012,vggnet2014}, and also aided by the increase in computational power and the availability of massive datasets. However, machine learning made its debut in astronomy in the late 1980s \citep[see][and references therein]{miller1993}, with artificial neural networks being the core of applications to star-galaxy classification \citep{odewahn1992,odewahn1993,bertin1996,bazell1998,anderson2000,qin2003}, galaxy morphology classification \citep{storrie1992,lahav1995,lahav1996,odewahn1996,cohen2003,madgwick2003}, photometric redshift estimation \citep{firth2003,tagliaferri2003,ball2004}, characterisation of stellar spectra \citep{klusch1993,hippel1994,bailer_jones1997}, quasar classification \citep{carballo2004,claeskens2006}, or cosmology \citep{auld2007,auld2008}. Moreover, in the early 2000s, decision trees and support vector machines began being used for galaxy morphology classification \citep{huertas2008,huertas2011}, photometric redshift estimation \citep{wadadekar2005}, or AGN/galaxy separation \citep{white2000,gao2008}. Within the SVO framework, the first studies using machine learning techniques emerged in the late 2000s, focusing on the automated supervised classification of eclipsing binary light curves \citep{sarro2006}, exoplanet light curves \citep{sarro2006b}, and variable star light curves \citep{debosscher2007,sarro2009}. This was followed by an important contribution to the use of machine learning techniques for the determination of physical parameters of ultracool dwarfs in the scope of the \textit{Gaia} mission \citep{sarro2013,bailerjones2013}. We refer the reader to \citet{ml_review_baron,ml_review_ball,huertas2023,smith2023} for a complete and extensive review of machine and deep learning techniques applied to astronomy.

Deep learning represents a new approach to data analysis in astronomy and in science in general, as it enables the development of unsupervised and self-supervised fully data-driven solutions that do not rely on laborious manual feature engineering or labelling. The simplest artificial neural network is the perceptron, originally introduced by \citep{rosenblatt1958}, which is equivalent to a single neuron node. This node consists of a set of numeric inputs $\textit{x}_{\mathrm{i}}$, which are multiplied by weights $\textit{w}_{\mathrm{i}}$ that represent the strength of the connection between each of the inputs and the neuron. The perceptron then sums the list of products, adds a bias term $\textit{b}$, which allows to the activation function to be shifted linearly, and passes the result to an activation function $\textit{f}$, which gives the final output $\textit{y}$:

\begin{equation}
    \textit{y} = \textit{f}\left ( \sum_{i=1}^{n} \textit{w}_i\textit{x}_i + \textit{b} \right ).    
    \label{eq:perceptron}
\end{equation}

\begin{figure}
    \centering
	\includegraphics[width=.75\columnwidth]{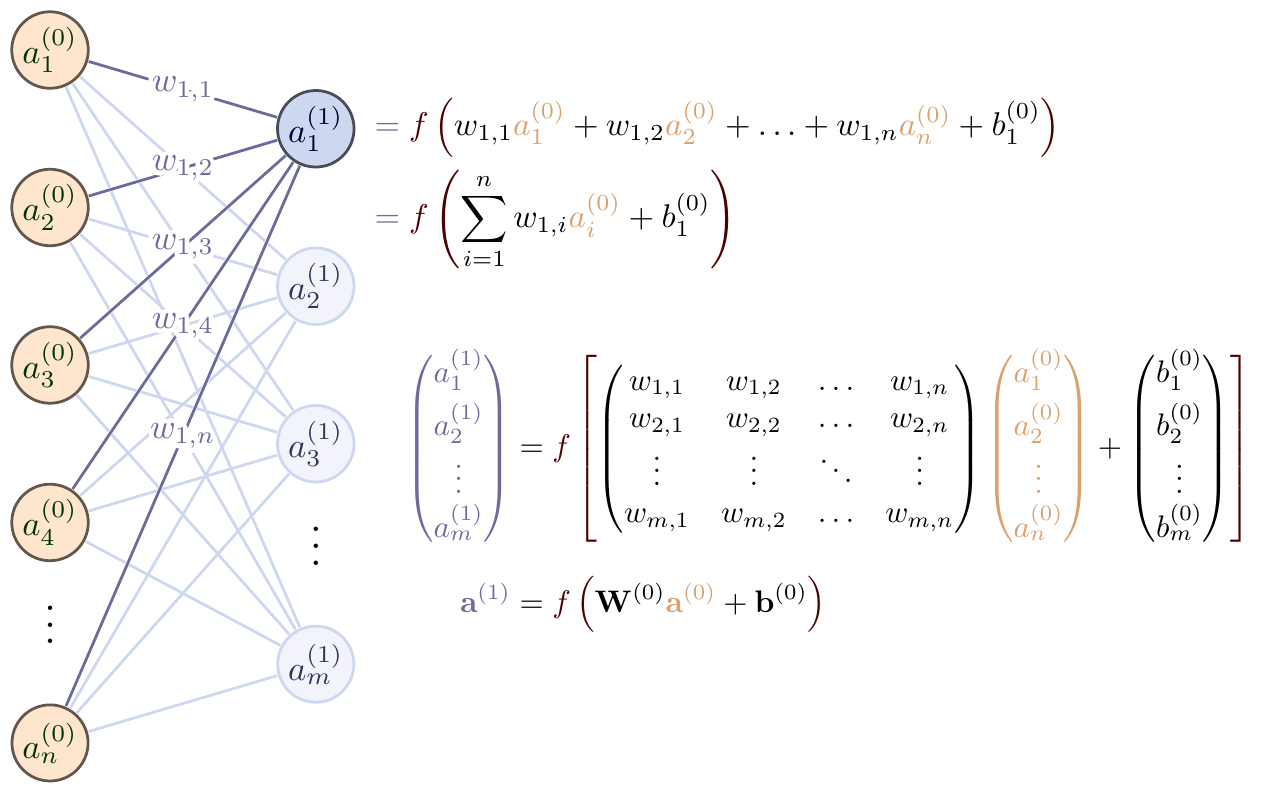}

    \caption{Schematic representation of the forward pass in a neural network node, as described in the text. Source: \url{https://tikz.net/neural_networks/}.}
    \label{fig:ann_activation}
\end{figure}

Feed-forward artificial neural networks, or multilayer perceptrons, are fully-connected multilayer stacks of individual nodes (see Figure \ref{fig:ann}) that compute non-linear input-output mappings. At each layer, the input of each individual node is obtained as a weighted sum of the outputs of the nodes of the previous layer, and passed to a non-linear activation function in a process known as forward pass (see Figure \ref{fig:ann_activation}). Typically, the rectified linear unit \citep[ReLU; ][]{nair2010}, $\textit{f}(\textit{x})=\max(\textit{x},0)$, is used as non-linear activation function due to its good scalability for networks with many layers and its ability to avoid vanishing gradients \citep{hochreiter1991}. In the training of the network, this forward pass is performed across all layers to compute the prediction of the neural network, which is passed to a loss function that computes the difference between this prediction and the ground truth, or expected output. Then, the gradient of the loss function with respect to the weights of the network is computed using the backpropagation procedure \citep{werbos1974,parker1985,lecun1985,rumelhart1986}, which propagates the gradients backwards from the last layer using the chain rule. Finally, the weights of the network are updated using gradient descent to minimise the loss function \textit{L}:

\begin{equation}
     \textit{w}_{i+1} = \textit{w}_i - \eta \frac{\partial \textit{L}}{\partial \textit{w}_i},  
    \label{eq:backpropagation}
\end{equation}

where $\eta$ is the learning rate, which controls how much the weights change. This process is repeated iteratively over multiple epochs, often using optimisation algorithms such as Adam \citep{adam}, until the network reaches a low enough loss.

\begin{figure}
    \centering
	\includegraphics[width=.6\columnwidth]{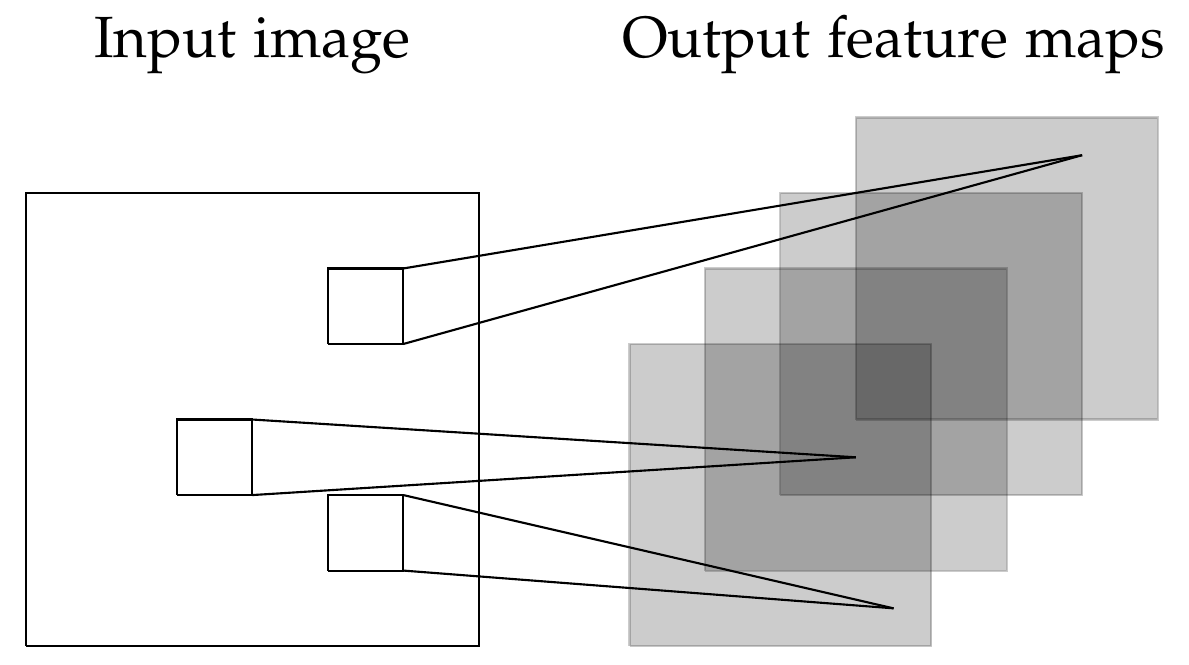}

    \vspace{5mm}

	\includegraphics[width=.6\columnwidth]{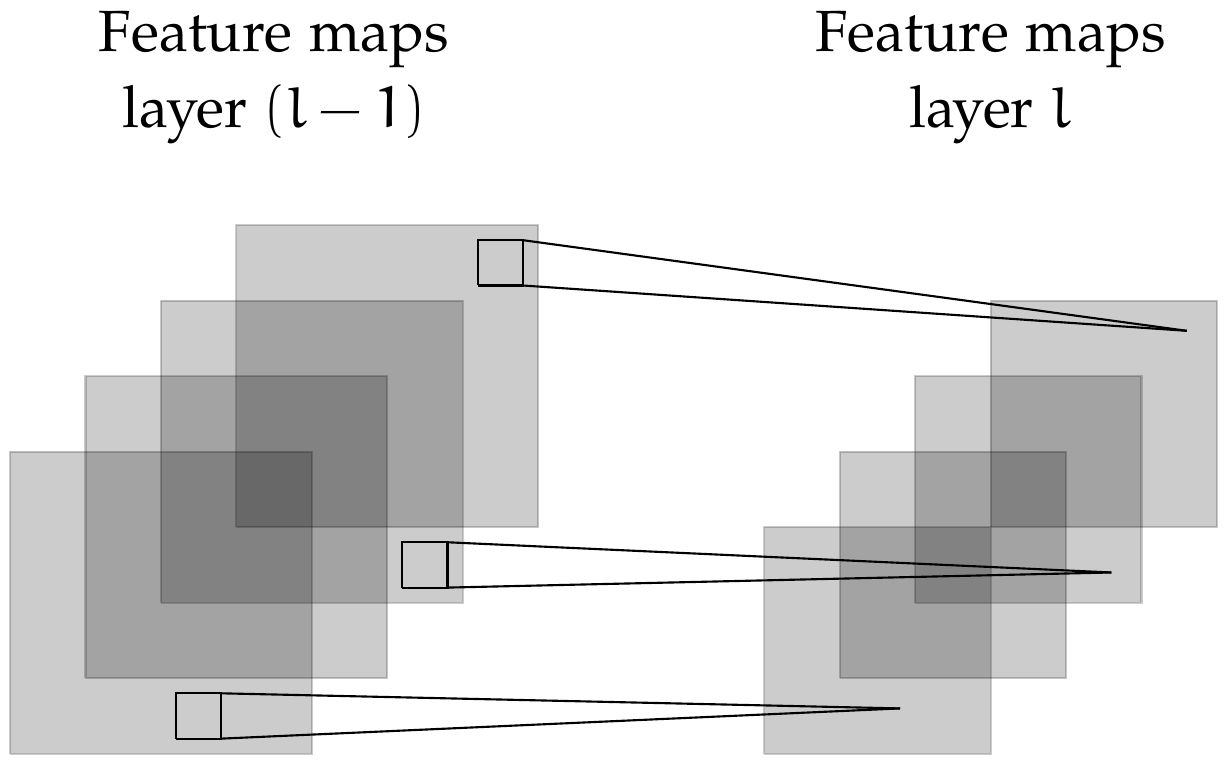}
    \caption{Schematic representation of the extraction of features in a convolutional layer. Source \url{https://davidstutz.de}.}
    \label{fig:cnn_intro}
\end{figure}

Inspired by the hierarchical structure of the human visual nervous system \citep[a precursor of convolutional neural networks; ][]{necognitron1980}, convolutional neural networks are a specific class of multilayered feedforward neural networks, initially developed for image classification and visual pattern recognition \citep{lecun1989,lecun1998}. The distinctive factor of convolutional neural networks is the use of convolution operations, in the convolutional layers, to automatically extract features from data. After the convolutional structure, the set of features is flattened and passed to a multilayer perceptron to predict the output of the layer. In each forward pass process, the input of each unit of the convolutional layer is obtained with an element-wise dot product between a set of weights known as convolution kernel or filter, and the output feature maps of the previous layer (see Figure \ref{fig:cnn_intro}). In a same layer, different units use different filters. The resulting arrays and a tunable bias are added up and passed through an activation function to obtain the output feature map of each unit. The set of weights of each kernel and the weights of the multilayer perceptron are adjusted in the training process, so that the different feature maps of the convolutional layers represent specific features detected in the input data\,\footnote{We refer the interested reader to \url{https://poloclub.github.io/cnn-explainer/} for an interactive visualisation of the internal workings of a convolutional neural network.}. This feature representation learnt by the network is hierarchical, preserving the generic learning in the lower layers (closer to the input) and the more specific features in the higher layers.

The deep learning explosion started in 2012. In the ImageNet Large Scale Visual Recognition Challenge competition \citep{russakovsky2014} of this year, a deep convolutional neural network called AlexNet \citep{alexnet2012} achieved incredible results, far outperforming its competitors\,\footnote{According to The Economist, ``Suddenly people started to pay attention, not just within the artificial intelligence community but across the technology industry as a whole.''.} thanks to the use of graphics processing units, ReLU activation functions, data augmentation, and a technique know as dropout that prevents the network from overfitting \citep{dropout}. This success initiated a revolution in the field of computer vision, and the pace of improvement in the following years of the ImageNet competition was dramatic \citep{vggnet2014,googlenet,resnet}. It did not take astronomers long to notice. Due to their nature, it is not surprising that early work using convolutional neural networks in astronomy focused on image classification, for pulsar identification \citep{zhu2014} and for galaxy morphological classification \citep{dieleman2015,huertas2015,aniyan2017}. Moreover, \citet{hala2014} pioneered the use of convolutional neural networks for spectral classification. These works signalled the beginning of the use of deep learning techniques in astrophysics, which has been growing at an overwhelming rate ever since.

As discussed in Section \ref{chp:vo_intro}, astronomical datasets are becoming increasingly large and complex, making the exploration of these archives almost impossible without the use of data discovery techniques. In this sense, machine learning has emerged as a powerful tool for visualising or detecting anomalies in vast datasets. Data visualisation is essential for the exploration of high-dimensional astronomical datasets, bringing the data to a lower dimensionality that allows it to be analysed in a more interpretable way. Dimensionality reduction techniques, such as principal component analysis \citep{hotelling:33}, t-SNE \citep{vandermaaten08}, UMAP \citep{McInnes2018}, or self-organising maps \citep[Kohonen networks; ][]{kohonen1982}, are widely used in this regard. Moreover, unsupervised and self-supervised representation deep learning, especially using autoencoder architectures \citep{serra1993} and more recently contrastive learning models \citep{chen2020}, are used to extract meaningful embeddings from high-dimensional astronomical data, that can be used as input for a downstream classification or regression task. These methodologies constitute also a vital tool for the detection of anomalies or outliers in big data surveys \citep{chalapathy2019}, which enables the discovery of rare or unexpected phenomena within massive datasets, where the combination of deep learning methods with the VO technology is extremely useful \citep{skoda2020}. Moreover, VO solutions are very helpful in further characterising these anomalous instances detected in surveys. This is particularly important for the field of transient astronomy, that will soon experience a revolution with the forthcoming LSST survey \citep{li2022} of the Vera C. Rubin Observatory.

The dependence of deep learning algorithms on massive training data is a crucial hurdle to overcome when a research scenario requires labelled data, as building a large annotated dataset can be incredibly complex and expensive. This is the case, for example, when a classification or regression task is to be performed on a small labelled sample. A straightforward, and widely used in astronomy, solution to this problem is the use of a data-rich labelled dataset similar to the target dataset or of synthetic data to train the deep learning models, but this may include a systematic error in the methodology if the training set is not identical to the observed data on which the inference is made \citep{pass20}. Transfer learning, in which knowledge is transferred from a rich source domain to a related but not identical target domain, plays a key role in solving the above problems. Knowledge transfer is typically performed by training a deep learning model on a data-rich dataset and then fine-tuning the neural network weights using the target dataset \citep{dominguez2018,walmsley2022,bello2023,masbuitrago2024}. Another approach to this problem is the use of active learning \citep{walmsley2020,stevens2021}, which reduces the number of required training samples by selecting the most informative data to label. For the simulation of data-rich labelled datasets, deep generative models (introduced in astronomy by \citealt{regier2015}) such as variational autoencoders, generative adversarial networks, score-based generative models, or diffusion models, can be leveraged to generate massive amounts of data similar to astronomical observations. Deep generative models enable data-driven simulation as they capture the underlying probability distribution of a given dataset, and use that knowledge to generate new, realistic synthetic data from it.

With the captivating title ``Attention is All You Need'', \citet{vaswani2017} presented the revolutionary transformer neural network, based on a mechanism known as attention that computes the relevance of each input token with respect to all others in a sequence and captures contextual relationships, and which is still largely unexploited in astronomy \citep{astromer,atat}. Transformer architectures are the pillars on which large language models, such as PaLM \citep{palm}, LLaMa \citep{llama}, or GPT-4 \citep{gpt4} are built, which have revolutionised the field of natural language processing in the last two years. Thanks to their versatility and their ability to handle multimodal data \citep{reed2022}, transformers can be harnessed to build what we know as foundation models, which are models that are trained on vast amounts of data using self-supervised learning for subsequent fine-tuning tailored to diverse specific downstream tasks. Interest in foundation models in astronomy is growing rapidly \citep{astrollama,rozanski2023,leung2024,parker2024}, since the natural evolution for the upcoming decades would be a transition from domain-specific deep learning models to fine-tuned versions of the same all-encompassing astronomical foundation model. The miscellaneous and rich nature of astronomical data generated from entirely different instruments, combined with the interoperability enabled by VO technology, represents a key opportunity in this regard. To this end, it is paramount that the astronomical community adopts a transparent, open-source, bazaar-style development, which has proven successful in large open-source projects such as Linux \citep{raymond2001}, with a strong commitment to interpretability (see \citet{ras2020} for a detailed discussion on explainable artificial intelligence). This open and democratised scenario would unlock the potential of state-of-the-art deep learning solutions for the entire astronomical community, solving the current inaccessibility of most astronomers to these models due to lack of resources.

The characterisation of M dwarfs and ultracool dwarfs is fundamental to advancing our understanding of stellar astrophysics, planetary formation and habitability, and the structure and kinematics of our Galaxy, yet their characterisation remains an ongoing challenge due to their intrinsic faintness and complex atmospheres. As astronomical datasets grow in size and complexity, the ability to efficiently mine and analyse these vast archives has become a necessity, with the VO playing a key role in enabling multi-wavelength data discovery and interoperability. In parallel, the rise of machine and deep learning has transformed how we extract knowledge from astronomical data at an unprecedented scale, offering new approaches for classification, parameter estimation, anomaly detection, and data-driven discovery. The synergy between VO technologies and machine learning has set the stage for a new era in astronomical research, one in which automated, scalable, and interpretable solutions will be essential for maximising the scientific return of upcoming large-scale surveys. Recent applications of machine and deep learning in astronomy, as exemplified above, illustrate how artificial intelligence is not only optimising data analysis but also driving new discoveries that would otherwise be unfeasible with traditional methods. The fusion of artificial intelligence and astronomy is no longer just an option--it is a necessity. 


\section{Aims and objectives of the thesis} \label{sec:thesis_obj}

The aim of this thesis is to explore the application of machine learning and deep learning techniques to spectroscopic and photometric surveys, with a particular focus on M dwarfs and ultracool dwarfs, demonstrating how these methodologies can enhance our understanding of low-mass stars and substellar objects, and push the boundaries of data-driven astronomical research. The thesis can be divided into two main objectives. The first, covered in Chapter \ref{chp:ucds_paper}, is to consolidate a methodology for identifying ultracool dwarfs in wide-field multi-filter photometric surveys, using data from the J-PLUS survey, driven by VO data mining techniques and tools. In view of the vast surveys with these characteristics that will come to light in the very near future, this thesis aims to demonstrate that a machine learning approach is able to significantly accelerate this process. A sub-objective derived from this first one is to leverage these surveys for the automatic detection of flares in M dwarfs (Chapter \ref{chp:flares_paper}), thanks to specific narrow-band filters located at specific spectral features. The second objective, which encompasses Chapters \ref{chp:autoencoders_paper} and \ref{chp:dtl_ucds}, is to develop an automatic and scalable deep learning-based methodology capable of determining the atmospheric parameters of M dwarfs and ultracool dwarfs from spectroscopic data. The strategy here starts with the use of M dwarf high-resolution spectra from CARMENES, and the subsequent adaptation to the ultracool domain is carried out with low-resolution spectra from the SpeX Prism Library.

\chapter{Ultracool Dwarfs in J-PLUS}
\label{chp:ucds_paper}




Ultracool dwarfs (UCDs) comprise the lowest mass members of the stellar population and brown dwarfs, from M7\,V to cooler objects with L, T, and Y spectral types. Most of them have been discovered using wide-field imaging surveys, for which the Virtual Observatory (VO) has proven to be of great utility. We aim to perform a search for UCDs in the entire Javalambre Photometric Local Universe Survey (J-PLUS) second data release (2\,176\,deg$^2$) following a VO methodology. We also explore the ability to reproduce this search with a purely machine learning (ML)-based methodology that relies solely on J-PLUS photometry. We followed three different approaches based on parallaxes, proper motions, and colours, respectively, using the \texttt{VOSA} tool to estimate the effective temperatures and complement J-PLUS photometry with other catalogues in the optical and infrared. For the ML methodology, we built a two-step method based on principal component analysis and support vector machine algorithms. We identified a total of 7\,827 new candidate UCDs, which represents an increase of about 135\,\% in the number of UCDs reported in the sky coverage of the J-PLUS second data release. Among the candidate UCDs, we found 122 possible unresolved binary systems, 78 wide multiple systems, and 48 objects with a high Bayesian probability of belonging to a young association. We also identified four objects with strong excess in the filter corresponding to the Ca~{\sc ii} H and K emission lines and four other objects with excess emission in the H$\alpha$ filter. Follow-up spectroscopic observations of two of them indicate they are normal late-M dwarfs. With the ML approach, we obtained a recall score of 92\,\% and 91\,\% in the  20$\times$20\,deg$^2$ regions used for testing and blind testing, respectively. We consolidated the proposed search methodology for UCDs, which will be used in deeper and larger upcoming surveys such as J-PAS and Euclid. We concluded that the ML methodology is more efficient in the sense that it allows for a larger number of true negatives to be discarded prior to analysis with \texttt{VOSA}, although it is more photometrically restrictive.

\section{J-PLUS} \label{jplus_intro}

J-PLUS is a multi-filter survey conducted from the Observatorio Astrofísico de Javalambre \citep[OAJ;][]{OAJ} in Teruel, Spain. Since it was primarily conceived to ensure the photometric calibration of J-PAS, it uses the second largest telescope at the OAJ, which is the 0.83 m Javalambre Auxiliary Survey Telescope (JAST80). J-PLUS is covering thousands of square degrees of the sky using the panoramic wide-field (2\,deg$^2$ field of view) camera T80Cam \citep{t80cam}, which is equipped with a CCD of 9.2k x 9.2k pixels and a pixel scale of 0.55\,arcsec\,pix$^{-1}$.

While J-PAS will use an unprecedented system of 56 narrow band filters in the optical, the J-PLUS filter system is composed of four broad  (\textit{gSDSS}, \textit{rSDSS}, \textit{iSDSS}, and \textit{zSDSS}), two intermediate (\textit{uJAVA} and \textit{J0861}) and six narrow (\textit{J0378}, \textit{J0395}, \textit{J0410}, \textit{J0430}, \textit{J0515}, and \textit{J0660}) band optical filters. The transmission curves, as well as additional information of these filters, can be found at the \textit{Carlos Rodrigo} Filter Profile Service maintained by the Spanish Virtual Observatory\,\footnote{\url{http://svo2.cab.inta-csic.es/theory/fps/index.php?&mode=browse&gname=OAJ&gname2=JPLUS}} \citep{fps}.

\begin{figure*}
    \centering
	\includegraphics[width=\columnwidth]{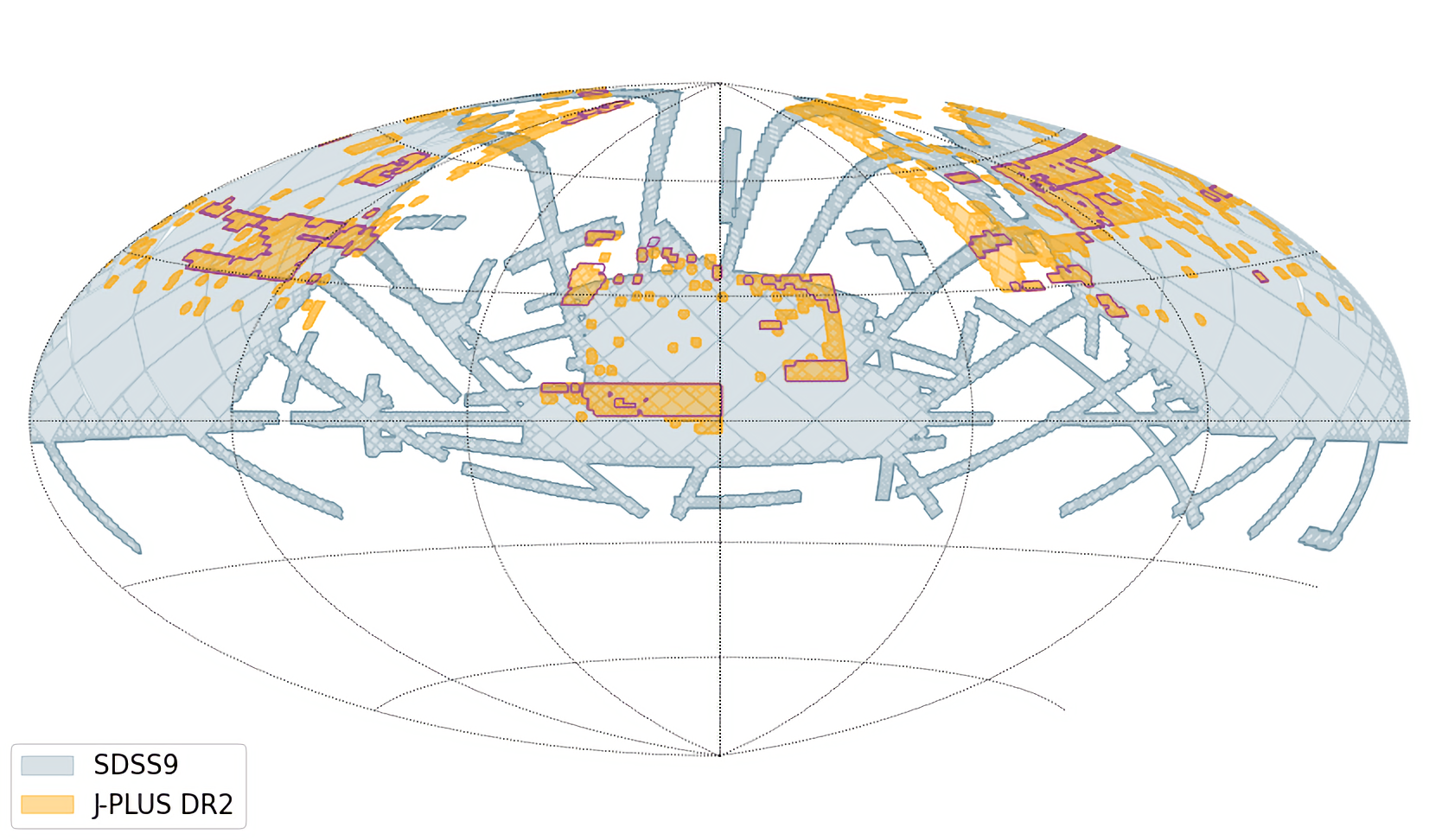}
    \caption{Sky coverage of J-PLUS DR2 (yellow) in $\alpha$, $\delta$ coordinates (centred on $\alpha = 0$ deg, $\delta = 0$ deg, with $\alpha$ rising to the left). The SDSS DR9 footprint is superimposed in blue. The purple line represents the border of the J-PLUS DR1 coverage map.}
    \label{fig:mocs}
\end{figure*}

The J-PLUS DR2, available since November 2020, comprises 1\,088 fields, covering 2\,176\,deg$^2$, observed in all the mentioned optical bands. Fig. \ref{fig:mocs} shows the sky coverage of this release. \citet{jpluscal} presents the updated photometric calibration for the DR2, that was improved by including the metallicity information from LAMOST DR5 in the stellar locus estimation. The limiting magnitudes of the 12 bands can be consulted in the Table 1 of the same paper.


\section{Methodology} \label{Methodology}

We divided the sky coverage of J-PLUS DR2 in 37 regions of 20$\times$20\,deg$^2$. To cope with the fact that queries to the J-PLUS archive are limited to 1 million objects, we decided to tessellate each region into smaller circular subregions of 1 deg radius. We made use of \texttt{TOPCAT}\footnote{\url{http://www.star.bris.ac.uk/~mbt/topcat/}} \citep{Taylor2005} to cross-match each tessellated region with the J-PLUS DR2 sky coverage in order to avoid searching regions of the sky that are not covered by it.

We used the package \texttt{STILTS}\footnote{\url{http://www.star.bris.ac.uk/~mbt/stilts/}} \citep{Taylor2006} to query the J-PLUS DR2 database through the Virtual Observatory TAP protocol. This allowed us to write ADQL\footnote{\url{https://www.ivoa.net/documents/REC/ADQL/ADQL-20081030.pdf}} code to search over all 20$\times$20\,deg$^2$ regions iteratively. A typical ADQL query example looks like this:

\newpage

\begin{code}
SELECT objs.filter_id,objs.alpha_j2000,
    objs.delta_j2000,objs.class_star,
    objs.mag_aper_6_0,objs.mag_err_aper_6_0,
    objs.mask_flags,imgs.aper_cor_6_0,
    imgs.aper_cor_err_6_0 
FROM jplus.MagABSingleObj as objs,
    jplus.TileImage as imgs 
WHERE objs.tile_id = imgs.tile_id
AND objs.alpha_j2000 between 2 and 5 
AND objs.delta_j2000 between 2 and 3 
AND objs.flags=0 
AND objs.filter_id between 1 and 4 
AND objs.class_star>0.1
\end{code}

In our case, we used the 6 arcsec diameter aperture photometry, since the aperture correction to pass 6 arcsec aperture magnitudes to total magnitudes for point-like sources is available in the J-PLUS DR2 database. We constrained the search to records with good photometric conditions by imposing  \texttt{flags=0} (no \texttt{SExtractor} flags\footnote{\url{https://sextractor.readthedocs.io/en/latest/Flagging.html}}). Since object detection is performed independently on each filter, this means that for each source the \texttt{flags=0} condition is applied at the filter level. We also required \texttt{class$\_$star > 0.1}. We were not very restrictive with \texttt{class$\_$star} (\texttt{SExtractor} stellarity index) in order not to loose faint sources that may appear as extended objects.

For each 20$\times$20\,deg$^2$ region, we concatenated the data for the corresponding circular subregions into a single table and removed duplicated instances (tessellated areas may overlap). As UCDs emit most of their flux at longer wavelengths, for the methodology described in Sects. \ref{parallax}, \ref{pm} and \ref{colordiagram}, we only considered the relevant filters for these objects, i.e., the reddest ones (filter IDs 1$-$4 and 10$-$12 in the J-PLUS DR2 database, see Table \ref{tab:filters}). Even so, we stored the data for all filters separately, as we required them for the flare detection workflow described in Sect. \ref{flares}. Finally, we used the CDS \texttt{X-Match} service\footnote{\url{http://cdsxmatch.u-strasbg.fr/}} in \texttt{TOPCAT} with \textit{Gaia} EDR3 J2016 (reference epoch 2016.0), using a 3 arcsec radius, to obtain the astrometric information. In those cases where more than one counterpart exists in the search region, only the nearest one was considered. In Sects. \ref{parallax}, \ref{pm} and \ref{colordiagram} we describe the analysis carried out for each 20$\times$20\,deg$^2$ region separately.

\begin{table}
\fontsize{11pt}{11pt}\selectfont
 \caption{J-PLUS filter information, taken from the J-PLUS DR2 database, sorted from shortest to longest wavelength.}
 \centering
 \label{tab:filters}
 \begin{tabular}{l c c}
 
  \hline\hline
  \noalign{\smallskip}

  Filter ID & Filter & $\lambda_{\mathrm{eff}}$ \\
  &  & [$\AA$] \\

  \noalign{\smallskip}
  \hline
  \noalign{\smallskip}

  5.0 & \textit{u} & 3542.20 \\
  6.0 & \textit{J0378} & 3793.38 \\
  7.0 & \textit{J0395} & 3938.55\\
  8.0 & \textit{J0410} & 4107.98\\
  9.0 & \textit{J0430} & 4298.36\\
  2.0\,$^{a}$ & \textit{g} & 4748.47\\
  10.0\,$^{a}$ & \textit{J0515} & 5139.67\\
  1.0\,$^{a}$ & \textit{r} & 6206.11\\
  11.0\,$^{a}$ & \textit{J0660} & 6606.67\\
  3.0\,$^{a}$ & \textit{i} & 7613.86\\
  12.0\,$^{a}$ & \textit{J0861} & 8610.16	\\
  4.0\,$^{a}$ & \textit{z} & 8940.28\\
  
  \noalign{\smallskip}
  \hline
 \end{tabular}
 \tablefoot{$^{(a)}$ Relevant filters (reddest ones) for UCDs search.}
\end{table}

\subsection{Parallax-based selection} \label{parallax}

From the cross-matched sample, we only kept sources with relative errors of less than 20\,\% in parallax and less than 10\,\% in both \textit{G} and $\textit{G}_{\rm RP}$ photometry. With these objects, we constructed a colour-magnitude diagram (see the left panel of Fig. \ref{fig:astro_diagrams}), where the absolute \textit{Gaia} magnitude in the \textit{G} band was estimated using

\begin{equation}
    M_G= \textit{G} + 5\log{\varpi} + 5,
	\label{eq:absoluteg}
\end{equation}

\noindent where \textit{G} is the \textit{Gaia} apparent magnitude and $\varpi$ is the parallax in arcseconds. To obtain a shortlist of candidate UCDs, we adopted a colour cut of $\textit{G} - \textit{G}_{\rm RP} > 1.3$\,mag, which corresponds to spectral types M5\,V or later according to the updated version of Table 5 in \citet{pecaut2013} \footnote{\label{mamajek}\url{http://www.pas.rochester.edu/~emamajek/EEM_dwarf_UBVIJHK_colors_Teff.txt}}, and an absolute magnitude limit of $M_G > 5$\,mag to leave aside the red giant branch.

\subsection{Proper motion-based selection} \label{pm}

Ultracool dwarfs may have photometric and morphological properties similar to those of objects such as giants, quasi-stellar objects (QSOs) or distant luminous red galaxies \citep[e.g.][]{Caballero2018, theissen2016, theyssen2017}. Assuming nearby objects will have high proper motions, reduced proper motion diagrams are a reliable tool for discriminating between nearby stellar populations and distant sources.

From the cross-matched sample introduced in Sect. \ref{Methodology}, we only kept sources with a relative error of less than 20\,\% in both proper motion components and less than 10\,\% in both \textit{G} and $\textit{G}_{\rm RP}$ photometry. Furthermore, we only took into account sources with non-zero proper motion, i.e., sources with, at least, one of the proper motion components greater (in absolute value) than three times the associated error.

\begin{figure*}
    \centering
    	\includegraphics[width=.48\columnwidth]{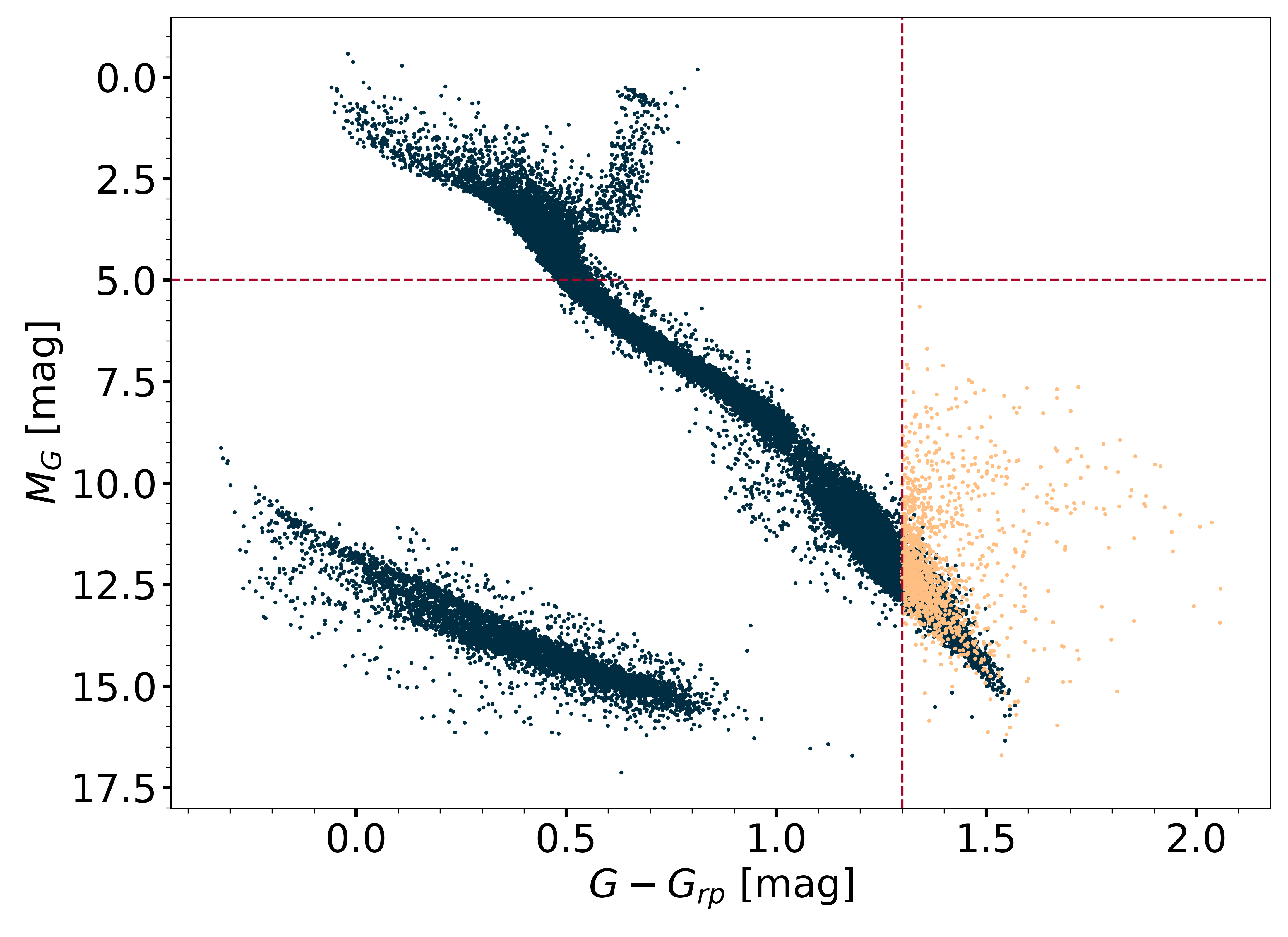}
    	\includegraphics[width=.48\columnwidth]{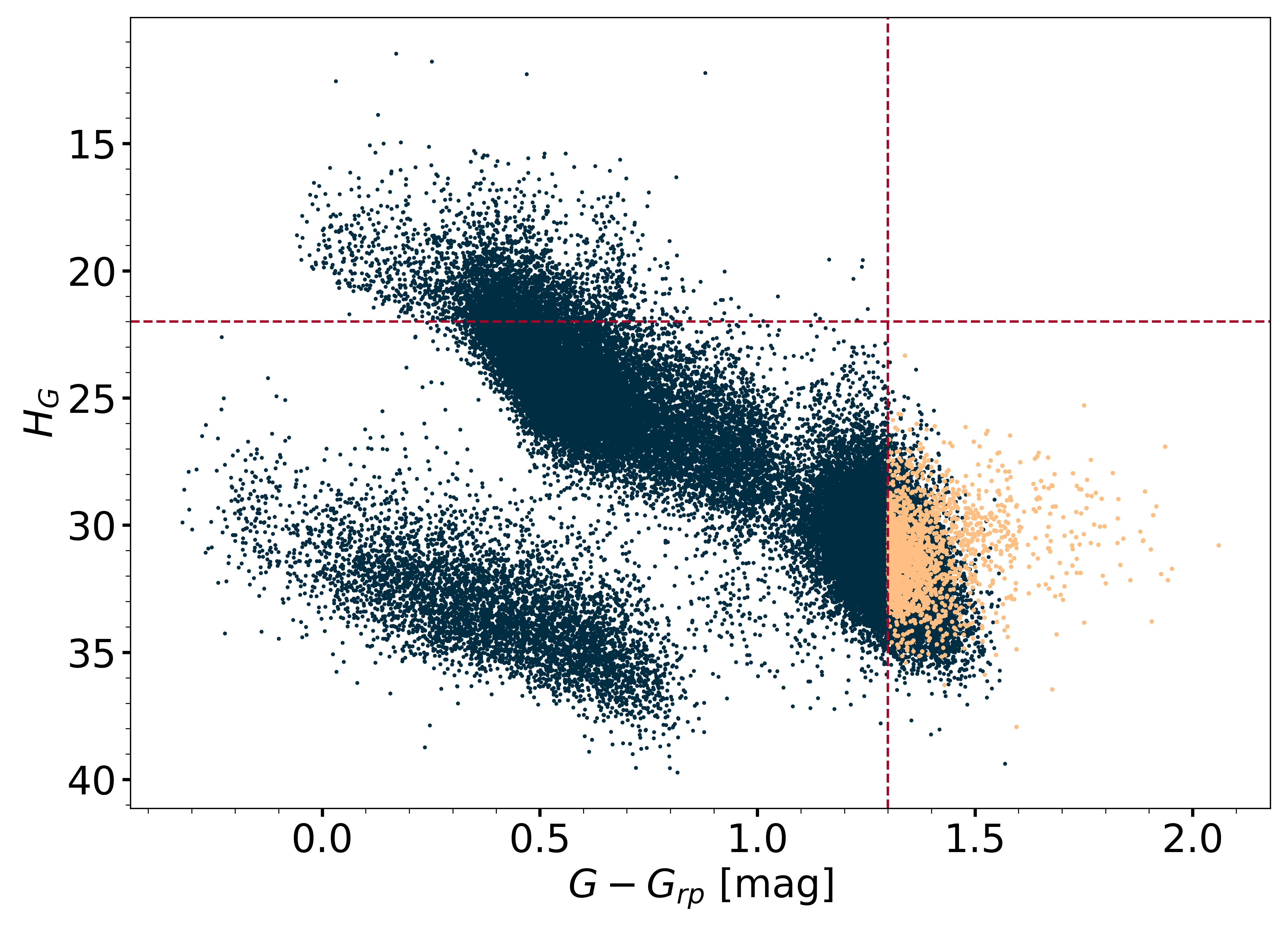}
    \caption{Location of the objects shortlisted as candidate UCDs via astrometric selection, for an example 20$\times$20\,deg$^2$ region, in a colour-magnitude (left) and a reduced proper motion (right) diagram built using \textit{Gaia} EDR3 sources with parallaxes larger tan 10 mas (dark blue dots). The vertical and horizontal red lines mark the boundaries for a source to be shortlisted as candidate UCD. Sources fulfilling these conditions are overplotted in yellow.}
    \label{fig:astro_diagrams}
\end{figure*}

The right panel of Fig. \ref{fig:astro_diagrams} shows the reduced proper motion diagram defined as:

\begin{equation}
    H_G=\textit{G} + 5\log{\mu} + 5,
	\label{eq:reducedpm}
\end{equation}

\noindent where \textit{G} is the \textit{Gaia} apparent magnitude and $\mu$ is the total proper motion in mas\,yr$^{-1}$. Of these sources, we filtered out those already pre-selected in the parallax-guided analysis described in Sect. \ref{parallax} and shortlisted as candidate UCDs those fulfilling the condition $\textit{G} - \textit{G}_{\rm RP} > 1.3$\,mag, and with a reduced proper motion $H_G > 22$\,mag to leave aside the red giant branch.

As discussed in Sect. \ref{Methodology}, the cross-match with \textit{Gaia} EDR3 J2016 is done using a 3 arcsec radius. Since J-PLUS DR2 is based on images collected from November 2015 to February 2020, we might miss some objects with a proper motion larger than 750\,mas\,yr$^{-1}$, as they could fall outside this 3\,arcsec radius. However, we decided not to increase the radius to avoid finding erroneous counterparts.

\subsection{Photometry-based selection} \label{colordiagram}

In the first two criteria (colour-magnitude and reduced proper motion diagrams) we are imposing parallax and proper motion constraints respectively, which makes these methods dependent on \textit{Gaia} astrometric information. This means that objects with good photometry but poor astrometry will be excluded from the lists of candidate UCDs. To solve this limitation, in this section we describe a method solely dependent on photometric information. This procedure consisted of two separate steps. First, we built a colour-colour diagram with the purpose of defining a colour cut to identify the UCD locus. Then, we applied this criterion to each 20$\times$20\,deg$^2$ region independently to obtain a shortlist of candidate UCDs.

To built the colour-colour diagram, we first searched in J-PLUS DR2 for true extended sources, defined as sources having \texttt{class$\_$star < 0.01}. Likewise, true point sources were defined as sources with \texttt{class$\_$star > 0.99}. Then, we performed a cross-match with 2MASS and built a $\textit{J}-\textit{K}_s$ (2MASS) vs. $\textit{r}-\textit{z}$ colour-colour diagram to separate the two types of sources.  As discussed in Sect. \ref{pm}, QSOs may have morphometric properties similar to those of UCDs, so it is crucial to also discriminate between these two types in the colour-colour diagram.

\begin{figure}
    \centering
	\includegraphics[width=.8\columnwidth]{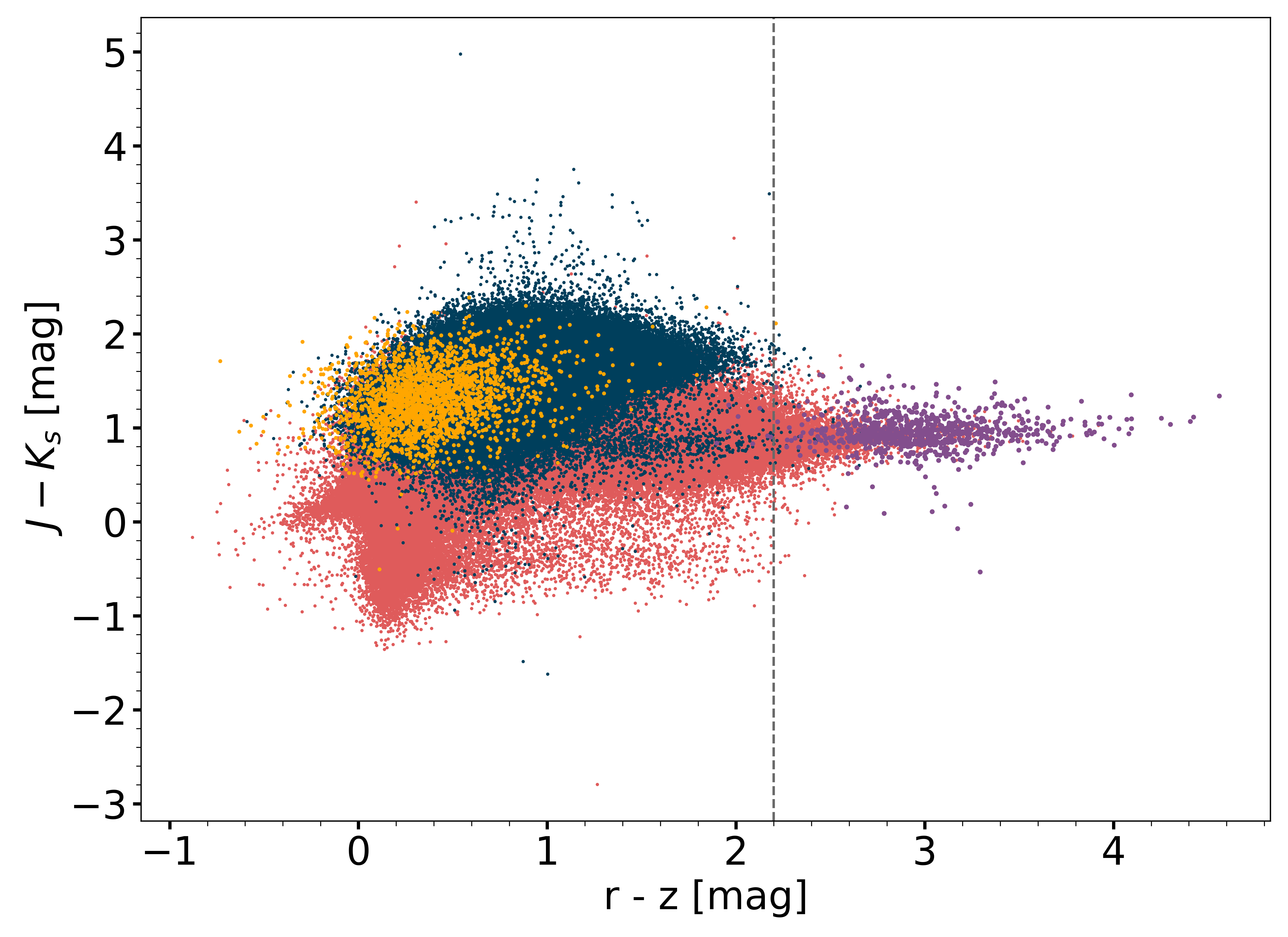}
    \caption{Colour-colour diagram built using true extended (dark blue) and true point (red) sources. Yellow dots represent the sample of 10\,481 QSOs. Purple dots represent the shortlisted candidate UCDs obtained by parallax-guided and proper motion-guided methods. The vertical grey line marks the $\textit{r}-\textit{z} > 2.2$\,mag limit for a source to be shortlisted as candidate UCD.}
    \label{fig:colour}
\end{figure}

Fig. \ref{fig:colour} shows the different types of objects in a colour-colour diagram. For the sample of QSOs, we cross-matched the SDSS-DR12 Quasar Catalog\footnote{\url{http://cdsarc.u-strasbg.fr/viz-bin/cat/VII/279}} with the J-PLUS DR2. To define the UCD locus, we overplotted in this diagram the candidate UCDs obtained by the methods described in Sects. \ref{parallax} and \ref{pm} for the region $\alpha$: 0 -- 20\,deg; $\delta$: 0 -- 20\,deg. As a compromise to balance the extended object contamination and the loss of candidate UCDs, we defined the UCD locus as the region fulfilling $\textit{r}-\textit{z} > 2.2$\,mag and applied this criterion to all the sources of each 20x20 deg$^2$ region. Of the sources fulfilling it, we filtered out those already pre-selected in the analysis described in Sects. \ref{parallax} and \ref{pm} and shortlisted the remaining ones as candidate UCDs.

\subsection{VOSA filtering} \label{vosa}

\begin{figure}
    \centering
	\includegraphics[width=.8\columnwidth]{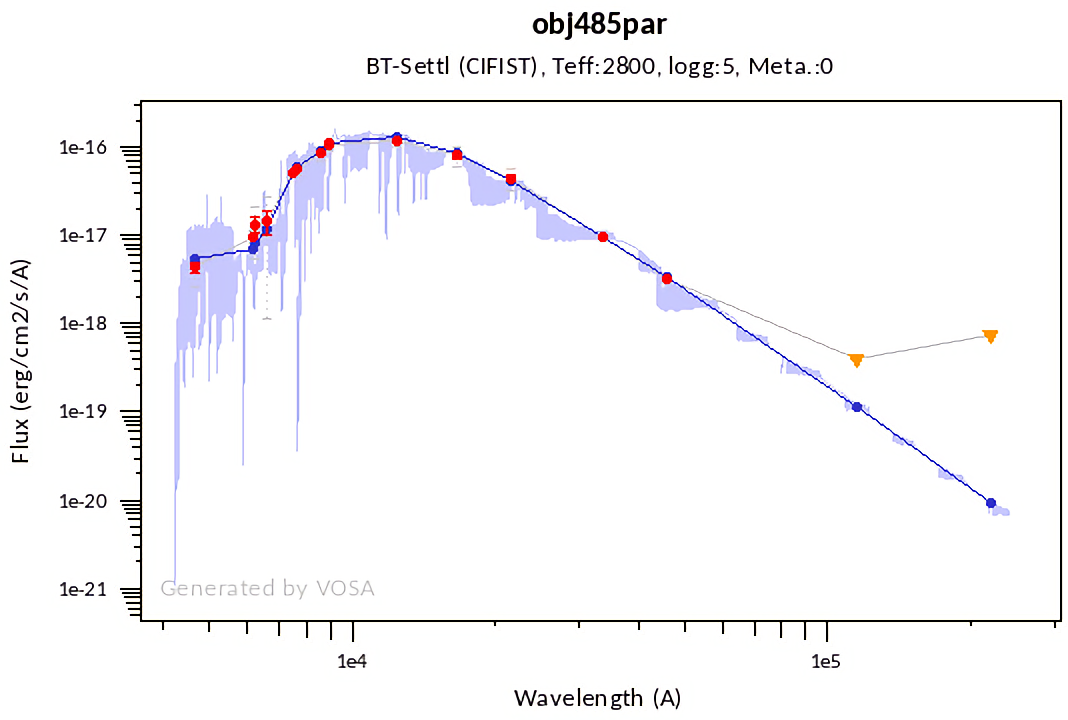}
    \caption{Example of an automatically generated SED fitting with \texttt{VOSA}. The blue spectrum represents the theoretical model that fits best, while red dots represent the observed photometry. The inverted yellow triangle indicates that the photometric value corresponds to an upper limit. These points are not considered in the fitting process.}
    \label{fig:vosa_sed}
\end{figure}

To estimate physical properties, such as effective temperature, luminosity or radius of the shortlisted objects described in the previous sections, we made use of the tool \texttt{VOSA}\footnote{\url{http://svo2.cab.inta-csic.es/theory/vosa/}} \citep{vosa}. This is a tool developed and maintained by the Spanish Virtual Observatory\footnote{\url{https://svo.cab.inta-csic.es}} which fits observational data to different collections of theoretical models. An example of \texttt{VOSA} Spectral Energy Distribution (SED) fitting can be found in Fig. \ref{fig:vosa_sed}. Before doing the fit, we built the observational SEDs using the J-PLUS photometric information as well as additional photometry from the 2MASS, UKIDSS, WISE, and VISTA infrared surveys, and from the SDSS data release 12 optical catalogue, available in \texttt{VOSA}.

In our analysis, we used the BT-Settl (CIFIST) collection of theoretical models \citep{allard2012, caffau2011}. Thus, the effective temperature estimated by \texttt{VOSA} is discretised due to the step adopted in the CIFITS grid of models (100\,K). We also assumed a surface gravity logg in the range 4.5 to 5.5 and solar metallicity. The limiting magnitude (5$\sigma$, 3\,arcsec diameter aperture) of J-PLUS DR2 is 20.5 [AB] in the \textit{z} band \citep{jpluscal}. If we take, for example, the object TVLM 891-15871, which is one of the objects in the UCD catalogue presented in \citet{Reyle2018} with the brightest absolute magnitude (11.36 [AB]) in the \textit{z} band, we see that it could be detected at a maximum distance of $\sim$680\,pc. This leads us to expect a maximum distance of about 650-700\,pc to find UCDs in the J-PLUS DR2.

Extinction plays a fundamental role in shaping the SED and, therefore, in the estimation of physical parameters \citep{extinction_laugalys, extinction_straizys}. Considering the maximum distance at which UCDs can be detected with J-PLUS, we adopted a range of values between $A_V=0$\,mag and $0.5$\,mag. We relied on the calibration described in Table 1 of \citet{solano2021} to adopt a temperature cutoff of 2\,900\,K for UCDs if the BT-Settl (CIFIST) models are used in \texttt{VOSA}. The goodness of fit of the SED in \texttt{VOSA} can be assessed with the vgfb parameter, a pseudo-reduced $\chi^2$ internally used by \texttt{VOSA} that is calculated by forcing $\sigma(F_{\rm obs}) > 0.1\times F_{\rm obs}$, where $\sigma(F_{\rm obs})$ is the error in the observed flux ($F_{\rm obs}$). Only sources with good SED fitting (vgfb < 12) were kept.

After applying these effective temperature and vgfb conditions, we used \texttt{TOPCAT} to remove the objects with a non-zero confusion flag (\texttt{cc\_flg}) in 2MASS, so as to ensure that objects are not contaminated or biased due to the proximity to a nearby source of equal or greater brightness. Moreover, we used the \texttt{Aladin} sky atlas \citep{aladin} to carry out a visual inspection of the coldest objects, in order to discard any problem related to blending or contamination by nearby objects. Finally, we ended up with 9\,810 final candidate UCDs. For the record, we checked that 204 of these objects have a renormalised unit weight error \citep[RUWE; ][]{lindegren2018} greater than 1.4 in \textit{Gaia} EDR3, which could mean that the source is affected by close binary companions. These objects were not removed since a binarity analysis is performed in Sect. \ref{binarity}.

\begin{figure}
    \centering
	\includegraphics[width=.8\columnwidth]{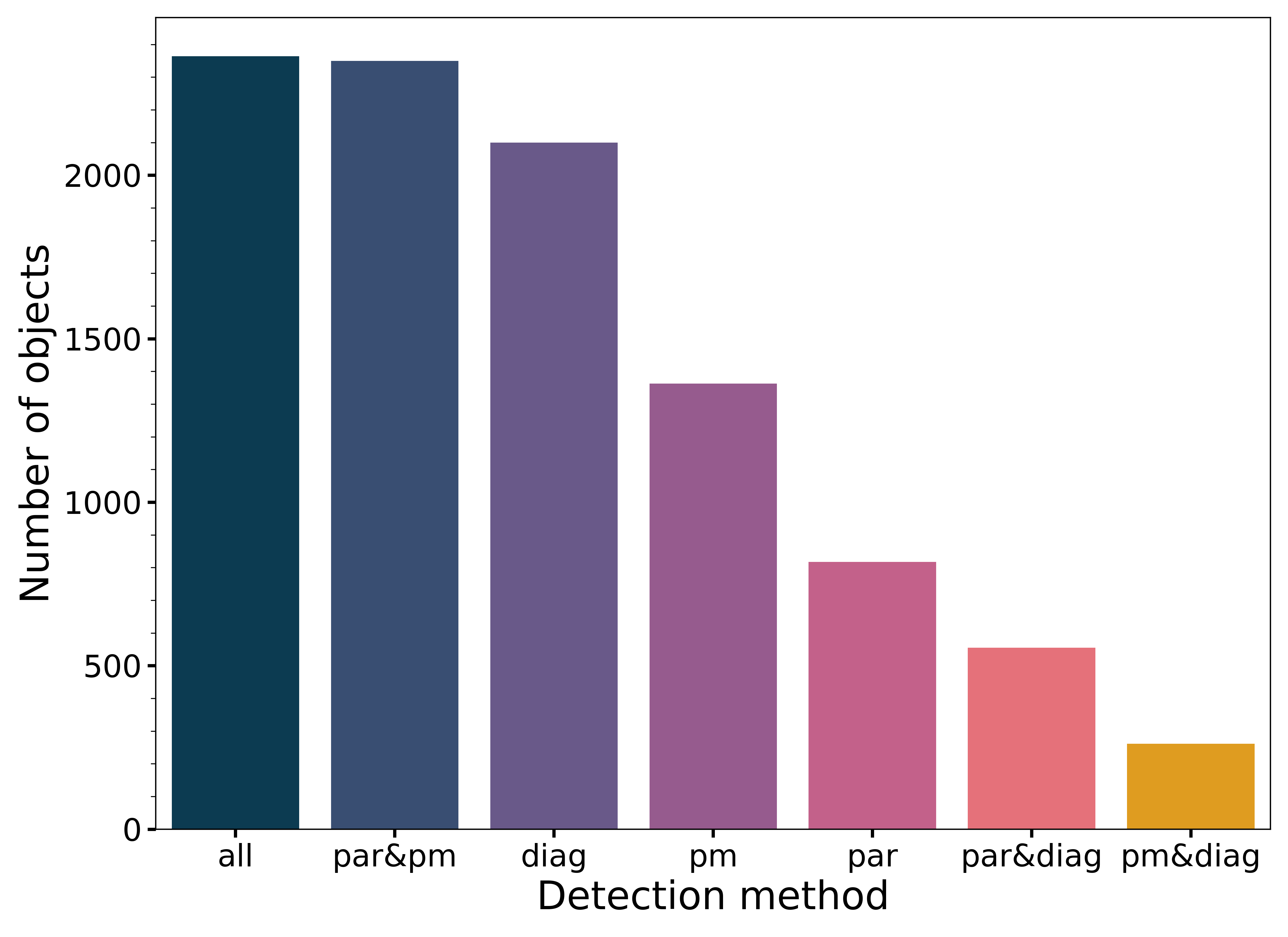}
    \caption{Breakdown of our candidate UCDs according to the methods by which they have been detected. `diag' label represents photometry-based selection, while `par' and `pm' labels represent selections based on parallax and proper motion, respectively. The `all' label comprises the candidate UCDs identified with the three approaches.}
    \label{fig:methods_both}
\end{figure}

As we use multiple detection methods in our methodology, distinct candidate UCDs may have been detected by different methods, or by several of them. Fig. \ref{fig:methods_both} shows the breakdown of the 9\,810 candidate UCDs according to the methods by which they have been detected. The fact that 2\,100 objects are only detected by the photometric methodology (`diag' bar in Fig. \ref{fig:methods_both}) and 4\,530 are only detected by the astrometric methodology (`par', `pm', and `par\&pm' bars in Fig. \ref{fig:methods_both}) argues for the complementary nature of both approaches. Considering each method separately, we detected 6\,086 candidates with parallax-based selection, 6\,338 with proper motion-based selection, and 5\,280 with photometry-based selection.

\section{Analysis} \label{Analysis}

\subsection{Temperatures and distances} \label{dist_temps}

\begin{figure}
    \centering
	\includegraphics[width=.8\columnwidth]{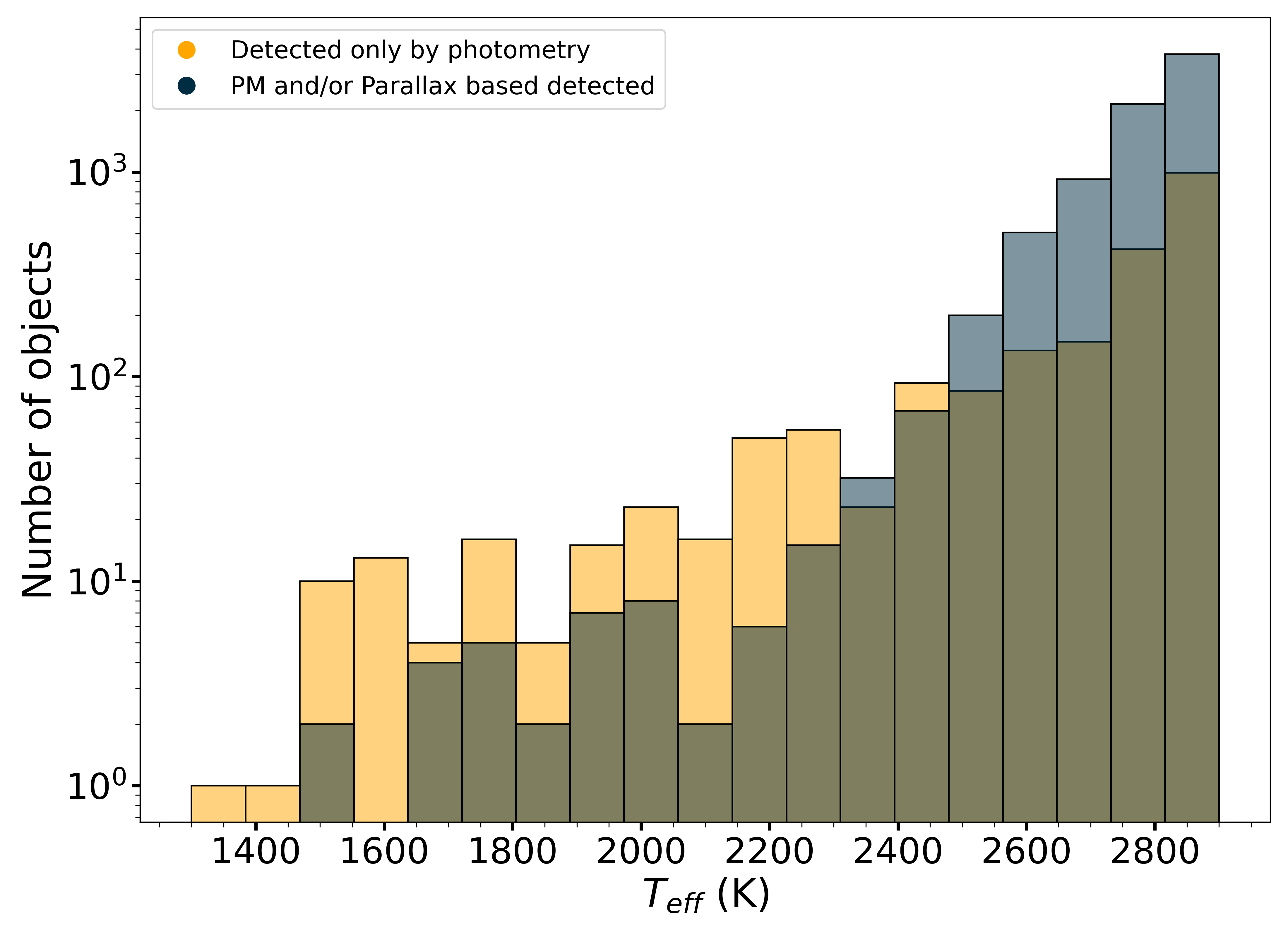}
    \caption{$T_{\rm eff}$ distribution for our candidate UCDs. In yellow we show the candidates that were only detected by photometry. In blue we show the candidates that were, at least, detected by astrometric methodology.}
    \label{fig:temphist}
\end{figure}

Fig. \ref{fig:temphist} shows that the distribution of effective temperatures for our candidate UCDs is not the same depending on whether they have been detected by astrometric methodology (blue) or not. To prove this, we performed a two-sample Kolgomorov-Smirnov test on the two samples, which returned a $p$ value = 3.66\,$\cdot\,10^{-15}$, rejecting the possibility that both samples are coming from the exact same distribution. The number of cold objects ($T_{\rm eff}\leq$ 2\,200\,K) is clearly higher in the only-photometry detected distribution (yellow). Most of our candidates ($\sim$86\,\%) have $T_{\rm eff}\geq$ 2\,700\,K, a clear consequence of the working wavelength, since UCDs peak in the near-infared, and J-PLUS covers only up to the z filter ($\lambda_{\rm eff}=8\,940.28$\,\AA).

\begin{figure}
    \centering
	\includegraphics[width=.8\columnwidth]{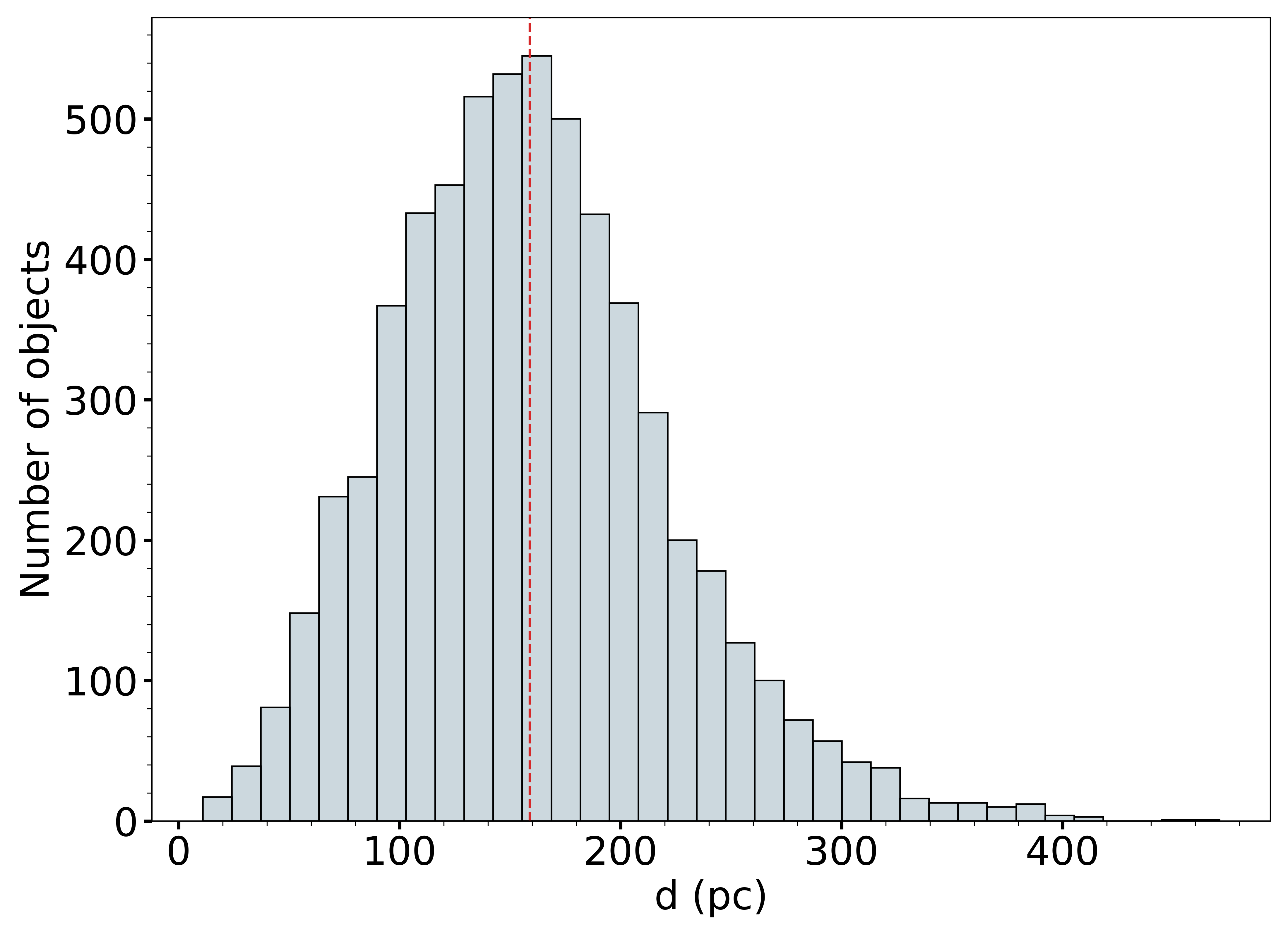}
    \caption{Distance distribution for our candidate UCDs with an error of less than 20\,\% in \textit{Gaia} EDR3 J2016 parallax. The mean value (red vertical line) of the distribution is 159\,pc, with the closest and farthest objects at 11\,pc and 471\,pc.}
    \label{fig:disthist}
\end{figure}

\begin{figure}
    \centering
	\includegraphics[width=.8\columnwidth]{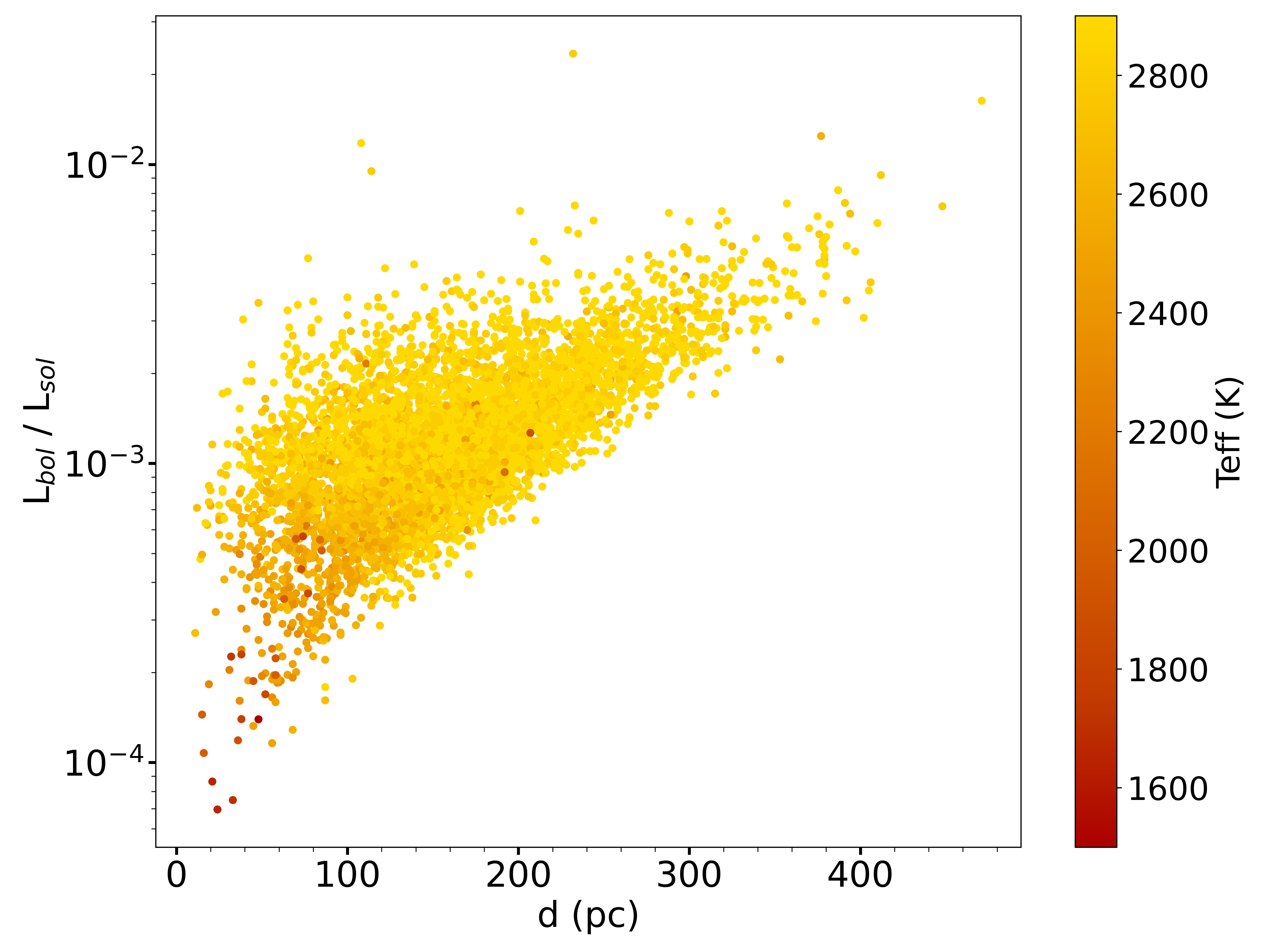}
    \caption{Bolometric luminosity (in solar units) vs. distance diagram of our candidate UCDs with good parallaxes. The points are colour-coded by temperature.}
    \label{fig:temp_lum}
\end{figure}

For the distance distribution of our candidate UCDs (Fig. \ref{fig:disthist}), we only considered the candidates with a relative error of less than 20\,\% in parallax (6\,086 objects), so we can rely on the inverse of the parallax as a distance estimator \citep{Luri2018}. In our case, as mentioned in Sect. \ref{Methodology}, the parallax are those of \textit{Gaia} EDR3. About 70\,\% of the objects lie in the $96 < {\rm D\,(pc)} < 222$ region (1$\sigma$ limits), with a maximum and minimun distance of 471\,pc and 11\,pc, respectively. This upper limit is consistent with the value estimated in Sect. \ref{vosa}. We found 68 nearby objects, at distances smaller than 40\,pc, that will be further discused in Sect. \ref{new_vs_known}. Fig. \ref{fig:temp_lum} gives a more in-depth view of the characteristics of our candidate UCDs. As expected, most of the cooler candidates are detected at closer distances and tend to have lower bolometric luminosity.

\subsection{Kinematics} \label{tangential_vel}

\begin{figure}
    \centering
	\includegraphics[width=.8\columnwidth]{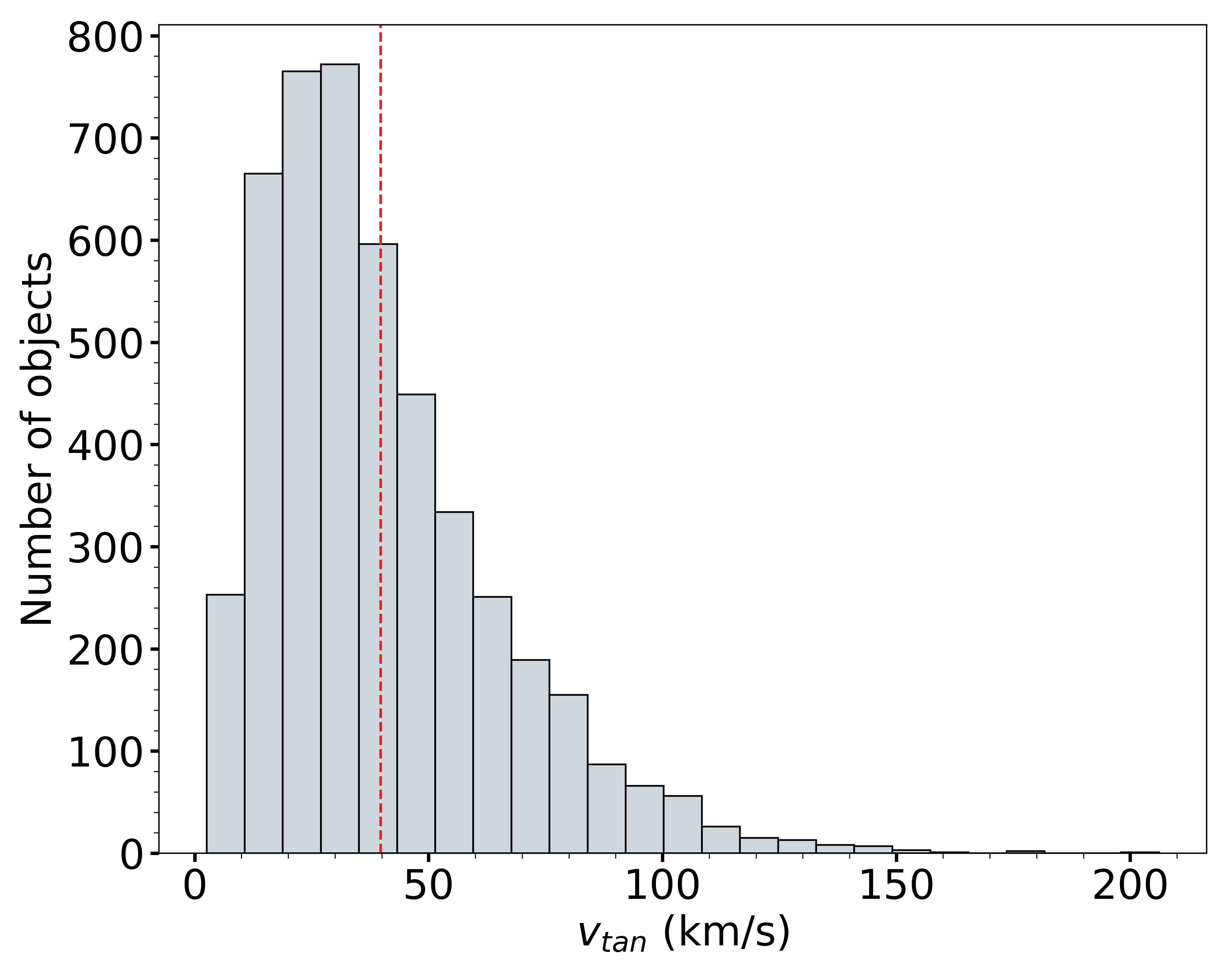}
    \caption{Tangential velocity distribution for our candidate UCDs with good parallax and pm conditions. The mean value (red vertical line) of the distribution is 39.78\,km\,s$^{-1}$, with a maximum value of 206\,km\,s$^{-1}$.}
    \label{fig:vtan_hist}
\end{figure}

Stellar kinematics is a reliable proxy for segregating large-scale galactic populations (thin disk, thick disk, and halo) \citep{Burgasser2015}. Using \textit{Gaia} EDR3 proper motions and parallaxes, we computed the tangential velocities of our candidate UCDs as $v_{\rm tan}=4.74 \mu d$, where $v_{\rm tan}$ is given in km\,s$^{-1}$, $\mu$ is the total proper motion in arcsec\,yr$^{-1}$ and $d$ is the distance in pc. For a correct estimation of the tangential velocity, we only considered candidates that met both conditions described in Sects. \ref{parallax} and \ref{pm} for good parallax and proper motion (4\,714). Fig. \ref{fig:vtan_hist} shows the distribution of tangential velocities for these candidates, with  a mean value of $v_{\rm tan}=39.78$\,km\,s$^{-1}$, a median value of $v_{\rm tan}=33.99$\,km\,s$^{-1}$, and a dispersion of $\sigma_{\rm tan}=24.85$\,km\,s$^{-1}$. Even taking into account objects located at the long tail of the distribution (134 objects, representing 2.8\,\% of the total, with $v_{\rm tan}>100$\,km\,s$^{-1}$), these values agree with previous calculations for UCDs \citep{faherty2009}.

\citet[][Fig. 10]{Torres2019} shows a breakdown of the tangential velocity based on the membership in the thin disk, the thick disk or the halo. Relying on these values, we can segregate our candidate UCDs into thin disk ($v_{\rm tan}\leq 85$\,km\,s$^{-1}$), thick disk ($85< v_{\rm tan}< 155$\,km\,s$^{-1}$), and halo ($v_{\rm tan}\geq 155$\,km\,s$^{-1}$) populations. We found 4\,441, 268 and five candidate UCDs in these intervals, respectively. According to \citet{Kilic2017}, the corresponding ages are 6.8-7.0\,Gyr (thin disk), 7.4-8.2\,Gyr (thick disk), and 12.5$^{+1.4}_{-3.4}$\,Gyr (halo).

\begin{figure}
    \centering
	\includegraphics[width=.8\columnwidth]{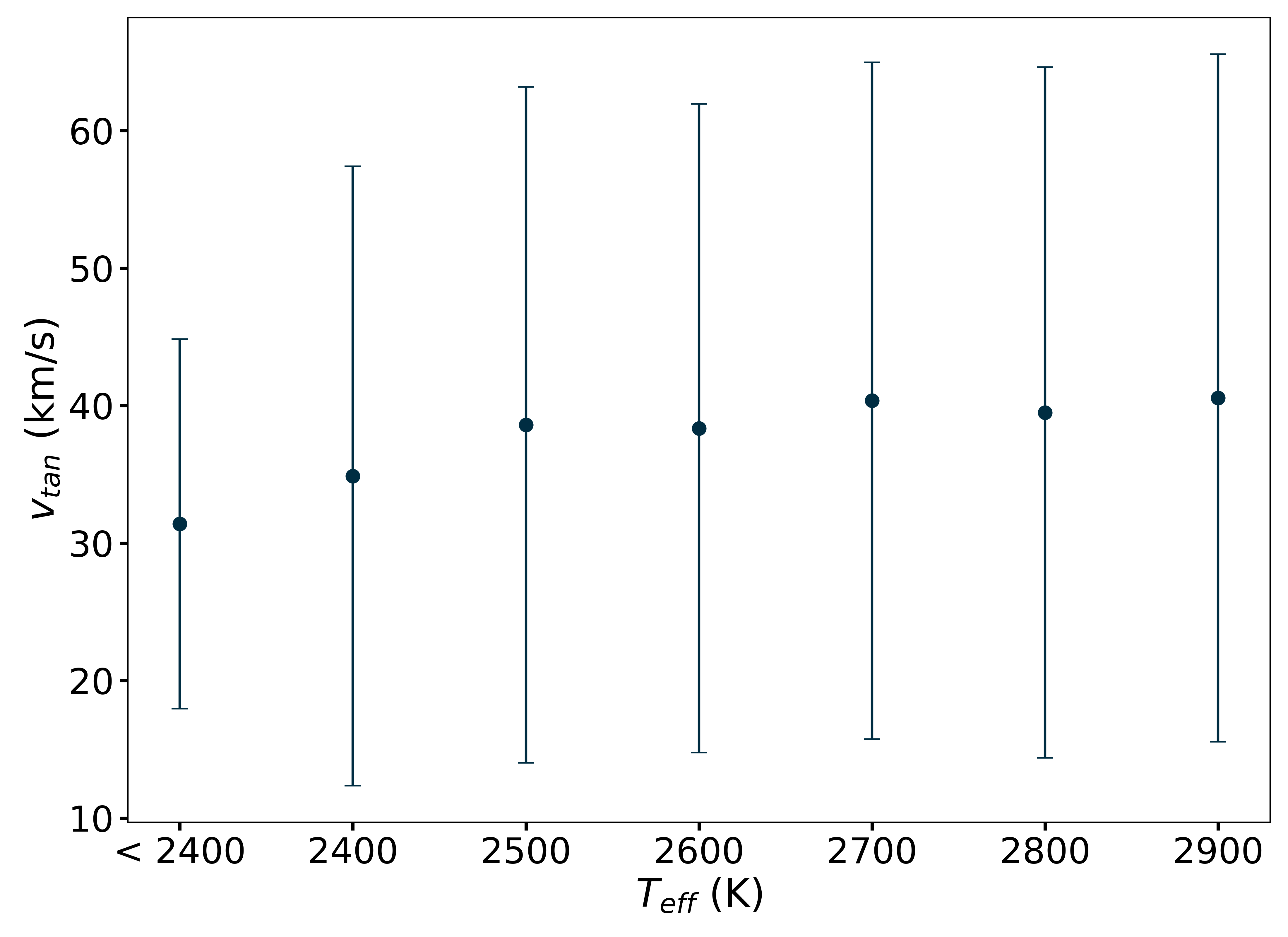}
    \caption{Mean tangential velocity for each value of effective temperature of our candidate UCDs with reliable parallax and proper motion. The error bars represent the standard deviation.}
    \label{fig:vtan_temp}
\end{figure}

Three of the potential halo members show a very high tangential velocity. Two of them, with Simbad identifiers 2MASS J18030236+7557587 and 2MASS J13155851+2814524, are not far from the thick disk-halo threshold, with tangential velocities of $v_{\rm tan}=176.25$\,km\,s$^{-1}$ and $v_{\rm tan}=177.47$\,km\,s$^{-1}$, respectively. Furthermore, one of the objects has $v_{\rm tan}=206.16$\,km\,s$^{-1}$, which significantly exceeds the limit. This object, at a distance of 179 pc, is reported as an M7 in the catalogue provided by \citet{ahmed2019} with the id J132625.03+333506.7. Due to its high tangential velocity, we conclude this object could be a potential member of the Galactic halo. We used the ($\textit{J}-\textit{K}_{\rm s}$, $\textit{i}-\textit{J}$) colour-colour diagram presented in \citet{lodieu2017} to study the metallicity of this object. With values of $\textit{J}-\textit{K}_{\rm s}=0.77$ and $\textit{i}-\textit{J}=3.29$, the object exhibits subdwarf behaviour (low metallicity). Fig. \ref{fig:vtan_temp} shows the mean and standard deviation of the tangential velocity for each value of the effective temperature. There is no evidence of correlation between effective temperature and tangential velocity among our candidates. 

To study the possible membership of our candidate UCDs to nearby young associatons, we relied on \texttt{BANYAN}~$\Sigma$\footnote{\url{http://www.exoplanetes.umontreal.ca/banyan/}} \citep{banyan}, a Bayesian analysis tool to identify members of young associations. Modelled with multivariate Gaussians in six-dimensional $\rm XYZUVW$ space, \texttt{BANYAN}~$\Sigma$ can derive membership probabilities for all known and well-characterised young associations within 150\,pc. As we found no radial velocity data available for any of the 4\,714 candidate UCDs with good parallax and proper motion, we introduced the sky coordinates, proper motion, and parallax of these objects as input parameters to the algorithm. 

For 4\,666 of the candidate UCDs, the algorithm predicted that most of them are field stars. However, it gave a high Bayesian probability for 48 objects to belong to a young association, in 30 of the cases with a probability greater than 95\,\%. In more detail, the algorithm mapped 34 candidate UCDs to the Pisces-Eridanus stellar stream \citep{PERI}, five to the Argus Association \citep{argus}, four to the AB Doradus Moving Group \citep{abdmg}, two to the Columba association \citep{columba}, and one each to the Tucana-Horologium \citep{tha}, $\beta$ Pictoris \citep{bpictoris}, and Carina-Near \citep{carn} associations. We verified all these 48 objects have tangential velocities typical of the thin disk, with mean $v_{\rm tan}=16.37$\,km\,s$^{-1}$ and standard deviation $\sigma = 6.17$\,km\,s$^{-1}$. As mentioned in \texttt{BANYAN}~$\Sigma$, a high membership probability in a young association does not guarantee that the star is a true member, or young, so further follow-up would be needed to demonstrate the youth of the object. Moreover, we note that the absence of radial velocity may cause the membership probabilities given by \texttt{BANYAN}~$\Sigma$ to be inflated.

\subsection{Binarity} \label{binarity}

We conducted a search for binary systems among our candidate UCDs in two ways. We searched for unresolved binaries using a methodology based purely on the photometry of our objects. Using the complementary photometry functionality of \texttt{VOSA}, we selected only the candidates fulfilling three conditions. First, with an excess detected by \texttt{VOSA} in any filter in the infrared. We discarded WISE $W3$ and $W4$ due to their poor angular resolution and sensitivity. Second, with good photometry in both 2MASS (\texttt{Qfl} = A) and WISE (\texttt{cc\_flags} = 0 and \texttt{ph\_qual} = A or B). Third, with at least three good photometric points in the infrared, apart from the detected excess.

After applying these conditions, we ended up with 291 objects with an excess in the infrared that could be ascribed to circumstellar material or to the presence of a close ultracool companion. Then, we used the binary fit functionality of \texttt{VOSA} to fit the observed SED of these 291 objects using the linear combination of two theoretical models. After this, we ended up with 122 candidate UCDs for which the infrared excess detected is nicely reproduced by performing a two-body fit, suggesting the existence of an unresolved companion.

In parallel to this, we looked for \textit{Gaia} companions of our candidate UCDs at large angular separations, using only those with reliable parallax and proper motion (4\,714). Firstly, we cross-matched these sources with \textit{Gaia} EDR3 J2016 to get all the objects separated a maximum of 180 arcsec in the sky (maximum separation allowed by the \texttt{X-match} service in \texttt{TOPCAT}) from each of our candidate UCDs. Then, we established a conservative upper limit of 100\,000\,au for the projected physical separation between a candidate and its companion. Finally, we relied on the conditions presented in \citet{smart_dwarfs} to ensure that the companion shares a parallax and proper motion similar to that of our candidate UCD:

\begin{itemize}

    \item $\Delta \varpi < max[1.0, 3\sigma_{\varpi}]$  
    \item $\Delta(\mu_{\alpha}\cos{\delta}) < 0.1\mu_{\alpha}\cos{\delta}$ 
    \item $\Delta \mu_{\delta} < 0.1\mu_{\delta}$ 

\end{itemize}

\noindent where $\varpi$ and $\mu$ are the parallax and proper motion of our candidate UCDs, respectively. After applying these criteria, we ended up with 73 candidate UCDs with one \textit{Gaia} companion and another five candidate UCDs with two \textit{Gaia} companions identified. Of these 78 objects, six are already tabulated as known binary systems by the Washington Double Star catalogue \citep[WDS;][]{WDS}. Table \ref{tab:binaries} lists the coordinates (J2000), parallaxes, proper motions, angular separations $\rho$ and projected physical separations $s$ of the six known systems. A table with the same information for the identified multiple systems that are not tabulated by the WDS is accesible through the catalogue described in Section \ref{data_av}.

A deeper knowledge of the \textit{Gaia} companion may allow us to infer properties, such as metallicity, of our candidate UCD. We only found spectral types in Simbad for two of the detected companions, with spectral types F2 and K3V. To obtain information about the rest of the companions, we first made use of \texttt{VOSA} to get an estimate of their effective temperature. Then, we relied on the updated version of Table 5 in \citet{pecaut2013} to map these effectives temperatures to the spectral types of the companions. As result, we ended up with four F-type, one G-type, 16 K-type and 42 M-type stars among the companions with good SED fitting in \texttt{VOSA}. For the rest of the companions, we obtained a bad SED fitting in \texttt{VOSA} (vgfb > 12), so we could not get an estimation of the effective temperature.


\section{Known ultracool dwarfs} \label{known}

\subsection{Recovered known UCDs} \label{recovered_known}
 
Here, we assess the number of known UCDs found in the J-PLUS DR2 field and the fraction of them that were recovered using our methodology. For this analysis, we used nine catalogues and services: SIMBAD\footnote{\url{http://simbad.u-strasbg.fr/simbad/}} \citep{Wenger00}, \citet{zhang2009}, \citet{zhang2010}, \citet{schmidt2010}. \citet{skrzypek2016}, \citet{smart2017}, \citet{Reyle2018}, \citet{panstarrs1}, and \citet{ahmed2019}. Using the SIMBAD TAP service\footnote{\url{http://simbad.u-strasbg.fr:80/simbad/sim-tap}} through \texttt{TOPCAT}, we selected objects with spectral types M7\,V, M8\,V, M9\,V or labelled as brown dwarfs. A total of 18\,282 objects were recovered. Also, from \citet{panstarrs1} we chose the 2\,090 objects having spectral type M7 or later. As all the 33\,665, 14\,915, 1\,886, 1\,361, 806, 484 and 129 objects in the \citet{ahmed2019}, \citet{Reyle2018}, \citet{smart2017}, \citet{skrzypek2016}, \citet{zhang2010}, \citet{schmidt2010}, and \citet{zhang2009} catalogues, respectively, are within our scope (spectral type M7 or later), we included them in their entirety. 

To select only the known UCDs that lie in the region of the sky covered by J-PLUS DR2 we made use of \texttt{TOPCAT} and its \texttt{nearMOC} functionality, which indicates whether a given sky position either falls within, or is within a certain distance of the edge of, a given MOC. The MOC\footnote{\url{https://www.ivoa.net/documents/MOC/}} (Multi-Order Coverage Map) is an encoding method dedicated to VO applications or data servers which allows to manage and manipulate any region of the sky, defining it by a subset of regular sky tessellation using the HEALPix method \citep{healpix}. Out of a total of 5\,817 objects lying in the J-PLUS DR2 field of view, we ended up with 4\,734 known UCDs with photometry in the relevant J-PLUS filters described in Sect. \ref{Methodology} (see Table \ref{tab:filters}), which are reduced to 4\,649 objects after removing those with non-zero confusion and contamination flags in 2MASS. From this set, 1\,983 were recovered using our methodology and 2\,666 were not. We conducted an in-depth analysis of the 2\,666 UCDs following the two methodologies (astrometric and photometric) separately, to see in which steps of the process these objects are discarded.

In short, of this 2\,666 unrecovered objects, 1\,520 are lost because they do not meet our parallax, proper motion, and photometry constraints, while another 119 are discarded in the $\textit{G} - \textit{G}_{\rm RP}$ and $\textit{r} - \textit{z}$ cuts. The remaining 1\,027 are lost in the temperature/vgfb cutoff after the analysis with \texttt{VOSA}, some due to a bad SED fitting (vgfb > 12) and most of them due to an estimated temperature higher than 2\,900\,K. We have checked the latter and the vast mayority of them are M7\,V from Simbad that lie at the temperature limit, with estimated temperatures of 3\,000 - 3\,100\,K.

\subsection{New candidate UCDs vs. previously known} \label{new_vs_known}

In this section, we analyse the differences between previously known UCDs and the remaining candidate UCDs among our sample. For this, we cross-matched our candidate UCDs with the known UCDs sample described in Sect. \ref{known}. As indicated above, of the 9\,810 candidates identified by the proposed VO methodology, only 1\,983 were previously reported as UCD. This amounts to a total of 7\,827 new candidate UCDs in the sky coverage of J-PLUS DR2, which represents an increase of about 135\,\% (7\,827/5\,817) in the number of UCDs for this area.

\begin{figure}
    \centering
	\includegraphics[width=.8\columnwidth]{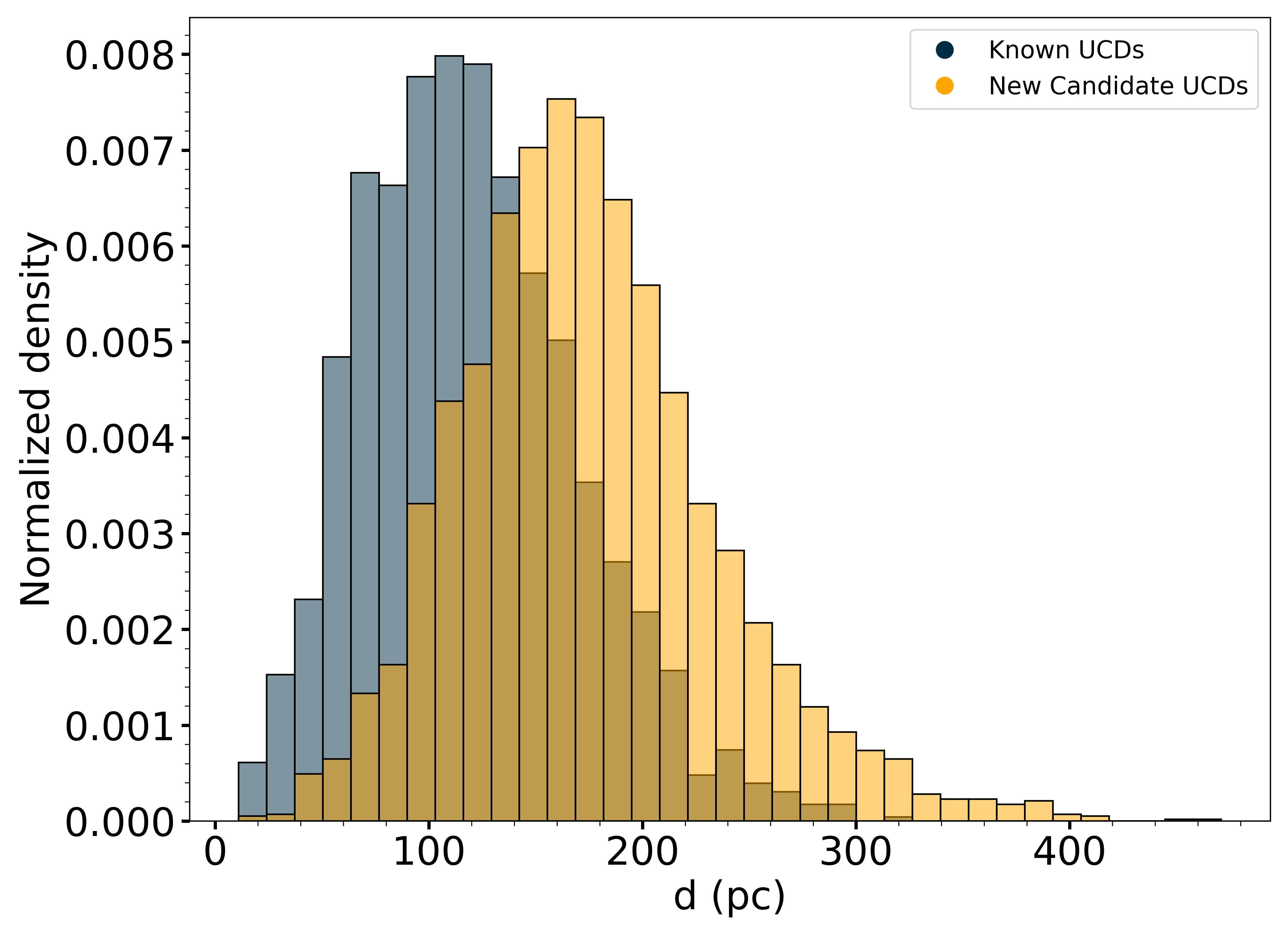}
    \caption{Distance distribution for previously reported (blue) and new (yellow) candidate UCDs with good parallax conditions.}
    \label{fig:disthistboth}
\end{figure}

\begin{figure}
    \centering
	\includegraphics[width=.8\columnwidth]{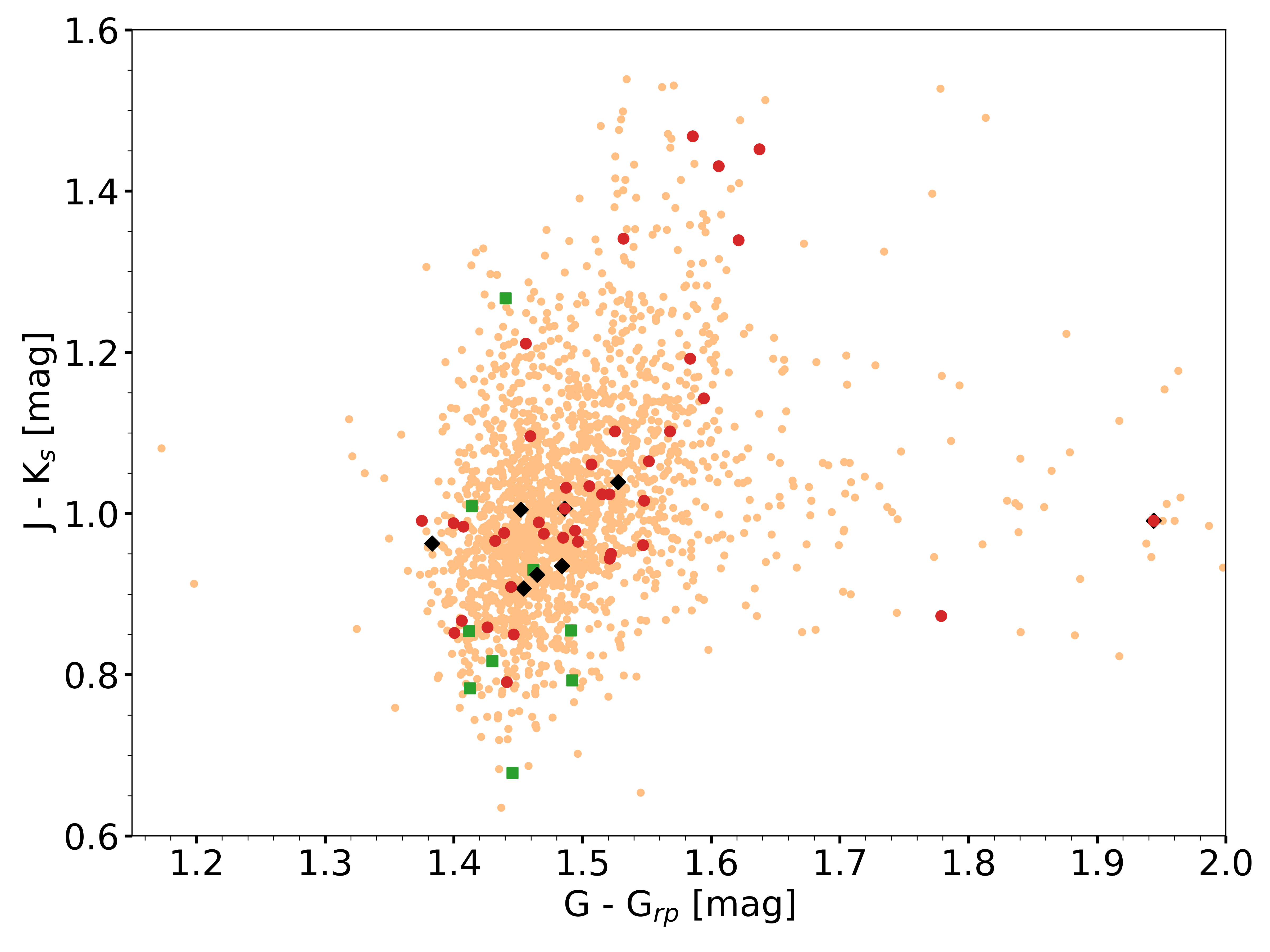}
    \caption{$\textit{J}-\textit{K}_s$ (2MASS) vs. $\textit{G}-\textit{G}_{\rm RP}$ (\textit{Gaia}) diagram of our candidate UCDs with good 2MASS photometric quality (Qflg=A) in \textit{J} and \textit{$K_s$} bands. Black diamonds represent our eight new nearby candidate UCDs at distances $d<40$\,pc. Green squares stand for new candidate UCDs with tangential velocities $v_{\rm tan}>100$
    \,kms\,$^{-1}$. Red circles represent candidate UCDs with a possible membership in a nearby young association.}
    \label{fig:2massdiag}
\end{figure}

Fig. \ref{fig:disthistboth} shows the distance distribution for our candidate UCDs, with good parallax conditions, discriminated by colour according to whether or not they were previously reported as UCD. It is clear that the new candidates detected are, on average, more distant, driven by the improvement of the quality of parallaxes with \textit{Gaia} EDR3. Of the 68 nearby objects found at distances smaller than 40\,pc, eight have not been previously reported as UCD. To check whether these objects could have been missed by other photometric surveys due to anomalies in their colours, we constructed a colour-colour diagram using \textit{$\textit{J}-\textit{K}_{\rm s}$} (2MASS) and $\textit{G}-\textit{G}_{\rm RP}$ (\textit{Gaia}) colours. Fig. \ref{fig:2massdiag} shows that this is not the case for any of these objects (black diamonds in the diagram).

\begin{figure}
    \centering
	\includegraphics[width=.8\columnwidth]{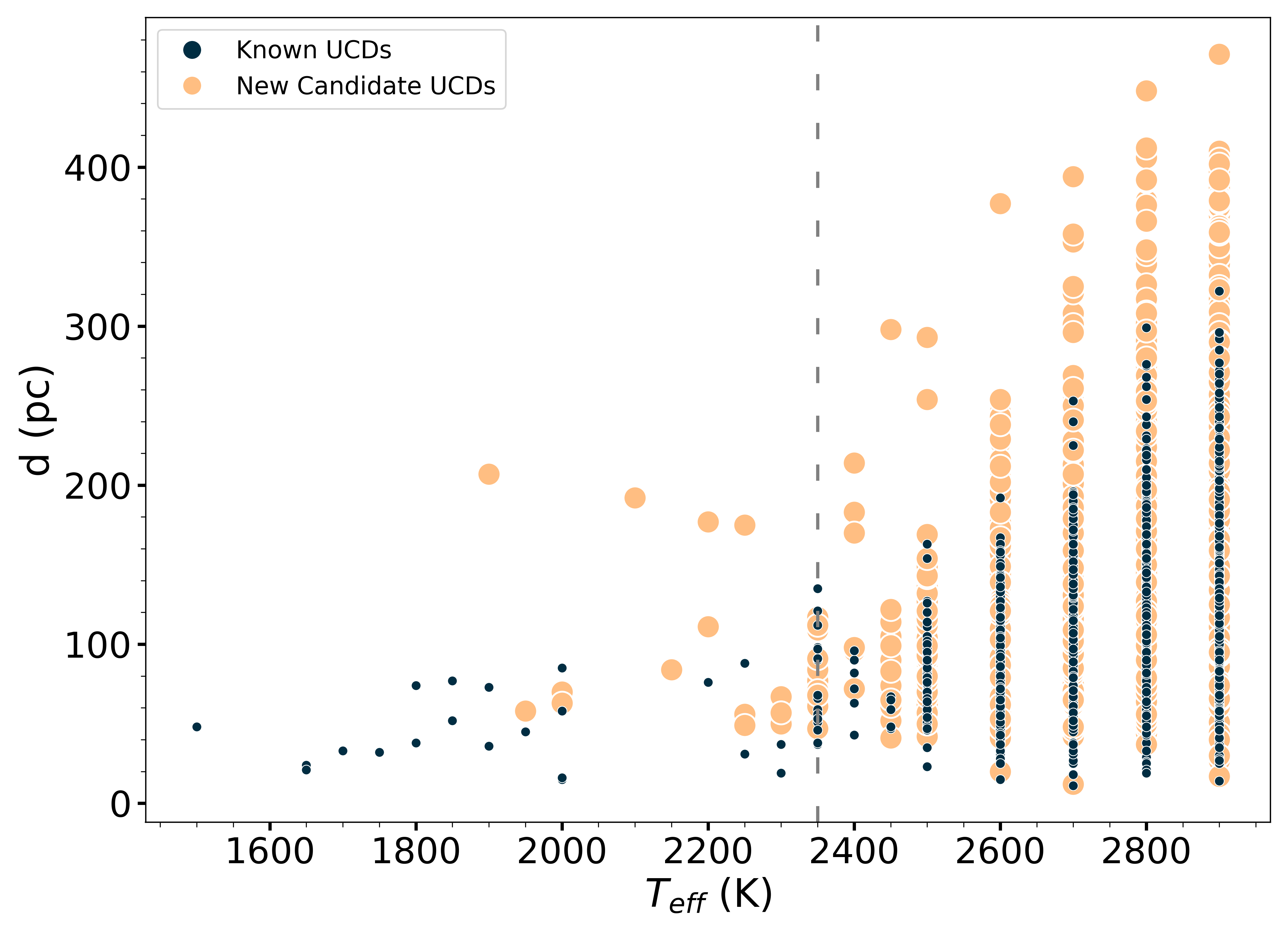}
    \caption{Distance vs. effective temperature diagram for previously reported (blue) and new (yellow) candidate UCDs with good parallax conditions. The vertical dashed line indicates the lower limit of effective temperature for M-type dwarfs (2\,359\,K) according to \citet{pecaut2013}.}
    \label{fig:distteff}
\end{figure}

A more in-depth view of this is the distance vs. effective temperature diagram shown in Fig. \ref{fig:distteff}. Here we can see how previously reported candidate UCDs tend to be at shorter distances for any value of the effective temperature. This trend is more clearly observed for higher temperature values, where the diagram shows how the new candidate UCDs cover the range of distances of the previously reported candidates and extend it to larger values, suggesting that our methodology allows us to go further in the search for new UCDs.

\begin{figure}
    \centering
	\includegraphics[width=.8\columnwidth]{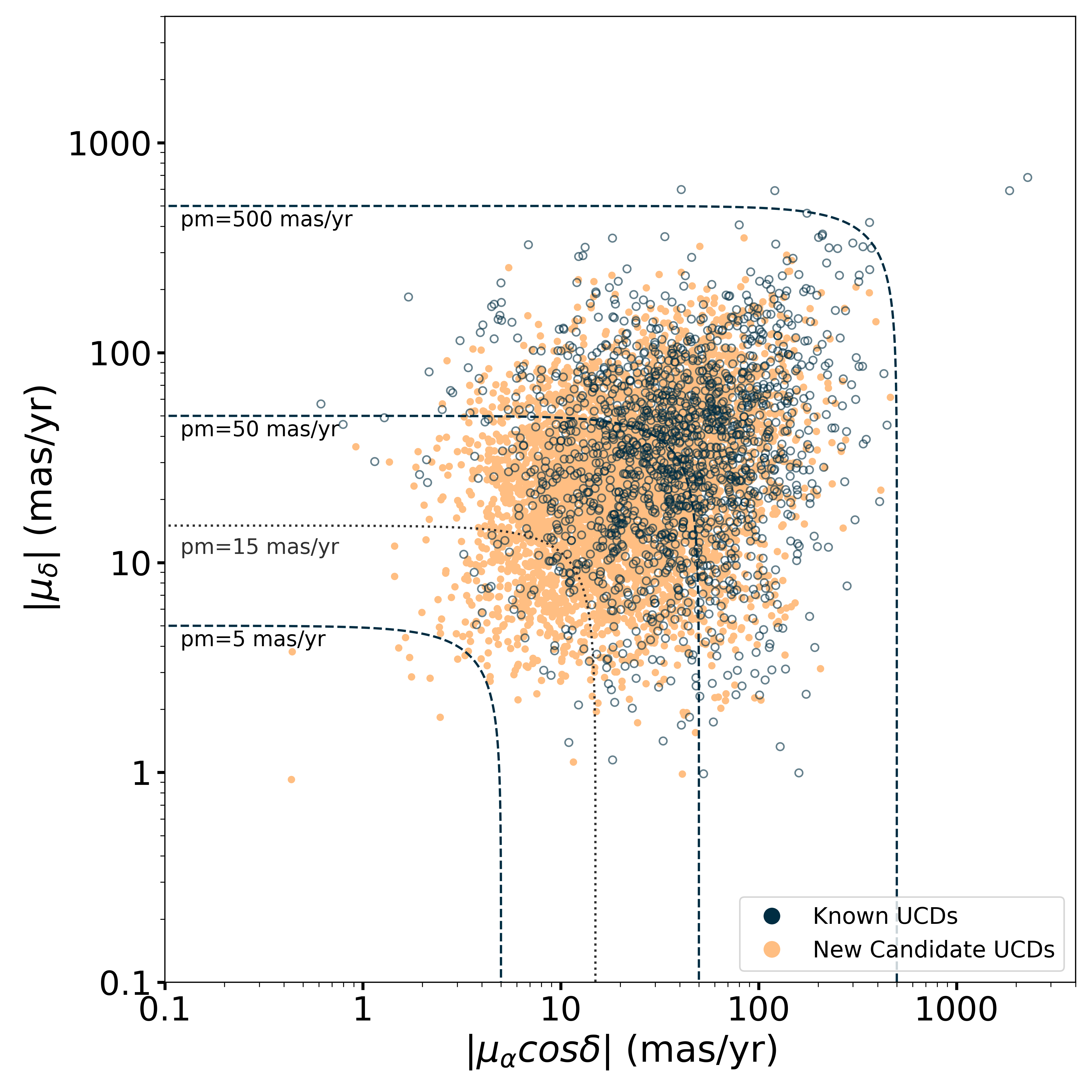}
    \caption{Absolute proper motion components for previously reported (blue) and new (yellow) candidate UCDs with good proper motion conditions.}
    \label{fig:pmanalysis}
\end{figure}

Going further, in Fig. \ref{fig:pmanalysis} we plot the absolute proper motions |$\mu_{\delta}$| and |$\mu_{\alpha}\cos{\delta}$| for our candidate UCDs with good proper motion conditions. It shows how the new candidate UCDs detected extend to smaller values of proper motion. Especially for values of proper motion of less than $15$\,mas\,yr$^{-1}$, the number of new candidates is significantly higher than the number of previously reported candidate UCDs, which reflects the improvement of the quality of proper motions with \textit{Gaia} EDR3.

\section{Machine learning analysis} \label{ml}

The filter system of J-PLUS offers a sufficiently high-dimensional space to reliably use ML techniques. We explored the ability to reproduce the presented search for candidate UCDs with a purely ML-based methodology that uses only J-PLUS photometry. Because the sample is strongly imbalanced, as a first step in the candidate UCDs identification, we proposed a filtering strategy to discard the objects that differ the most from the UCDs using the PCA algorithm. Then, with the reduced sample, SVM models were trained and fine-tuned to maximise the identification of candidate UCDs.

Principal component analysis~\citep{hotelling:33}, one of the most popular linear dimensionality reduction algorithms, is a non-parametric method that aims to reduce a complex data set to a lower dimension by identifying the axes that account for the largest amount of variance. The unit vectors defining each of these axes are called principal components. PCA works on the assumption that principal components with larger associated variance encompass the underlying structure of the data set in order to find the best basis for re-expressing it. The expectation behind this method, as with any method of dimensionality reduction, is that the entire data set can be well characterised along a small number of dimensions (principal components). By projecting the data set onto the hyperplane defined by these principal components, you ensure that the projection will preserve as much variance as possible. 

The selection of PCA in our approach instead of other non-linear dimensionality reduction techniques, such as t-Distributed Stochastic Neighbor Embedding~\citep[t-SNE;][]{vandermaaten08} or Uniform Manifold Approximation and Projection~\citep[UMAP;][]{McInnes2018}, is mainly based on (1) the computational efficiency, since PCA allows projecting new data along the new axes without having to reapply the algorithm, and (2) the deterministic nature of the PCA solution, i.e., different runs of PCA on a given dataset will always produce the same results. These properties of PCA are crucial in our proposal, since we use the 2D representation of PCA to perform the filtering as a first step in our ML task.

Support vector machine is a supervised (requires labelled training data) ML algorithm that has been widely used in classification and regression problems~\citep{2013A&A...550A.120S,2017MNRAS.465.4556G}. The origin of this algorithm dates back to the late 70s, when \citet{vapnik} delved into the statistical learning theory. The idea behind SVM is to find a hyperplane that separates data into two classes while maximising a marging, defined as the distance from the hyperplane to the closest point across both classes. Thus, the SVM chooses the best separating hyperplane as the one that maximises the distance to these points, so the decision surface is fully specified by a subset of points on the inner edge of each class, known as support vectors. The SVM is a linear classifier, so if the data is not linearly separable in the instance space, we can gain linear separation by mapping the data to a higher dimensional space. To do so, different kernels are used, such as the polynomial or the radial basis function (RBF), since the kernel trick allows us to define a high-dimensional feature space without actually storing these features.

\subsection{PCA cut} \label{pca_cut}

In our methodology, we used J-PLUS DR2 data from one of the 20$\times$20\,deg$^2$ mentioned in Sect. \ref{Methodology}. We selected as features seven different J-PLUS colours built with the most relevant filters for UCDs, i.e., the reddest ones (see Table \ref{tab:filters}): $\textit{i}-\textit{z}$, $\textit{r}-\textit{i}$, $\textit{i}-\textit{J0861}$, $\textit{J0861}-\textit{z}$, $(\textit{i}-\textit{z})^2$, $(\textit{r}-\textit{i})^2$, and $\textit{r}-\textit{z}$. We discarded the filter \textit{J0660} because the available photometry in this filter is less abundant than in the others. Thus, we first built these variables from the J-PLUS photometry, discarding objects with no information in any of the required filters, and labelled the instances as positive or negative class using the candidate UCDs obtained with the previous methodology. After this, we ended up with a sample composed of 317 UCDs and 495\,274 non-UCD objects.

\begin{figure}
    \centering
	\includegraphics[width=.8\columnwidth]{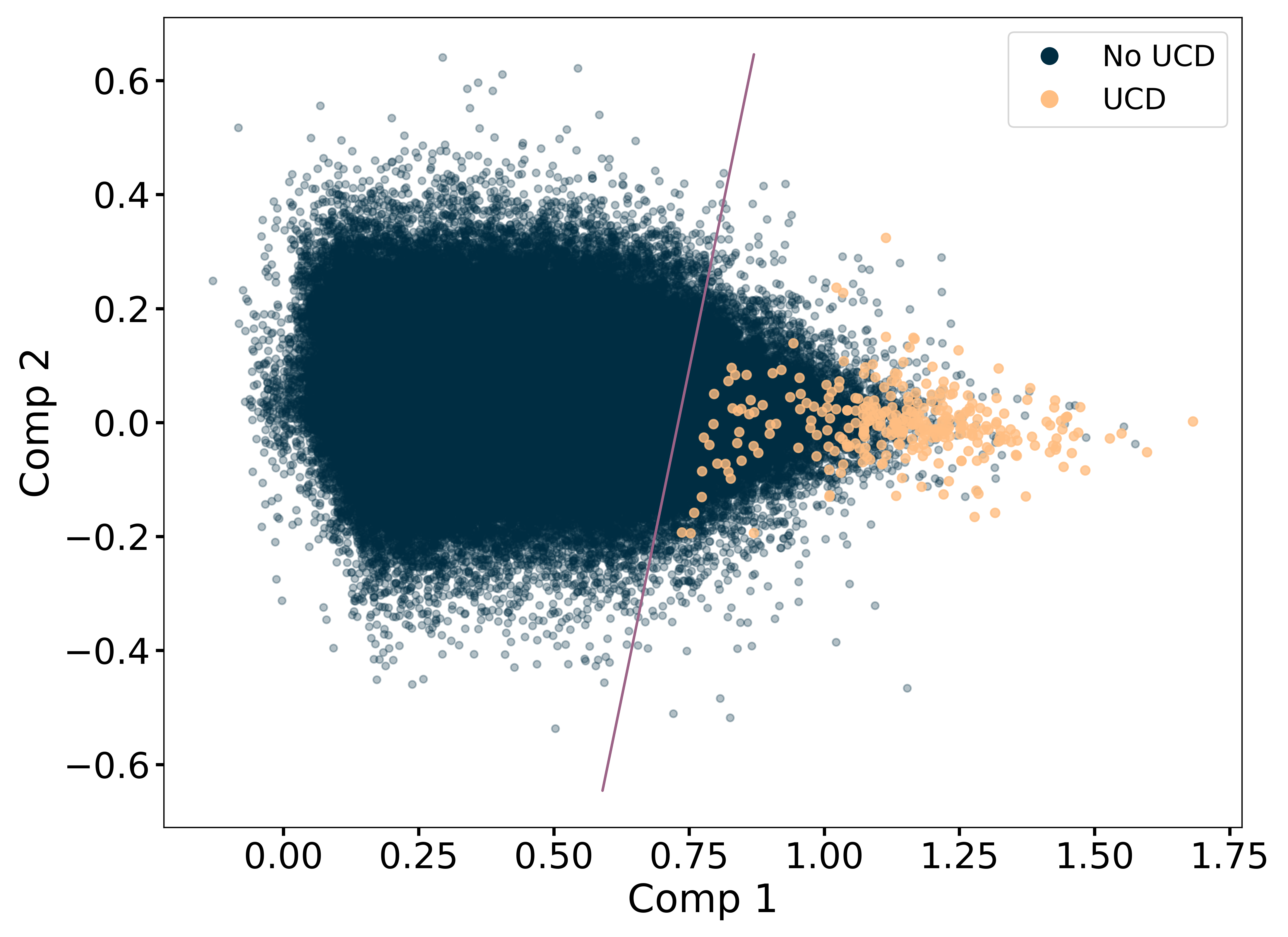}
    \caption{Projection of the sample used in the ML methodology onto the hyperplane defined by the first two principal components, with an explained variance ratio of 93\,\%. Points are colour-coded according to their class, UCD (yellow) or non-UCD (dark blue). The purple line represents the decision threshold used to make a first cut at identifying UCDs, keeping only the objects that fall on the UCD side.}
    \label{fig:pca}
\end{figure}

To perform the PCA, we first divided the sample into training (70\,\%) and test (30\,\%) sets using stratified sampling to ensure that these sets are representative of the overall population (have the same percentage of samples from each target class as the complete set). Thus, we trained the PCA model using the training set, obtaining that 93\,\% of the sample's variance lied along the two first principal components. Projecting the training data onto the hyperplane defined by these two principal components, the vast majority of non-UCD objects are clearly separated from the UCDs. Thus, it is possible to make a first cut in the identification of UCDs with this 2D projection, by defining a decision threshold (purple line in the Figure) and keeping only the objects that fall on the UCD side. Fig. \ref{fig:pca} shows the same projection for the entire sample (training + test). After this cut, we reduced our sample to 317 and 29\,732 UCD and non-UCD objects, respectively, achieving a 94\,\% reduction on the negative class. Despite still being strongly imbalanced, this reduced sample has a better balance between the negative and positive class, which facilitates better results when using the SVM.

\subsection{SVM model} \label{svm}

To develop the SVM model, we used the reduced sample obtained in the PCA filtering, keeping the same training and test set structure. We used the test set for the validation of the classification model. The seven J-PLUS colours described in Sect. \ref{pca_cut} were used as features in the training step.

Then, we conducted a search for the SVM's optimal hyperparameters on the training test. To do this, we created a grid for the SVM kernel and hyperparameters and did an exhaustive search over this parameter space using the \texttt{GridSearchCV} class from the \texttt{scikit-learn} package, which optimises the hyperparameters of an estimator by k-fold cross-validation using any score to evaluate the performance of the model. In our case, we used the recall score, which measures the ability of the classifier to find all the positive instances, since our priority is to identify as many candidate UCDs as possible. For the \texttt{GridSearchCV} class, we used ten k-folds and set the hyperparameter \texttt{class\_weight} to `balanced' to address the imbalance by adjusting the weights inversely proportional to the class frequencies. In the grid of hyperparameters, we tested the regularisation parameter $C$ for values of 1, 10, 100 and 1000, and the kernel scale $\gamma$ of the RBF kernel for 0.001, 0.01, 0.1, 1, 10 and 100.

\begin{figure}
    \centering
	\includegraphics[width=.7\columnwidth]{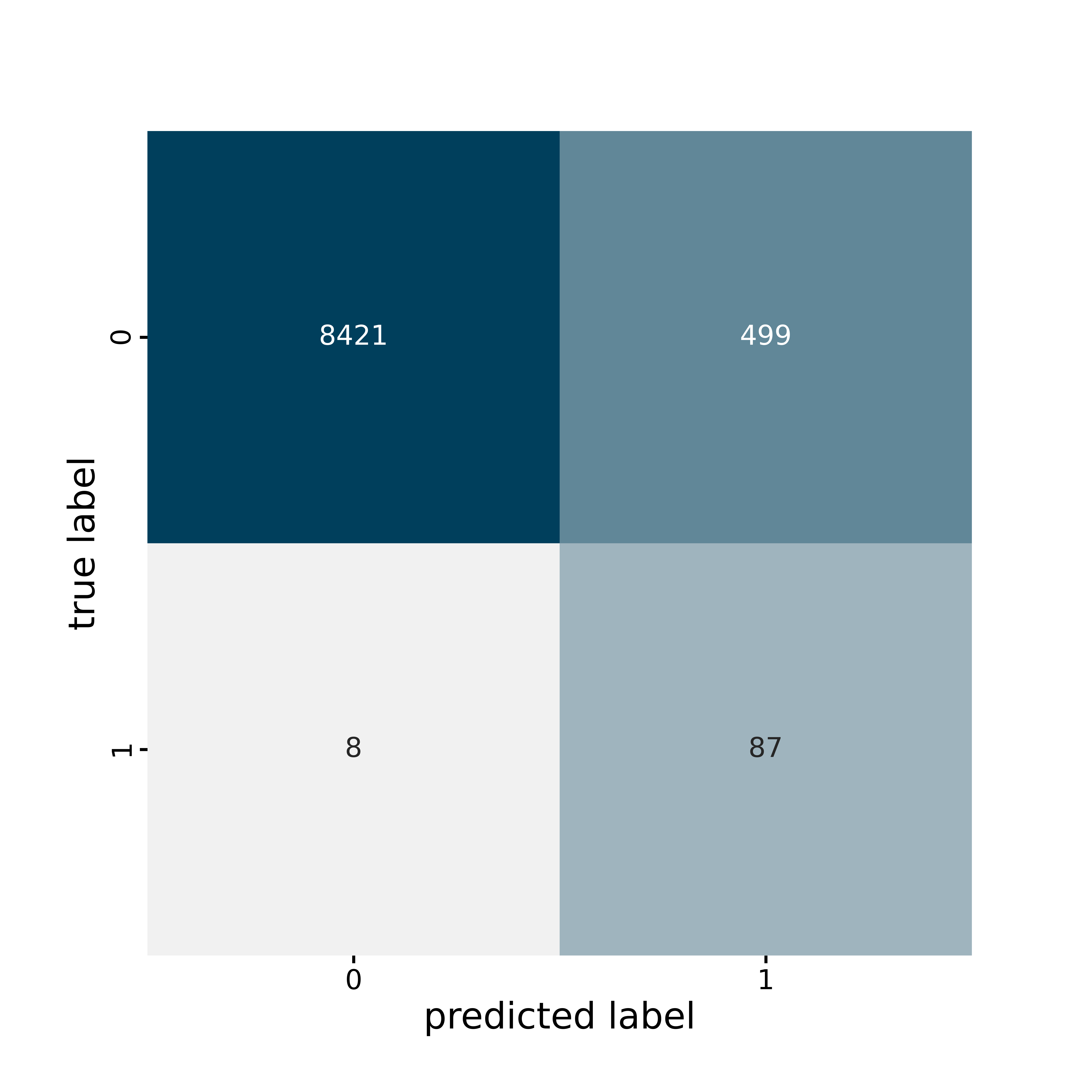}
    \caption{Confusion matrix for the test set, with a recall of 92\,\%.}
    \label{fig:conf_mat_tr}
\end{figure}  

After this search for the optimal hyperparameters, we obtained the best recall score with an RBF kernel and hyperparameters $C=10$ and $\gamma=0.001$, with a total recall, precision, and F1 score of 92\,\%, 15\,\%, and 26\,\%, respectively, on the test set.  Fig. \ref{fig:conf_mat_tr} shows confusion matrix on the test set. The confusion matrix is a performance measurement in machine learning classification that compares the labels predicted by the model (x-axis) with the ground-truth labels in the data set (y-axis). The most important thing to note here is that the SVM model manages to recover nearly all positive instances, which is our main priority, as we do not want to lose any candidate UCD in the process. Also, the SVM performs very well at identifying True Negatives (TN, negative instances predicted as negative). In conclusion, the model allows us to filter out the vast majority of non-UCD objects, while keeping almost all the candidate UCDs. However, the class imbalance of the data causes the number of False Positives (FP, negative instances predicted as positive) to be larger than the number of True Positives (TP, positive instances predicted as positive). This makes the analysis with VOSA still necessary to differentiate the final candidate UCDs.

\subsection{Blind test} \label{blind_test}

\begin{figure}
    \centering
	\includegraphics[width=.7\columnwidth]{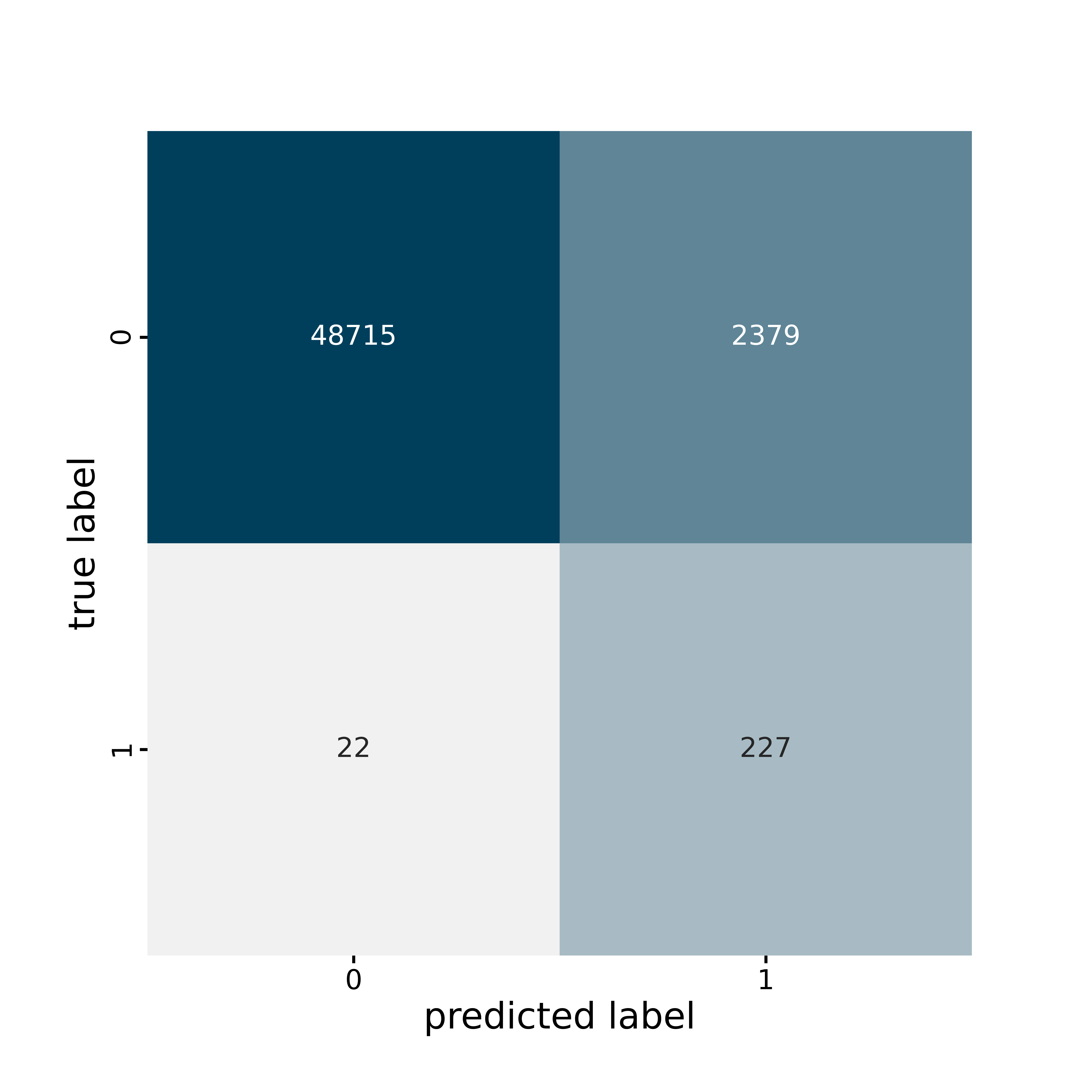}
    \caption{Confusion matrix for the blind test, with a recall of 91\,\%.}
    \label{fig:conf_mat_new}
\end{figure}

To validate the classifier's performance on unseen data, we applied our ML methodology on the J-PLUS DR2 data from another of the 20$\times$20\,deg$^2$ regions containing 607\,801 objects with good photometry in all relevant filters. Firstly, we used the same PCA model fitted with the previous region to perform the PCA filtering on this new region, reducing the total number of instances to 51\,343. We used the previously fitted SVM model to predict over this reduced set, obtaining a recall, precision, and F1 score of 91\,\%, 9\,\%, and 16\,\%, respectively. Fig. \ref{fig:conf_mat_new} shows the confusion matrix for this blind test. Thus, we ended up with 2\,606 (2\,379 + 227) objects to be analysed with VOSA for the final UCD identification, which means the SVM model achieved to discard $\sim$95\,\% (1 - 2\,379/51\,094) of the non-UCD objects that pass the PCA filtering.

We used the objects analysed with \texttt{VOSA} in the VO methodology for this same region to make a thorough analysis of our ML method. Thus, we found that, of these objects, the PCA filter removes those with $T_{\rm eff}\gtrsim4\,100$\,K, so this first cut is able to purge the initial set of the hottest objects. The ML methodology is more restrictive in terms of photometric quality, as it is only applicable to the objects with photometry in all the filters used to build the input features. This means that all the final candidate UCDs with no photometry in any of these filters (around 50\,\% for this region), obtained with the VO methodology, are not captured by the ML procedure. In summary, we concluded that the ML methodology is more efficient in the sense that it allows for a greater number of true negatives (non-UCD objects) to be discarded prior to analysis with \texttt{VOSA}, although it is a more restrictive method as it relies only on the photometry of the J-PLUS filters used. Another advantage of the proposed ML approach is that it consists of a single process instead of the three separate ones required in the VO methodology.


\section{Detection of strong emission line emitters} \label{flares}

Strong emission lines have been detected serendipitously in UCD optical spectra, both as transient flaring  phenomena  \citep{Liebert1999,Liebert2003,Martin2001,Schmidt2007} as well as steady  features  \citep{Schneider1991,Mould1994,Martin1999,Burgasser2011}. Stellar flares, events powered by the sudden release of magnetic energy, that is converted to kinetic energy of electrons and ions due to magnetic reconnection in the stellar atmosphere, are a common phenomenon around M dwarfs. Works such as those presented in \citet{flare_xray}, \citet{Berger2010} and \citet{arecibo} have confirmed that optical, radio and X-ray flares do occur in UCDs. 

We decided to focus our search for strong emission on the H$\alpha$ and Ca~{\sc ii} H and K lines, important chromospheric activity indicators \citep{cincunegui}, which correspond to filters 11.0 ($J0660$) and 7.0 ($J0395$) in the J-PLUS filter system, respectively. Since this is a rare phenomenon, we decided to conduct this search on a larger sample of objects, including all the objects that met the $\textit{G} - \textit{G}_{\rm RP}$ and $\textit{r}-\textit{z}$ colour criteria presented in Sects. \ref{parallax}, \ref{pm} and \ref{colordiagram}. Therefore, since we did not apply the effective temperature cutoff, the search also covered spectral types hotter than those of the UCDs.

With this purpose, we developed a Python algorithm that detects any drop in magnitude in filters \textit{J0395} and \textit{J0660}. Firstly, the algorithm joins the J-PLUS DR2 photometry obtained in the search described in Sect. \ref{Methodology} to the shortlisted objects obtained with the methodology described in Sect. \ref{parallax}, \ref{pm} and \ref{colordiagram}. Then, object by object, it computes the magnitude ratio between the filter of interest and all its neighbours. We chose as neighbours the filters 6.0, 8.0 and 9.0 for the filter 7.0 (Ca~{\sc ii} H and K) and the filters 1.0 and 3.0 for the 11.0 (H$\alpha$). If this ratio is lower than a fixed threshold value (entered by the user) for any neighbouring filter, the algorithm recognises a  possible strong line emitter and plots the photometry of the object. For the object to be recognisable, we need at least photometry in one of the neighbouring filters, so we can detect this emission peak. The algorithm receives as input a file with the candidate UCDs photometry and returns both the plotted photometry of the objects with possible strong emission and a table with the computed magnitude drop for each of them. We were permissive with the fixed threshold, so as not to discard any interesting object, and imposed a value of 0.96. Then, we visually inspected all the possible strong emitters detected by the algorithm given this threshold.

\begin{figure}
    \centering
	\includegraphics[width=.8\columnwidth]{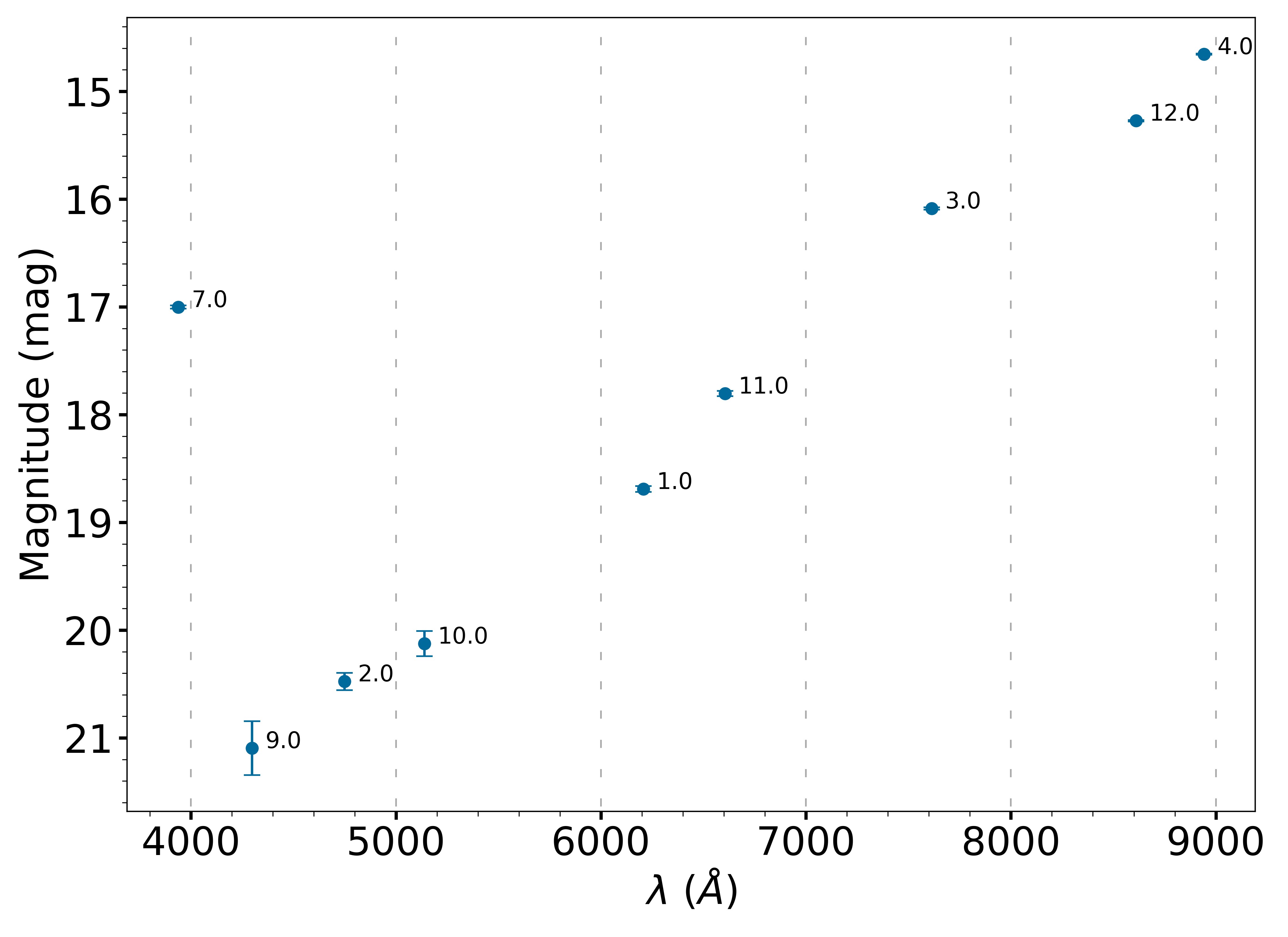}
    \caption{Example of a detected strong excess in the filter 7.0 (\textit{J0395}), which corresponds to Ca~{\sc ii} H and K emission lines. The Figure shows the J-PLUS photometry of the first object in Table \ref{tab:flares}, with error bars representing the error in the magnitude. The algorithm detects that the magnitude in the filter \textit{J0395} is 0.8 times the magnitude in the filter 9.0 (\textit{J0430}), and recognises this as a possible emission line. In this case, the threshold value for the excess detection was 0.96.}
    \label{fig:flares}
\end{figure}

\begin{figure*}
    \centering
    \includegraphics[width=\linewidth/6]{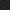}
    \hspace{0.1cm}
    \includegraphics[width=\linewidth/6]{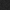}
    \hspace{0.1cm}
    \includegraphics[width=\linewidth/6]{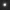}
    \hspace{0.1cm}
    \includegraphics[width=\linewidth/6]{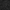}
    \hspace{0.1cm}
    \includegraphics[width=\linewidth/6]{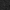}
    \\[\smallskipamount]
    \includegraphics[width=\linewidth/6]{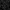}
    \hspace{0.1cm}
    \includegraphics[width=\linewidth/6]{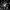}
    \hspace{0.1cm}
    \includegraphics[width=\linewidth/6]{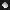}
    \hspace{0.1cm}
    \includegraphics[width=\linewidth/6]{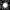}
    \hspace{0.1cm}
    \includegraphics[width=\linewidth/6]{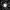}

    \caption{Images from the J-PLUS DR2 archive with the photometry in different filters for two of the strong line emitters detected. The first row corresponds to an excess in the Ca~{\sc ii} H and K (filter \textit{J0395}) emission lines (first object in Table \ref{tab:flares}), with images in the filters \textit{u}, \textit{J0378}, \textit{J0395}, \textit{J0410}, and \textit{J0430} (from left to right). The second row shows an excess in the H$\alpha$ (filter \textit{J0660}) emission line (seventh object in Table \ref{tab:flares}), with images in the filters \textit{J0515}, \textit{r}, \textit{J0660}, \textit{i}, and \textit{J0861} (from left to right). For both objects, all images shown were taken within a time interval of about 40 minutes.}
    \label{fig:flares_imgs}
\end{figure*}

Finally, we ended up with eight objects that exhibit significant emission peaks in the filters of interest, that are presented in Table \ref{tab:flares}. We used \texttt{VOSA} to estimate the effective temperature of these objects and found only one UCD, with $T_{\rm eff}=2\,500$\,K, among the eight objects (fifth object in Table \ref{tab:flares}). The remaining seven objects have estimated effective temperatures (see Table \ref{tab:flares}) typical of mid-M dwarfs \citep{zhang_2018_midm}. Fig. \ref{fig:flares} shows the photometry of the object with the highest line emission excess (first object in Table \ref{tab:flares}). Also, in Fig. \ref{fig:flares_imgs} we include images from the J-PLUS DR2 archive with the emission in different filters for the object with highest excess activity in the Ca~{\sc ii} H and K (first object in Table \ref{tab:flares}) and H$\alpha$ (seventh object in Table \ref{tab:flares}) emission lines. With this analysis, we underline the possibility of systematically detecting strong emission lines in UCDs and earlier M-type stars with photometric surveys such as J-PLUS.

For the fifth  object listed in Table \ref{tab:flares}, namely LP 310-34, we carried out a follow-up optical spectroscopy monitoring study. Five exposures of half an hour integration time each were obtained on January 12th, 2020 in service time (proposal 60-299, PI Martín) with ALFOSC attached to the Nordic Optical Telescope in La Palma. The grism number 4 and the slit with of 1.0 arcsec were selected providing a dispersion of 3.75 \AA pixel$^{-1}$ and a resolving power of R=700. Our spectra confirm that it is a very late M dwarf (dM8) with H$\alpha$ in emission \citep{Schmidt2007}. We measured an H$\alpha$ equivalent width of -14.6 \AA , using the gaussian profile integration option available in the IRAF task splot applied to the co-added spectrum of the five exposures. Individual measurements of the equivalent width in each spectrum ranged from -7.0 to -20.7 \AA , suggesting variability in the strength of the H$\alpha$ emission. This level of H$\alpha$ emission is not uncommon among late-M dwarfs \citep{Martin2010,Pineda2016}. No other emission lines were detected in our spectra. 

One of the new strong line-emission candidates (sixth object in Table \ref{tab:flares}) was observed on April 21st, 2022 with the long-slit low-resolution mode of the SpeX instrument \citep{Rayner2003} at the NASA Infrared Telescope Facility (IRTF, program 2022A011, PI A. Burgasser). Preliminary analysis of the data indicates that the near-infrared spectrum is well matched by a M5 dwarf template (A. Burgasser, private communication). Further details of these observations and additional spectroscopic follow-up of the J-PLUS candidates presented in this work is planned for a future paper.  

This study suggests that our J-PLUS search for strong emission lines may be revealing previously unknown sporadic very strong activity in otherwise normal late-M dwarfs. It is worth noting that our search for strong line emitters has detected as many objects with Ca~{\sc ii} H and K excess than with H$\alpha$ excess, and no object showing both excesses simultaneously. Events of strong Ca~{\sc ii} H and K line emission in normal late-M dwarfs may have important implications for studies of 
exoplanetary space weather and habitability \citep{Yamashiki2019}.

\begin{table}
\fontsize{11pt}{11pt}\selectfont
 \caption{Objects with strong flux excess in H$\alpha$ (filter \textit{J0660}) or Ca~{\sc ii} H and K (filter \textit{J0395}) emission lines, identified with our Python algorithm.}
 \label{tab:flares}
 \centering          
 \begin{tabular}{c c c c c c c}
  \hline\hline
  \noalign{\smallskip}
  
  $\alpha$ & $\delta$ & Filter & Magnitude\,$^{(a)}$ & Ratio\,$^{(b)}$ & Simbad ID & Estimated $T_{\rm eff}$\\
  
  [deg] & [deg] &  of interest &  &  & & [K]\\
  
  \noalign{\smallskip}
  \hline
  \noalign{\smallskip}
  
  18.53140 & 7.94229 & \textit{J0395} & 16.811 & 0.799 & \ldots & 3\,200\\
  
  36.68415 & 34.75973 & \textit{J0660} & 18.100 & 0.884 & \ldots & 3\,300\\
  
  107.18550 & 71.90704 & \textit{J0660} & 18.199 & 0.915 & \ldots & 3\,200\\

  116.14374 & 40.14576 & \textit{J0395} & 19.333 & 0.946 & \ldots & 3\,100\\
  
  121.85651 & 32.21826 & \textit{J0395} & 17.002 & 0.806 & LP 310-34 & 2\,500\\

  135.92497 & 34.80495 & \textit{J0395} & 18.319 & 0.895 & LP 259-39 & 3\,200\\

  138.52385 & 23.87355 & \textit{J0660} & 17.443 & 0.850 & \ldots & 3\,200\\

  199.02058 & 56.12370 & \textit{J0660} & 20.233 & 0.911 & \ldots & 3\,300\\

  \noalign{\smallskip}
  \hline
 \end{tabular}
 \tablefoot{$^{(a)}$ In the filter of interest. $^{(b)}$ Ratio of magnitudes between filter of interest and neighbour filter.}
\end{table}

\section{Conclusions} \label{conclusions}

Using a Virtual Observatory methodology, we provide a catalogue of 9\,810 candidate UCDs over the entire sky coverage of J-PLUS DR2. With 7\,827 previously not reported as UCD, we show there is still room for the discovery of these objects even with a small telescope such as the JAST80. Our main goal is to consolidate and further develop a search methodology, introduced in \citet{solano2019}, to be used for deeper and larger surveys such as J-PAS and Euclid, both being an ideal scenario for the study and discovery of UCDs thanks to their unprecedented photometric system of 54 narrow-band filters and excellent sensitivity, respectively. Further confirmation by spectroscopy of the UCD nature of these candidates goes beyond the scope of this study. However, the candidate UCDs that are reported in Simbad, but are not in our sample of known UCDs (see Sect. \ref{known}), mostly present spectral type M6\,V or are left out because they lack the luminosity class, so we expect the degree of contamination to be small.

The use of different approaches based on astrometry and photometry tends to minimise the drawbacks and biases associated to the search of ultracool objects: photometric-only selected samples may leave out peculiar UCDs not following the canonical trend in colour-colour diagrams and they can also be affected by extragalactic contamination. Proper motion searches may ignore objects with small values of projected velocity in the plane of the sky. Regarding parallax-based searches, they will be limited to the brightest objects with parallax values from \textit{Gaia}.

Based on our kinematics study, almost all our candidate UCDs can be considered thin disk members, with 268 of them being potential members of the thick disk. Also, five of the candidates are likely to belong to the Galactic halo. Using the \texttt{BANYAN}~$\Sigma$ tool, we find 48 candidate UCDs with a high Bayesian probability of belonging to seven different young moving associations, in 30 of the cases with a probability greater than 95\,\%. A further spectroscopic follow-up will be required to search for spectral signatures of youth. In the binarity analysis, we find 122 possible unresolved companions among our candidate UCDs. Searching for wide \textit{Gaia} companions of our candidate UCDs, we find 78 possible multiple systems (73 binary + 5 triple), six of them already tabulated by the WDS. We use \texttt{VOSA} to get an estimation of the effective temperature of the wide \textit{Gaia} companions identified in all the systems, finding that most of them are M-type stars.

Among the non-recovered known UCDs that lie in the sky coverage of J-PLUS DR2, we find that more than half are lost due to lack of photometric or astrometric information with enough quality. The remaining objects are discarded due to our conservative temperature cutoff at 2\,900\,K or a bad SED fitting (vgfb$>$12). Compared to previously reported candidates, the new ones are on average more distant and extend to smaller values of proper motion.

We achieve promising results when reproducing the search for UCDs with a purely ML-based methodology. In this approach, we find crucial the preliminar PCA filtering to deal with the strong imbalance of the data and discard the hottest objects. This allows us to significantly reduce the negative class and improve the classification capability of the posterior SVM model. Using the developed ML methodology to predict on unseen data, we are able to recover 91\,\% of the candidate UCDs found with the VO methodology, discarding a larger number of true negatives (non-UCD objects) before the analysis with VOSA in a faster way. This is a significant achievement, since the main bottleneck of the VO methodology is the high number of objects to be analysed with \texttt{VOSA}.

In this line, the real turning point would be to develop a ML methodology that more significantly filters the number of objects we need to analyse with \texttt{VOSA} for the final UCD identification. This is not a straightforward task due to the imbalance of the data and because the analysis with \texttt{VOSA} is based on complex theoretical models. To this end, we are exploring the use of independent component analysis in the initial filtering and ensemble learning in the classification step.

Finally, we develop an algorithm capable of detecting strong emission line emitters in the optical range. We identify four objects with strong excess in the filter corresponding to the Ca~{\sc ii} H and K emission lines and four other objects with excess emission in the H$\alpha$ filter.
\chapter{Detection of Flaring M dwarfs with multi-filter Photometry}
\label{chp:flares_paper}




Understanding and characterising the magnetic activity of M dwarfs is of paramount importance in the search for Earth-like exoplanets orbiting around them. Energetic stellar activity phenomena, such as flares or coronal mass ejections, which are common in these stars, are deeply connected with the habitability and atmospheric evolution of the surrounding exoplanets. We present a follow-up of a sample of M dwarfs with strong H$\alpha$ and Ca~{\sc ii} H and K emission lines identified with J-PLUS photometry in a previous work. We collected low-resolution NOT/ALFOSC and GTC/OSIRIS spectra, measuring the PC3 index for the spectral type determination. We used two-minutes cadence TESS calibrated light curves to identify and characterise multiple flares, and to calculate the rotation period of the two active M dwarfs found in our sample. We confirmed that the strong emission lines detected in the J-PLUS photometry are caused by transient flaring activity. We found clear evidence of flaring activity and periodic variability for LP 310-34 and LP 259-39, and estimated flare energies in the TESS bandpass between $7.4\times10^{30}$ and $2.2\times10^{33}$\,erg for them. We characterised LP 310-34 and LP 259-39 as very rapidly rotating M dwarfs with Ca~{\sc ii} H and K and H$\alpha$ in emission, and computed a rotation period of 1.69\,d for LP 259-39 for the first time. This work advocates the approach of exploiting multi-filter photometric surveys to systematically identify flaring M dwarfs, especially to detect episodes of strong Ca~{\sc ii} H and K line emission that may have important implications for exoplanetary space weather and habitability studies. Our results reveal that, apart from the already known H$\alpha$ flares, flare events in Ca~{\sc ii} H and K can also be detected using optical narrow-band filters in common M dwarfs.

\section{Observations}\label{sec:obs}

\subsection{Sample selection}\label{sec:sample}

The sample studied in this work is the result of the search for strong emission lines performed in our previous work \citep{masbuitrago2022}, using multi-filter optical photometry from the Javalambre Photometric Local Universe Survey \citep[J-PLUS;][]{Cenarro2019}. For this, we developed a Python algorithm capable of detecting excess in the J-PLUS filters corresponding to the H$\alpha$ ($J0660$) and Ca~{\sc ii} H and K ($J0395$) emission lines. Following this approach, we identified eight M dwarfs with emission excess in these filters (four of them in each of the filters and none showing both excesses simultaneously). In the end, one of these objects was discarded for spectroscopic follow-up because it was not bright enough, resulting in a final sample of seven M dwarfs. Table~\ref{tab:targets} lists the selected targets.

The J-PLUS spectral energy distribution (SED) of each target star is provided in Appendix \ref{app:app_flares}. The excess emission in the $J0395$ filter is evident for J-PLUS0114, J-PLUS0744, J-PLUS0807, and J-PLUS0903. On the other hand, the SEDs of J-PLUS0226, J-PLUS0708, and J-PLUS0914 show strong emission in the $J0660$ filter. We attribute this behaviour to the fact that the star experiences flaring activity during the corresponding J-PLUS observing block, in which all filters are observed sequentially. The strategy for each J-PLUS observing block is to obtain, for the same pointing, three consecutive exposures per filter, with a total exposure time of approximately one hour \citep{Cenarro2019}. Flaring phenomena during the exposures for the filters of interest would explain the SED behaviour found. Given the low probability of observing a flare during the exposures for the filters of interest, it is easier to detect the less energetic and shorter-lived flares, which are more frequent and last a few minutes as we confirm in Section \ref{sec:flares}.

The estimated effective temperatures for these objects, obtained with the tool \texttt{VOSA}\footnote{\url{http://svo2.cab.inta-csic.es/theory/vosa/}} \citep{vosa}, locate them as mid-M dwarfs except for one, namely LP 310-34, with a $T_{\rm eff}=2\,500$\,K. As mentioned in \citet{masbuitrago2022}, we already carried out an spectroscopic follow-up for the latter that confirmed it as a late M dwarf (dM8) with H$\alpha$ in emission \citep{Schmidt2007}.

\begin{table*}
\fontsize{9pt}{9pt}\selectfont
 \caption{Targets selected for spectroscopic observation.}
 \label{tab:targets}
 \centering          
 \begin{tabular}{l l c c c c c}
  \hline\hline
  \noalign{\smallskip}
  
  Object\,$^{(a)}$ & SIMBAD & $\alpha\,^{(b)}$ & $\delta\,^{(b)}$ & $T$\,$^{(c)}$ & \texttt{VOSA} $T_{\rm eff}$ & Excess \\
   & & [J2016.0] & [J2016.0] & [mag] & [K] & \\
  
  \noalign{\smallskip}
  \hline
  \noalign{\smallskip}
  
  J-PLUS DR2 J0114+07 & \ldots & 01:14:07.54 & 07:56:32.2  & $15.78\pm0.01$ & 3\,200 & Ca~{\sc ii} HK\\
  
  J-PLUS DR2 J0226+34 & \ldots & 02:26:44.20 & 34:45:35.0  & $18.91\pm0.05$ & 3\,300 & H$\alpha$\\
  
  J-PLUS DR2 J0708+71 & \ldots &  07:08:44.51 & 71:54:25.3  & $18.19\pm0.01$ & 3\,200 & H$\alpha$\\

  J-PLUS DR2 J0744+40 & \ldots & 07:44:34.50 & 40:08:44.7  & $14.79\pm0.01$ & 3\,100 & Ca~{\sc ii} HK\\
  
  J-PLUS DR2 J0807+32 & LP 310-34 & 08:07:25.60 & 32:13:06.0  & $14.72\pm0.01$ & 2\,500 & Ca~{\sc ii} HK\\

  J-PLUS DR2 J0903+34 & LP 259-39 & 09:03:41.95 & 34:48:18.6  & $13.83\pm0.01$ & 3\,200 & Ca~{\sc ii} HK\\

  J-PLUS DR2 J0914+23 & \ldots & 09:14:05.74 & 23:52:24.9  & $18.39\pm0.08$ & 3\,200 & H$\alpha$\\

  \noalign{\smallskip}
  \hline
 \end{tabular}
 \tablefoot{$^{(a)}$ Hereafter we use J-PLUSHHMM as an abbreviation. $^{(b)}$ From \textit{Gaia} Data Release 3 \citep[DR3;][]{gaiadr3}. $^{(c)}$ TESS magnitude from \citet{tic8_2}.  } 
\end{table*}

\subsection{Observational details}
We collected low-resolution optical spectra of our seven targets with The Alhambra Faint Object Spectrograph and Camera (ALFOSC) mounting on the 2.56-m Nordic Optical Telescope (NOT) with proposal number 66-208 (P.I. ELM). Also, we observed two bright targets (J-PLUS DR2 J0807+32 and J-PLUS DR2 J0903+34 in Table \ref{tab:targets}) with the Optical System for Imaging and low-Intermediate-Resolution Integrated Spectroscopy (OSIRIS) mounting on the 10.4-m Gran Telescopio Canarias (GTC), at the Roque de los Muchachos Observatory on the island of La Palma, Spain, with programme GTCMULTIPLE2I-22B (P. I. ELM).

ALFOSC is equipped with a Teledyne e2v CCD231-42-g-F61 back illuminated, deep depletion, astro multi-2 detector. The detector dimension is 2\,048$\times$2\,064 pixels with a scale of 0.2138 arcsec/pix. The NOT/ALFOSC observation was executed under visitor mode during the nights of January 26-27, 2023 (observers PMB \& JYZ). We used a 1.0-arcsec slit, and \#4 grism, which provide a wavelength range from 3\,200 \AA\ to 9\,600 \AA\ with a resolution power $R\approx360$.

OSIRIS is a commonly used instrument of GTC. It covers the wavelength range $3\,650- 10\,050$\,\AA\ and has an effective field of view of 7.5$\times$6.0 arcmin. OSIRIS has two Marconi CCD44-82 (2\,048$\times$4\,096 pixels) detectors with gap in between. The 2$\times$2 binned pixel size is 0.254 arcsec/pix. In the mode of long-slit spectroscopy, the object is centred on the slit at the coordinate $\rm{X}
=250$ of the CCD2. The GTC/OSIRIS observation was executed under service mode. We requested as conditions a maximum seeing of 1.2\,arcsec, cloud free sky, and grey moon phase, using the R1000B grism and a 1.2-arcsec slit under the parallactic angle. This configuration yields a wavelength coverage from 3\,600\,\AA\ to 7\,900\,\AA\ with a resolution power $R\approx500$. Table~\ref{tab:obs} shows the record of observations.

\begin{table}
\fontsize{11pt}{11pt}\selectfont
 \caption{Record of observations.}
 \raggedright
 \label{tab:obs}
 \centering          
 \begin{tabular}{l c c c c c c c}
  \hline\hline
  \noalign{\smallskip}
  
  Object  &  Date  &  Configuration & Slit width & Grism & Exposures\\
  
  \noalign{\smallskip}
  \hline
  \noalign{\smallskip}
      
    J-PLUS0114 & 26 Jan. 2023 & ALFOSC Long Slit & 1.0" & \#4 & 1800s$\times$2\\
    
    J-PLUS0114 & 27 Jan. 2023 & ALFOSC Long Slit & 1.0" & \#4 & 1800s$\times$1\\
    
    J-PLUS0226 & 27 Jan. 2023 & ALFOSC Long Slit & 1.0" & \#4 & 2000s$\times$5\\
    
    J-PLUS0708 & 26 Jan. 2023 & ALFOSC Long Slit & 1.0" & \#4 & 2000s$\times$5\\
    
    J-PLUS0744 & 27 Jan. 2023 & ALFOSC Long Slit & 1.0" & \#4 & 1500s$\times$3\\
    
    J-PLUS0807 & 29 Oct. 2022 & OSIRIS Long Slit & 1.2" & R1000B & 180s$\times$6\\
    
    J-PLUS0807 & 26 Jan. 2023 & ALFOSC Long Slit & 1.0" & \#4 & 1800s$\times$2\\
    
    J-PLUS0807 & 27 Jan. 2023 & ALFOSC Long Slit & 1.0" & \#4 & 1800s$\times$1\\
    
    J-PLUS0903 & 29 Oct. 2022 & OSIRIS Long Slit & 1.2" & R1000B & 90s$\times$6\\

    J-PLUS0903 & 27 Jan. 2023 & ALFOSC Long Slit & 1.0" & \#4 & 1000s$\times$3\\
    
    J-PLUS0914 & 26 Jan. 2023 & ALFOSC Long Slit & 1.0" & \#4 & 2000s$\times$4\\
    
    J-PLUS0914 & 27 Jan. 2023 & ALFOSC Long Slit & 1.0" & \#4 & 2000s$\times$1\\

  \noalign{\smallskip}
  \hline
 \end{tabular}
 \tablefoot{Targets are ordered first by right ascension and then observation date.}
\end{table}


\subsection{Data reduction}
We reduced both the GTC/OSIRIS and NOT/ALFOSC data using v1.12 of \texttt{PypeIt} \citep{pypeit:zenodo,pypeit:joss_pub}, a community-developed open-source Python package for semi-automated reduction of spectroscopical data in astronomy. \texttt{PypeIt} supports a long list of spectrographs and provides the code infrastructure to automatically process the image, identify the slit in a given detector, extract the object spectra, and perform wavelength calibration. We observed the standard stars HD~19445 and Feige~110 for flux calibration of NOT/ALFOSC and GTC/OSIRIS data, respectively.


\section{Results and discussion}
\label{sec:results}

\subsection{Reduced spectra}
\label{sec:reduced_sp}

\begin{figure}
    \centering
    	\includegraphics[width=.48\columnwidth]{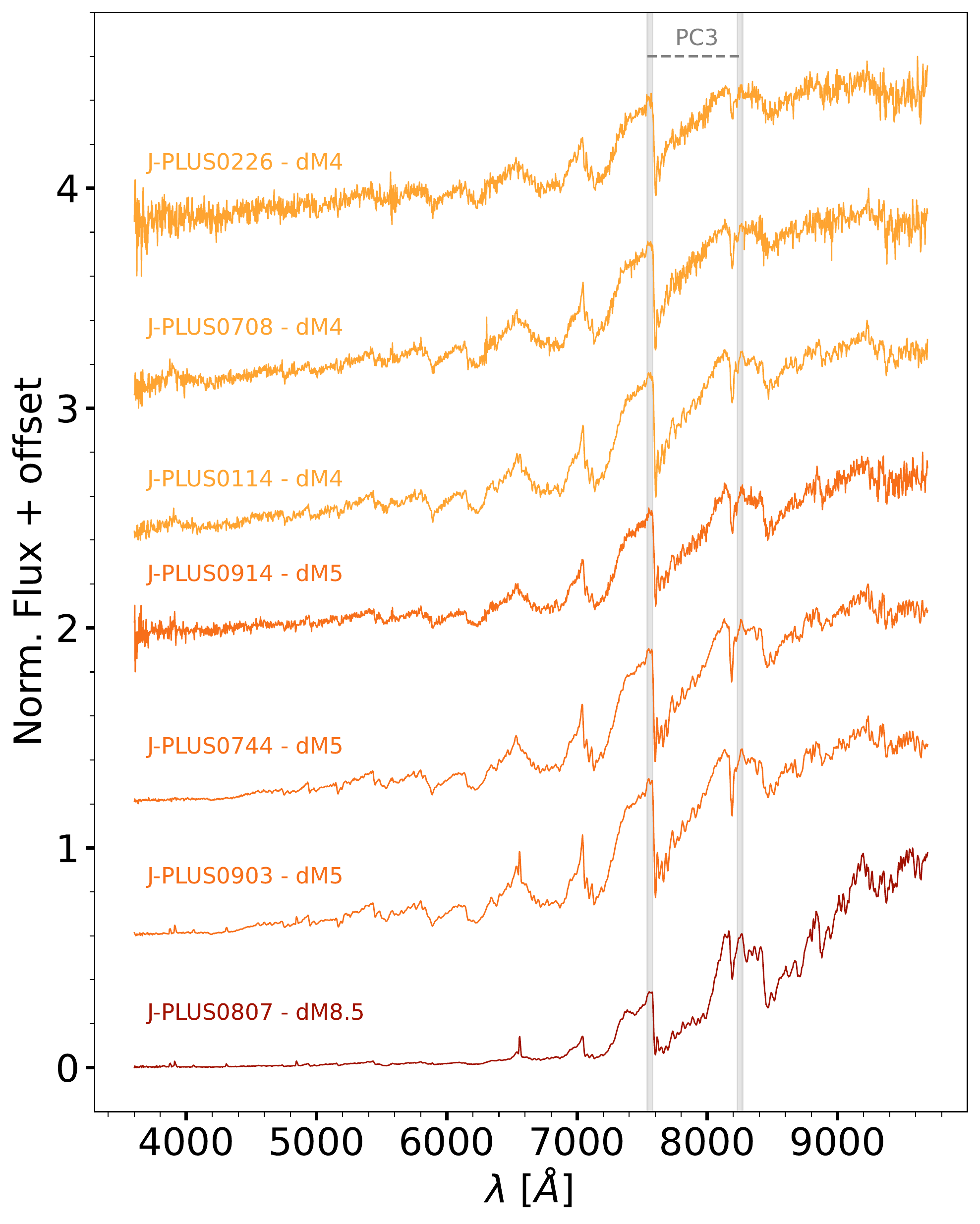}
        \includegraphics[width=.48\columnwidth]{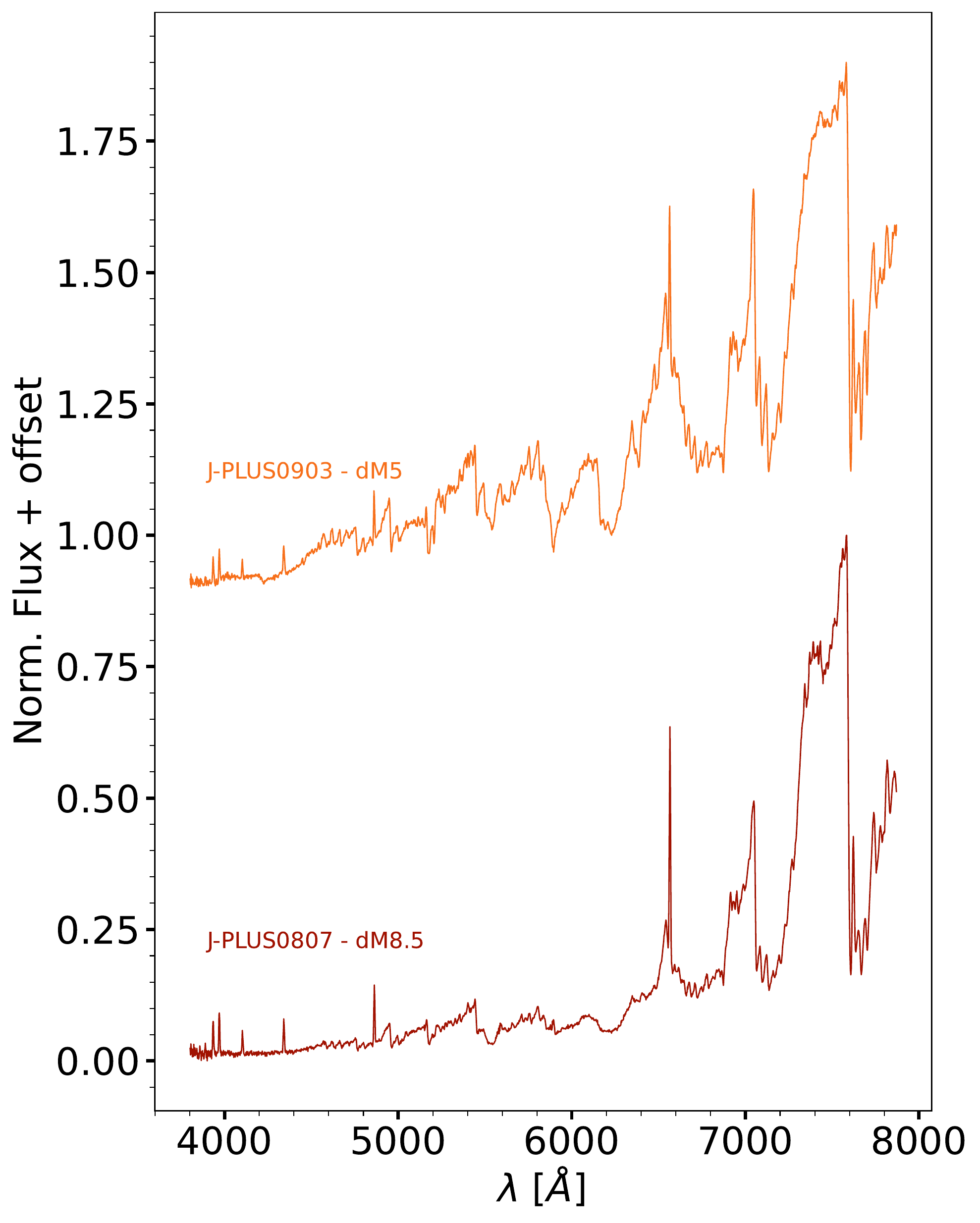}
    \caption{Co-added spectra observed with NOT/ALFOSC (\textit{left panel}) and GTC/OSIRIS (\textit{right panel}), sorted from top to bottom by the derived PC3 index (see Table \ref{tab:pc3}). The grey shaded bands in the \textit{left panel} show the spectral regions used to derive the PC3 index.}
    \label{fig:spectra_all}
\end{figure}

Figure \ref{fig:spectra_all} shows the co-added spectra for the observed targets. We note that no intense steady line emission is observed in the spectra (for examples of strong line emission in low-resolution spectra, see Fig. 1 in \citealp{Burgasser2011} or Figs. 10 and 11 in \citealp{Schmidt2007}), confirming that the excess emission detected in the J-PLUS photometry is not steady and is indeed caused by transient flaring activity. Moreover, objects with excess emission in the J-PLUS Ca~{\sc ii} H and K filter show no apparent differences in their spectral features compared to objects with excess emission in the J-PLUS H$\alpha$ filter (see Table \ref{tab:targets}), suggesting that the flaring activity detected in Ca~{\sc ii} H and K is not particular to a specific type of star. Hence, it follows that common M dwarfs experience two types of flares, those already well-known in H$\alpha$ and those in Ca~{\sc ii} H and K revealed in this work.

Several spectroscopic indices have been explored for the spectral classification of M dwarfs \citep{lepine2003} and, in particular, for late-M dwarfs using low-resolution optical spectra \citep{kirkpatrick1995, martin1996, martin1999b}. To derive spectral types for our sample, we measured the PC3 index \citep{martin1999b}, which is a reliable indicator of spectral type in the [M2.5, L1] range and has been used consistently in the literature \citep{crifo2005,martin2006,Martin2010,phanbao2006,reyle2006,phanbao2008}. The PC3 index is a pseudo-continuum spectral ratio between the $8230-8270$\,\AA\,(numerator) and $7540-7580$\,\AA\,(denominator) intervals, which can be used to derive spectral types between M2.5 and L1 following the calibration presented by \citet{martin1999b}:

\begin{equation}    
    \rm{SpT}=-6.685+11.715\times (\rm PC3)-2.024\times (\rm PC3)^2.
    \label{eq:ffd}
\end{equation}

Table \ref{tab:pc3} lists the PC3 index and the adopted spectral type, with an uncertainty of $\pm0.5$ subclasses, for our targets. The classification obtained for J-PLUS0807 is consistent with that provided in \citet{Schmidt2007}, who derived a spectral type of dM8, with an uncertainty of $\pm0.5$ subclasses, by visual comparison of the spectra to spectral standards. These results confirm the rest of our sample, still spectroscopically unclassified in the literature, as mid-M dwarfs.

\begin{table}
\fontsize{11pt}{11pt}\selectfont
 \caption{PC3 index and adopted spectral type for our targets.}
 \label{tab:pc3}
 \centering          
 \begin{tabular}{l c l}
  \hline\hline
  \noalign{\smallskip}
  
  Object & PC3 & SpT\\
  
  \noalign{\smallskip}
  \hline
  \noalign{\smallskip}
  
  J-PLUS0114 & 1.12 & dM4 \\
  
  J-PLUS0226 & 1.09 & dM4 \\
  
  J-PLUS0708 & 1.12 &  dM4 \\

  J-PLUS0744 & 1.29 & dM5 \\
  
  J-PLUS0807 & 1.94 & dM8.5 \\

  J-PLUS0903 & 1.29 & dM5 \\

  J-PLUS0914 & 1.29 & dM5 \\

  \noalign{\smallskip}
  \hline
 \end{tabular}
\end{table}

\begin{figure}
    \centering
    	\includegraphics[width=.48\columnwidth]{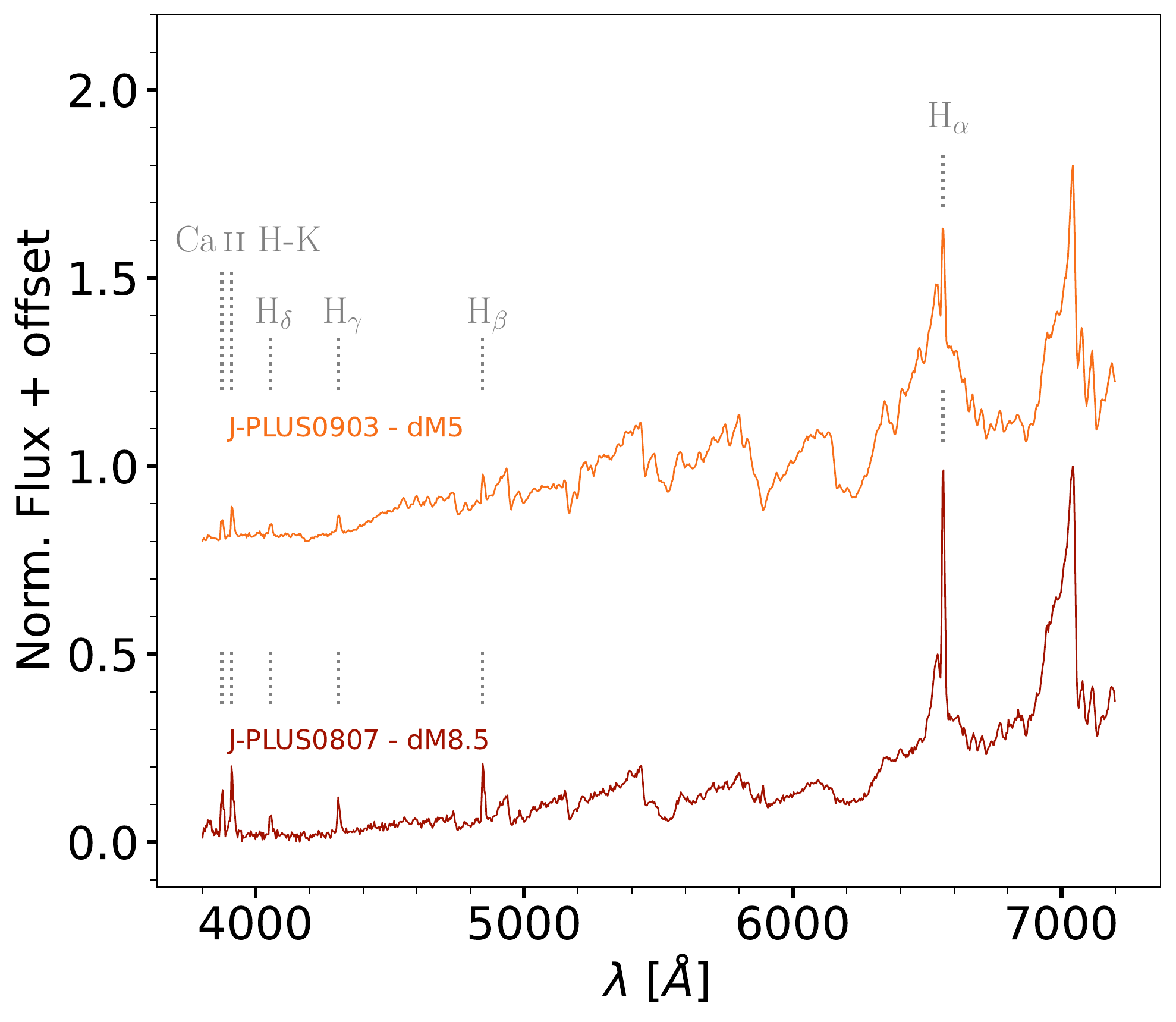}
        \includegraphics[width=.48\columnwidth]{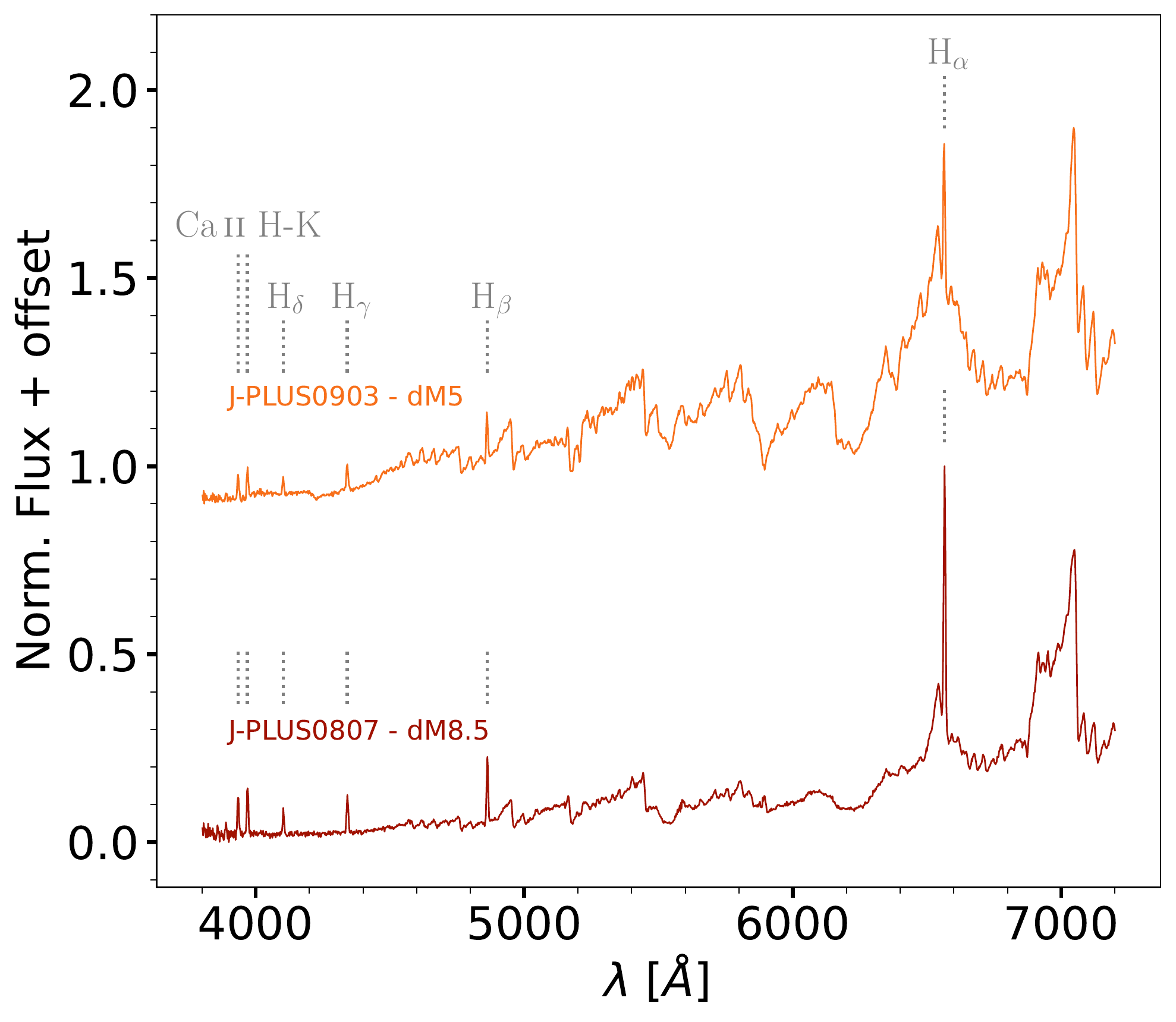}
    \caption{Zoom-in for the NOT/ALFOSC (\textit{left panel}) and GTC/OSIRIS (\textit{right panel}) co-added spectra of J-PLUS0807 and J-PLUS0903. The grey dashed lines mark Ca~{\sc ii} H and K, H$\delta$, H$\gamma$, H$\beta$, and H$\alpha$ emission lines.}
    \label{fig:spectra_red}
\end{figure}

The obtained spectra confirm both J-PLUS0807 and J-PLUS0903 as active M dwarfs with Ca~{\sc ii} H and K and H$\alpha$ in emission, while the rest of the targets show no signs of activity. Figure \ref{fig:spectra_red} shows a close-up view of the spectral region of interest for these stars, with prominent Ca~{\sc ii} H and K, H$\delta$, H$\gamma$, H$\beta$ and H$\alpha$ emission lines. We quantified the H$\alpha$ emission using the \texttt{specutils}\footnote{\url{https://specutils.readthedocs.io/en/stable/index.html}} \citep{specutils} Python package, obtaining an H$\alpha$ equivalent width of $-16.80\,\AA$ and $-5.90\,\AA$ for the co-added NOT/ALFOSC spectra, and of $-18.36\,\AA$ and $-6.08\,\AA$ for the co-added GTC/OSIRIS spectra of J-PLUS0807 and J-PLUS0903, respectively. These results correspond to levels of H$\alpha$ emission that are not uncommon among this type of stars \citep{Schmidt2007,Martin2010}. We found no significant differences between the equivalent width measurements for the individual spectra of each exposure. 

\subsection{Light curve analysis}
\label{sec:tess}

We queried the Mikulski Archive for Space Telescopes (MAST\footnote{\url{https://mast.stsci.edu/portal/Mashup/Clients/Mast/Portal.html}}) to fetch high-cadence photometric data for our sample. We found two-minutes cadence TESS calibrated LCs for the two closest stars, J-PLUS0807 and J-PLUS0903, with TIC IDs 461654150 and 166597074, respectively. Table \ref{tab:TESS_data} shows the details of the retrieved LCs, which are processed using the pipeline developed by the Science Processing Operations Centre \cite[SPOC;][]{spoc}. The contamination ratio, \texttt{Rcont}, listed in the TESS Input Catalog \citep{tic8_2} is 10\% and 0.22\% for J-PLUS0807 and J-PLUS0903, respectively. To further study a possible contamination of the TESS photometry for these two stars, we used the \texttt{tpfplotter}\footnote{\url{https://github.com/jlillo/tpfplotter}} \citep{tpfplotter} tool to explore the target pixel files (TPFs) of the fields of our targets. Thus, we only found a $\sim1\%$ contamination, obtained from the difference in \textit{Gaia} magnitudes, from \textit{Gaia} sources within the photometric apertures selected by the SPOC pipeline to process the LCs of the two stars.

\begin{table}
\fontsize{11pt}{11pt}\selectfont
 \caption{Details of two-minutes cadence TESS LCs used in this work.}
 \label{tab:TESS_data}
 \centering  
 \begin{tabular}{l c c}
  \hline\hline
  \noalign{\smallskip}
  
  TIC ID & TESS Sectors & Observation length\\
   & & [d]\\
  
  \noalign{\smallskip}
  \hline
  \noalign{\smallskip}
  
  461654150 & 20, 44, 45, 46, and 47 & 112.41 \\
  
  166597074 & 21 & 23.96 \\

  \noalign{\smallskip}
  \hline
 \end{tabular}
\end{table}

We identified clear evidence of flaring activity and periodic variability in the retrieved two-minutes cadence LCs for J-PLUS0807 and J-PLUS0903, which are analysed in detail in Sections \ref{sec:flares} and \ref{sec:periods}. For the remaining five stars in our sample, that do not have processed, short-cadence TESS data, we used the Python package \texttt{lightkurve} \citep{lightkurve} to manually extract LCs from the TESS Full Frame Images cutouts, but did not find any flare events or periodic variability signals. We also searched for time-resolved UV data from the NASA Galaxy Evolution Explorer \citep[GALEX;][]{galex} mission for our targets, using the \texttt{gPhoton} \citep{gphoton} database and software, but we did not find any.


\subsubsection{Flares}
\label{sec:flares}

\begin{figure}
\centering
	\includegraphics[width=.85\columnwidth]{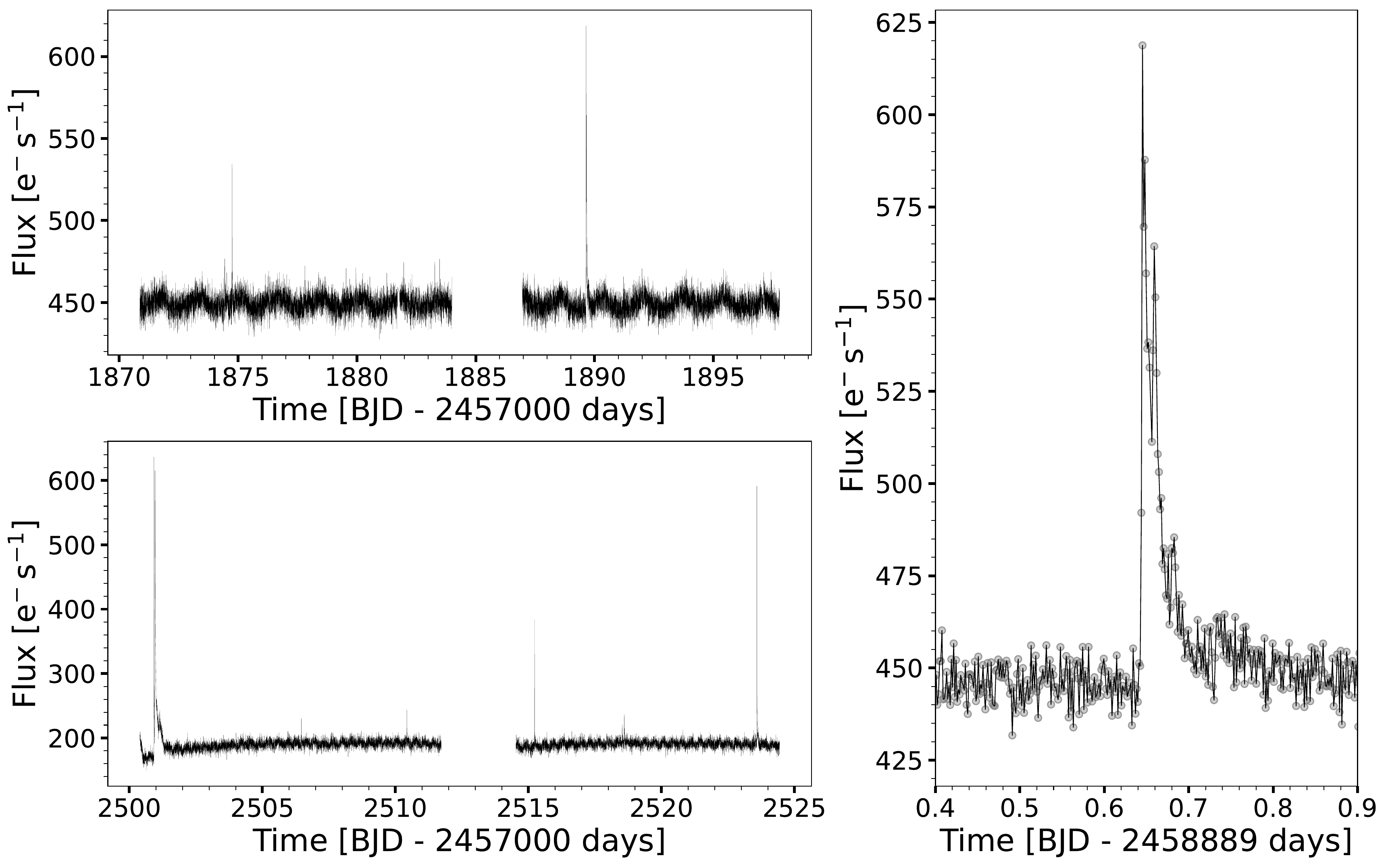}
    \caption{LCs of J-PLUS0903 (sector 21, \textit{top left panel}) and J-PLUS0807 (sector 44, \textit{bottom left panel}). The \textit{right panel} shows the largest flare event of the J-PLUS0903 LC.}
    \label{fig:flares_lc}
\end{figure}

For our analysis, we used the Pre-search Data Conditioning Simple Aperture Photometry \citep[PDCSAP;][]{smith2012,stumpe2012,stumpe2014} flux, available in the TESS LCs retrieved for J-PLUS0807 and J-PLUS0903, which is already corrected from long-term trends, instrumental effects, and excess flux due to starfield crowding. We removed all data points with non-zero quality flags, after visually verifying that points with 512 (`Impulsive outlier removed before cotrending') or 1024 (`Cosmic ray detected on collateral pixel row or column') quality flags were not actually part of a real flare. We identified several flare events in all the J-PLUS0807 and J-PLUS0903 TESS LCs. For example, Figure \ref{fig:flares_lc} shows the LC of J-PLUS0903 (top left panel) and one of the LCs of J-PLUS807 (sector 44, bottom left panel), with multiple flaring episodes observed in both of them. Moreover, the right panel provides a zoomed-in view of the flare event occurring around day 1890 (BJD - 2457000 days) in the J-PLUS0903 LC.

We used the open-source Python software  \texttt{AltaiPony}\footnote{\url{https://altaipony.readthedocs.io/en/latest/}} \citep{davenport2016,ilin2021} to automatically identify and characterise flares in the LCs. Prior to flare detection, we detrended the LCs using a Savitzky-Golay filter \citep{Savitzky1964} to remove rotational modulation trends. For flare detection, we followed the same procedure as \citet{davenport2014,doyle2019,doyle2022,kumbhakar2023}, identifying flares as two or more consecutive points that are $2.5\sigma$ above the local scatter of the data \citep{chang2015}. As reported by \citet{vida2017,doyle2018}, we found no obvious relationship between flaring activity and rotational phase. \texttt{AltaiPony} automatically determines several flare properties, such as start and end times, flare amplitude, and equivalent duration (ED), which is the area under the flare light curve in units of seconds. Using the observed NOT/ALFOSC spectra and the tool \texttt{Specphot}\footnote{\url{http://svo2.cab.inta-csic.es/theory/specphot/}} \citep{specphot}, developed and maintained by the Spanish Virtual Observatory\footnote{\url{http://svo2.cab.inta-csic.es}}, we obtained the star quiescent flux in the TESS bandpass. We relied on the calculated flux and \textit{Gaia} distances of our targets to derive the quiescent stellar luminosity and multiplied it by the ED to obtain the flare energy in the TESS bandpass. We obtained $L_{\rm TESS}=2.3\times10^{29}$~erg\,s$^{-1}$ and $L_{\rm TESS}=2.2\times10^{30}$~erg\,s$^{-1}$ for the quiescent luminosity in the TESS bandpass of J-PLUS0807 and J-PLUS0903, respectively. Table \ref{tab:flares} details the flare properties for each target.

\begin{table}
\fontsize{11pt}{11pt}\selectfont
 \caption{Detailed flare properties of our targets with TESS processed LCs.}
 \label{tab:flares}
 \centering          
 \begin{tabular}{l c c c c c}
  \hline\hline
  \noalign{\smallskip}
  
  Object & Sector & Number of flares & $\log(E)$ range & Duration range & Flare rate \\
   & & & [erg] & [min] & [d$^{-1}$]\\
  
  \noalign{\smallskip}
  \hline
  \noalign{\smallskip}
  
  J-PLUS0807 & 20 & 5 & 30.9--32.1 & 4.0--32.0 & 0.22\\
  
  J-PLUS0807 & 44 & 6 & 31.2--33.4 & 6.0--416.0 & 0.28\\
  
  J-PLUS0807 & 45 & 5 & 30.9--32.9 & 6.0--54.0 & 0.23\\

  J-PLUS0807 & 46 & 7 & 31.1--32.6 & 4.0--36.0 & 0.30\\

  J-PLUS0807 & 47 & 4 & 31.0--33.3 & 6.0--368.0 & 0.17\\

  J-PLUS0903 & 21 & 4 & 31.6--33.0 & 4.0--60.0 & 0.17\\

  \noalign{\smallskip}
  \hline
 \end{tabular}
\end{table}
   
\begin{figure}
    \centering
	\includegraphics[width=.75\columnwidth]{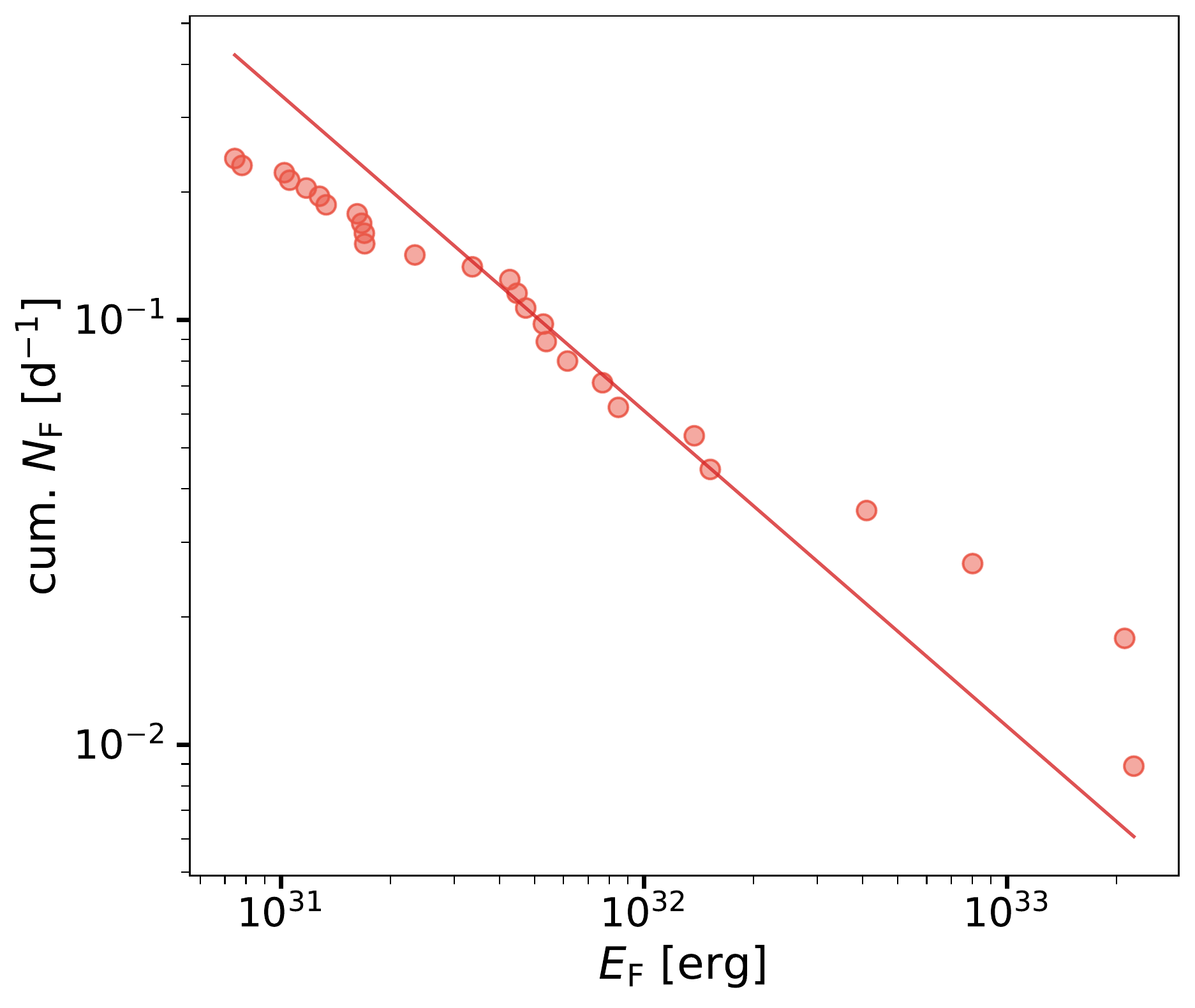}
    \caption{Cumulative FFD for J-PLUS0807. The red solid line represents the power-law fit obtained with \texttt{AltaiPony}.}
    \label{fig:ffd}
\end{figure}

The flare energy and rate obtained for our targets are typical of active, fast rotating mid- and late-M dwarfs \citep{doyle2019,ramsay2020,stelzer2022}. With the observed flares for J-PLUS0807, we built the cumulative flare frequency distribution (FFD) to study the flare rate as a function of flare energy. This was not possible for J-PLUS0903 due to the low number of events available. FFDs can be expressed as a power-law relation \citep{stelzer2007,lin2019}:

\begin{equation}    
    \frac{d \nu}{d E_{\rm F}}\sim E_{\rm F}^{-\alpha},
    \label{eq:ffd}
\end{equation}

where $\nu$ is the cumulative flare rate for a given flare energy $E_{\rm F}$, and $1-\alpha$ is the slope of a linear fit to a log-log representation. To fit the FFD, we relied on \texttt{AltaiPony}'s \texttt{FFD.fit\_powerlaw()} method, which fits the power-law parameters simultaneously using the Markov Chain Monte Carlo method described in \citet{wheatland2004}. Since the detection probability decreases in the low-energy regime, where flares may go undetected due to the noise present in the LC, we discarded the low-energy tail of the FFD in the fit of the power-law \citep{hawley2014,chang2015,ilin2021}. Figure \ref{fig:ffd} shows how the power-law relation breaks down around $E_{\rm F}=10^{31.5}$~erg, which is the threshold we applied to consider flares in the FFD fitting. Following this methodology, we obtained $\alpha=1.74_{-0.17}^{+0.20}$ for J-PLUS0807, which is in agreement with what \citet{lin2019}, \citet{raetz2020}, and \citet{murray2022} found for their samples of 548, 56 and 85 flaring M dwarfs, respectively. We found that the less energetic flares, which are more frequent as illustrated in the FFD, are also shorter in duration and the easiest to detect with the J-PLUS observation strategy (see Section \ref{sec:sample}).

\subsubsection{Rotation periods}
\label{sec:periods}

\begin{figure}
    \centering
    	\includegraphics[width=.48\columnwidth]{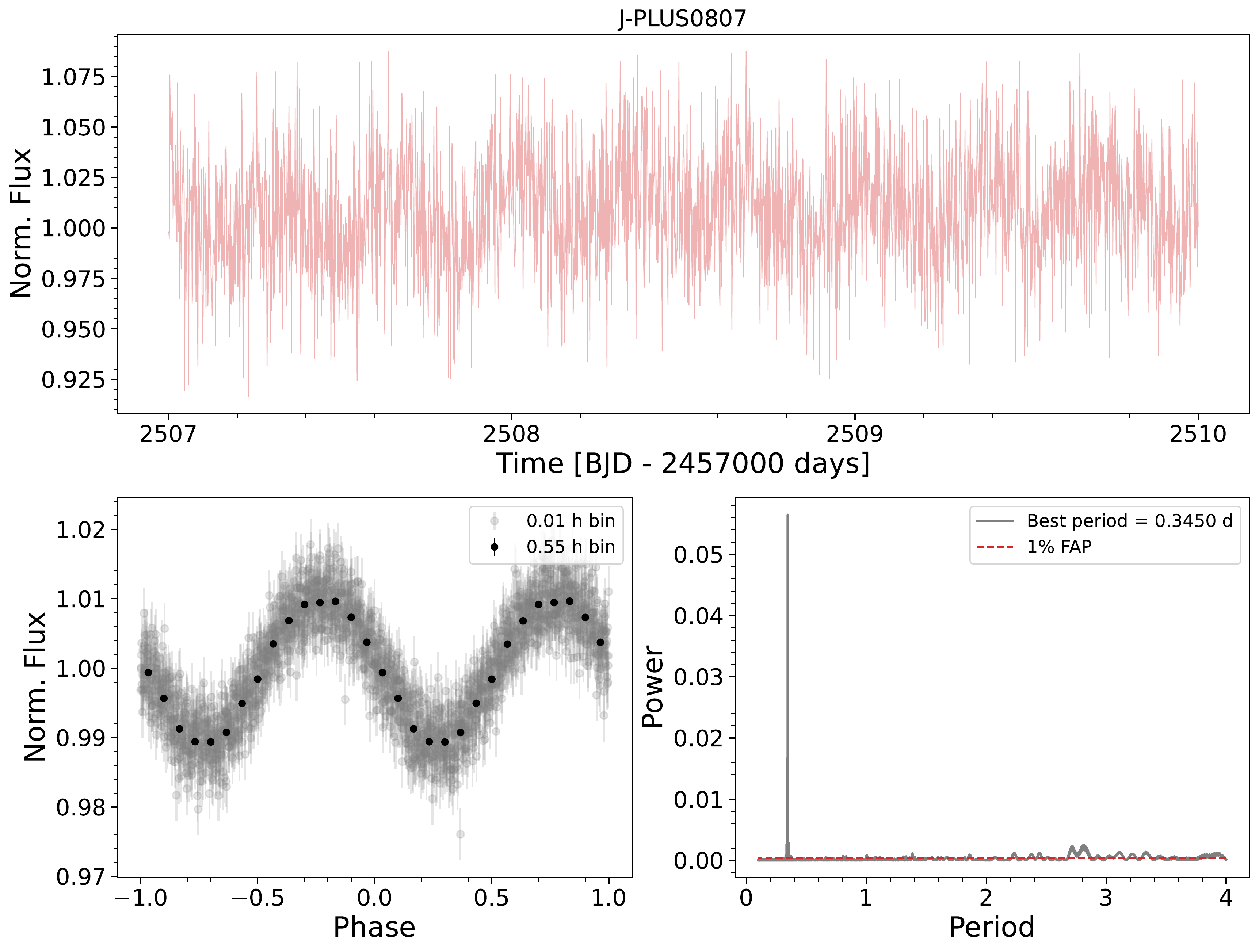}
            \includegraphics[width=.48\columnwidth]{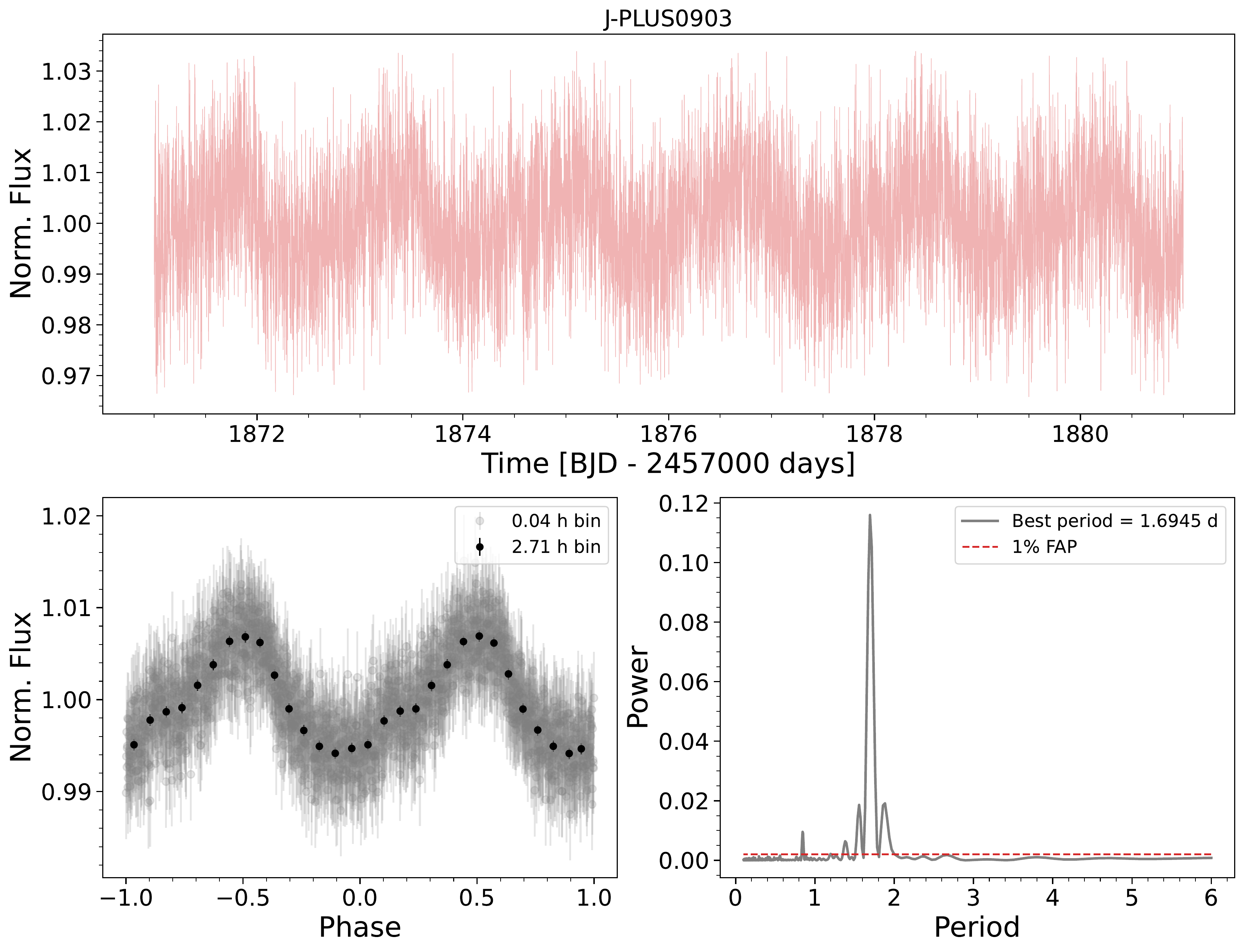}
    \caption{TESS LCs (\textit{top panels}), phase-folded LCs (\textit{bottom left panels}), and periodograms (\textit{bottom right panels}) for our two targets. The \textit{top panel} shows only a small section of the LCs for better visibility. Two different bin sizes are shown for the binned phase-folded LCs, with grey and black dots. In the periodograms, the red dashed line represents the 1\% FAP level.}
    \label{fig:periods}
\end{figure}

All TESS LCs retrieved for J-PLUS0807 and J-PLUS0903 show a clear periodic variability, which usually arises due to co-rotating star-spots that appear and disappear from the line of sight. Therefore, we relied on a Lomb-Scargle periodogram \citep{lomb,scargle}, using the \texttt{astropy} Python package \citep{astropy}, to search for the rotation period of each of our targets. Figure \ref{fig:periods} shows the periodogram for each of the  targets and the phase-folded LCs with the chosen periods, which are very prominent in the periodograms.

For J-PLUS0807, we computed the period using the data from all available sectors and obtained $P_{\rm rot}=0.3450$\,d, which is consistent with the values reported by \citet{guangwei2024} using TESS data from sectors 20, 45, 46 and 47, \citet{seli2021} using only TESS data from sector 20, and \citet{newton2016}, who relied on photometry from the MEarth Project \citep{berta2012}. For J-PLUS0903, we obtained $P_{\rm rot}=1.69$\,d, which is the first estimation for the rotation period of this object. As a measure of the uncertainty of the peak position in the periodogram, we used the standard deviation of all the values with a power greater than the half height of the periodogram peak, obtaining 0.0002\,d and 0.02\,d for J-PLUS0807 and J-PLUS0903, respectively.

We confirmed that our targets were the sources of the detected variability using the \texttt{TESS\_localize}\footnote{\url{https://github.com/Higgins00/TESS-Localize}} \citep{tess_localize} Python package.

The computed rotation periods place our two targets as very fast rotators \citep{irwin2011}, which is deeply interlinked with the activity level observed. After reaching the main sequence, low-mass stars slowly spin-down due to the loss of angular momentum by stellar winds, thus undergoing a decrease in their magnetic activity over time \citep{yang2017,davenport2019,raetz2020} that may also be dependent on stellar metallicity \citep{see24}, which makes obtaining robust age estimations for low-mass stars notoriously difficult. To explore this, we relied on \texttt{stardate}\footnote{\url{https://stardate.readthedocs.io/en/latest/}} \citep{stardate}, a Python tool that combines isochrone fitting with gyrochronology for measuring stellar ages. In our case, we included magnitudes from the Two-Micron All Sky Survey \citep[2MASS;][]{2mass}, parallax and magnitudes from \textit{Gaia} DR3, magnitudes from the Sloan Digital Sky Survey \citep[SDSS;][]{sdss}, and the rotation periods obtained in this work as input parameters. Following this procedure, we obtained an age of $0.79^{+0.62}_{-0.09}$\,Gyr and $1.94^{+1.74}_{-1.26}$\,Gyr for J-PLUS0807 and J-PLUS0903, respectively, which is in agreement with the values found in the literature for fast rotators \citep{newton2016,doyle2019}. Here, the chosen value and uncertainties correspond to the median and $\pm1\sigma$ thresholds of the Markov Chain Monte Carlo samples computed by \texttt{stardate}.


\section{Planetary habitability}
\label{sec:habitability}

Understanding the impact of the magnetic activity of M dwarfs on a planet's evolution and habitability is of crucial interest in the search for Earth-like planets. The common flaring activity and CMEs, together with the nearby habitable zone of these stars, can lead to substantial alteration of planetary atmospheres or even their erosion. It is unclear whether stellar flares are beneficial or detrimental to the habitability of exoplanets. It is possible that UV radiation emitted during flare events can trigger the development of prebiotic chemistry \citep{rimmer2018,airapetian2020}. Although abiogenesis would potentially be slower compared to prebiotic Earth due to the lower emission of M dwarfs at these wavelengths \citep{rugheimer2015,ranjan2017}, flares could provide the lacking UV energy \citep{buccino2007,jackman2023}. In this line, the flare events in Ca~{\sc ii} H and K emission lines revealed in this work may play an important role.

Continued exposure to $E_{\rm bol}>10^{34}$\,erg flares would make the presence of ozone layers impossible on any habitable zone terrestrial exoplanet orbiting an M dwarf \citep{tilley2019,chen2021}. Moreover, \citet{berger2024} recently demonstrated that the 9\,000\,K blackbody commonly assumed for flares underestimated the FUV emission for 98\% of their sample, which would significantly increase the number of stars with sufficient flaring activity to fall into the ozone depletion zone from previous studies. Following the relation provided by \citet{seli2021}, we converted the TESS energies of the detected flares to bolometric flare energies. Thus, we obtained a rate of 0.02\,day$^{-1}$ for $E_{\rm bol}>10^{34}$\,erg flares for J-PLUS0807, which is an order of magnitude lower than the rate found by \citet{tilley2019} for the ozone layer to be eroded in habitable zone terrestrial exoplanets around M dwarfs. For J-PLUS0903, none of the flares exceeded this energy threshold.


\section{Conclusions} \label{sec:conclusions}

This work serves as a follow-up study of the sample of M dwarfs with strong excess emission in the J-PLUS filters corresponding to Ca~{\sc ii} H and K and H$\alpha$ emission lines, identified in our previous work \citep{masbuitrago2022}. Using low-resolution spectra collected with NOT/ALFOSC and GTC/OSIRIS, we measured the PC3 spectral index of our targets and spectroscopically confirmed the mid-M dwarf nature of six of them for the first time. We confirmed that the strong excess emission detected in the J-PLUS photometry is caused by transient flare events, suggesting that two types of flares are detected using narrow-band optical photometry in common M dwarfs, those already well-known in H$\alpha$ and those in Ca~{\sc ii} H and K presented in this work. Work dedicated to the study of flares in large M dwarf samples usually focuses only on H$\alpha$ flare events, which could lead to an underestimation of the number of flaring M dwarfs. In the future, multi-wavelength simultaneous observations will be essential to further study the flaring activity of M dwarfs.

We analysed two-minutes cadence TESS LCs for J-PLUS0807 and J-PLUS0903 and performed a thorough characterisation of the multiple flare events observed in them. We estimated the flare energies in the TESS bandpass and found them to be in the range of $7.4\times10^{30}-2.2\times10^{33}$\,erg. We found clear signs of a periodic variability in the TESS LCs, confirming the previously reported ultra-fast rotating nature of J-PLUS0807 with data from sectors 20, 44, 45, 46, and 47. Also, we computed for the first time a rotation period of 1.69\,d for J-PLUS0903.

This work demonstrates the potential of multi-filter photometric surveys such as J-PLUS or the upcoming J-PAS to systematically detect flare events in M dwarfs, especially episodes of strong Ca~{\sc ii} H and K line emission that may have important implications for exoplanetary space weather and habitability studies. Using a detection algorithm such as the one developed in \citet{masbuitrago2022}, it is possible to identify a sample of candidates that can be confirmed and analysed with spectroscopic follow-up and high-cadence photometric LCs from TESS or similar missions such as K2. It also highlights the fundamental role of stellar flares in shaping the habitability of exoplanets. A high frequency of energetic flares implies that planets around these stars may experience significant atmospheric erosion and elevated levels of surface radiation, although it could also trigger the development of prebiotic chemistry. 


\chapter{Autoencoders and Deep Transfer Learning in CARMENES}
\label{chp:autoencoders_paper}




Deep learning (DL) techniques are a promising approach among the set of methods used in the ever-challenging determination of stellar parameters in M dwarfs. In this context, transfer learning could play an important role in mitigating uncertainties in the results due to the synthetic gap (i.e. difference in feature distributions between observed and synthetic data). We propose a feature-based deep transfer learning (DTL) approach based on autoencoders to determine stellar parameters from high-resolution spectra. Using this methodology, we provide new estimations for the effective temperature, surface gravity, metallicity, and projected rotational velocity for 286 M dwarfs observed by the CARMENES survey. Using autoencoder architectures, we projected synthetic PHOENIX-ACES spectra and observed CARMENES spectra onto a new feature space of lower dimensionality in which the differences between the two domains are reduced. We used this low-dimensional new feature space as input for a convolutional neural network to obtain the stellar parameter determinations. We performed an extensive analysis of our estimated stellar parameters, ranging from 3050 to 4300\,K, 4.7 to 5.1\,dex, and $-$0.53 to 0.25\,dex for $\textit{T}_{\rm eff}$, log \textit{g}, and [Fe/H], respectively. Our results are broadly consistent with those of recent studies using CARMENES data, with a systematic deviation in our $\textit{T}_{\rm eff}$ scale towards hotter values for estimations above 3\,750\,K. Furthermore, our methodology mitigates the deviations in metallicity found in previous DL techniques due to the synthetic gap. We consolidated a DTL-based methodology to determine stellar parameters in M dwarfs from synthetic spectra, with no need for high-quality measurements involved in the knowledge transfer. These results suggest the great potential of DTL to mitigate the differences in feature distributions between the observations and the PHOENIX-ACES spectra.


\section{Context}


The precise determination of the stellar parameters of M dwarfs is crucial to improve our understanding of planetary formation and evolution, which depends fundamentally on the thorough characterisation of their host stars \citep{cifuentes2020}. However, well-established photometric and spectroscopic methods for determining these parameters encounter particular challenges, mainly due to the inherent faintness of M dwarfs and their frequent manifestation of strong stellar activity. Specifically for spectroscopic analyses, establishing the spectral continuum can be a difficult task.
Despite these problems, numerous efforts have been devoted to estimating photospheric parameters in M dwarfs, including effective temperature ($\textit{T}_{\rm eff}$), surface gravity (log \textit{g}), and metallicity ([M/H]). Several methods have proven successful in inferring these parameters, such as fitting synthetic spectra, as in  \citet[][hereafter Pass19]{pass2019} and  \citet[][hereafter Mar21]{mar21}, pseudo-equivalent widths (pEWs) \citep[e.g.][]{Mann2013,Mann2014,Neves2014}, 
spectral indices \citep[e.g.][]{RojasAyala2010,Rojas2012}, empirical calibrations \citep[e.g.][]{casagrande08,Neves2012}, interferometry \citep[e.g.][]{Boyajian2012,Rabus2019}, and machine learning  \citep[e.g.][hereafter Pass20]{Antoniadis2020,pass20}.

The approaches based on pEWs, measurements of the strength of absorption lines in a spectrum, and spectral indices, calculated from carefully chosen spectral regions --and  often derived from absorption lines or bands--, leverage their sensitivity and correlation with stellar parameters (mainly, $\textit{T}_{\rm eff}$ and [Fe/H]). As a recent example of these approaches, \citet{Khata2020} determined $\textit{T}_{\rm eff}$ and  metallicities, among other parameters, for 53 M dwarfs using \textit{H}- and \textit{K}-band pEWs and H$_{2}$O indices.
Another approach relies on empirical calibrations based on observations of M dwarfs that have an F, G, or K binary companion with known metallicity. This is grounded in the idea that the metallicity of an M dwarf is comparable to that of the hotter primary star, assuming the system originated from the same proto-stellar cloud \citep{Neves2012,montes2018,duque24}. For example, \citet{Rodriguez2019} employed the relationships of \citet{Newton2015} and \citet{Mann2013b} to derive $\textit{T}_{\rm eff}$ and metallicity, respectively, from moderate-resolution spectra of 35 M dwarfs from the \textit{K2} mission. 
Numerous spectral indices have also been empirically calibrated. For instance, \citet{Veyette2017} determined $\textit{T}_{\rm eff}$, [Fe/H], and [Ti/H] from high-resolution \textit{Y}-band spectra of 29 M dwarfs by combining spectral synthesis with empirically calibrated indices and pEWs using FGK+M systems \citep{bonfils2005,Mann2013}.

Interferometric measurements have also proven useful for deriving index-based calibrations for $\textit{T}_{\rm eff}$ \citep{Mann2013b}, performing empirical calibrations for $\textit{T}_{\rm eff}$ \citep{maldonado2015,Newton2015}, or determining $\textit{T}_{\rm eff}$ from interferometric observations in combination with parallaxes and bolometric fluxes \citep{Boyajian2012,vonBraun2014,Rabus2019}. However, their application is limited to a relatively small number of stars due to the requirement that they must be bright and nearby.

The fitting of synthetic spectra relies on a minimisation algorithm to find the synthetic spectrum that best matches the observed spectrum. Variations exist in terms of the synthetic grid employed (e.g. BT-Settl, PHOENIX-ACES, MARCS), using high or low spectral resolution, and the number and wavelength of features selected for comparison.
For example, the BT-Settl models \citep{allard2012,Allard2013} were used by \citet{GaidosMann2014} and \citet{mann2015} to derive $\textit{T}_{\rm eff}$ values for M dwarfs with low-resolution visible SNIFS (Supernova Integral Field Spectrograph) spectra, and by \citet{Rajpurohit2018} to compute $\textit{T}_{\rm eff}$, log \textit{g}, and [Fe/H] for 292 M dwarfs using high-resolution CARMENES spectra \citep{reiners2018}. \citet{Kuznetsov2019} applied BT-Settl models to intermediate-resolution spectra from the visible arm of VLT/X-shooter \citep[intermediate resolution, high-efficiency spectrograph,][]{Vernet2011} to determine $\textit{T}_{\rm eff}$, log \textit{g}, [Fe/H], and $\textit{v}\sin{i}$ for 153 M dwarfs. More recently, \citet{Hejazi2020} derived $\textit{T}_{\rm eff}$, $\log{g}$, 
metallicity [M/H], and alpha-enhancement [$\alpha$/Fe] of 1\,544 M dwarfs and subdwarfs from low- to medium-resolution spectra collected at the Michigan-Dartmouth-MIT observatory, Lick Observatory, Kitt Peak National Observatory, and Cerro Tololo Interamerican Observatory. Additionally, \citetalias{mar21} determined $\textit{T}_{\rm eff}$, log \textit{g}, and [Fe/H] for a sample of 343 M dwarfs observed with CARMENES using a Bayesian implementation of the spectral synthesis technique, the \texttt{SteParSyn}\footnote{\url{https://github.com/hmtabernero/SteParSyn}} code \citep{Tabernero2022}.

Based on the PHOENIX-ACES library \citep{Husser2013}, \citet{Birky2017} derived $\textit{T}_{\rm eff}$, log \textit{g}, and [Fe/H] for late-M and early-L dwarfs from high-resolution near-infrared APOGEE spectra \citep{Wilson2010}. Similarly, \citet{pass18} and  \citet[][hereafter Schw19]{schw19} determined these parameters for M dwarfs observed with CARMENES in the visible wavelength region. Building upon these works,  \citetalias{pass2019} extended the analysis by determining $\textit{T}_{\rm eff}$, log \textit{g}, and [Fe/H]  not only from the visible range covered with CARMENES but also from the near-infrared and the combination of visible and near-infrared data. The comparison conducted in \citetalias{pass2019} led to the conclusion that utilising both spectral ranges for parameter determination maximises the amount of available spectral information while minimising possible effects caused by imperfect modelling.
The MARCS model atmospheres \citep{Gustafsson2008} have also been employed to compute photospheric parameters. For instance, in a recent study by \citet{Souto2020}, $\textit{T}_{\rm eff}$, log \textit{g}, and [Fe/H] were determined for 21 M dwarf mid-resolution APOGEE \textit{H}-band spectra using MARCS models and the {\ttfamily turbospectrum} code \citep{Plez2012} through the {\ttfamily bacchus} wrapper \citep{Masseron2016}. 
Similarly, \citet{Sarmento2021} derived $\textit{T}_{\rm eff}$, log \textit{g}, [M/H], and microturbulent velocity $v_{\rm mic}$ for 313 M dwarfs from APOGEE \textit{H}-band spectra using MARCS models, {\ttfamily turbospectrum}, and {\ttfamily iSpec} python code \citep{BlancoCuaresma2014}.

As large surveys release extensive databases containing thousands of stars, there is a need for flexible and automated methods capable of handling vast amounts of data to infer stellar atmospheric parameters. In this sense, machine learning (ML) techniques have also been used for determining photospheric parameters for M dwarfs from stellar spectra. For example, \citet{Sarro2018} proposed an automated procedure based on genetic algorithms to identify pEWs and integrated flux ratios from BT-Settl models that yield good estimations of $\textit{T}_{\rm eff}$, log \textit{g}, and [M/H] for spectra from the NASA Infrared Telescope Facility (IRTF). Also based on pEWs, \citet{Antoniadis2020} present an ML tool, named {\ttfamily ODUSSEAS}, to derive $\textit{T}_{\rm eff}$ and [Fe/H] of M dwarf stars from 1D spectra for different resolutions. In \citet{Birky2020}, {\ttfamily The Cannon} \citep{Ness2015,Casey2016}, a data-driven spectral-modelling and parameter-inference framework, is used to estimate $\textit{T}_{\rm eff}$ and [Fe/H] for 5\,875 M dwarfs in the APOGEE \citep{Abolfathi2018} and {\it Gaia} DR2 \citep{GaiaDR2} surveys. Using the Stellar LAbel Machine \citep[SLAM,][]{Zhang2020}, \citet{Li2021} trained a model with APOGEE stellar labels and synthetic spectra from the BT-Settl model, resulting in the determination of $\textit{T}_{\rm eff}$ and [M/H] for M dwarfs from the LAMOST DR6\footnote{\url{http://dr6.lamost.org/}} catalogue.

This study extends previous works on applying deep learning (DL) to predict stellar parameters from high-resolution spectra observed with CARMENES. \citetalias{pass20} presented a DL approach where convolutional neural networks (CNNs) were trained on synthetic PHOENIX-ACES models to estimate $\textit{T}_{\rm eff}$, log \textit{g}, [M/H], and $\textit{v}\sin{i}$ for 50 M dwarfs observed with CARMENES. After a thorough analysis of their methodology, in which different architectures and spectral windows were tested, they found that all DL models were able to estimate stellar parameters from synthetic spectra in a precise and accurate way. However, when testing these models on the CARMENES spectra, they found significant deviations for the metallicity because of the synthetic gap \citep{fabbro2018,Tabernero2022}, which is the difference in feature distributions between synthetic and observed data.
In a more recent study, \citet[][hereafter Bello23]{bello2023} employed a deep transfer learning (DTL) approach to mitigate the uncertainties associated with the synthetic gap (see their Figs. 1 and 2). Following the training of DL models on a large set of synthetic spectra from the PHOENIX-ACES model, the models underwent fine-tuning based on external knowledge about stellar parameters. This external knowledge included 14 stars from the CARMENES survey with interferometric angular diameters measured by \citet{Boyajian2012}, \citet{vonBraun2014}, and references therein. Additionally, it was supplemented with five mid-to-late M dwarf stars from \citet{passegger2022}. They achieved the determination of new $\textit{T}_{\rm eff}$ and [M/H] values for 286 M dwarfs from the CARMENES survey, and although this approach improved the estimation of $\textit{T}_{\rm eff}$ and [M/H] for M dwarfs from high-resolution spectra obtained with CARMENES, the lack of sufficiently large number of reference stars to transfer knowledge is a limitation for the technique. If the reference dataset is limited in size, diversity, or representation across the parameter space, the models may not generalise well to a broader range of M dwarfs.

In this work, we present a novel transfer learning approach for estimating photospheric parameters in M dwarfs based on their stellar spectra. The primary goal of the proposed method is to address the aforementioned limitation identified by \citetalias{bello2023} by eliminating the requirement for interferometric values in the knowledge transfer process.
To achieve this, instead of employing a model-based transfer learning approach, as in \citetalias{bello2023}, where the transferred knowledge is encoded into model parameters, priors or model architectures, we propose a feature-based transfer learning. In this approach, the knowledge to be transferred can be considered as the learned feature representation. The idea is to learn a `good' feature representation so that, by projecting data onto the new representation, the differences between domains (source and target, i.e. synthetic and observed spectra in our case) can be reduced. This allows the source domain labelled data (synthetic spectra with known parameters) to be used to train a precise model for the target domain constituted by the observed spectra \citep{yang2020}.



\section{Data} \label{acs_sec:data}

The proposed approach was tested using the same sample spectra as \citetalias{pass2019}. This sample, listed in their Table B.1, comprise 282 M dwarfs observed with CARMENES. Additionally, four more stars from an independent interferometric sample, as described by \citetalias{bello2023}, were included.

CARMENES is installed at the Calar Alto Observatory, located in Spain, and stands as one of the leading instruments in the quest for searching for Earth-like planets within the habitable zones around M dwarfs. It comprises two separate spectrographs: one for the visible (VIS) wavelength range (from 520 to 960\,nm) and the other for the near-infrared (NIR) range (from 960 to 1710\,nm), each offering high-spectral resolutions of R\,$\approx$\,94\,600 and 80\,500, respectively \citep{Quirrenbach20,reiners2018}. 

A detailed description of the data reduction procedure is available in \citet{Zechmeister14}, \citet{Caballero2016}, and \citetalias{pass2019}.
Similar to the latter, we used a high signal-to-noise (S/N) template spectra for each star. These templates are generated as byproducts of the CARMENES radial-velocity pipeline, known as {\tt serval} 
\citep[SpEctrum Radial Velocity AnaLyser;][]{Zechmeister2018}. In the standard data flow, the code constructs a template for each target star from a minimum of five individual spectra to derive the radial velocities through least-square fitting to the template. 
The S/N of the observed CARMENES sample used in this work was above 150. Concerning the wavelength window, we adopted the range 8\,800--8\,835\,\AA, consistent with \citetalias{bello2023}, as this window displayed the smallest mean squared error among all the investigated windows in \citetalias{pass20}.

To train the neural network models, we utilised the PHOENIX-ACES spectra library\footnote{\url{https://phoenix.astro.physik.uni-goettingen.de/}} \citep{Husser2013}. This library is chosen for its consideration of spectral features present in cool dwarfs. Furthermore, the use of synthetic models enables the generation of a large number of spectra with known parameters, eliminating the need for limited samples of observations with well-known stellar parameters. We used the same PHOENIX-ACES grid as in previous works (\citetalias{pass20}; \citetalias{bello2023}), which was generated by linearly interpolating between the existing grid points using {\ttfamily pyterpol} \citep{Nemravov2016}. The complete dataset contains a grid of 449\,806 synthetic high-resolution spectra between 8\,800\,{\AA} and 8\,835\,{\AA} with $\textit{T}_{\rm eff}$ between 2\,300 and 4\,500\,K (step 25\,K), log \textit{g} between 4.2 and 5.5\,dex (step 0.1\,dex), [M/H] between -1.0 and 0.8\,dex (step 0.1\,dex), and $\textit{v}\sin{i}$ between 1.5 and 60.0\,km\,s$^{-1}$ (with a variable step of 0.5, 1.0, 2.0 or 5.0; see Table 1 in \citetalias{pass20}). A degeneracy between $\textit{T}_{\rm eff}$, log \textit{g}, and [Fe/H] was described by \citet{pass18}, who found exceptionally high values of log \textit{g} and [Fe/H] for well-fitting PHOENIX-ACES models. This degeneracy was further underscored by \citetalias{pass2019} and \citetalias{pass20} during the application of DL models to the observed CARMENES spectra, and the latter imposed additional constraints to the grid leveraging the PARSEC v1.2S evolutionary models \citep{parsec2012,Chen2014,Chen2015,Tang2014}. Degeneracies between stellar parameters are often found when fitting synthetic spectra, and some authors have explored several ways to help break them \citep{buzzoni2001,brewer2015}. The refinement performed by \citetalias{pass20} aimed to exclude parameter combinations for M dwarfs that do not fit the main sequence, as discussed in Section 4.2 of their work. Notably, \citetalias{pass20} demonstrated that the imposition of these constraints on the synthetic model grid used in the training of the DL models is capable of breaking the observed parameter degeneracy. After applying these restrictions, the grid includes 22\,933 PHOENIX-ACES spectra.

Due to the negligible presence of telluric features in the investigated range, telluric correction was not applied to the VIS spectra.
For normalisation, we employed the Gaussian Inflection Spline Interpolation Continuum ({\tt GISIC}\footnote{\url{https://pypi.org/project/GISIC/}}), the same method and routine used by \citetalias{pass20} and developed by D.\,D.~Whitten, designed for spectra with strong molecular features. 
Following the same approach as \citetalias{bello2023}, we applied this procedure to both observed and synthetic spectra within the spectral window 8800--8835\,\AA{} with an additional 5\,\AA{} on each side to mitigate potential edge effects. Moreover, the observed spectra underwent radial velocity correction to align with the rest frame of the synthetic spectra, achieved through cross-correlation \citep[\texttt{crosscorrRV} from PyAstronomy,][]{Czesla2019} between a PHOENIX model spectrum and the observed spectrum. To ensure a universal wavelength grid, essential for applying the proposed method, the wavelength grid of the observed spectra was linearly interpolated with the grid of the synthetic spectra.

\begin{figure}
    \centering
	\includegraphics[width=.8\columnwidth]{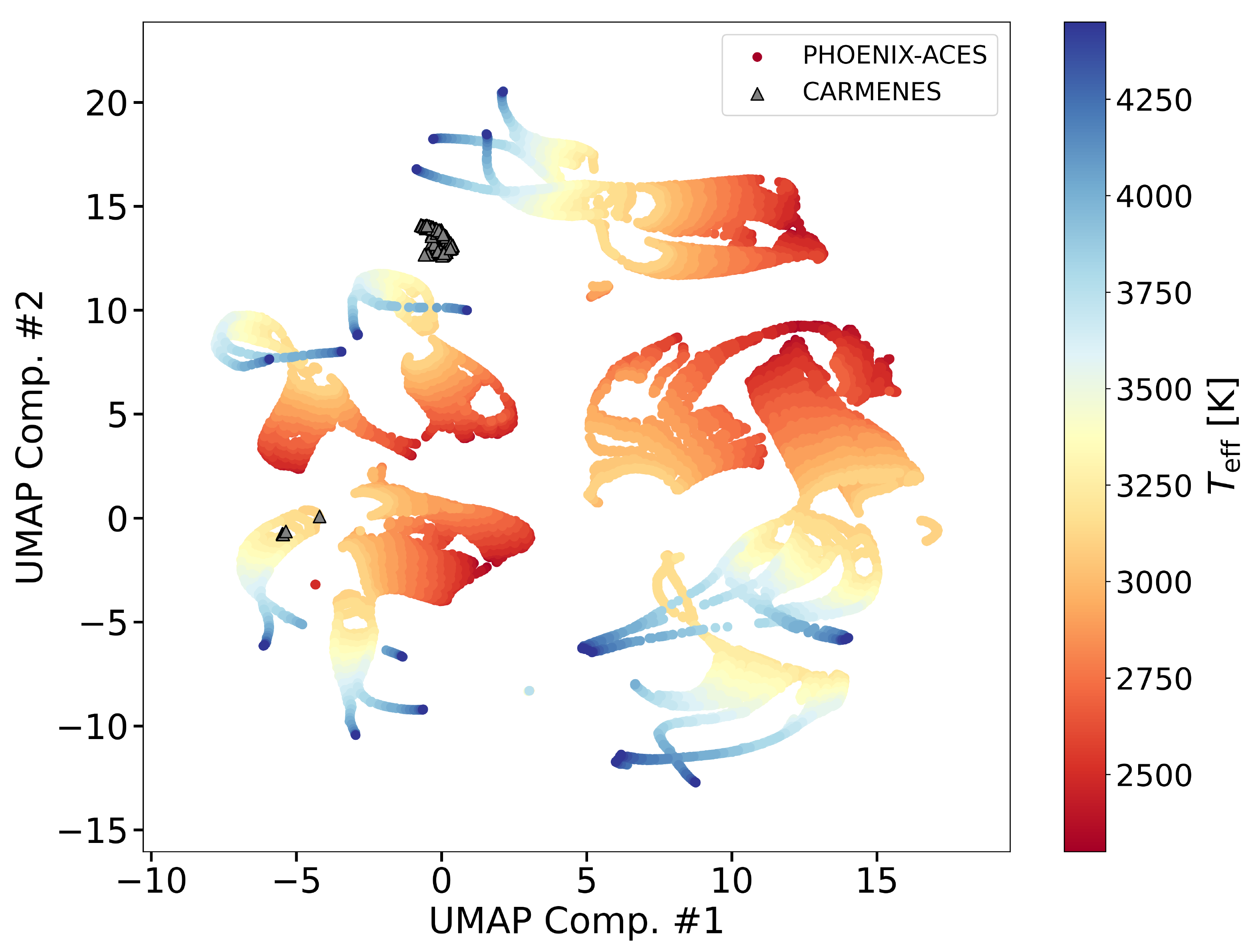}
    \caption{Two-dimensional UMAP projection of PHOENIX-ACES (dots colour-coded by $\textit{T}_{\rm eff}$) and CARMENES (grey triangles) spectra from the 8\,800--8\,835\,$\AA$ window. Almost all CARMENES spectra are isolated from the PHOENIX-ACES family feature space.}
    \label{fig:umap_flux}
\end{figure}

In spite of the performed spectra preparation, differences in the feature distributions of the synthetic and observed sets of spectra (i.e. synthetic gap) were identified. We used the Uniform Manifold Approximation and Projection \citep[UMAP;][]{McInnes2018}, with a metric that considers the correlation between the spectra, to project the high-dimensional input space (3\,500 flux values for each spectrum) into a two-dimensional space while preserving inter-distances. As shown in Fig. \ref{fig:umap_flux}, akin to \citetalias{pass20} and \citetalias{bello2023}, most of the CARMENES spectra (grey triangles) do not align precisely within the synthetic spectra (colour-coded dots). Thus, a transfer learning approach appears appropriate to extend the applicability of the regression models trained with the synthetic spectra to the observed spectra. 

\section{Methodology} \label{acs_sec:methodology}

The DTL approach proposed in this paper can be summarised as follows. Initially, we extract a low-dimensional representation of synthetic spectra based on the PHOENIX-ACES library using autoencoders (AEs), a special kind of neural network initially proposed for dimensionality reduction \citep{hinton2006b}. Then, the knowledge transfer process is performed by fine-tuning these AEs with high-resolution spectra observed with the CARMENES instrument. It must be noted that no stellar parameters were used during this re-training. With the low-dimensional representations of the synthetic spectra resulting from the initial step, we trained CNNs. Finally, using these CNNs, we estimated the stellar parameters ($\textit{T}_{\rm eff}$, log \textit{g}, [M/H], and $\textit{v}\sin{i}$) for 286 CARMENES M dwarfs by using their low-dimensional representations obtained after the fine-tuning step.

\subsection{Feature extraction using an autoencoder} \label{acs_sec:ac}

In this study, we explore unsupervised feature extraction from stellar spectra using AEs to facilitate feature-based transfer learning and leverage the new representations for estimating photospheric parameters. 
Belonging to representation learning --a subfield of machine learning--, AEs have the capability to capture the underlying factors hidden in the observed data \citep{Bengio13,Goodfellow16}. 
They have been succesfully used in various astrophysical applications, including unsupervised feature learning from galaxy spectral energy distribution \citep{FronteraPons17}, learning of non-linear representations from rest-frame spectroscopic data for redshift estimation \citep{FronteraPons19}, galaxy classification \citep{Cheng21}, astrophysical component separation \citep{Milosevic21}, reconstruction of missing magnitudes from observed objects before classifying them into stars, galaxies, and quasars \citep{Khramtsov21}, and telluric correction \citep{Kjarsgaard23}. 
In addition, some authors have used AEs to estimate stellar atmospheric parameters from spectra \citep{Yang2015,Li2017}. However, their approach is different from our proposal since the training of the models was performed in a supervised manner: spectra from SDSS/SEGUE DR7 \citep{Abazajian2009} were used, and $\textit{T}_{\rm eff}$, log \textit{g}, and [Fe/H] were obtained from the SDSS/SEGUE Spectroscopic Parameter Pipeline \citep[SSPP;][]{SSPP01,SSPP02,SSPP03,SSPP04} for stars in the temperature range 4\,088-9\,747\,K (earlier than our CARMENES targets). 
In our case, we are interested in the use of AEs to enable transfer learning, as representation learning enables the transfer of knowledge when there are features useful for different settings or tasks that correspond to underlying factors appearing in more than one setting \citep{Goodfellow16}.

\begin{figure*}
    \centering
	\includegraphics[width=\columnwidth]{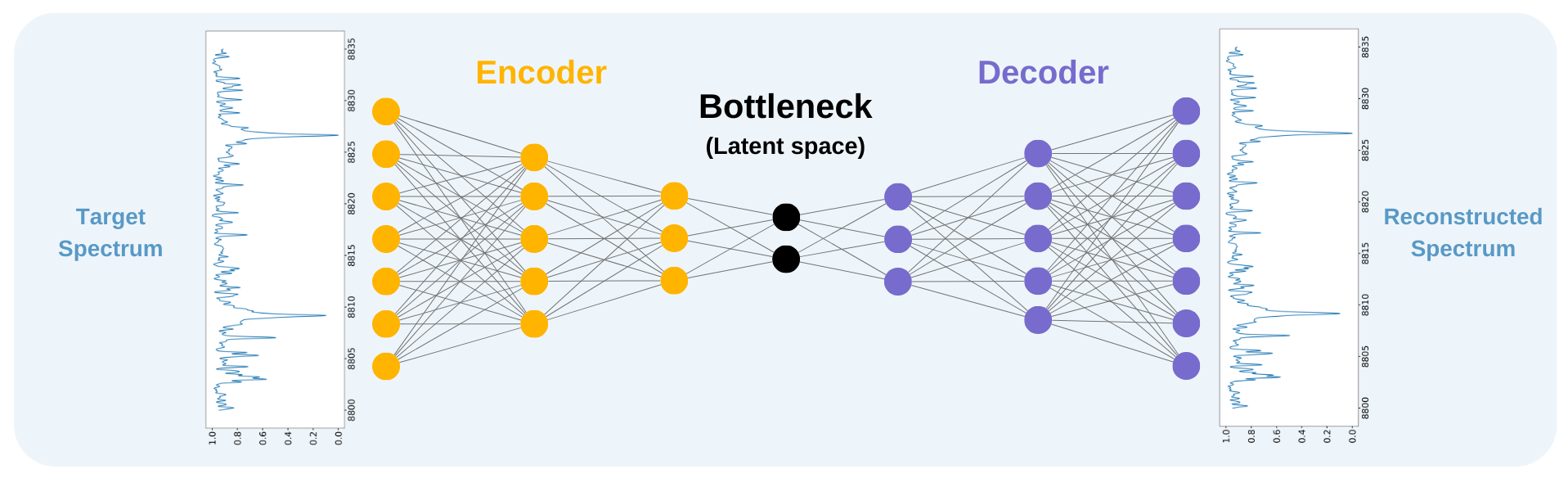}
    \caption{Schematic representation of the AE architecture used in this work.}
    \label{fig:ac_info}
\end{figure*}

The rationale behind the first step of our methodology is to find a meaningful low-dimensional representation, referred to as the latent space, of the synthetic spectra. To accomplish this, we employed an AE, which consists of an `encoder' trained to transform the high-dimensional spectrum into a low-dimensional code, and a `decoder' trained to reconstruct the original spectrum as accurate as possible from its lower-dimensional latent space (see Fig. \ref{fig:ac_info}).

First, we divided the grid of synthetic spectra into a training set (70\,\%) and a test set (30\,\%). We considered multiple AE architectures, developing a python code to create a flexible AE structure. The number of neurons on each layer, the L1 regularisation term for the dense layers (used to prevent overfitting), and the learning rate for the Adam optimisation \citep[a computationally efficient stochastic gradient descent method,][]{adam} were passed as parameters. For this code, we relied on the \texttt{Keras}\footnote{\url{https://keras.io/about/}} \citep{keras} deep learning API, which runs on top of the \texttt{Tensorflow}\footnote{\url{https://www.tensorflow.org/}} \citep{tensorflow} machine learning platform. Next, we created a grid for these hyperparameters and performed an exhaustive search using the \texttt{GridSearchCV} class from the \texttt{scikit-learn}\footnote{\url{https://scikit-learn.org/stable/}} package, which optimises the hyperparameters of an estimator through k-fold cross-validation, using any scoring metric to evaluate the model. In our case, we used 4-fold cross-validation and the mean squared error between the reconstructed and the original validation data as the scoring metric. To integrate our python code into a \texttt{scikit-learn} workflow, we used the \texttt{KerasRegressor} wrapper from the \texttt{scikeras}\footnote{\url{https://adriangb.com/scikeras/stable/}} python package.

After this search for the best hyperparameter combinations, we only kept those with a mean cross-validation score below the median, evaluated using the entire grid. We trained an AE for each of these architectures, adding a contractive regularisation term in the loss function, consisting of the squared Frobenius norm of the Jacobian matrix of the encoder activations with respect to the input:

\begin{equation}
    \left \|\,J_{f}\,(x)\, \right \|_{F}^{2}=\sum_{ij}\left ( \frac{\partial h_{j}\,(x)}{\partial x_i} \right)^2,
	\label{eq:contractive_loss}
\end{equation}

\noindent where $f$ represents the encoding function that maps the input $x$ to the hidden representation $h$. The main idea of contractive AEs is to make the feature extraction more robust to small perturbations in the training data. In the overall loss function optimisation, the trade-off between the reconstruction and the L1 regularisation terms will retain the important variations in the latent space for the reconstruction of the input \citep{rifai2011}.

\begin{figure*}
    \centering
	\includegraphics[width=\columnwidth]{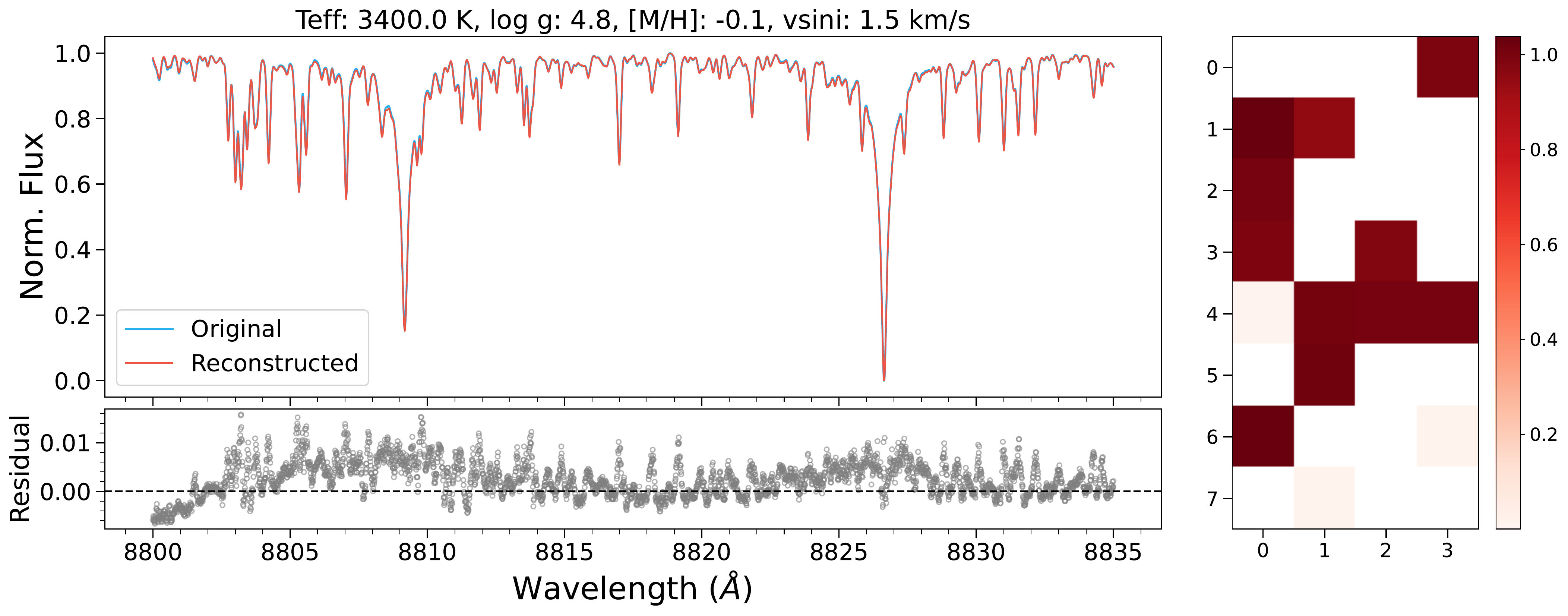}
    \caption{Reconstructed spectrum (\textit{left}) and latent representation (\textit{right}) of a PHOENIX-ACES synthetic spectrum for one of the trained AEs. \textit{Left panel:} comparison of the original (blue) and reconstructed (red) spectrum. Both spectra overlap as they are almost similar. The title shows the stellar parameters of the synthetic spectrum. Reconstruction residuals (original$-$reconstructed) are shown in the \textit{bottom panel}. \textit{Right panel:} 32-dimensional latent space of the input spectrum obtained by the encoder, reshaped to a 8$\times$4 matrix only for a better visibility. The colour scale indicates the strength of the features. The decoder uses this compressed representation to obtain the reconstructed spectrum.}
    \label{fig:ph_rec}
\end{figure*}

We only kept the AEs with a learning rate equal to 0.0001, as we found that some of them with a higher learning rate were not able to converge properly, leading to a poor latent representation of the spectra. With this, we ended up with 26 final AE architectures and evaluated them on the test set, obtaining mean squared reconstruction errors $\sim 5\cdot10^{-5}$. Fig. \ref{fig:ph_rec} shows the reconstruction and the latent space of a PHOENIX-ACES synthetic spectrum for one of the AEs. Using the encoder networks of the AEs, we obtained 26 sets (one for each AE) of 32-dimensional compressed representations for the grid of synthetic spectra.

\subsection{Deep transfer learning} \label{acs_sec:dtl}

The dependence of DL algorithms on massive training data is a crucial hurdle to overcome when a research scenario requires labelled data. In some fields, such as astrophysics, building a large, annotated data set can be incredibly complex and expensive. A straightforward and widely used solution to this problem is the use of synthetic data to train the DL models, but this may include a systematic error in the methodology if the synthetic gap (see Section \ref{acs_sec:data}) is significant, as is the case in this work.

Transfer learning (TL) plays a key role in solving the above problems, as it allows knowledge to be transferred from a rich source domain to a related but not identical target domain. The transition from TL to DTL, with incomplete DTL as an intermediate stage \citep[deep neural networks are only used as feature extractors in TL models;][]{yu2022}, came with the integration of DL techniques into the TL paradigm.

In the context of TL, a domain can be represented as $D=\left \{\mathcal{X},P(X)  \right \}$, where $\mathcal{X}$ denotes a feature space and $P(X)$ represents the marginal probability distribution for $X=\left \{x_1,...,x_n \right \} \in \mathcal{X}$. Also,  a task can be represented as $T=\left \{Y, f(\cdot) \right \}$, where $Y$ denotes a label space and $f(\cdot)$ is a predictive function. According to the definition provided by \citet{pan2010}, given a source domain $D_{\rm S}$ and task $T_{\mathrm{S}}$, and a target domain $D_{\mathrm{T}}$ and task $T_{\mathrm{T}}$, TL aims to enhance the performance of a predictive function $f_{\mathrm{T}}(\cdot)$ in $D_{\mathrm{T}}$, using the knowledge available in $D_{\mathrm{S}}$ and $T_{\mathrm{S}}$, where $D_{\mathrm{S}}\neq D_{\mathrm{T}}$ and/or $T_{\mathrm{S}}\neq T_{\mathrm{T}}$. In our work, the source domain is represented by the grid of synthetic PHOENIX-ACES spectra, while the target domain is built from the 286 CARMENES observed spectra. Moreover, the predictive function is defined as the encoder network of the AE architecture, responsible for compressing the input spectra into the low-dimensional latent representation.

The purpose of this step in the methodology is to adopt a DTL-based strategy, in particular the fine-tuning approach \citep{brian2016,yosinski2014}, using the AE architectures we already trained in the source domain to obtain a meaningful low-dimensional latent representation of our data-poor target domain. In this process, we kept the weights frozen in all encoder layers until the last one, leaving only the deepest encoder layer, the bottleneck (i.e. the latent space or compressed representation of the spectrum, as illustrated in Fig. \ref{fig:ac_info}), and the decoder network to be re-trained. The motivation for keeping the lower layers frozen is to prevent generic learning from being overwritten, thus preserving the knowledge acquired by the network to recognise relevant spectral features, while the more specific features are tailored to the target domain \citep{sadr2020}.

\citet{pan2018} already explored the possibility of finding a low-dimensional latent space in which source and domain data are close to each other, and using it as a bridge to transfer the knowledge from the labelled source domain to the unlabelled target domain. In our case, the ultimate goal of this process is to find a low-dimensional representation of the observed spectra that is closer to the synthetic latent representation than in the initial high-dimensional space of the spectra (see Fig. \ref{fig:umap_flux}). Furthermore, we want for these target representations to be as meaningful as possible, since we intend to use them later as a starting point for estimating the stellar parameters.

First, we divided the target set of 286 CARMENES spectra into a training set (80\,\%) and a test set (20\,\%), with the latter being used to assess the reconstruction error across the target domain. Then, we fine-tuned the 26 AE architectures, following the process explained above, obtaining mean squared reconstruction errors $\sim4\cdot 10^{-4}$ on the test set, in contrast to the reconstruction errors ($\sim3\cdot 10^{-3}$) obtained on the CARMENES set using the AEs pre-trained on the PHOENIX-ACES spectra. It must be noted that no stellar parameters were used during this re-training.

\begin{figure}
    \centering
	\includegraphics[width=\columnwidth]{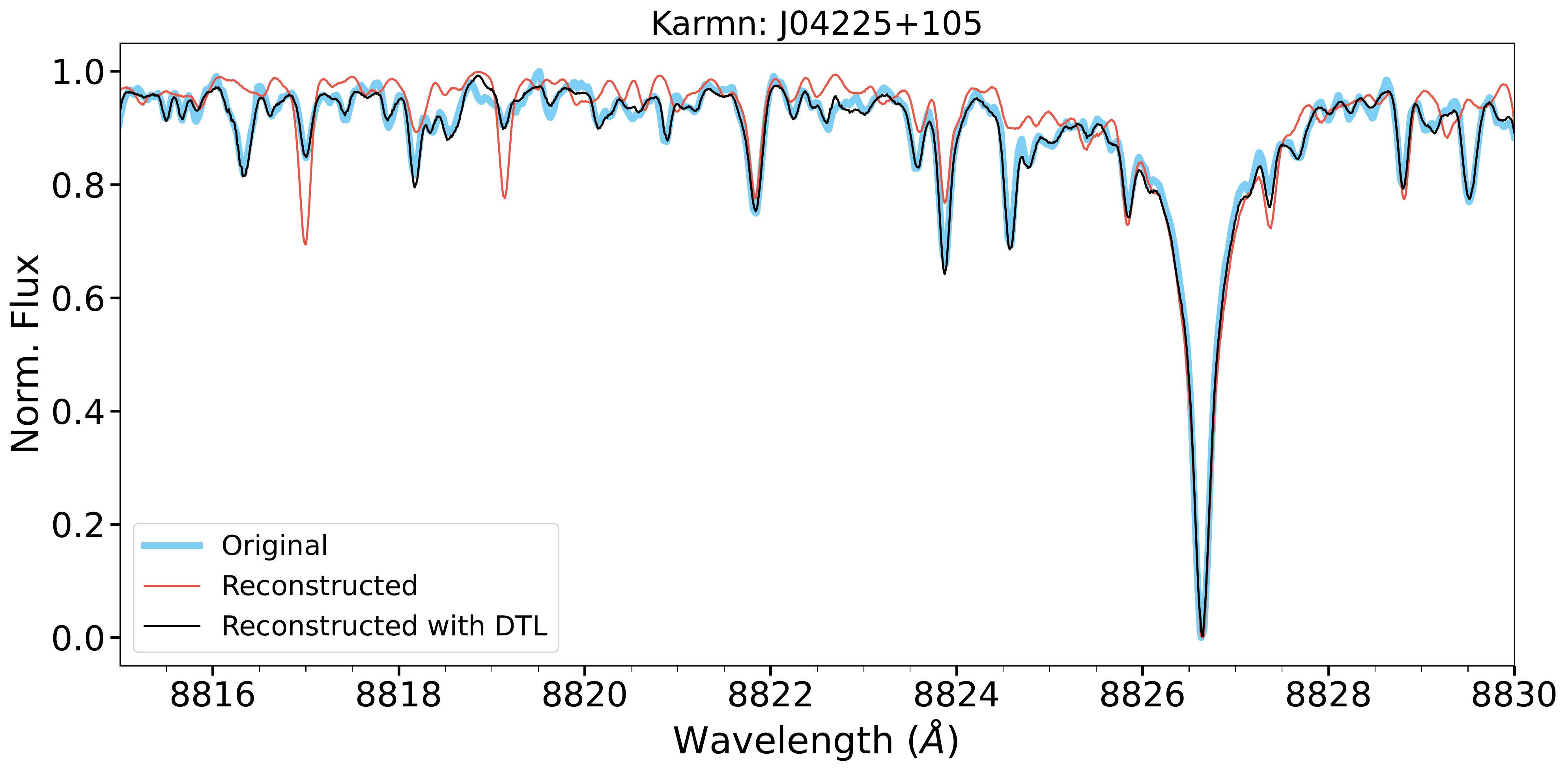}
    \caption{Original (blue) vs. reconstructed CARMENES spectrum for LSPM J0422+1031 (Karmn J04225+105, M3.5\,V). The Figure only shows a section of the spectrum for better visibility, with the unique purpose of emphasising how the reconstruction after fine-tuning (black) captures much more detailed spectral features than the reconstruction with the initial training (red).}
    \label{fig:vsdtl}
\end{figure}

Fig. \ref{fig:vsdtl} illustrates the importance of this step for the AE to effectively adapt to our specific target domain, ensuring that the compressed representations provided by the fine-tuned encoders will be more meaningful than those we would have obtained with the initial training. Using these fine-tuned encoder networks, we obtained the final 26 sets of 32-dimensional representations for the observed CARMENES spectra.

\begin{figure}
    \centering
	\includegraphics[width=.8\columnwidth]{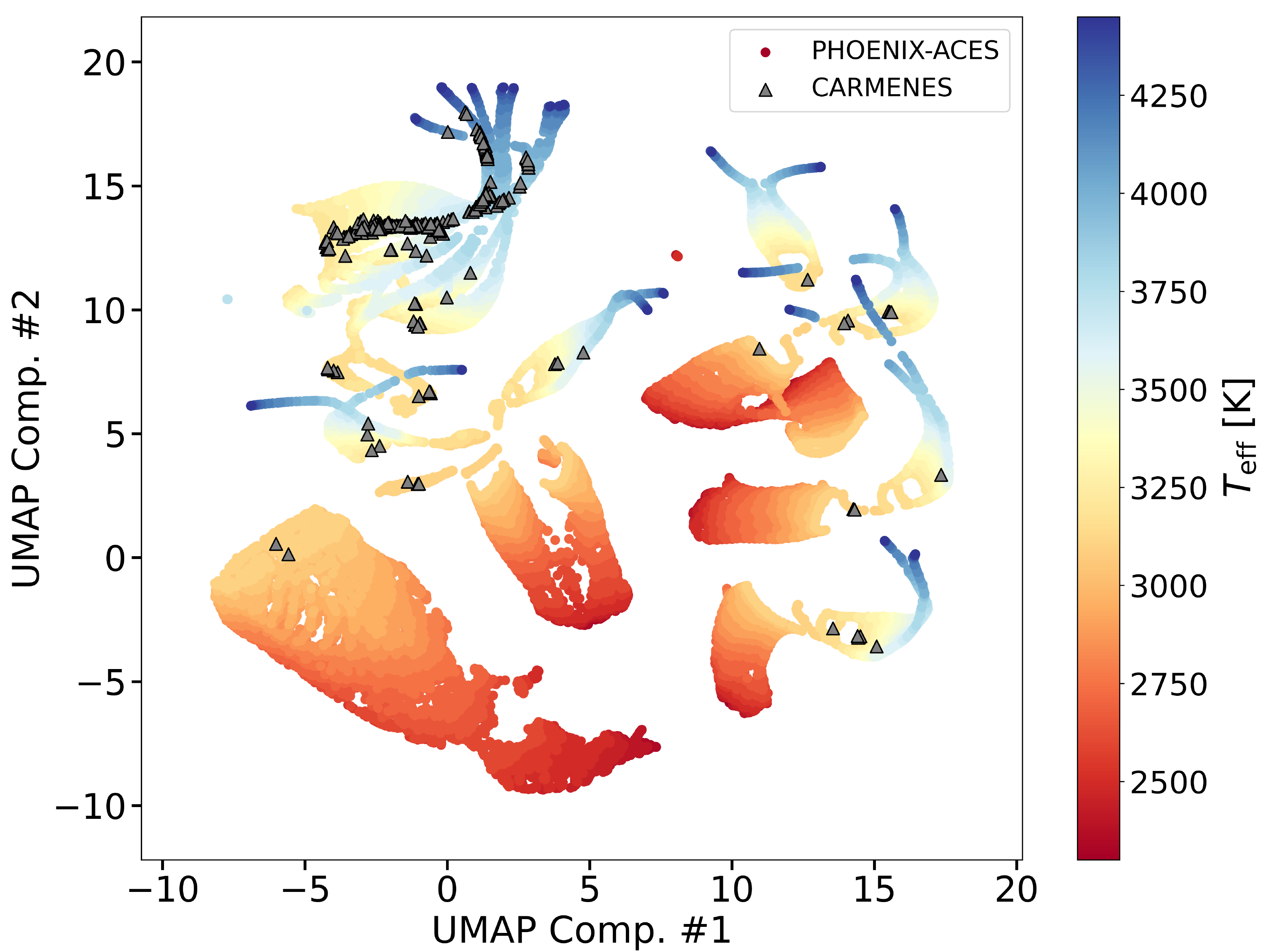}
    \caption{Two-dimensional UMAP projection of one of the 26 sets of PHOENIX-ACES (dots colour-coded by $\textit{T}_{\rm eff}$) and CARMENES (grey triangles) compressed representations. PHOENIX-ACES encodings are obtained with the initially trained AE and CARMENES encodings with the fine-tuned network.}
    \label{fig:umap_enc}
\end{figure}

While our goal was to preserve the meaningfulness of the low-dimensional representations of the synthetic and observed spectra, we aimed, above all, to minimise the disparity between the observed and synthetic compressed representations. For instance, Fig. \ref{fig:umap_enc} illustrates a UMAP two-dimensional projection, using the same metric as in Fig. \ref{fig:umap_flux}, for one of the 26 sets of PHOENIX-ACES and CARMENES representations. In contrast to Fig. \ref{fig:umap_flux}, in this case, the CARMENES objects are integrated over the space occupied by the PHOENIX-ACES family of projections, leading to a significant reduction of the differences in feature distributions between the two domains. Consequently, we calculated the minimum Euclidean distance from each CARMENES instance to the synthetic grid in both the initial high-dimensional space and the new low-dimensional feature space. While the mean distance is $2.72$ when evaluated in the initial feature space (Fig. \ref{fig:umap_flux}), it is reduced to a mean value of $0.086$ for the encoded representations (Fig. \ref{fig:umap_enc}), averaged over the 26 sets.
In this manner, a latent space that encodes the shared knowledge from both domains was learned, effectively bridging the gap between them.

\subsection{Stellar parameter estimation} \label{acs_sec:cnn}

In the final step of our methodology, we employed CNNs, one of the oldest deep learning approaches \citep{lecun1998}, to estimate the stellar parameters of the 286 CARMENES stars. As a starting point for this process, we used the 26 sets of encoded representations for the PHOENIX-ACES and CARMENES spectra obtained in the previous steps of our work. 

Inspired by the hierarchical structure of the human visual nervous system \citep[a precursor of CNNs; ][]{necognitron1980}, CNNs are therefore generally used to deal with image data. They are a specific class of multilayered feedforward neural networks, initially developed for image classification and visual pattern recognisition \citep{lecun1998,alexnet2012,vggnet2014}. The distinctive factor of CNNs is the use of convolution operations, in the convolutional layers, to automatically extract features from data. After the convolutional structure, the set of features is flattened and passed to an artificial neural network (ANN) to perform the classification or regression task.

In each forward-propagation process, the input of each neuron of the convolutional layer is obtained with an element-wise dot product between a convolution kernel (or filter), with trainable coefficients, and the outputs of the previous layer. The resulting arrays and a tunable bias are added up and passed through an activation function to obtain the output feature map of the neuron. The set of kernels is tuned during the training process, as the weights of the deep ANN layers are adjusted, so that the different feature maps of the layer represent specific features detected in the input data. \citet{surveycnn} provided a detailed review of CNNs.

\begin{figure}
    \centering
	\includegraphics[width=.95\columnwidth]{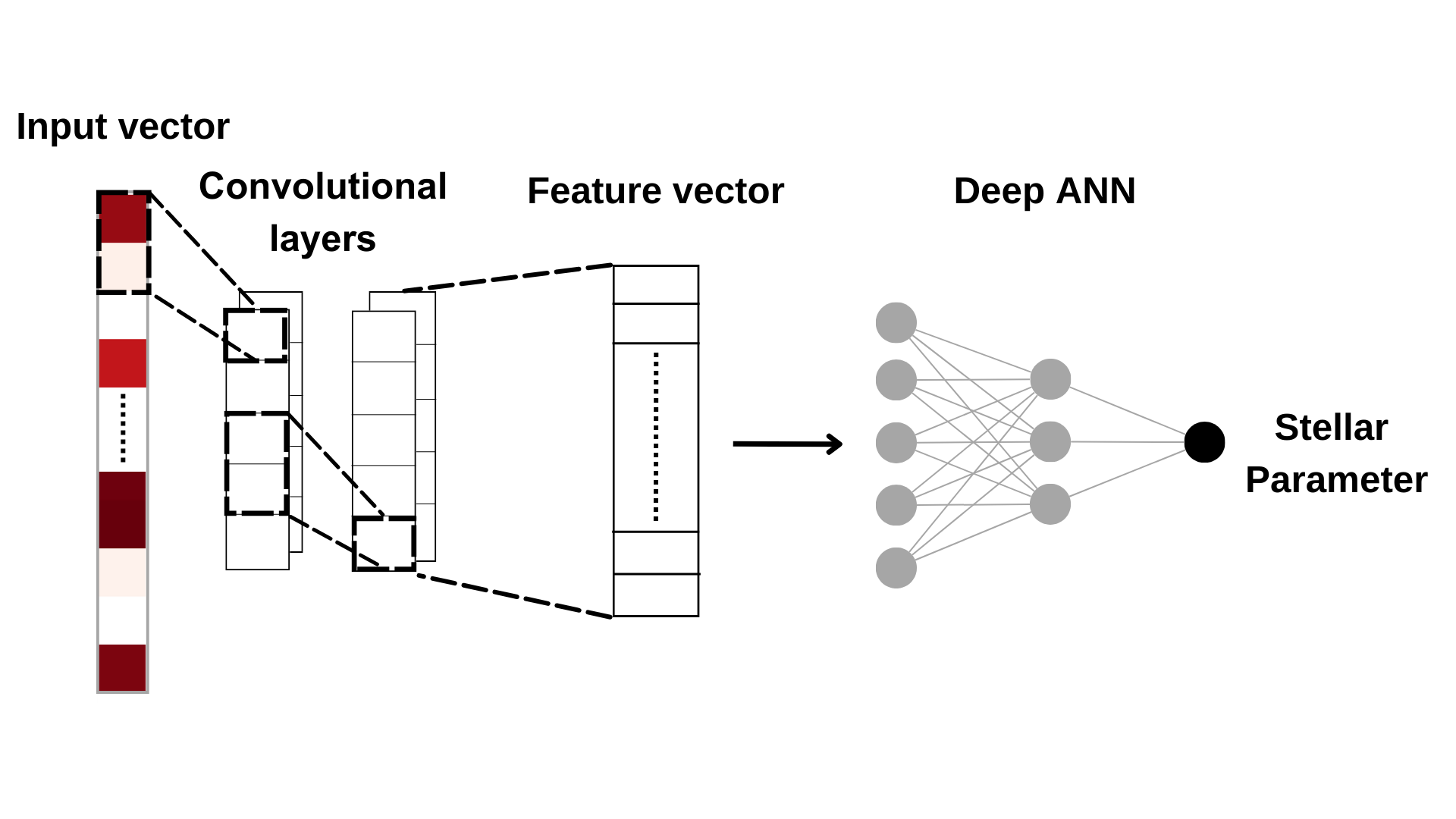}
    \caption{Schematic representation of a one-dimensional CNN architecture.}
    \label{fig:cnn}
\end{figure}

In one-dimensional (1D) CNNs (see Fig. \ref{fig:cnn}), the convolution kernel slides along a sequence of non-independent values to extract relevant features, and they have proven to be highly performant in several applications during the recent years \citep{kiranyaz2019}. \citet{sharma2020} presented a semisupervised learning approach to handle the scarcity of labelled samples, using AE and 1D CNN architectures for stellar spectral classification. \citet{zheng2020} explored how the generation of stellar spectra to balance the training data set can significantly improve the performance of a 1D CNN classifier.

Since we used 32-component vectors as input data for the stellar parameter estimation, we built a 1D CNN architecture. This architecture consists of two convolutional layers (Conv1D) with a variable number of filters (see Table \ref{tab:cnn_arc}), followed by four fully-connected (Dense) layers. A flattening step is incorporated between the convolutional and the ANN components to reshape the output of the final convolutional layer (number of outputs $\times$ number of filters) into a one-dimensional vector. This vector is then fed into the dense layers. We used a rectified linear unit (ReLU) activation function in all layers except the output layer, with a linear activation. We estimated $\textit{T}_{\rm eff}$, log\,$\textit{g}$, [M/H], and $\textit{v}\sin{i}$ independently, searching for the optimal hyperparameters of the 1D CNN architecture (same procedure as in Section \ref{acs_sec:ac}) in the estimation of each parameter. Table \ref{tab:cnn_arc} describes in detail the CNN architectures used. We followed the same procedure in the independent estimation of the different stellar parameters. To have a significant number of final estimates and to assess the robustness of our methodology, we built five CNN models for each of the 26 sets of encoded representations, thus obtaining a total of 130 regressors for each of the parameters.

To train the CNN models, we use stratified sampling to create the indices of the traning (70\,\%) and test (30\,\%) sets from the PHOENIX-ACES low-dimensional representations, ensuring that the distribution of the target parameter is representative of the overall distribution in both sets. For this, we relied on the \texttt{StratifiedShuffleSplit} class of the \texttt{scikit-learn} python package, which automatically performs stratification based on a target variable and generates indices to split data into training and test set. We trained the CNN models using the synthetic compressed representations, with a mean squared error loss function, and evaluated them on the test set. As final regressors, we kept the 80 models with the lowest mean squared error in the test set, obtaining an upper value of 353\,K, 0.0042\,dex, 0.0016\,dex, and 0.054\,km\,s$^{-1}$ for $\textit{T}_{\rm eff}$, log\,$\textit{g}$, [M/H], and $\textit{v}\sin{i}$, respectively. Using these models, we obtained 80 final parameter estimates for each of the CARMENES stars.

\begin{figure*}
    \centering
    	\includegraphics[width=\columnwidth]{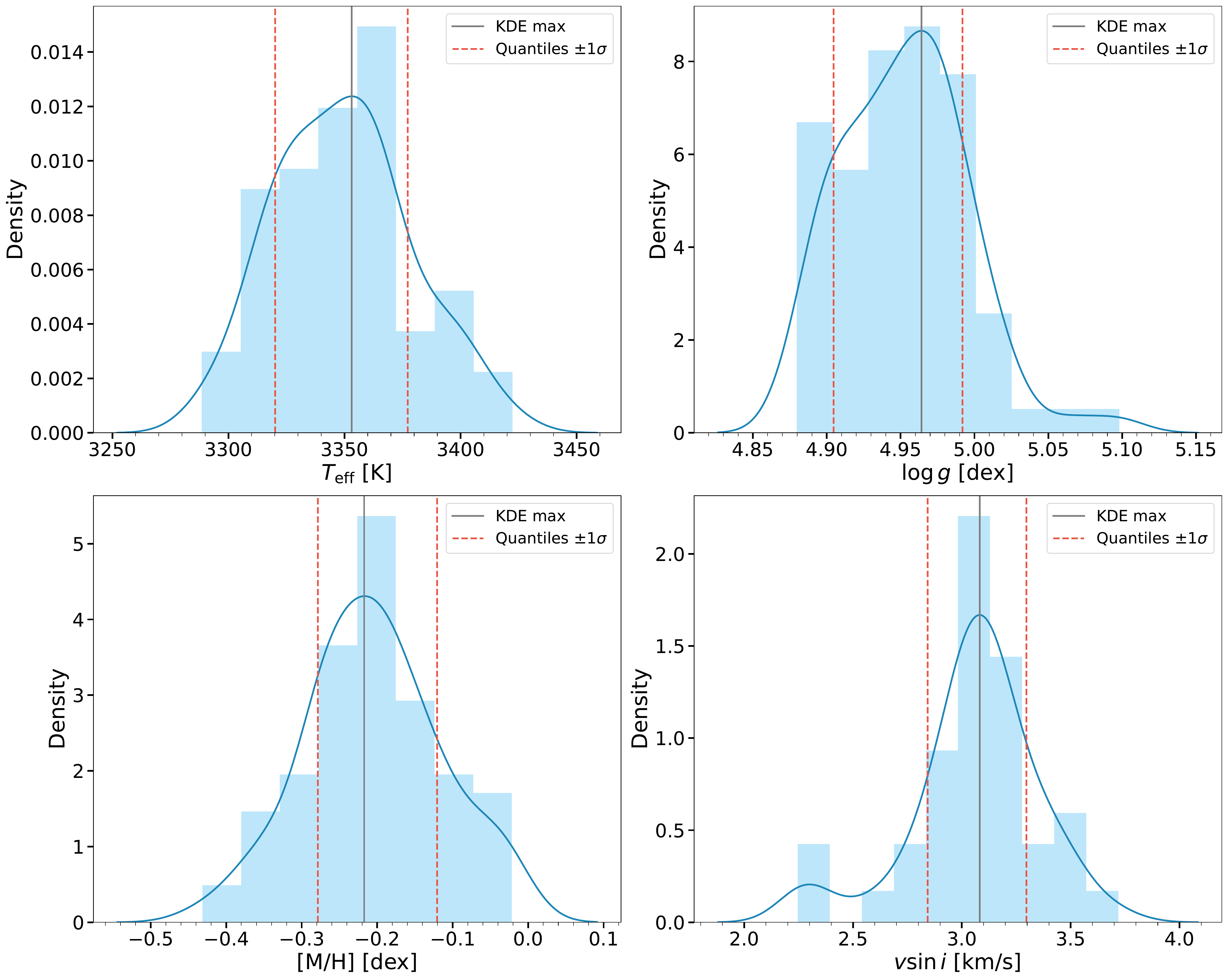}
    \caption{Distribution of stellar parameter estimations of J17578+046 \citep[Barnard's star, M3.5\,V;][]{alonsofloriano2015}. The blue solid line represents the KDE, with the maximum marked with a grey solid line. The red dashed lines represent the $\pm1\sigma$ uncertainties.}
    \label{fig:kde_plots}
\end{figure*}

We followed the same strategy used by \citetalias{pass20} and \citetalias{bello2023} for the uncertainty estimation of the stellar parameters. For each star, we gathered the 80 estimations and approximated the probability density function using the Kernel Density Estimate \citep[KDE; ][]{chen1997, poggio2021} technique. We took the maximum of this probability density function as the confident estimation for the stellar parameter, together with the 1$\sigma$ thresholds as the corresponding uncertainties. Here, the final stellar parameter is derived from a distribution of parameter estimates which come from 26 different sets of input features, together with the five CNN models built for each set. Therefore, the uncertainties provided should be understood as an intrinsic error of our methodology. Fig. \ref{fig:kde_plots} shows an example of the results for a single star.

\section{Results and discussion} \label{acs_sec:results}

\subsection{Stellar parameters analysis} \label{acs_sec:par_analysis}

\begin{figure*}
    \centering
    	\includegraphics[width=.48\columnwidth]{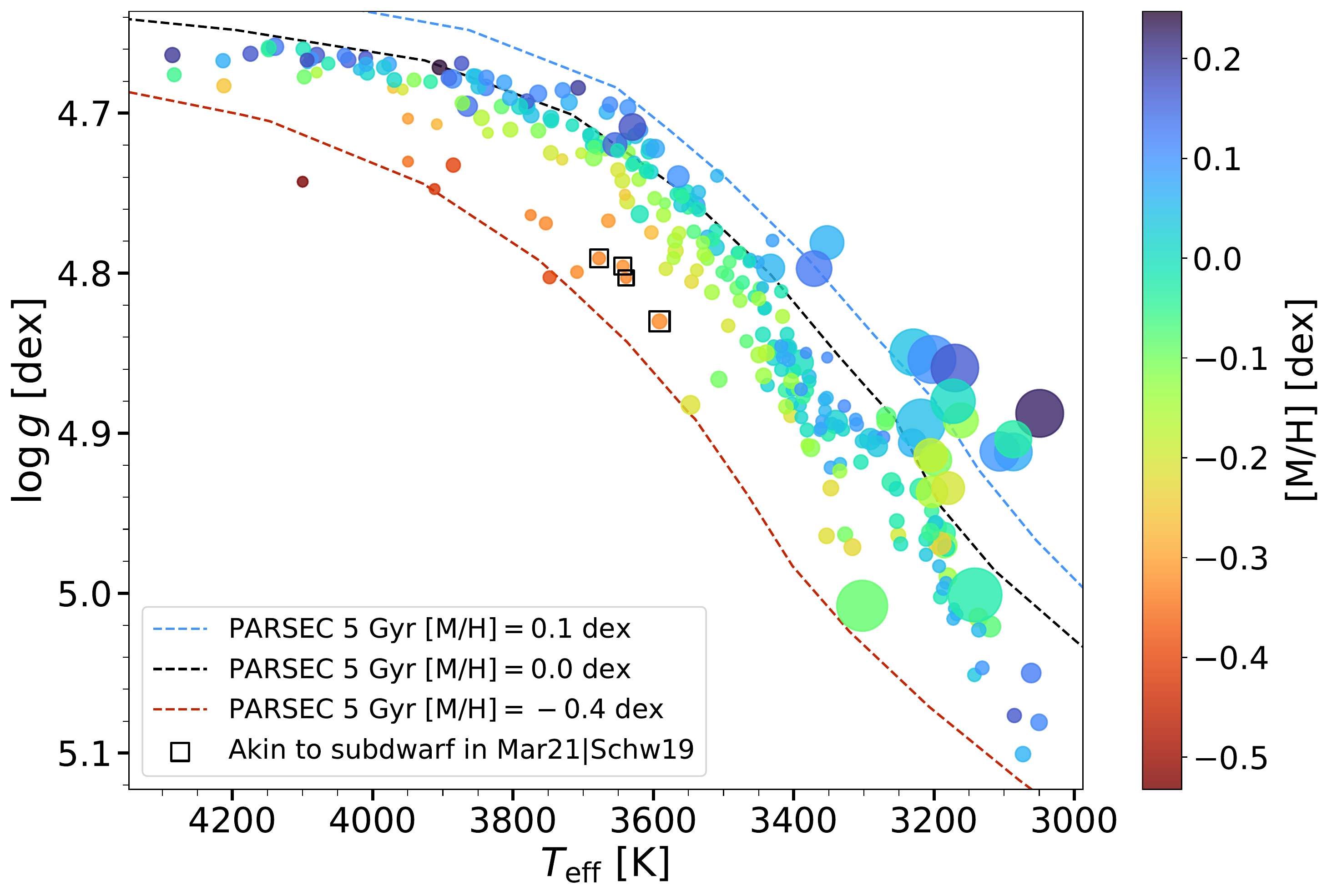}
    	\includegraphics[width=.48\columnwidth]{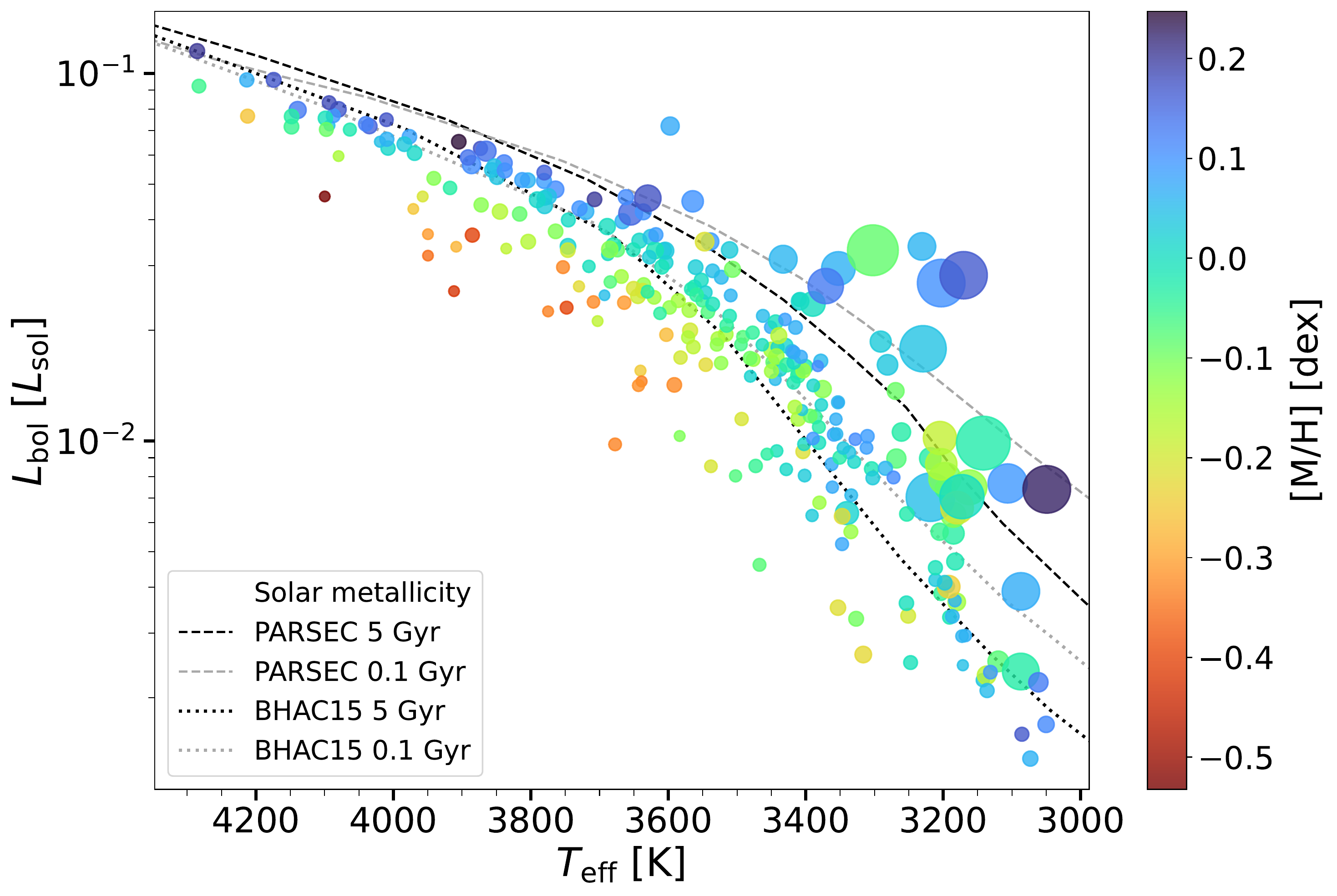}
            \\
    	\includegraphics[width=.48\columnwidth]{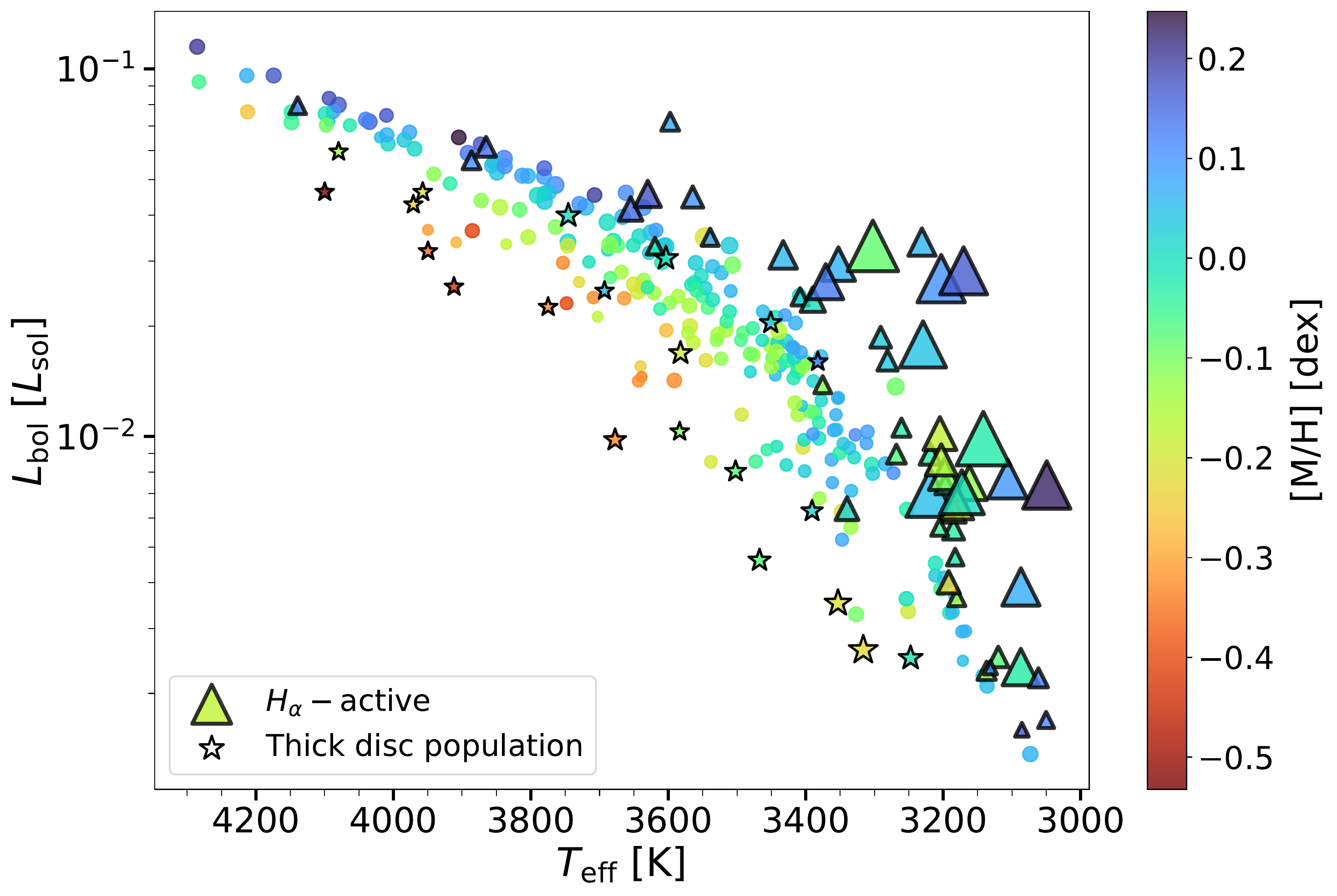}
            \includegraphics[width=.48\columnwidth]{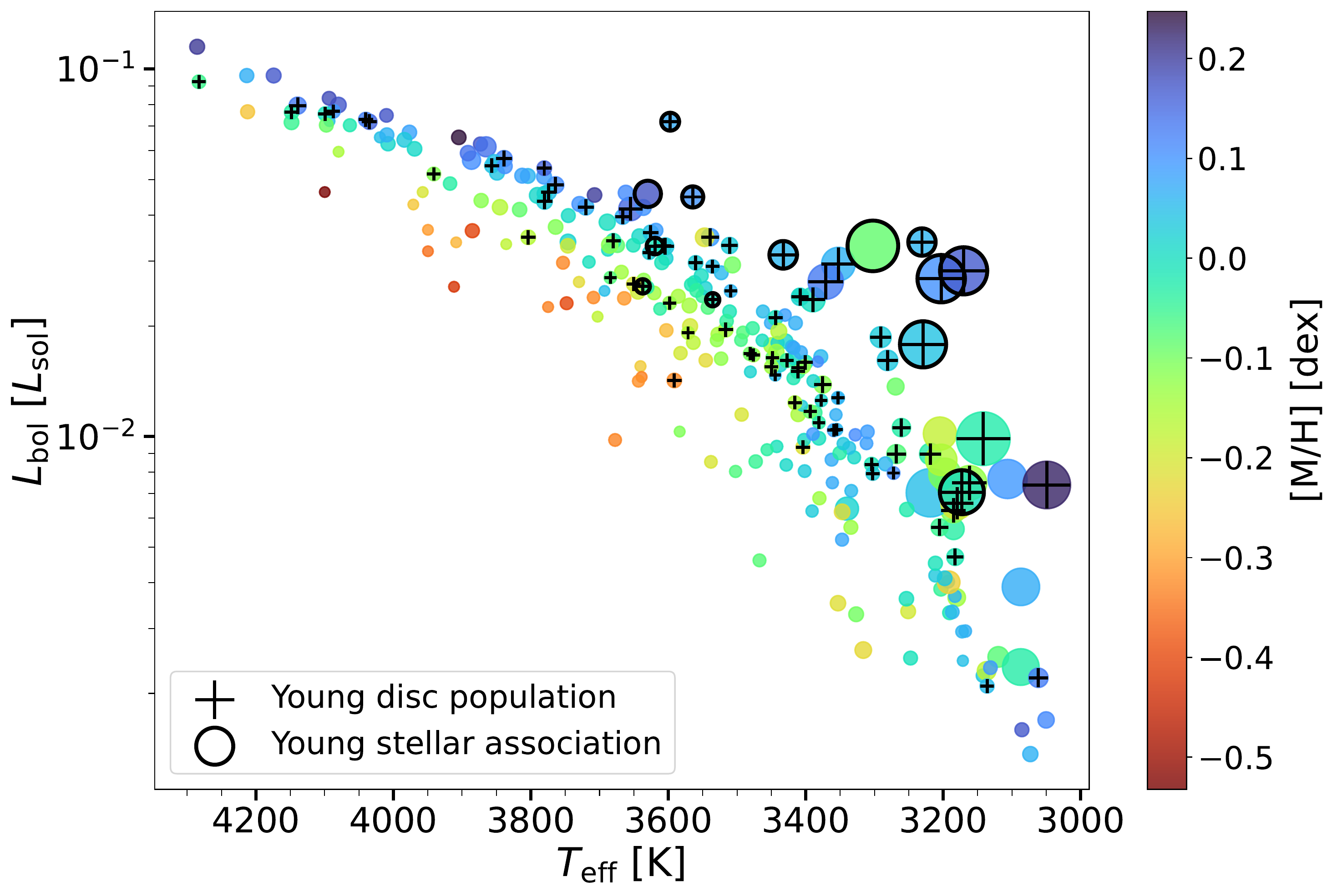}
        \caption{Analysis of the stellar parameters derived with our methodology. The dots are colour-coded according to the estimated metallicity. The size of the dots is proportional to the estimated projected rotational velocity. The \textit{top left panel} shows a Kiel diagram, with the red, black, and blue dashed lines corresponding to 5 Gyr PARSEC isochrones with [M/H] $=-0.4, 0.0$ and $0.1$\,dex, respectively. Empty squares represent the stars reported to have a behaviour akin to subdwarfs both in \citetalias{mar21} and \citetalias{schw19} (same for \textit{bottom left panel}). \textit{Top right:} black and grey dashed lines correspond to solar metallicity PARSEC isochrones for 5 and 0.1 Gyr, respectively. Black and grey dotted lines correspond to solar metallicity \citet{baraffe2015} isochrones for 5 and 0.1 Gyr, respectively. \textit{Bottom left:} triangles represent stars identified as H$\alpha$ active in \citet{schofer2019}. Empty stars depict members of the thick disc Galactic population (Cortés-Contreras et al., in prep.).  \textit{Bottom right:} plus symbols correspond to stars identified as members of the young disc Galactic population by Cortés-Contreras et al. (in prep.). Empty circles represent stars with a possible membership in a young stellar associaton, as explained in Section \ref{acs_sec:par_analysis}.}
    \label{fig:par_diags}
\end{figure*}

Table \ref{tab:pars} presents the stellar atmospheric parameters determined with our methodology. The top left panel in Fig. \ref{fig:par_diags} shows a Kiel diagram that relates all our estimated parameters, along with isochrones based on the  PAdova and TRieste Stellar Evolution Code \citep[PARSEC release v1.2S;][]{parsec2012} for 5\,Gyr and [M/H] $=-0.4, 0.0,$ and $0.1$\,dex. The results obtained with our methodology follow the trend set by the isochrones and the structure observed in the estimated metallicities is also consistent with them. The remaining three panels in Fig. \ref{fig:par_diags} show a Hertzsprung-Russell diagram (HRD) of our results, with different features highlighted in each of them. We computed the bolometric luminosities, $\textit{L}_{\mathrm{bol}}$, as \citet{cifuentes2020} using the latest astrometry and photometry from {\it Gaia} DR3 \citep{gaiadr3}. Theoretical isochrones, for solar metallicity, from PARSEC v1.2S and from evolutionary models presented by \citet{baraffe2015} are overplotted in the top right panel for 0.1 and 5\,Gyr. Both the Kiel diagram and the HRD reveal a clear outlier region at the lowest temperatures \citep[mid M-dwarf regime;][]{cifuentes2020,pecaut2013}, populated mostly by the stars with a high estimated projected rotational velocity ($\textit{v}\sin{i}$). These fast rotators in our sample are located at the expected M-dwarf regime, following the relation between the spectral types from the CARMENES input catalogue \citep[Carmencita; ][]{alonsofloriano2015,caballero2016a} and the $\textit{v}\sin{i}$ values calculated by \citet{reiners2018} \citepalias[see Fig. 2 in][]{mar21}.

The bottom panels in Fig. \ref{fig:par_diags} help to understand the outliers that deviate from the main sequence. The bottom left panel shows that almost all the overluminuous outliers in the HRD are identified as H$\alpha$ active stars by \citet{schofer2019}, considered as such if the pseudo-EW of the H$\alpha$ line satisfies pEW$'$(H$\alpha)<-0.3$\,\AA~(H$\alpha$ flag from Table B.1 in \citetalias{mar21}). As found in previous works \citep[e.g. ][]{jeffers2018,reiners2018}, the fraction of H$\alpha$ active stars is higher at later spectral types. There are clear patterns in the HRD which arise from the kinematic membership of the targets. For instance, and in agreement with \citet{jeffers2018}, most H$\alpha$ active and rapidly rotating stars are kinematically young (dots marked with a + in the bottom right panel).

To study the possible membership of our sample to nearby young stellar associatons, we relied on \texttt{BANYAN}~$\Sigma$\footnote{\url{http://www.exoplanetes.umontreal.ca/banyan/}} \citep{banyan}, a Bayesian analysis tool to identify members of young associations. Modelled with multivariate Gaussians in six-dimensional $\rm XYZUVW$ space, \texttt{BANYAN}~$\Sigma$ can derive membership probabilities for all known and well-characterised young associations within 150\,pc. In our case, we used the python version of \texttt{BANYAN}~$\Sigma$\footnote{\url{https://github.com/jgagneastro/banyan_sigma}}, and included the \textit{Gaia} DR3 sky coordinates, proper motion, radial velocity, and parallax of our target stars as input parameters to the algorithm. The classifier gave a high probability (>80\,\%) for 9 objects to belong to a young stellar association, in 7 of the cases with a probability greater than 95\,\%. Table \ref{tab:young} lists the details of these objects. All these stars with a possible membership in a young stellar associaton are represented with a thick open circle in the bottom right panel of Fig. \ref{fig:par_diags}. Here, we also considered four extra stars, namely J09133+688 (G\,234-057), J12156+526 (StKM\,2-809), J15218+209 (GJ\,9520), and J18174+483 (TYC\,3529-1437-1), which \citetalias{schw19} mentioned as young age-based outliers.

\begin{table}
\fontsize{11pt}{11pt}\selectfont
 \caption{Stars in our sample classified by \texttt{BANYAN}~$\Sigma$ with a high probability of belonging to a young stellar association.}
 \label{tab:young}
 \centering          
 \begin{tabular}{l c c c}
  \hline\hline
  \noalign{\smallskip}
  
  Karmn &  \texttt{BANYAN}~$\Sigma$  Prob.\,$^{(a)}$ & Young association\,$^{(b)}$ & Association reference\\
  
  \noalign{\smallskip}
  \hline
  \noalign{\smallskip}

  J02088+494 & 99.94\,\% & AB Doradus & \citet{abdmg} \\
  
  \noalign{\smallskip}

  J02519+224 & 99.79\,\% & $\beta$ Pictoris & \citet{bpictoris} \\
  
  \noalign{\smallskip} 

  J03473-019 & 99.94\,\% &  AB Doradus & \citet{abdmg} \\
  
  \noalign{\smallskip}
  
  J05019+011\,$^{(c)}$ & 99.91\,\% & $\beta$ Pictoris & \citet{bpictoris} \\

  \noalign{\smallskip} 

  J05062+046\,$^{(c)}$ & 99.79\,\% & $\beta$ Pictoris & \citet{bpictoris} \\
  
  \noalign{\smallskip} 
  
  J09163-186 & 95.01\,\% &  Argus & \citet{argus} \\
  
  \noalign{\smallskip}   
  
  J10289+008 & 99.97\,\% &  AB Doradus & \citet{abdmg} \\
  
  \noalign{\smallskip} 
  
  J19511+464 & 94.17\,\% & Argus & \citet{argus} \\

  \noalign{\smallskip} 
  
  J21164+025 & 85.20\,\% & Argus & \citet{argus} \\  
  
  \noalign{\smallskip}  
  \hline
 \end{tabular}
 \tablefoot{$^{(a)}$ The probability that this object belongs to the young stellar association. $^{(b)}$ Most probable Bayesian hypothesis (including the field). $^{(c)}$ Already mentioned in \citetalias{schw19} as candidate members of the corresponding young stellar association.}
\end{table}

The bottom left panel in Fig. \ref{fig:par_diags} shows that outliers below the main sequence are typically members of the thick disc Galactic population \citep[Cortés-Contreras et al., in prep.; ][]{tesis_miriam}. Furthermore, four of these outliers are reported to have a behaviour akin to subdwarfs (empty squares in top and bottom left panels) both by \citetalias{mar21} and \citetalias{schw19}. Table \ref{tab:subd} details all the outliers we identified with low-metallicity behaviour, along with the metallicity estimations found in the literature. As discussed by \citet{jao2008}, with the decrease in the metallicity of these objects the TiO opacity also strongly decreases, and this less blanketing from the TiO bands causes more continuum flux to radiate from the deeper and hotter layer of the stellar atmosphere, so that these stars appear bluer than their solar metallicity counterparts (see Fig. 1 in \citealt{jao2008}). Our [M/H] determinations for these stars are, in general, in good agreement with the literature.

\begin{figure}
    \centering
	\includegraphics[width=.6\columnwidth]{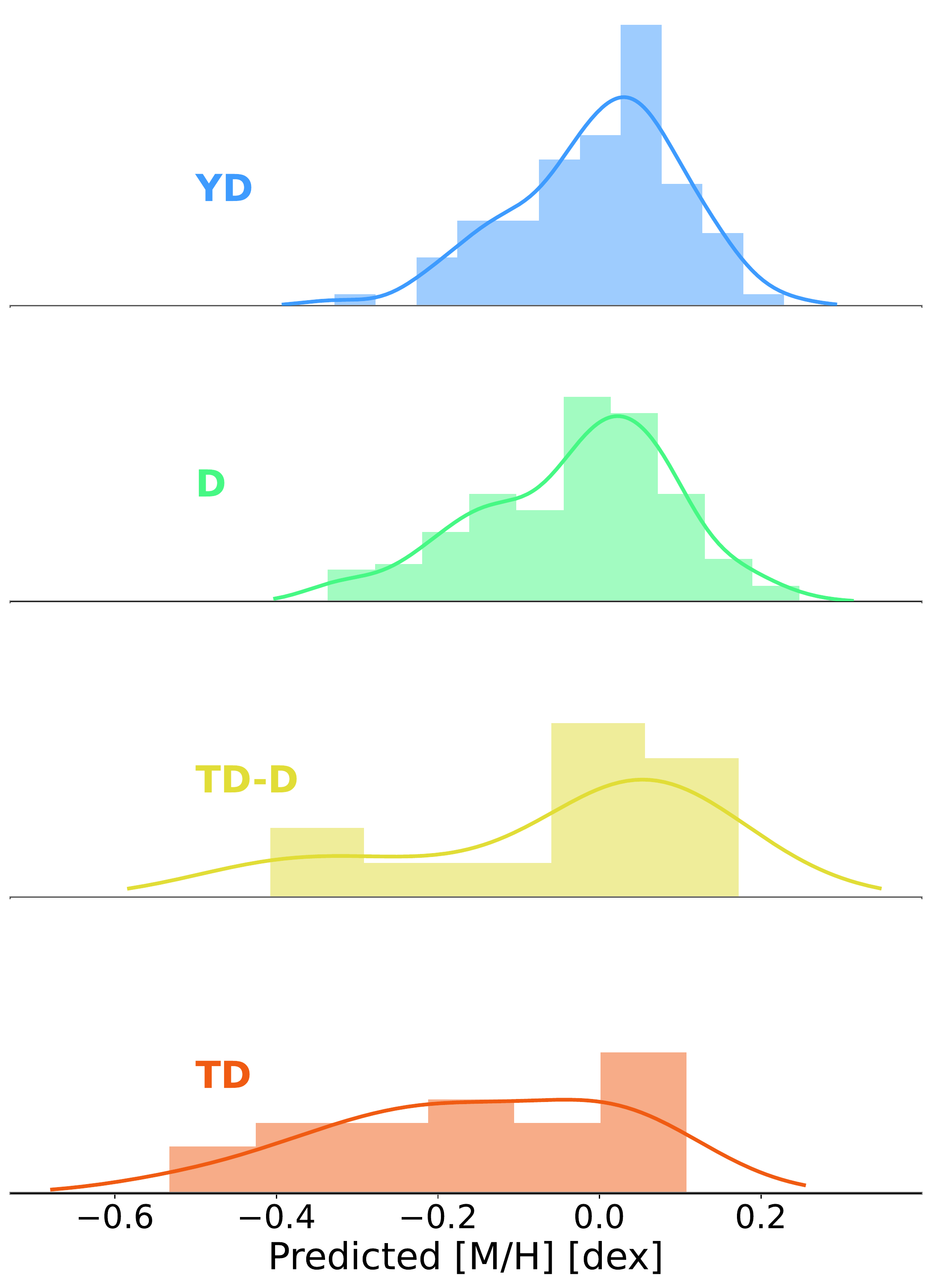}
    \caption{Distribution of our predicted metallicities broken down by kinematic membership in the the thick disc (TD), the thick disc-thin disc transition (TD-D), the thin disc (D), and the young disc (YD) Galactic populations \citep[Cortés-Contreras et al., in prep.; ][]{tesis_miriam}. The bins are normalised so that the total area of the histogram equals one, and the solid lines represent the KDE.}
    \label{fig:pop_mh_hist}
\end{figure}

Fig. \ref{fig:pop_mh_hist} shows the distribution of our predicted metallicities broken down by kinematic membership in the thick disc (TD), thick disc-thin disc transition (TD-D), thin disc (D), and young disc (YD) Galactic populations \citep[Cortés-Contreras et al., in prep.; ][]{tesis_miriam}. This breakdown reveals the distinction between metal-rich thin disc stars and metal-poor stars in the older thick disc \citep{bensby2005,gaiamh2023}, with the TD-D transition as an intermediate step. To prove this, we performed a two-sample Kolmogorov-Smirnov test \citep{kolmogorov33,smirnov48} on the thin and thick disc samples, which returned a $p\,\rm{value} = 0.0071$, rejecting the hypothesis that both samples come from the same distribution.

\begin{figure}
    \centering
	\includegraphics[width=.8\columnwidth]{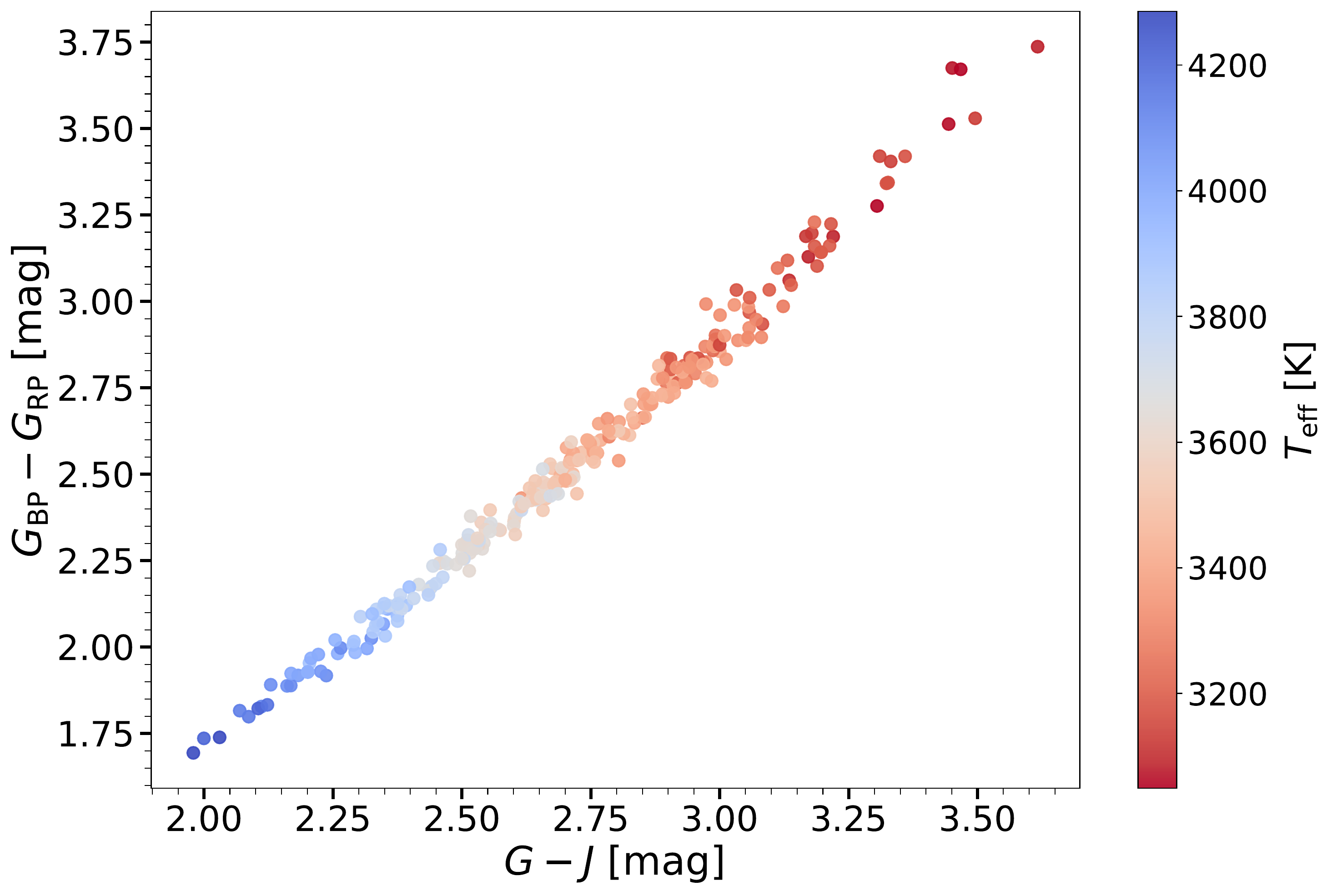}
    \caption{2MASS-\textit{Gaia} $\textit{G}_{\mathrm{BP}} - \textit{G}_{\mathrm{RP}}$ versus $\textit{G}-\textit{J}$ diagram of our target stars with good photometric quality (2MASS \texttt{Qflg}=A and a relative error of less than 10\,\% in \textit{Gaia}\,DR3 photometry). The points are colour-coded according to the effective temperatures derived in this work.}
    \label{fig:col_diag}
\end{figure}

Also, the 2MASS-\textit{Gaia} $\textit{G}_{\mathrm{BP}} - \textit{G}_{\mathrm{RP}}$ versus $\textit{G}-\textit{J}$ colour-colour diagram in Fig. \ref{fig:col_diag} shows how the evolution in our estimated effective temperatures is coherent with the colour-colour relationship (see Fig. 14 in \citealt{cifuentes2020}). For this diagram, we only considered stars with reliable 2MASS $J$-band and \textit{Gaia} DR3 $\textit{G}_{\mathrm{BP}}$ and $\textit{G}_{\mathrm{RP}}$ photometry.

\subsection{Comparison with the literature}
\label{acs_sec:lit_comp}

We compared our results with different collections found in the literature. Whereas this section focuses on the latest studies using CARMENES data, namely \citetalias{bello2023}, \citetalias{mar21}, \citetalias{pass2019}, \citetalias{pass20}, and \citetalias{schw19}, a more extensive compilation of literature, together with the uncertainties of the estimations, is provided in Appendix \ref{app:appb_lit}. For \citetalias{pass2019}, we considered the parameters derived from VIS spectra. Table \ref{tab:comparison_lit} lists the mean difference ($\overline{\Delta}$; literature$-$this work), root mean squared error (rmse), and Pearson correlation coefficient ($r_{\rm p}$) for the comparison with each of the literature collections. An interactive version of the results presented in this section is available to the astronomical community\,\footnote{\url{https://cab.inta-csic.es/users/pmas/}}.

\begin{table}
\fontsize{11pt}{11pt}\selectfont
 \caption{Mean difference ($\overline{\Delta}$; literature$-$this work), root mean square error (rmse), and Pearson correlation coefficient ($r_{\rm p}$) for the comparison between our results and the literature.}
 \label{tab:comparison_lit}
 \centering
 \begin{tabular}{l c c c c c c c c c c c c}
 
  \hline\hline
  \noalign{\smallskip}

  Reference & \multicolumn{3}{c}{$\textit{T}_{\rm eff}$} & \multicolumn{3}{c}{log\,$\textit{g}$} & \multicolumn{3}{c}{[Fe/H]} & \multicolumn{3}{c}{$\textit{v}\sin{i}$} \\
  
  & \multicolumn{3}{c}{[K]} & \multicolumn{3}{c}{[dex]} & \multicolumn{3}{c}{[dex]} & \multicolumn{3}{c}{[km\,s$^{-1}$]} \\

  & $\overline{\Delta}$ & rmse & $r_{\rm p}$ & $\overline{\Delta}$ & rmse & $r_{\rm p}$ & $\overline{\Delta}$ & rmse & $r_{\rm p}$ 
  &$\overline{\Delta}$ & rmse & $r_{\rm p}$ \\
  
  \noalign{\smallskip}
  \hline
  \noalign{\smallskip}

  \citetalias{bello2023} & -117 & 180 & 0.87 & \ldots & \ldots & \ldots & 0.01 & 0.14 & 0.60 & \ldots & \ldots & \ldots \\
  \citetalias{mar21} & -19 & 102 & 0.94 & 0.12 & 0.18 & 0.39 & -0.11 & 0.16 & 0.65 & \ldots & \ldots & \ldots \\
  \citetalias{pass2019} & -80 & 117 & 0.96 & 0.00 & 0.05 & 0.86 & 0.06 & 0.15 & 0.52 & \ldots & \ldots & \ldots \\
  \citetalias{pass20} & -35 & 51 & 0.99 & -0.04 & 0.06 & 0.93 & 0.23 & 0.25 & 0.76 & 1.64 & 1.94 & 0.99 \\
  Rein18$^{\,(a)}$ & \ldots & \ldots & \ldots & \ldots & \ldots & \ldots & \ldots & \ldots & \ldots & -0.86 & 1.51 & 0.98 \\
  \citetalias{schw19} & -40 & 93 & 0.96 & 0.13 & 0.14 & 0.89 & 0.00 & 0.10 & 0.63 & \ldots & \ldots & \ldots \\
  
  \noalign{\smallskip}
  \hline
 \end{tabular}
 \tablefoot{$^{(a)}$ From \citet{reiners2018}.}
\end{table}

\begin{figure*}
    \centering
    	\includegraphics[width=.45\columnwidth]{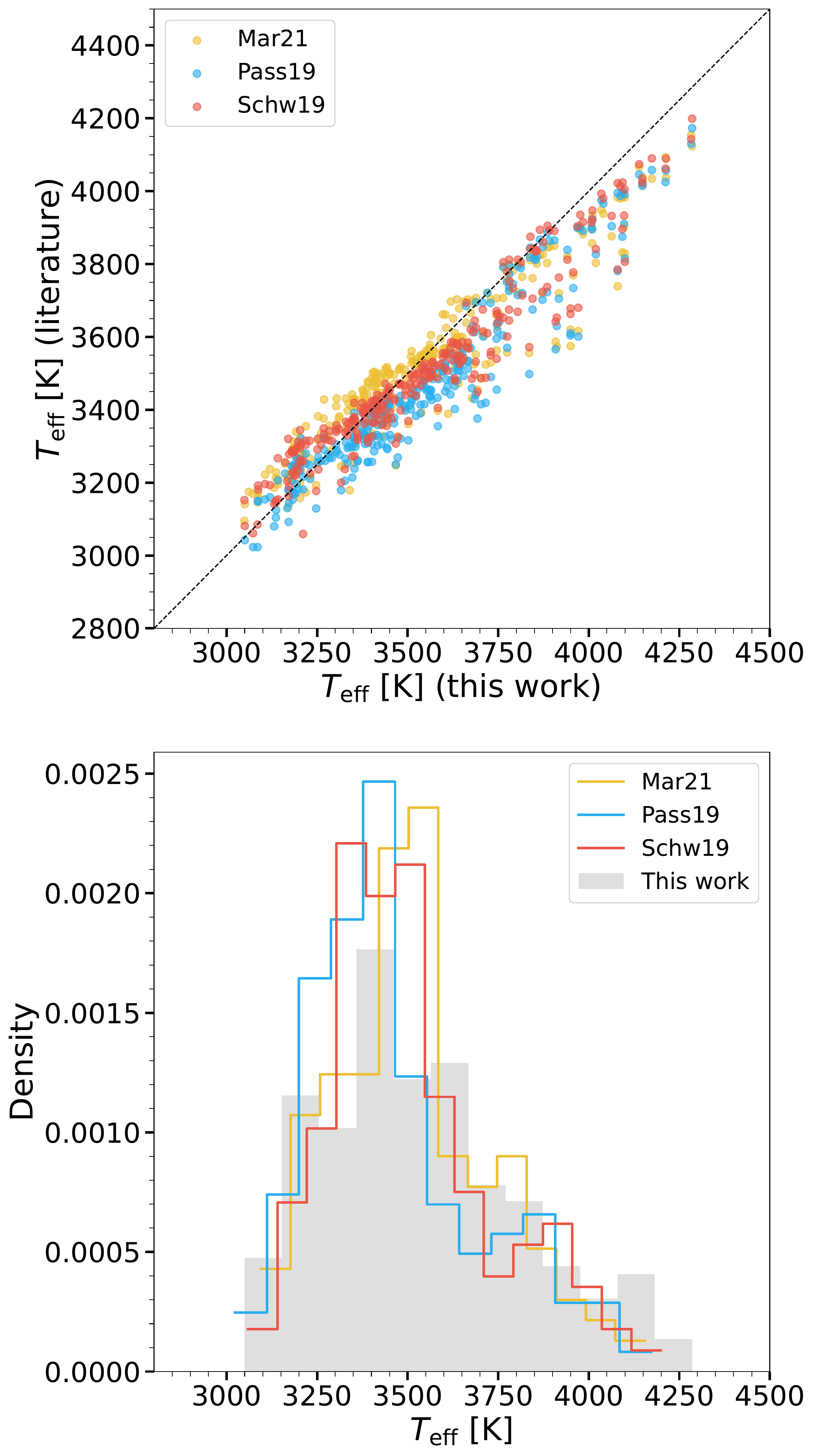}
            \includegraphics[width=.45\columnwidth]{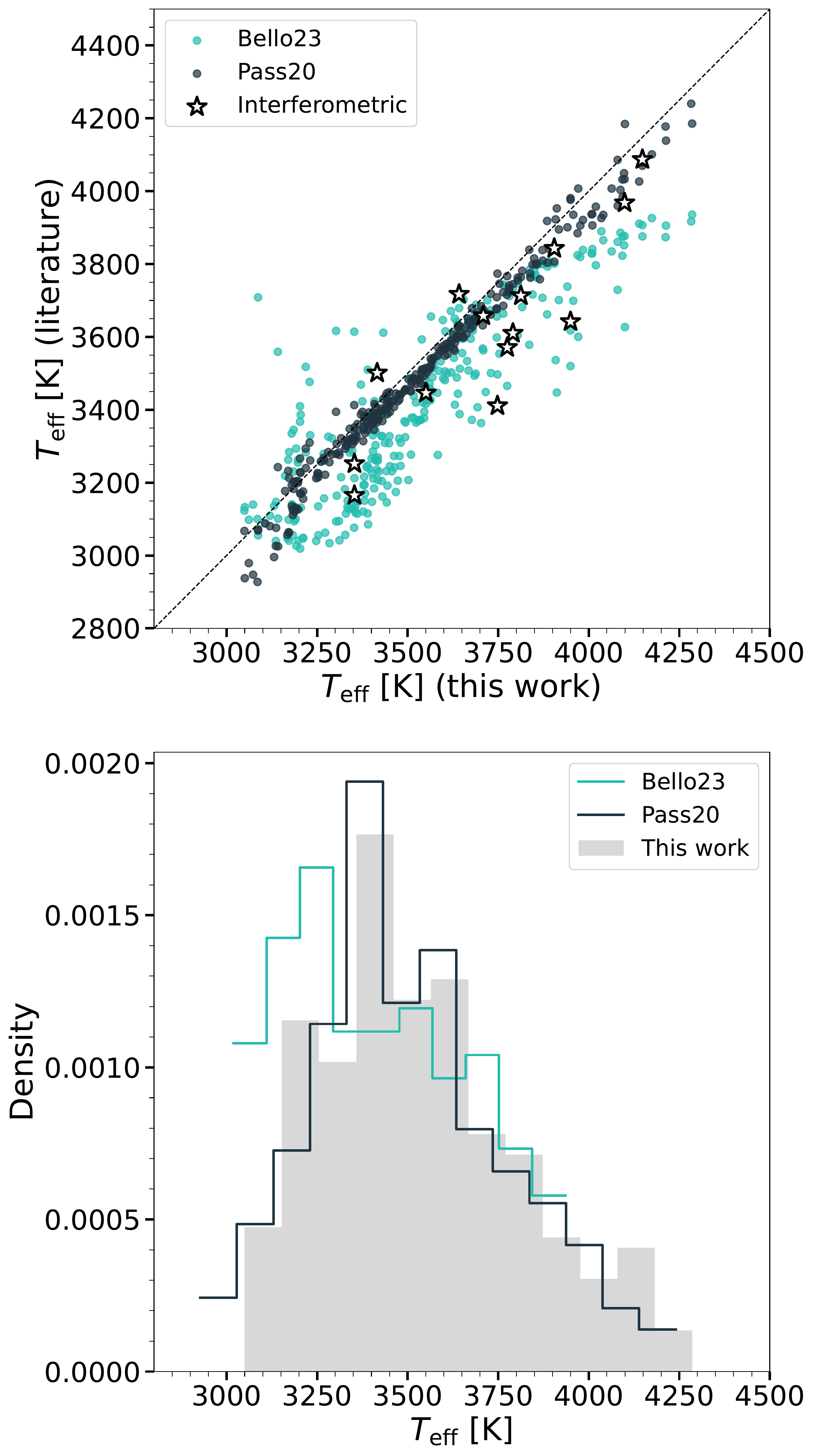}
    \caption{Comparison between our derived $\textit{T}_{\rm eff}$ values and the literature. The \textit{left panels} include the results from \citetalias{mar21} (yellow), \citetalias{pass2019} (blue), and \citetalias{schw19} (red). The \textit{right panels} include the work from \citetalias{bello2023} (cyan) and the results obtained following the DL methodology described by \citetalias{pass20} (dark blue). The black stars in the \textit{top right panel} correspond to the interferometrically derived $\textit{T}_{\rm eff}$ values from \citetalias{bello2023}. The dashed black lines in the \textit{top panels} correspond to the 1:1 relation. For the bin width in the histograms shown in the \textit{bottom panels}, we used the default parameters of the \texttt{seaborn histplot} function.}
    \label{fig:scatter_teff}
\end{figure*}

Figure~\ref{fig:scatter_teff} depicts the comparison with literature values for $\textit{T}_{\rm eff}$. The left panels show a similar linear trend among \citetalias{mar21}, \citetalias{pass2019}, and \citetalias{schw19} with our values, all of them with a slope of less than one, for the region $\textit{T}_{\rm eff}$ (this work) $\lesssim 3\,750$\,K. From this value onwards, where the number of stars in our training set is smaller, the dispersion increases significantly and our $\textit{T}_{\rm eff}$ estimations deviate towards hotter values, resulting in a mean difference of $\overline{\Delta}=-19$\,K, $\overline{\Delta}=-80$\,K, $\overline{\Delta}=-40$\,K for \citetalias{mar21}, \citetalias{pass2019}, and \citetalias{schw19}, respectively. The figures provided in Appendix \ref{app:appb_lit} show that the uncertainties intrinsic to our methodology are also larger for estimations above 3\,750\,K. The right panels show how the agreement with the values obtained following the approach described by \citetalias{pass20} is excellent, which is expected since their methodology is the closest to the one presented in this work. Moreover, the comparison with the results from \citetalias{bello2023} reveals the same structure, but inverted, as shown in Fig. 9 of their work, with a larger dispersion than that observed for the other literature collections. The black stars in the top right panel represent the 14 interferometrically derived $\textit{T}_{\rm eff}$ values (see Table 1 in \citetalias{bello2023}), which are on average cooler than the temperatures obtained with our methodology ($\overline{\Delta}_{\rm interf}=-119$\,K). The $r_{\rm p}$ values listed in Table~\ref{tab:comparison_lit} show a strong correlation with all the collections.

\begin{figure*}
    \centering
    	\includegraphics[width=.45\columnwidth]{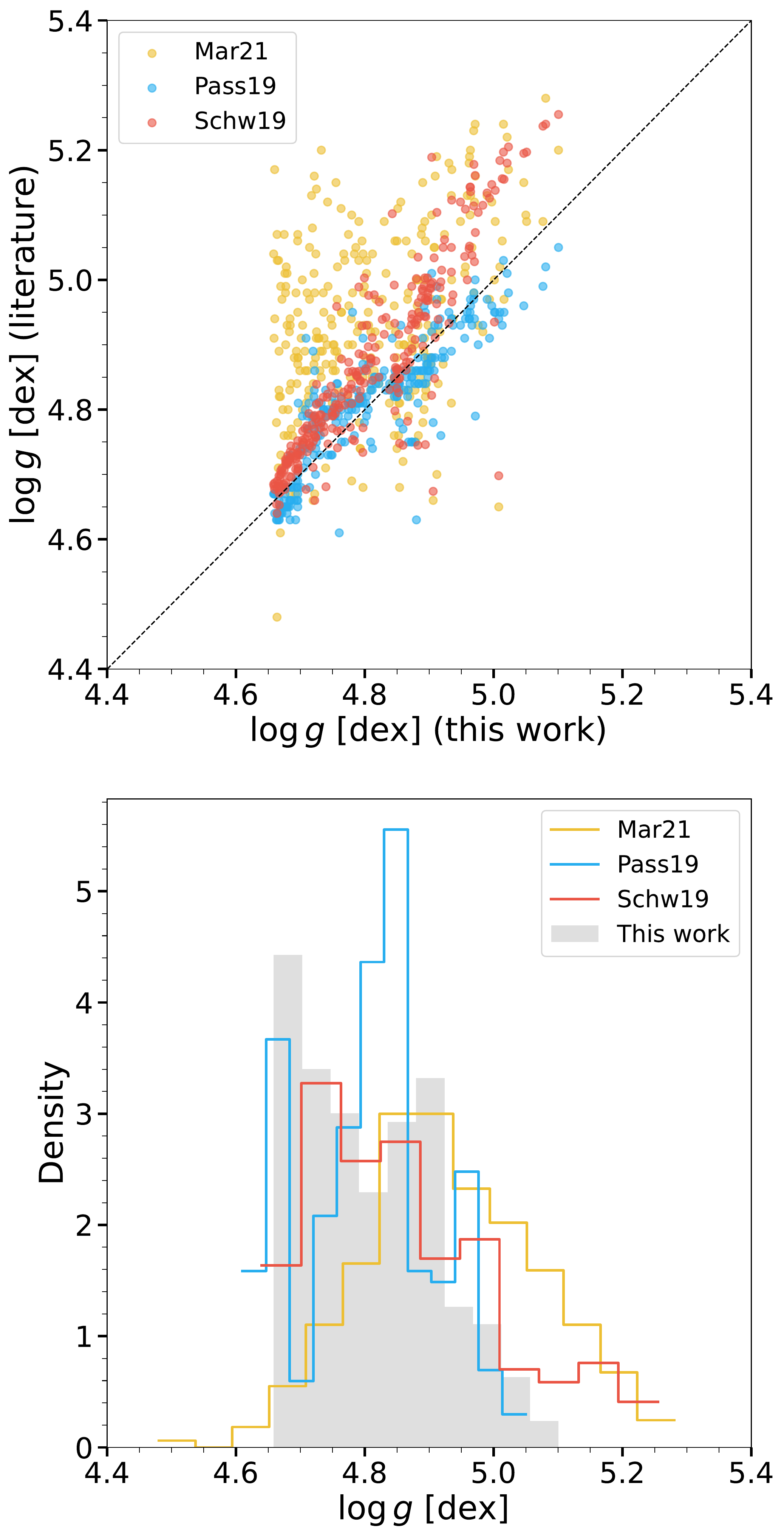}
            \includegraphics[width=.45\columnwidth]{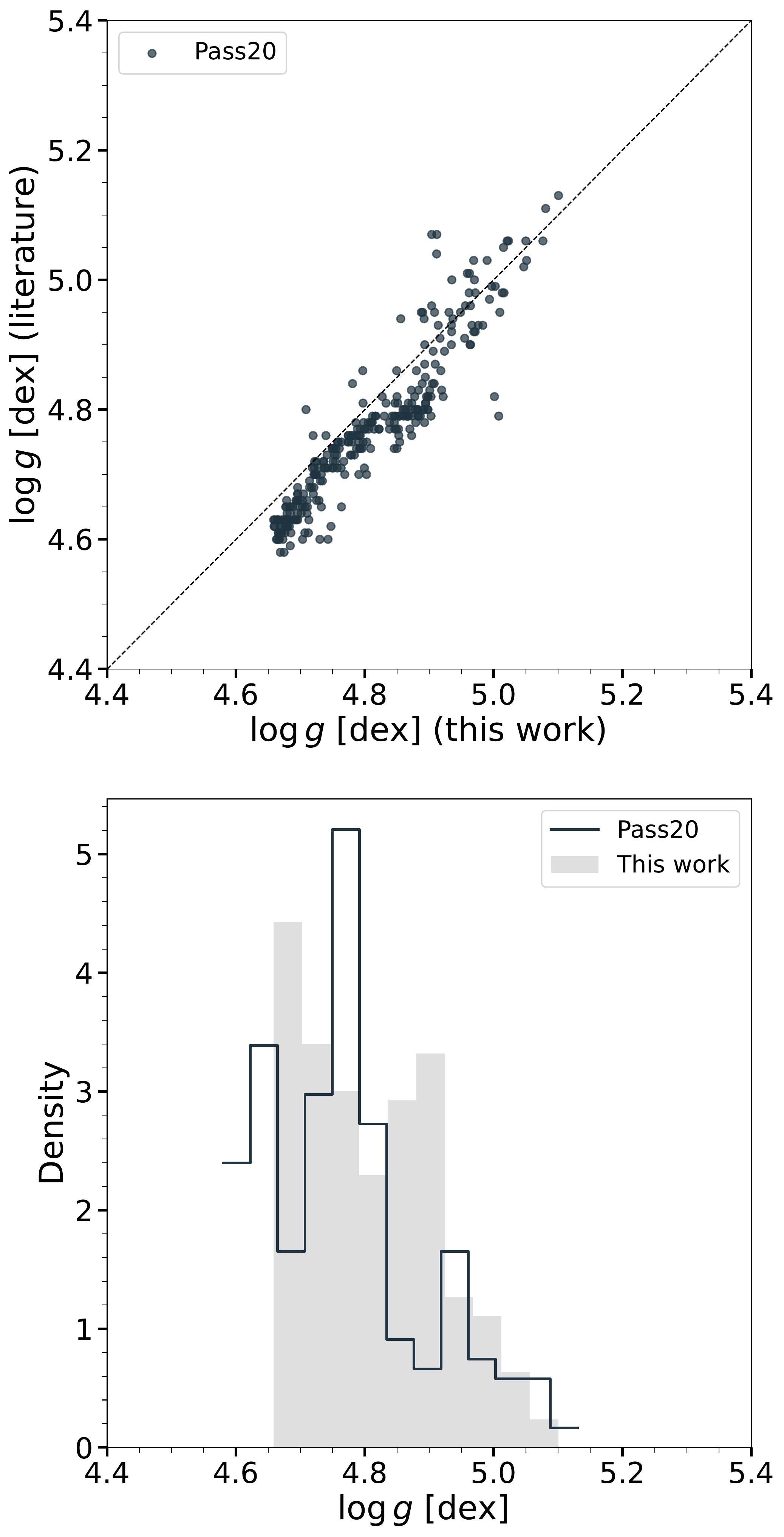}
    \caption{Comparison between our derived log\,$\textit{g}$ values and the literature. Colours and symbols are the same as in Fig. \ref{fig:scatter_teff}.}
    \label{fig:scatter_logg}
\end{figure*}

Figure \ref{fig:scatter_logg} shows a similar literature comparison for log\,$\textit{g}$. For \citetalias{schw19}, we considered the values derived using their mass-radius relation and the Stefan-Boltzmann's law. The log\,$\textit{g}$ values from \citetalias{mar21} show a large dispersion ($r_{\rm p}=0.39$), as already mentioned in their work, and are generally spread towards higher values ($\overline{\Delta}=0.12$\,dex). While the results from \citetalias{pass2019} cover the same range and are similar on average to our obtained log\,$\textit{g}$ ($\overline{\Delta}=0.00$\,dex), those from \citetalias{schw19} extend to higher values and are on average higher than ours ($\overline{\Delta}=0.13$\,dex). It should be noted that, while \citetalias{pass2019} and \citetalias{schw19} fix log\,$\textit{g}$ using theoretical isochrones, \citetalias{mar21} has log\,$\textit{g}$ as a free parameter.
Moreover, our results show a good correlation ($r_{\rm p}=0.93$) with those obtained following the methodology described by \citetalias{pass20}, although the latter are deviated to lower values ($\overline{\Delta}=-0.04$\,dex).

\begin{figure*}
    \centering
    	\includegraphics[width=.45\columnwidth]{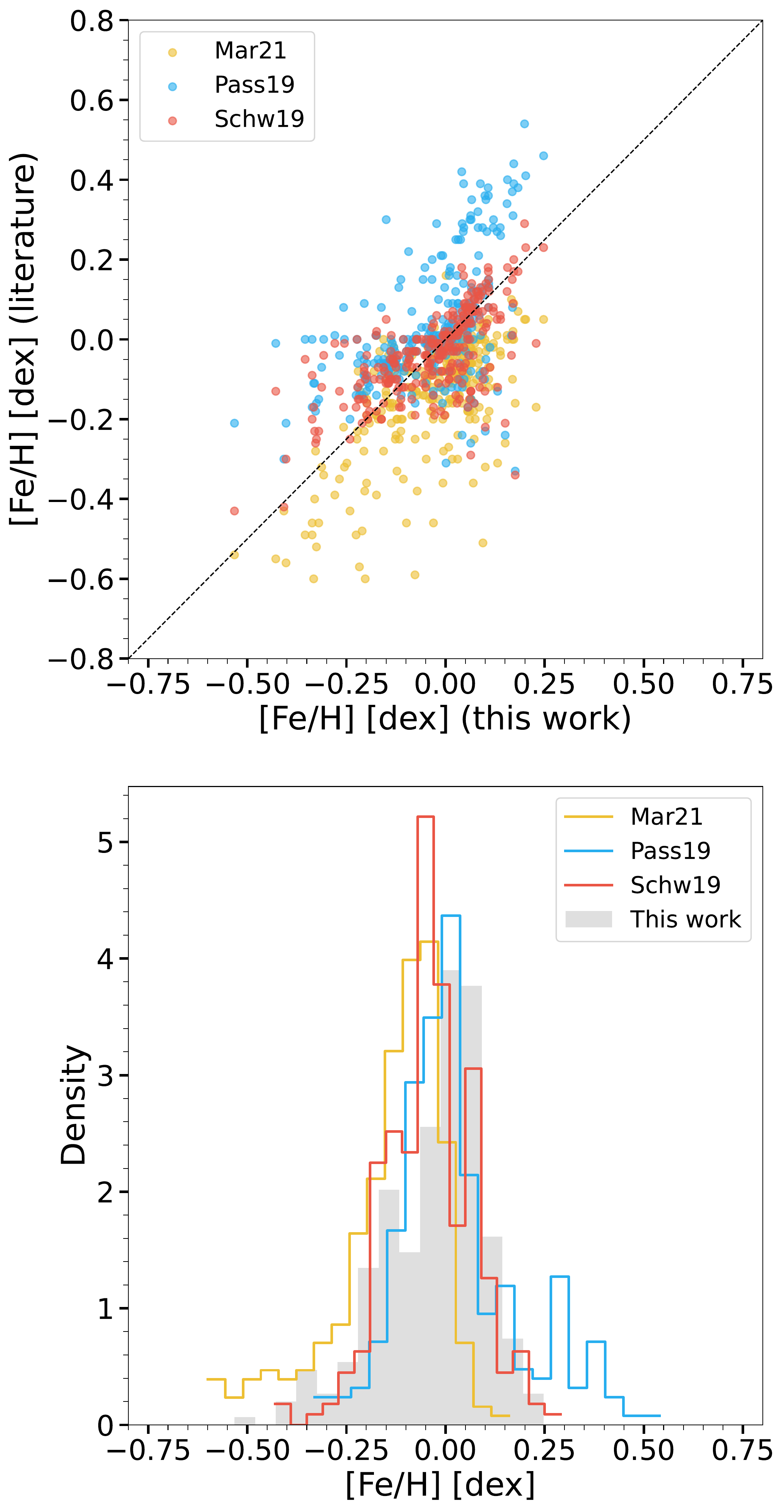}
            \includegraphics[width=.45\columnwidth]{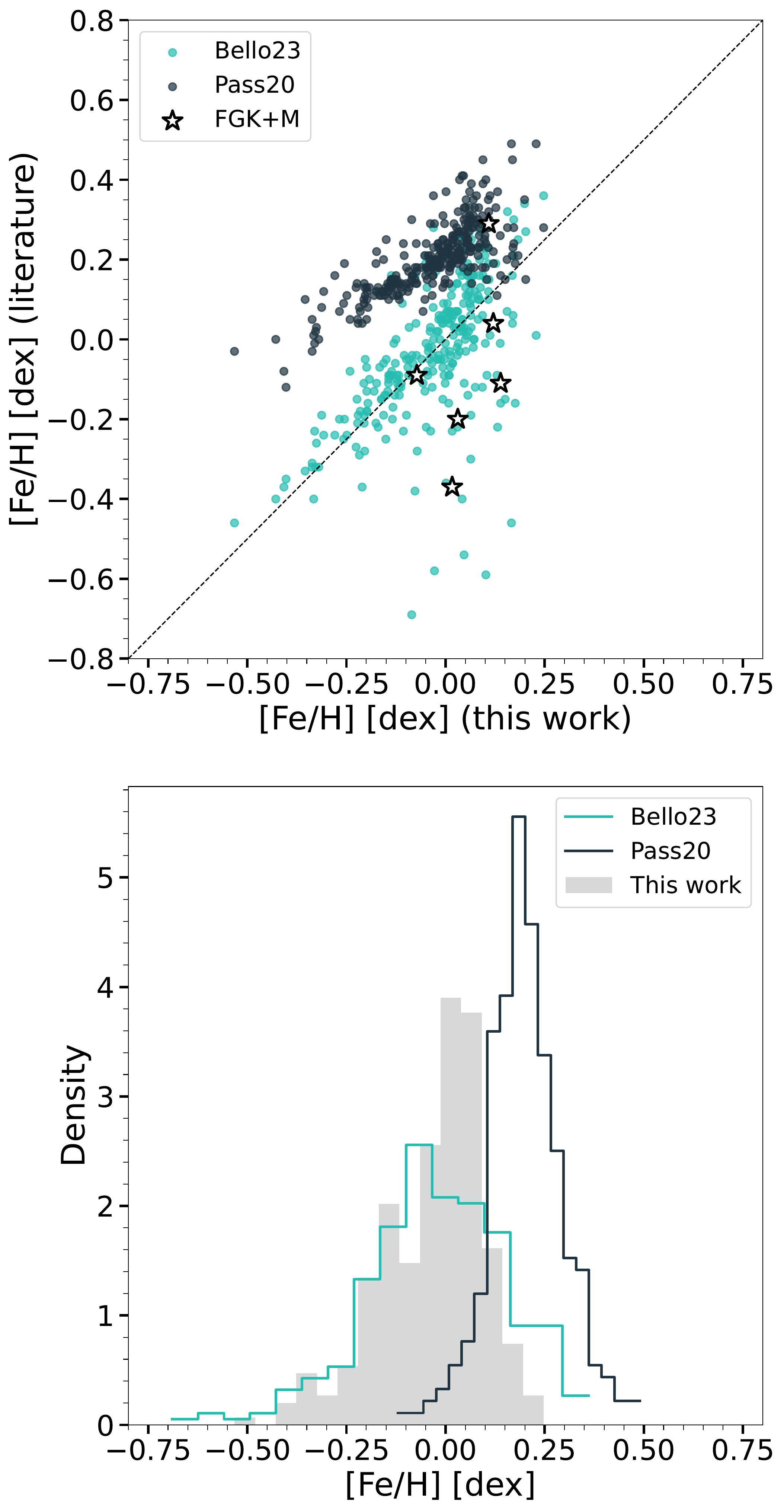}
    \caption{Comparison between our derived [Fe/H] values and the literature. Colours and symbols are the same as in Fig. \ref{fig:scatter_teff}. The black stars in the \textit{top right panel} correspond to the spectroscopically determined [Fe/H] values from FGK+M systems presented in \citetalias{bello2023}.}
    \label{fig:scatter_mh}
\end{figure*}

As discussed in \citet{passegger2022}, several discrepancies can be found when comparing metallicities of M dwarfs obtained with different methodologies. Figure \ref{fig:scatter_mh} shows the comparison with literature values for our [M/H] estimations, which directly translate into [Fe/H] values \citep{pass20,passegger2022}. For \citetalias{mar21}, we considered the values corrected for alpha enhancement. Our results are similar on average to those from \citetalias{schw19} ($\overline{\Delta}=0.00$\,dex), while \citetalias{pass2019} and \citetalias{mar21} results tend to be higher and lower, with $\overline{\Delta}=0.06$ and $\overline{\Delta}=-0.11$\,dex, respectively. As already mentioned in \citet{passegger2022}, the results from the DL methodology described by \citetalias{pass20} are deviated towards more metal-rich values, with $\overline{\Delta}=0.23$\,dex. We note that this deviation, which is attributed to the synthetic gap by \citetalias{pass20}, does not appear in the DTL methodologies presented by \citetalias{bello2023} and here. \citetalias{bello2023} metallicities cover more or less the same range as our results, and the spectroscopically determined [M/H] values from FGK+M systems (see Table 3 in \citetalias{bello2023}) (black stars in the top right panel) are systematically lower ($\overline{\Delta}=-0.13$\,dex).

\begin{figure}
    \centering
	\includegraphics[width=.7\columnwidth]{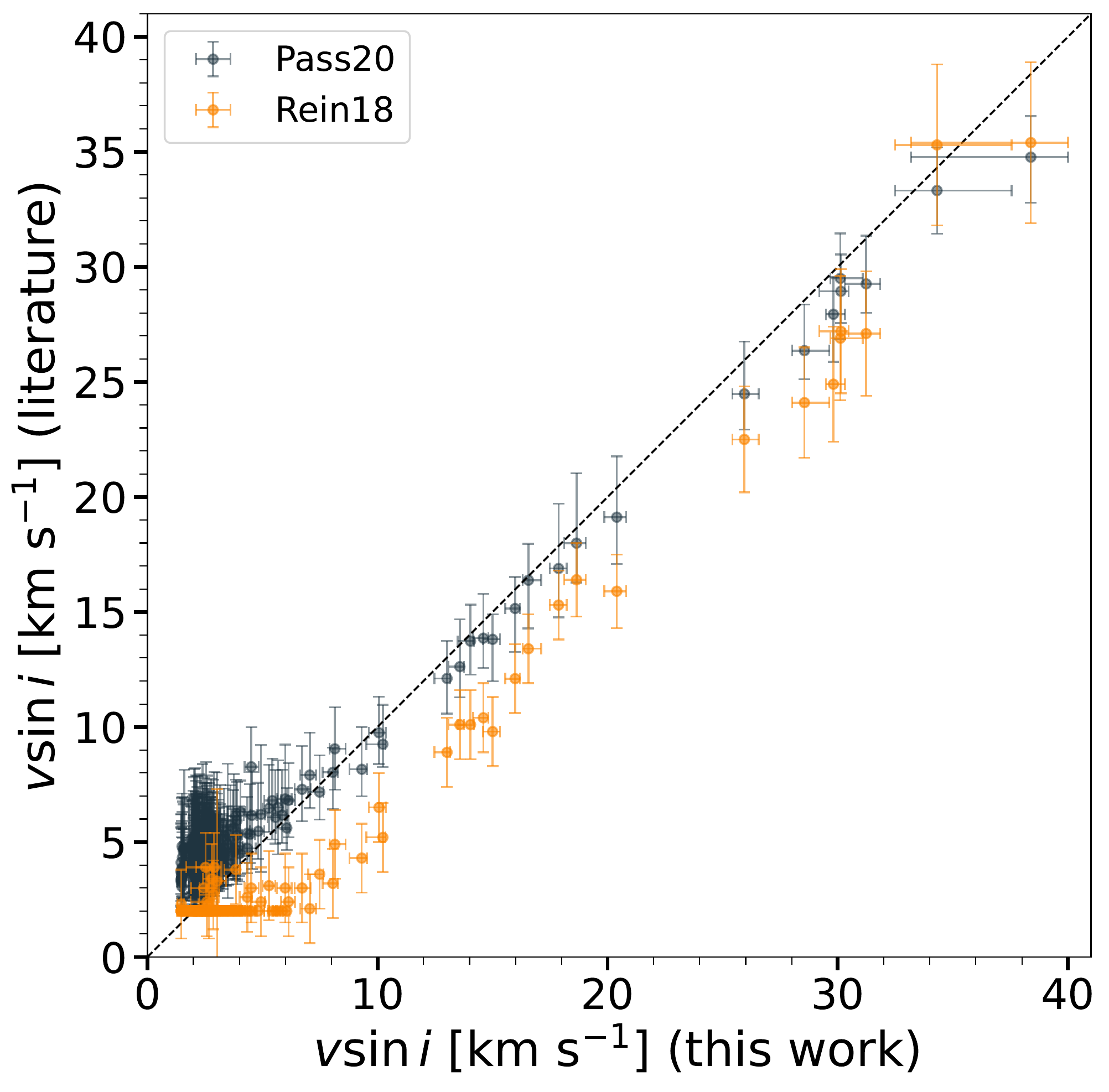}
    \caption{Comparison between our derived $\textit{v}\sin{i}$ values and the literature. The `Rein18' label stands for the results presented in \citet{reiners2018}.}
    \label{fig:scatter_vsini}
\end{figure}

We also compared our $\textit{v}\sin{i}$ determinations with the ones derived by \citet{reiners2018} using the cross-correlation method and with those obtained following the DL methodology described by \citetalias{pass20}. Fig. \ref{fig:scatter_vsini} shows how our derived $\textit{v}\sin{i}$ are mostly consistent with the literature within their errors. Both \citetalias{pass20} and \citet{reiners2018} results show a good correlation with our values ($r_{\rm p}=0.99$ and 0.98, respectively). Since most of the objects are located at lower $\textit{v}\sin{i}$ values, it is convenient to split the analysis provided in Table \ref{tab:comparison_lit} at a cut-off value of $v\sin{i}\,{\rm (this\,work)}=12$\,km\,s$^{-1}$. Below this value, \citetalias{pass20} presents $\overline{\Delta}=1.83$\,km\,s$^{-1}$ and ${\rm rmse}=1.97$\,km\,s$^{-1}$, with $\overline{\Delta}=-1.22$\,km\,s$^{-1}$ and ${\rm rmse}=1.45$\,km\,s$^{-1}$ for faster rotators. Similarly, for \citet{reiners2018}, we obtained $\overline{\Delta}=-0.68$\,km\,s$^{-1}$ and ${\rm rmse}=1.24$\,km\,s$^{-1}$ for values below the threshold, and $\overline{\Delta}=-3.47$\,km\,s$^{-1}$ and ${\rm rmse}=3.71$\,km\,s$^{-1}$ for values above.


\section{Conclusions}\label{acs_sec:conclusions}

This work serves as an extension of a series of papers (\citetalias{pass20}; \citetalias{bello2023}) dedicated to exploring the use of DL for stellar parameter estimation of CARMENES M dwarfs, based on synthetic spectra. \citetalias{bello2023} developed a model-based DTL technique to bridge the significant differences in flux features between the two spectral families, reported by \citetalias{pass20}. Here, we propose a parallel feature-based DTL strategy that addresses the limitations mentioned in their work regarding the need for high-quality stellar parameter estimations in the knowledge transfer process.

Using a methodology that combines the use of AEs and CNNs, we derived new estimations for the stellar parameters $\textit{T}_{\rm eff}$, log\,$\textit{g}$, [M/H], and $\textit{v}\sin{i}$ of 286 M dwarfs observed with CARMENES. The AE models were trained on PHOENIX-ACES synthetic spectra and then fine-tuned using the CARMENES high-S/N, high-resolution spectra. In the fine-tuning process, no data other than the observed spectra are required, which gives our methodology great flexibility, as no measured stellar parameters are involved in the knowledge transfer. We used the low-dimensional representations of the synthetic and observed spectra, resulting from the initial training and the fine-tuning steps, respectively, as input to the CNNs for the estimation of the stellar parameters. In this way, parameter estimation is conducted using a dataset in which no significant differences in the feature distributions between the synthetic and observed data are evident.

We performed an in-depth analysis of our estimated stellar parameters, using the diagram shown in Fig. \ref{fig:par_diags} to study the objects that deviate from the main sequence. We found that almost all the overlumimuous outliers are identified as H$\alpha$ active stars by \citet{schofer2019}, while outliers located below the main sequence are typically metal-poor stars from the thick disc Galactic population. In particular, using the \texttt{BANYAN}~$\Sigma$ tool, we found 9 objects with a high Bayesian probability of belonging to five different young stellar associations, in 7 of these cases with a probability of more than 95\,\%. Together with the low-metallicity objects already reported in \citetalias{mar21} and \citetalias{schw19}, we identified eight more stars that exhibit the same behaviour.

We also conducted a comparative study between our results and the latest studies using CARMENES data, finding good consistency with the literature in most cases. Both our $\textit{T}_{\rm eff}$ and log\,$\textit{g}$ determinations are, in general, strongly correlated with the results from the literature, with a systematic deviation in our $\textit{T}_{\rm eff}$ scale towards hotter values for estimations above 3\,750\,K. As expected, our parameter determinations are in very good agreement with \citetalias{pass20}, since their methodology is the most similar to the one presented in this paper. More importantly, the deviation in metallicity attributed to the synthetic gap in their work is not observed in ours thanks to the DTL approach. This, together with the work presented by \citetalias{bello2023}, demonstrates the great potential of DTL-based strategies to bridge the synthetic gap in stellar parameter estimation from synthetic spectra.

\chapter{Characterisation of Ultracool Dwarfs with Deep Transfer Learning} \label{chp:dtl_ucds}

The future is bright for the field of ultracool dwarfs. The Visible Instrument (VIS) and the Near-Infrared Spectrometer and Photometer (NISP) aboard the ESA \textit{Euclid} mission will provide a unique combination of wide-area ($\sim15\,000$\,deg$^2$) coverage, high-spatial resolution, and unprecedented sensitivity, with a low-resolution near-infrared spectroscopic survey that will enable the spectral characterisation of a huge number of previously undiscovered ultracool dwarfs. This was recently demonstrated by \citet{jerry2024}, who highlighted the reliability that the data provided by the slitless spectroscopic mode of the NISP instrument will deliver for the spectral characterisation of ultracool dwarfs in both the deep and wide surveys. In the summer of 2027, the NASA Nancy Grace Roman Space Telescope will join Euclid to explore the infrared sky as never before possible, with a much deeper and more precise core survey, but over a smaller area ($\sim2\,000$\,deg$^2$). These surveys will be complemented in the optical by the LSST, carried out in the Vera C. Rubin Observatory, expected to start operations in mid-2025. The LSST will provide a high-spatial resolution, high-cadence, and high-sensitivity multi-band photometric survey over the entire Southern Hemisphere sky, that will supersede the previous SDSS and Pan-STARRS optical datasets. The combination of all these upcoming surveys will lead to a quantum leap of over an order of magnitude in the number of ultracool dwarfs detected \citep{solano2021,martin2023}, enabling the study of more distant ultracool dwarf populations than ever before.

The low-resolution near-infrared spectroscopic survey conducted by \textit{Euclid} represents a key opportunity for developing a flexible, automated, and reliable methodology capable of harnessing these vast amounts of data to spectroscopically classify ultracool dwarfs and determine their effective temperature. In this line, we leveraged of the deep transfer learning framework introduced by \citet{masbuitrago2024} and explored its adaptation to the low-resolution, ultracool dwarf domain, with a view to its further application to the \textit{Euclid} dataset.

\section{Testbed environment with SpeX}

\begin{figure}
    \centering
	\includegraphics[width=.48\columnwidth]{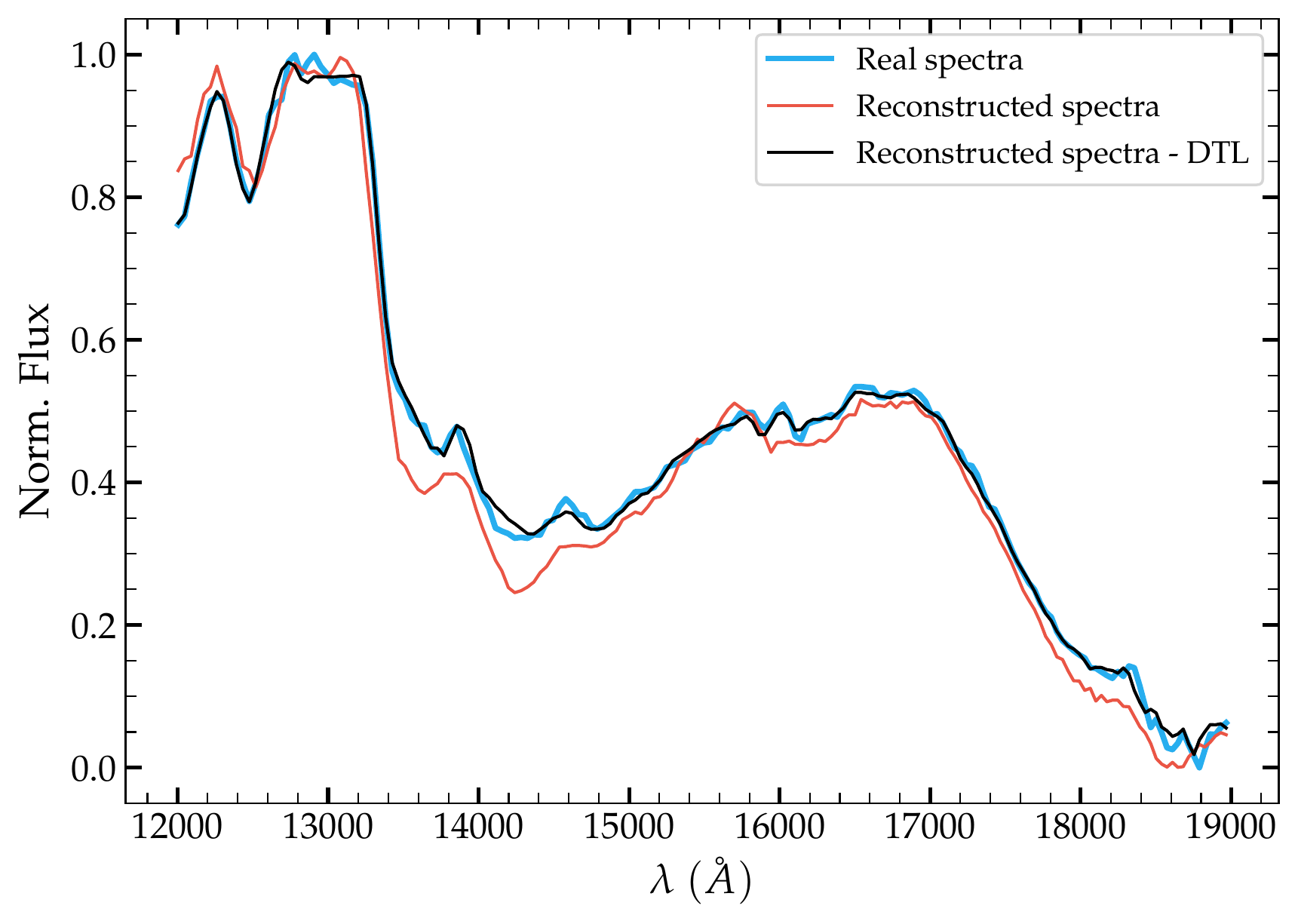}
	\includegraphics[width=.48\columnwidth]{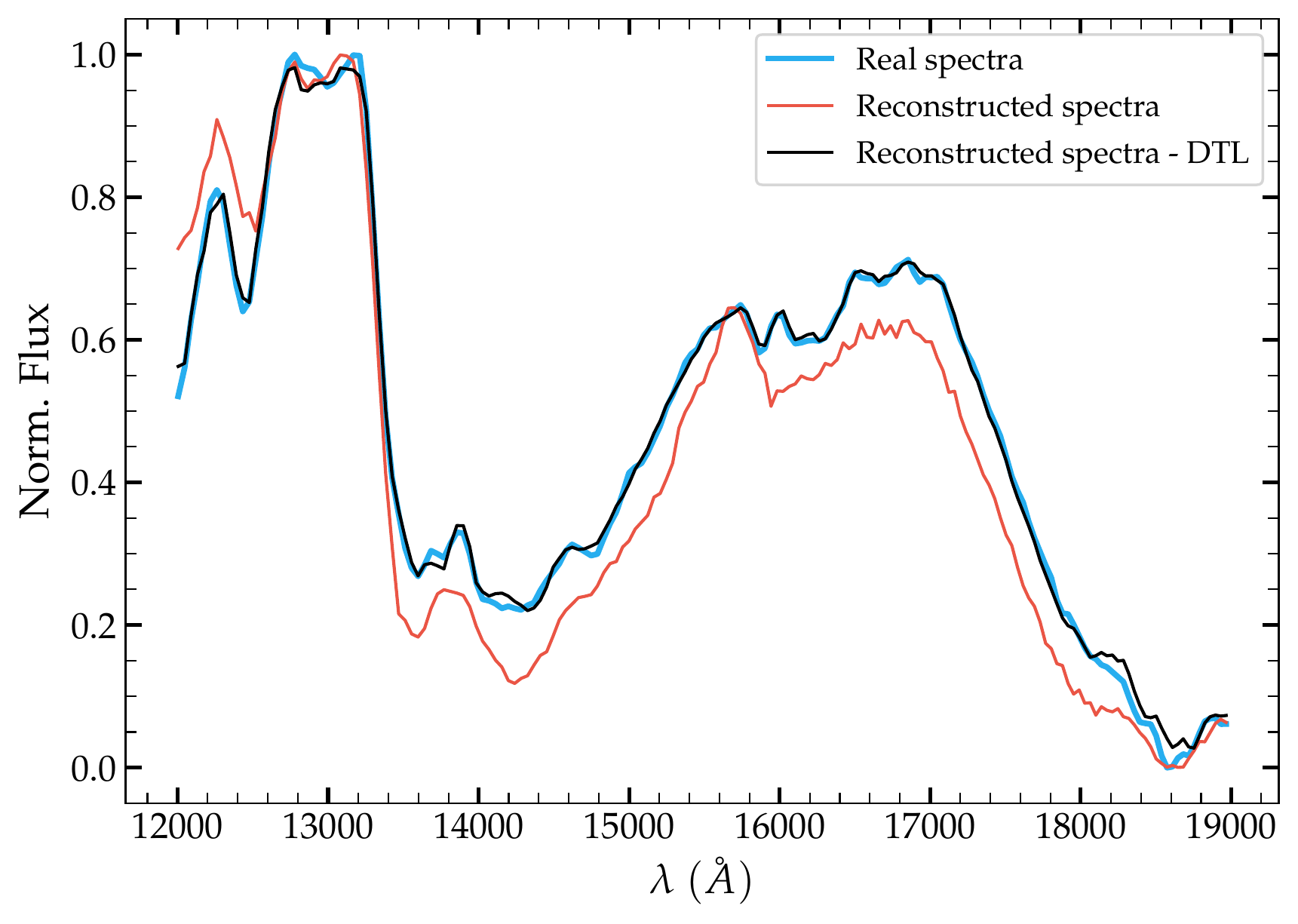}
    \caption{Original (blue) vs. reconstructed SpeX spectra of 2MASS J02540582-1934523 (M8) and 2MASSW J0015447+351603 (L2). The figure shows how the reconstruction after fine-tuning (black) captures much more detailed spectral features than the reconstruction with the initial training (red).}
    \label{fig:spex_reco}
\end{figure}

To adjust our methodology and prepare it to the arrival of the first spectroscopic data from \textit{Euclid}, we created a testbed environment using low-resolution spectra from the SpeX Prism Library, an online repository of over 3\,000 low-resolution, near-infrared spectra, primarily of ultracool dwarfs. The spectra available in the SpeX Prism Library were observed with the prism mode of the SpeX spectrograph \citep{Rayner2003} of the NASA Infrared Telescope Facility, with a resolving power $\sim200$ across $0.8-2.5$\,$\mu$m when using the 0.8\,arcsec slit. This repository is easily accessible using \texttt{SPLAT} \citep{splat}, a python-based access and analysis package designed to search for spectral data in the SpeX Prism Library and perform comprehensive spectral analysis. To obtain our sample of spectra, or target domain, we cross-matched the sample of high-quality ultracool dwarfs from the UltracoolSheet catalogue, presented in Section \ref{sec:ucds_intro}, with the SpeX Prism Library and obtained a final sample of 692 spectra with a spectroscopic classification in SpeX covering spectral types from M6 to T9.

As source domain for our deep transfer learning methodology, we built a grid of synthetic spectra based on the recent Sonora Elf Owl \citep{elfowl} substellar atmosphere models, which present developments in atmospheric chemistry compared to earlier model collections such as Sonora Bobcat \citep{bobcat} or Sonora Cholla \citep{cholla}. For this, we adjusted the Sonora Elf Owl models to the resolution and wavelength solution of SpeX\,\footnote{The adapted models are available in the \texttt{ucdmcmc} package of Dr. Adam Burgasser: \url{https://github.com/aburgasser/ucdmcmc/tree/main}}, and added three different random Gaussian noise values to each spectrum to enrich the dataset, ending up with a final synthetic grid of 31\,050 spectra (see Table 2 in \citealt{elfowl} for the grid of parameters). For both the synthetic and the observed spectra, we only considered the wavelength interval $12\,000-19\,000$\,\AA, since this will be the range covered by the \textit{Euclid} wide survey \citep{euclid2023}. Moreover, given the temperature constraints of the Sonora Elf Owl models, we retained only the SpeX spectra corresponding to spectral types M8 or later, and only kept the highest SNR spectrum when several were available for the same source. Doing this, we ended up with a sample of 585 SpeX spectra.

\section{Ultracool dwarf characterisation}

To determine the effective temperature of our sample of low-resolution, near-infrared ultracool dwarf spectra, we replicated the deep transfer learning methodology presented by \citet{masbuitrago2024}. First, we trained the autoencoder neural networks using the grid of synthetic spectra, obtaining reconstruction errors $\sim10^{-4}$ on the test set. Since the number of input features is significantly smaller than in \citet{masbuitrago2024} due to the lower resolution of the data, we adjusted the number of neurons in the input layer and reduced the number of hidden layers to two in the autoencoder architectures (see Fig. \ref{fig:ac_info}). For the knowledge transfer process, we fine-tuned the autoencoder neural networks with the SpeX spectra, tailoring the high-level features of the encoder network to our target domain. Figure \ref{fig:spex_reco} shows the importance of this step to adapt the autoencoder to our target domain, ensuring that the compressed representations obtained with the fine-tuned autoencoders are more meaningful than those obtained without the transfer learning process.

\begin{figure}
    \centering
	\includegraphics[width=.48\columnwidth]{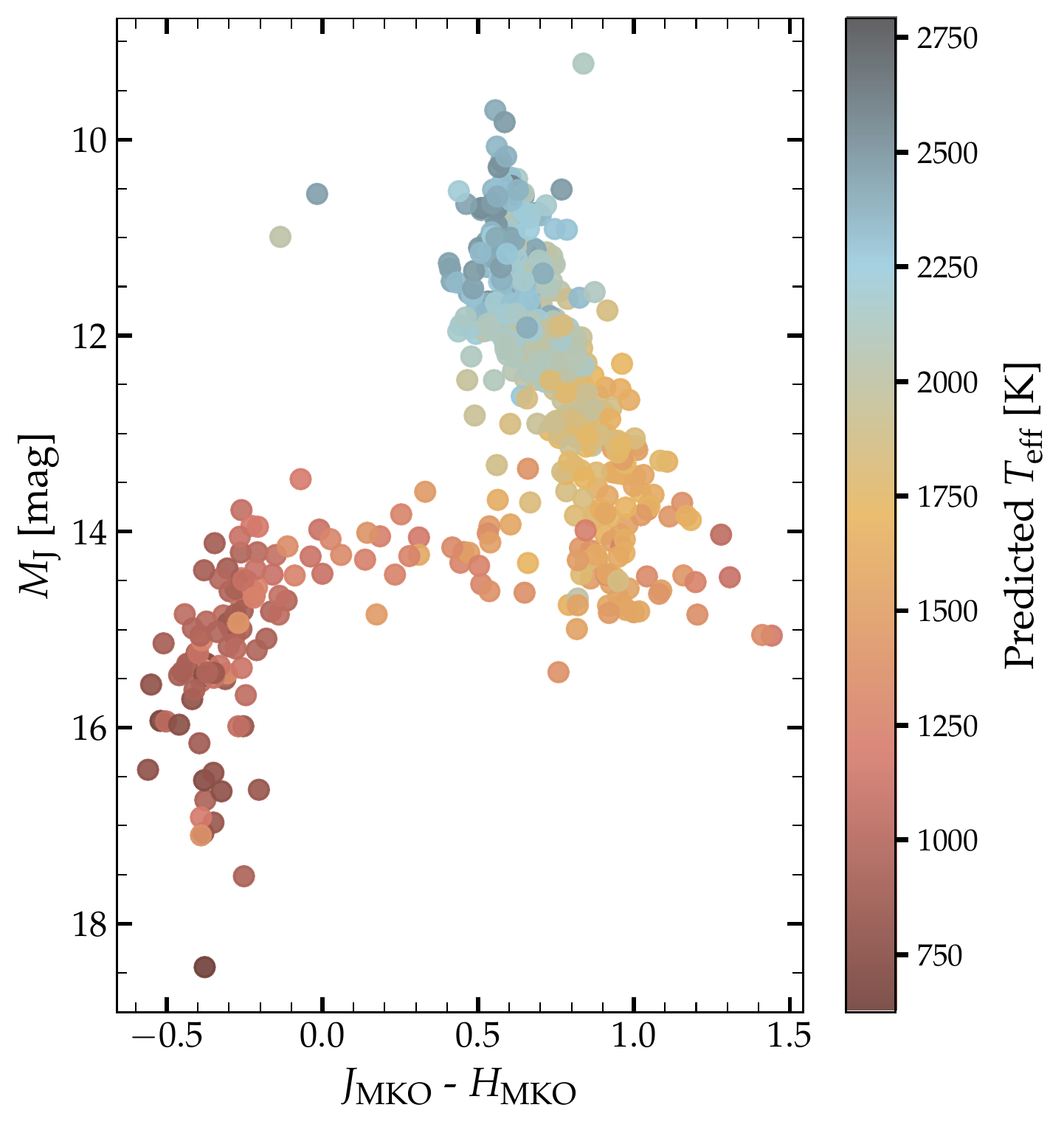}
	\includegraphics[width=.48\columnwidth]{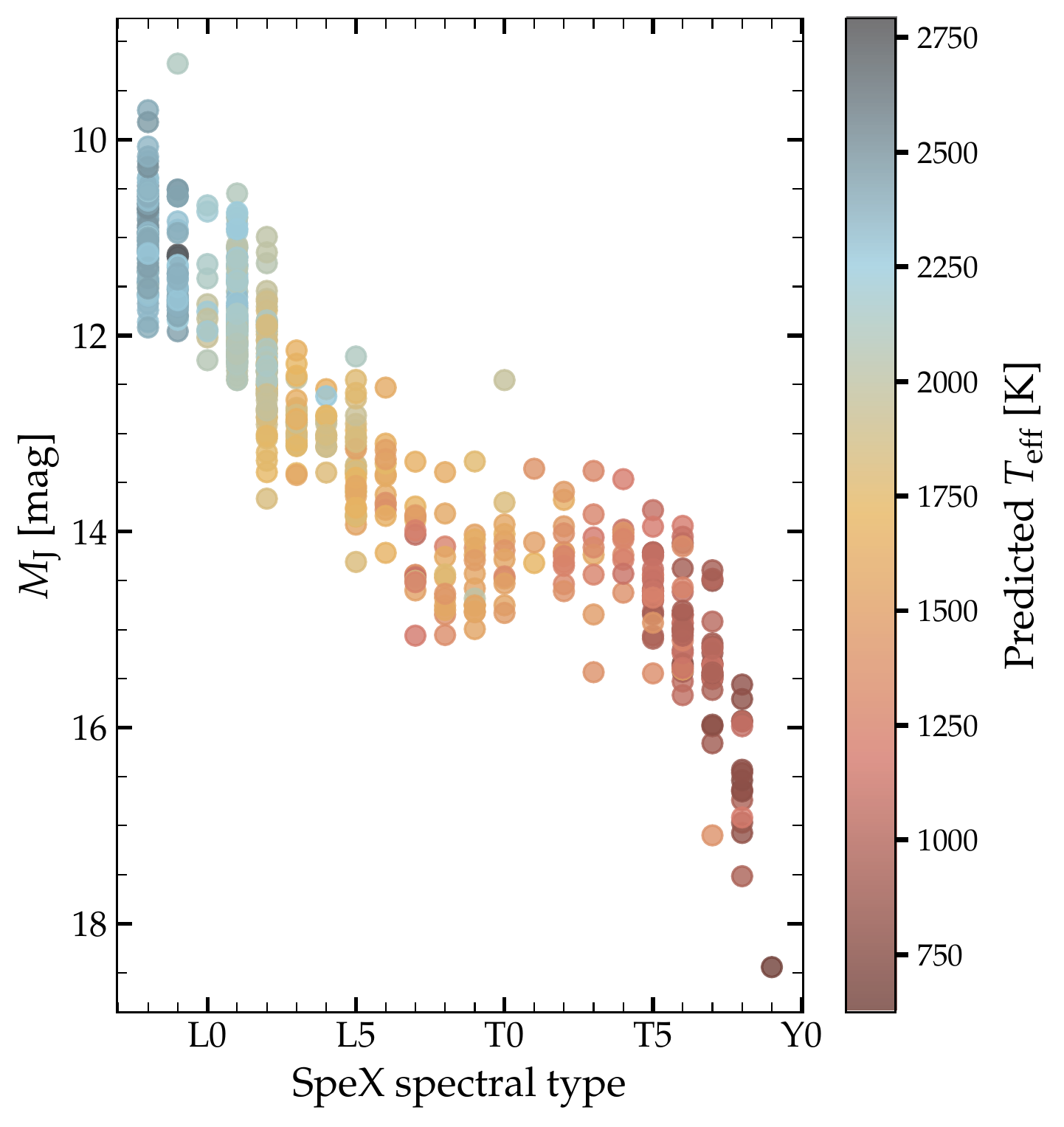}
    \caption{\textit{Left panel:} $M_{\mathrm{J}}$ vs. $\textit{J}-\textit{H}$ colour-magnitude diagram, with the dots colour-coded by spectral type, of our sample of ultracool dwarfs. \textit{Right panel:} Evolution of $M_{\mathrm{J}}$ with the spectral type for the same sample. The dots are colour-coded by the $\textit{T}_{\rm eff}$ determined in this work. The magnitudes and the spectral types have been taken from the UltracoolSheet and SpeX, respectively.}
    \label{fig:ucds_teff}
\end{figure}

We used the low-dimensional representations of the Sonora Elf Owl and SpeX spectra, resulting from the initial training and the fine-tuning of the autoencoders, respectively, as input to the convolutional neural networks (see Fig. \ref{fig:cnn}) for the estimation of the effective temperature of our target sample. We calculated the minimum Euclidean and correlation distances  from each SpeX instance to the synthetic grid in both the initial high-dimensional space and the new low-dimensional feature space, obtaining a reduction of over an order of magnitude for the compressed low-dimensional representations, averaged over all the sets obtained from the different autoencoder architectures. In this way, we effectively bridge the gap between the two domains, and parameter estimation is conducted using a dataset in which discrepancies in feature distributions between the synthetic and observed data are reduced.

\begin{figure}
    \centering
	\includegraphics[width=.48\columnwidth]{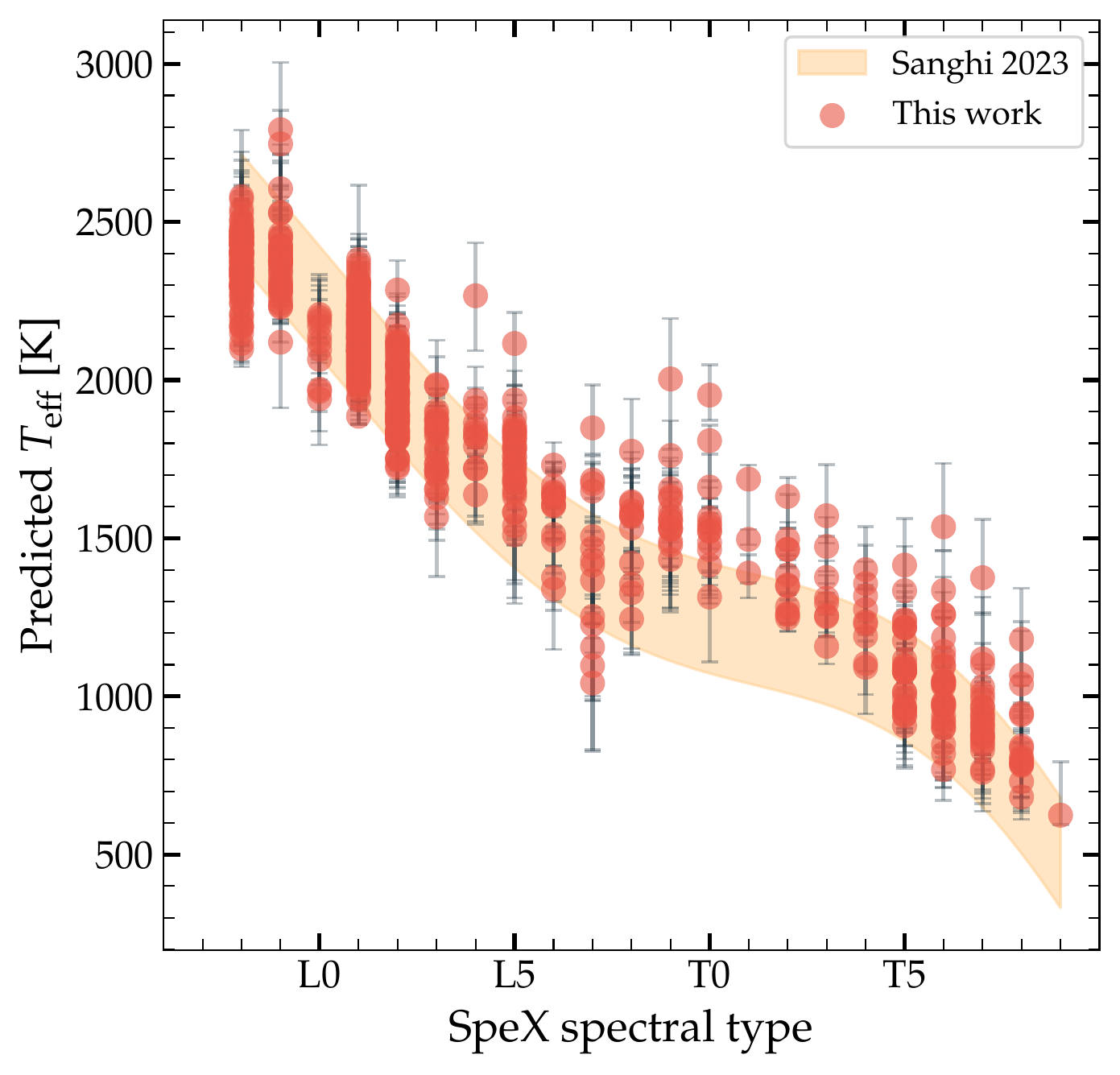}
	\includegraphics[width=.48\columnwidth]{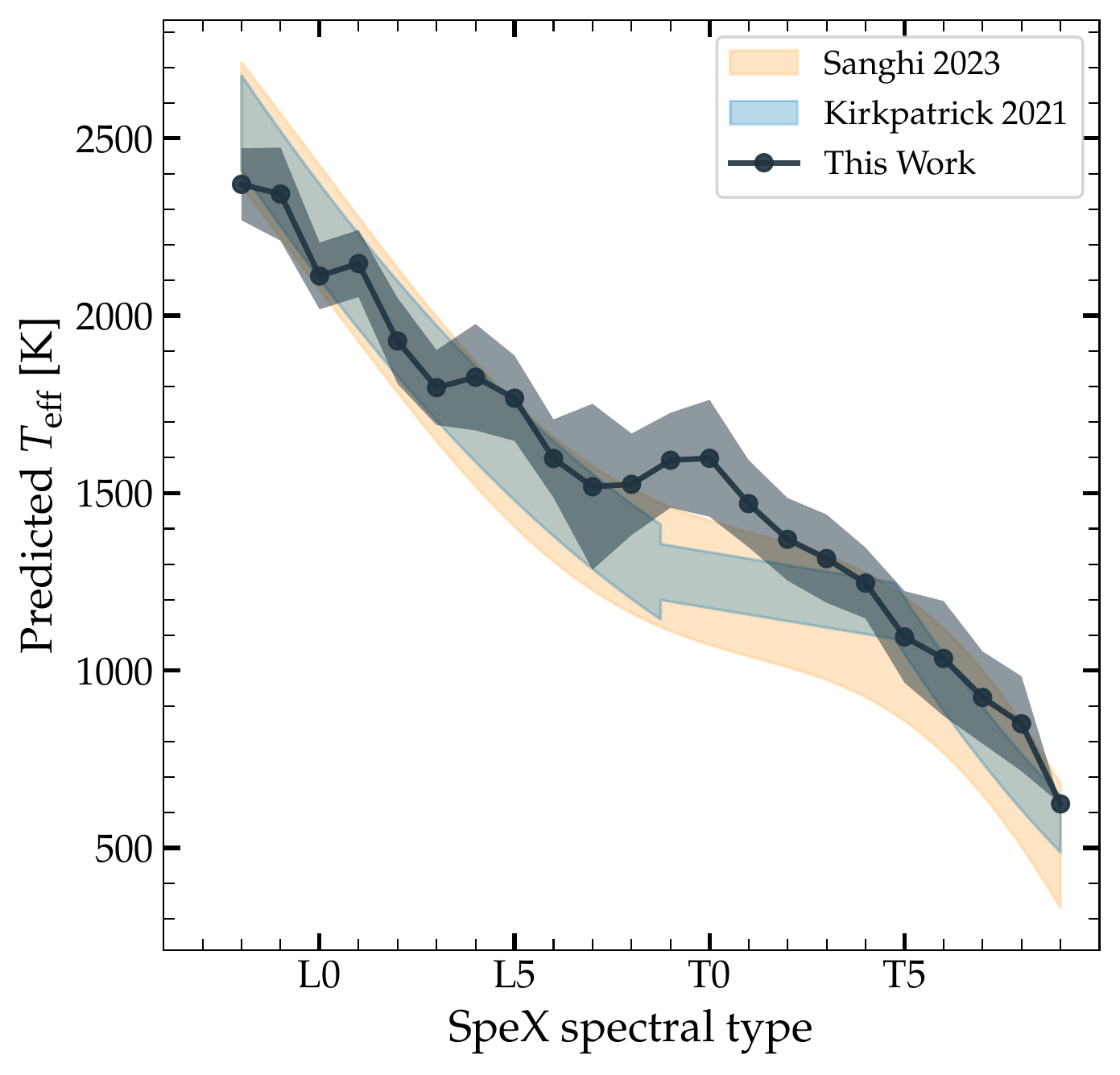}
    \caption{\textit{Left panel:} Effective temperatures derived in this work, for our sample of ultracool dwarfs (red dots with grey error bars), as a function of the spectral type listed in SpeX. The shaded orange area indicates the semi-empirical relation by \citet{sanghi2023}. \textit{Right panel:} Mean effective temperatures (black dots), weighted using the uncertainties derived in this work, for each of the spectral types. The shaded black area marks the standard deviation for each of the spectral types. The shaded orange and blue areas indicate the semi-empirical relations by \citet{sanghi2023} and \citet{kirkpatrick2021}, respectively.}
    \label{fig:spex_spts_teff}
\end{figure}

Figure \ref{fig:ucds_teff} reproduces the diagrams presented in Fig. \ref{fig:ucds_diag}, colouring the dots with the effective temperatures determined for our target sample to illustrate the temperature evolution of ultracool dwarfs. The near-infrared colour-magnitude diagram in the left panel shows how, when the trend changes abruptly to bluer $\textit{J}-\textit{H}$ values, the effective temperature of the ultracool dwarfs remains roughly constant at $\sim1450$\,K. During this L/T transition, visible in both panels as a plateau in $M_{\mathrm{J}}$, the effective temperature evolves very slowly \citep{golimovsky2004,kirkpatrick2021}, decreasing only $\sim200$\,K throughout the entire transition. Table \ref{tab:ucds_teffs} lists all the effective temperatures determined for our sample of ultracool dwarfs.

\begin{figure}
    \centering
	\includegraphics[width=.7\columnwidth]{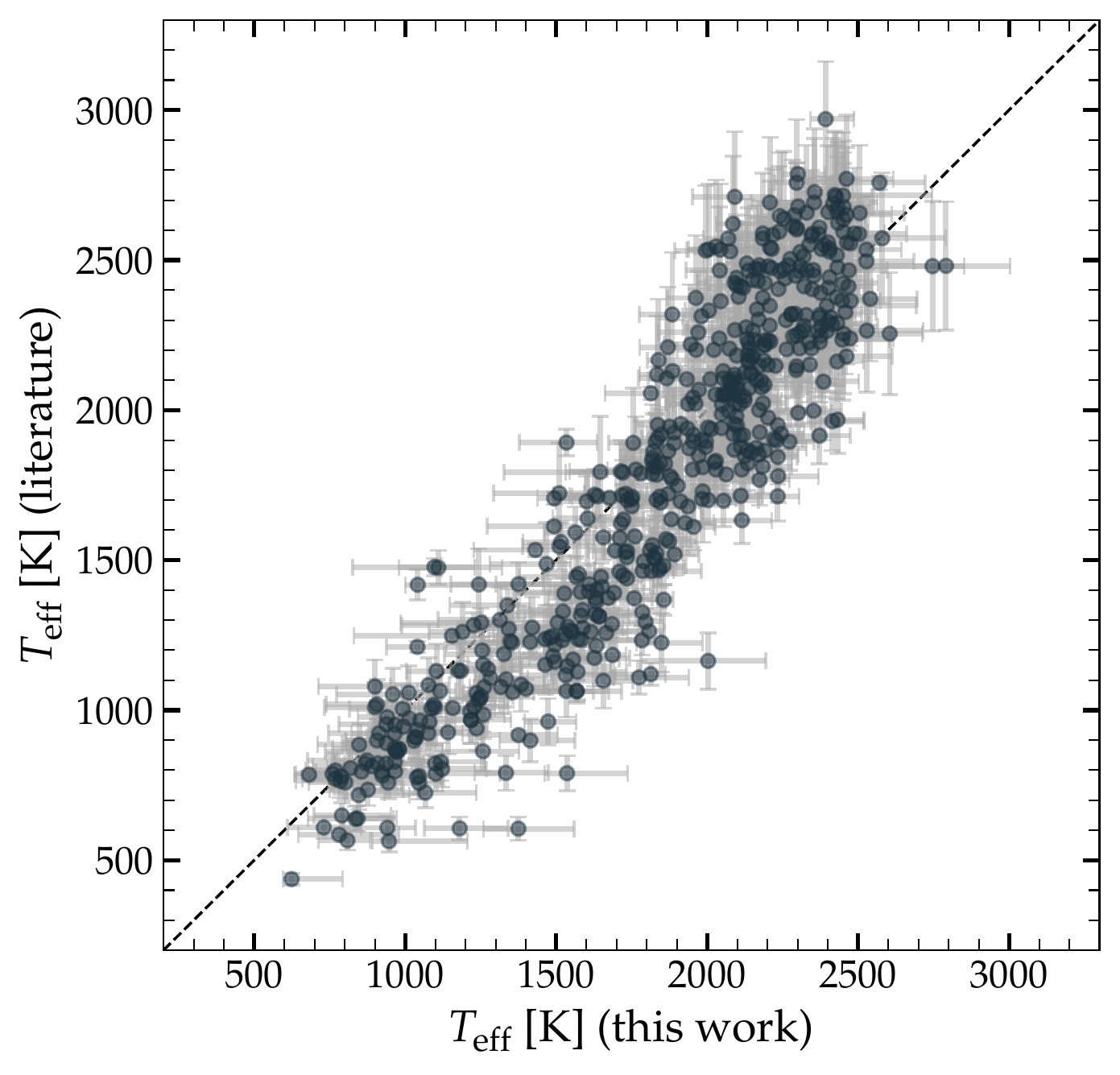}
    \caption{Comparison between our derived effective temperatures (X axis) and those in \citet{sanghi2023}.}
    \label{fig:teffs_comp_sanghi}
\end{figure}

Figure \ref{fig:spex_spts_teff} shows the determined effective temperatures as a function of spectral type, together with the mean weighted with the uncertainties derived in this work and standard deviation for each of the spectral types (right panel), which are listed in Table \ref{tab:spt_teff_rel}. Both panels demonstrate how the temperature decreases steeply for spectral types M8-L7 and $\sim$T2-T9, with the well-known narrow range of effective temperature throughout the L/T transition. The right panel shows how the calculated effective temperatures are in general in very good agreement with the semi-empirical relations from \citet{kirkpatrick2021} and \citet{sanghi2023}. Our values are in average higher than the aforementioned relations during the L/T transition. Since \citet{sanghi2023} uses also uses a sample of ultracool dwarfs extracted from the UltracoolSheet catalogue, we can directly compare our effective temperature determinations with their semi-empirical values. Figure \ref{fig:teffs_comp_sanghi} illustrates this comparison, confirming a good consistency between the two sets and a deviation towards higher values in our temperatures for the L/T transition. This transition is still a less understood phase of ultracool dwarf evolution. The increase of cloud opacity from early-L to late-L dwarfs, and the evolution to cloudless T dwarfs, hugely complicates the modelling of these atmospheres. In the future, a better treatment of clouds for this transition in atmospheric models will be the key to mitigating this effect.

\begin{table}
\fontsize{11pt}{11pt}\selectfont
 \caption{Relation between spectral type and effective temperature for ultracool dwarfs derived in this work.}
 \label{tab:spt_teff_rel}
 \centering          
 \begin{tabular}{l c c c}
  \hline\hline
  \noalign{\smallskip}
  
  Spectral type & Weighted mean $\textit{T}_{\rm eff}$\,$^{(a)}$ & Standard deviation & Number of objects\,$^{(b)}$\\

    & [K] & [K] & \\
  
  \noalign{\smallskip}
  \hline
  \noalign{\smallskip}
  
  M8 & 2370 & 101 & 86\\
  
  \noalign{\smallskip}
  
  M9 & 2343 & 131 & 36\\
  
  \noalign{\smallskip}

  L0 & 2112 & 94 & 11\\
  
  \noalign{\smallskip}

  L1 & 2147 & 94 & 132\\
  
  \noalign{\smallskip}

  L2 & 1929 & 121 & 62\\
  
  \noalign{\smallskip}

  L3 & 1798 & 105 & 25\\
  
  \noalign{\smallskip}
  
  L4 & 1827 & 150 & 12\\
  
  \noalign{\smallskip}

  L5 & 1768 & 120 & 32\\
  
  \noalign{\smallskip}

  L6 & 1598 & 110 & 13\\
  
  \noalign{\smallskip}

  L7 & 1518 & 234 & 14\\
  
  \noalign{\smallskip}

  L8 & 1525 & 143 & 12\\
  
  \noalign{\smallskip}

  L9 & 1593 & 134 & 15\\
  
  \noalign{\smallskip}

  T0 & 1598 & 165 & 12\\
  
  \noalign{\smallskip}
  
  T1 & 1471 & 123 & 3\\
  
  \noalign{\smallskip}

  T2 & 1370 & 116 & 11\\
  
  \noalign{\smallskip}

  T3 & 1316 & 124 & 8\\
  
  \noalign{\smallskip}

  T4 & 1247 & 100 & 9\\
  
  \noalign{\smallskip}

  T5 & 1095 & 129 & 26\\
  
  \noalign{\smallskip}
  
  T6 & 1034 & 162 & 28\\
  
  \noalign{\smallskip}

  T7 & 924 & 130 & 22\\
  
  \noalign{\smallskip}

  T8 & 850 & 134 & 15\\
  
  \noalign{\smallskip}

  T9 & 624 & $\cdots$ & 1\\

  \noalign{\smallskip}
  \hline
 \end{tabular}
 \tablefoot{$^{(a)}$ The uncertainties derived in this work are used as weights. $^{(b)}$ Number of objects within each spectral type.}
\end{table}

\begin{figure}
    \centering
	\includegraphics[width=.71\columnwidth]{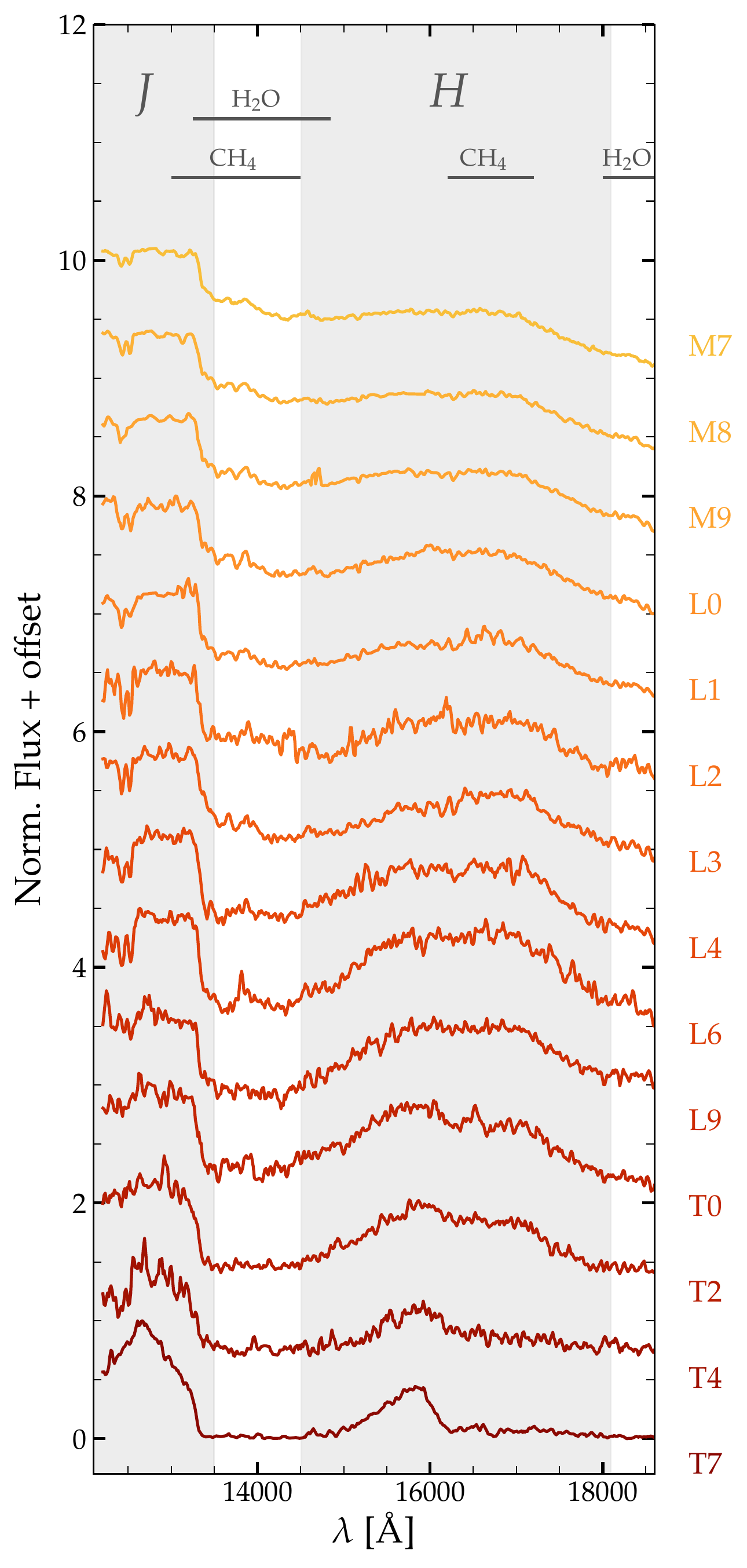}
    \caption{Spectral sequence for a sample of confirmed ultracool dwarfs using \textit{Euclid} spectra. The spectra were classified by comparing them to the standard templates published by SPLAT (see Dominguez-Tagle et al. in press.). The relevant bands, as discussed in Section \ref{sec:ucds_intro}, are highlighted.}
    \label{fig:euclid}
\end{figure}

The results obtained in this study indicate that the methodology presented by \citet{masbuitrago2024}, developed for the determination of stellar parameters of M dwarfs from high-resolution spectra, can be successfully adapted to the low-resolution domain to estimate the effective temperature of ultracool dwarfs. In this line, the methodology consolidated in this chapter will serve as a basis for the characterisation of ultracool dwarfs in the promising surveys to come in the next years, which envisage a scientific leap in this field, starting with its direct application to the wide-field \textit{Euclid} low-resolution spectroscopic survey. We are already making progress in this regard, working with the first spectroscopic data from \textit{Euclid}, and have successfully tailored the procedure to its wavelength solution. Doing this, we have applied the methodology to near-infrared, low-resolution \textit{Euclid} spectra of a sample of confirmed ultracool dwarfs (see Figure \ref{fig:euclid}), and determined the effective temperatures of these objects, which are in excellent agreement with the spectral types derived by comparing them to the standard templates published by SPLAT (see Dominguez-Tagle et al. in press.).

\chapter{General conclusions and future work}
\label{chp:general_intro}

This thesis delves into the discovery and characterisation of low-mass objects from a data-driven perspective, providing a rich catalogue of ultracool dwarf candidates and a deep transfer learning methodology for the estimation of stellar parameters of M dwarfs that we hope will be of great value for the astronomical community to exploit. In recent years, astronomy is undergoing a paradigm shift driven by an exponential growth in observational data, with new-generation surveys that have produced vast amounts of information that have pushed traditional methods of data analysis to their limits. We address this challenge by exploring the application of machine and deep learning techniques, in combination with Virtual Observatory technologies, for the development of methodologies to advance our understanding of M dwarfs and ultracool dwarfs in the years to come. The results obtained reinforce the growing role of machine learning in astronomy, highlighting its transformative potential for handling large astronomical datasets, and advocate data-driven approaches that combine Virtual Observatory technologies with machine and deep learning techniques as the way forward for the future of observational astronomy. These results have led to the publication of three scientific papers in the course of this thesis: \citet{masbuitrago2022}, \citet{masbuitrago2024}, and \citet{masbuitrago2025}.

\section{Summary of the Thesis}

The primary contributions of this work are as follows:

\begin{itemize}

    \item This thesis has demonstrated how Virtual Observatory data mining technologies can be harnessed to streamline the discovery and characterisation of ultracool dwarfs. Combining multi-filter photometry from several surveys and astrometric data, we consolidated a Virtual Observatory methodology to efficiently identify ultracool dwarf candidates in wide-field surveys and subsequently characterise them. Using this approach, we provided a catalogue of ultracool dwarfs over the entire sky coverage of the J-PLUS second data release, increasing the number of ultracool dwarfs reported in this region by $\sim135$\,\%. We demonstrated how a machine learning approach could accelerate this process, which is an important achievement considering the application of this methodology to larger and deeper surveys such as J-PAS and \textit{Euclid}. In this sense, future work could be focused in mitigating the main limitations of the developed methodology, which is based on a combination of principal component analysis and support vector machines, namely the significant number of false positives obtained prior to the determination of the effective temperature. An approach involving cost-sensitive learning techniques \citep{cost_sensitive} could be the way forward.

    \item We consolidated a deep transfer learning approach, based on autoencoder neural networks, to determine atmospheric stellar parameters of M dwarfs from high-resolution spectra. Using this methodology, we provided new estimations for the effective temperature, surface gravity, metallicity, and projected rotational velocity for 286 M dwarfs observed by the CARMENES survey, mitigating the deviations in previous works attributed to the differences between synthetic and observed data. Since no other data than the observed spectra are required in the transfer learning process, our methodology proves to be very flexible and represents a significant step forward in bridging the synthetic gap in stellar parameter estimation from synthetic spectra. We further demonstrated this by successfully adapting the procedure to the low-resolution domain to estimate the effective temperature of ultracool dwarfs using near-infrared spectra from SpeX Prism Library.

    \item We demonstrated the potential of multi-filter photometric surveys to systematically detect flare events in M dwarfs. Combining Virtual Observatory capabilities to query huge amount of data and a flexible detection algorithm developed for this end, we managed to analyse millions of spectral energy distributions and obtain a sample of flaring M dwarfs. We confirmed and studied the flaring nature of these objects using low-resolution spectra collected with NOT/ALFOSC and GTC/OSIRIS, and high-cadence photometric data from TESS. This procedure, which can easily be used in other multi-filter photometric surveys, allowed the detection of episodes of strong Ca~{\sc ii} H and K line emission, which are not usually taken into account in the study of flares in large M dwarf samples and may have important implications for exoplanetary space weather and habitability studies.

\end{itemize}

The results of this thesis have broad implications for both stellar and substellar astrophysics. This work has contributed to expanding the census of ultracool dwarfs, which are among the least understood populations due to their intrinsic faintness and complex atmospheres, by employing a flexible and scalable Virtual Observatory approach that integrates multi-filter photometry and astrometric data. The newly identified 7\,827 candidates constitute valuable targets for follow-up observations and further refinement of ultracool dwarf population statistics. By identifying Ca~{\sc ii} H and K flaring M dwarfs, and a methodology to detect them in multi-filter photometric surveys, this thesis contributes to the ongoing discussion of how stellar activity affects the long-term viability of planetary systems around these stars, which are ubiquitous in the solar neighbourhood and are prime targets for exoplanet searches.

A persistent challenge in stellar astrophysics is the discrepancy between synthetic models and observed spectra. By using a deep transfer learning approach to project synthetic and observed data into a common feature space, this thesis consolidates a novel approach to overcoming the gap between them, improving the reliability of parameter estimation from synthetic spectra in low-mass objects with a flexible and scalable methodology that can be easily applied to the large surveys expected in the years to come. As machine learning becomes more widespread, it will continue to contribute to the development of astronomical data analysis methodologies, superseding or complementing traditional methods, enabling discoveries that would otherwise be difficult or impossible to achieve.

\section{Future Directions} 

The methodologies developed in this thesis open up several promising avenues for future research. In this sense, future efforts should focus on scaling up these approaches and integrate them into the workflow of next-generation astronomical surveys such as J-PAS, \textit{Euclid} or LSST, which will dramatically increase the amount of data available. Since machine learning pipelines will play a crucial role in managing the vast datasets produced by these missions, the techniques developed in this thesis can be adapted to upcoming surveys to automate the discovery and characterisation of low-mass stellar and substellar objects. 

The most direct application of the work developed in this thesis is the use of the deep transfer learning methodology presented in Chapters \ref{chp:autoencoders_paper} and \ref{chp:dtl_ucds} for the discovery and characterisation of ultracool dwarfs in the first data release of \textit{Euclid}, which will be publicly available in mid-2026. In this sense, we will go a step further in the methodology, taking advantage of the power of autoencoder architectures for outlier detection, to identify ultracool dwarfs using the whole dataset of low-resolution spectra from the \textit{Euclid} wide-field spectroscopic survey. This can be achieved by using autoencoder neural networks trained with a grid of synthetic spectra, and fine-tuned with real \textit{Euclid} spectra corresponding to well-known ultracool dwarfs. By analysing the reconstruction error of this system, we will study the use of a limiting value to discard all objects for which a significantly higher reconstruction error is obtained, which will make it possible to analyse huge amounts of spectroscopic data in a very efficient way, retaining only the ultracool objects of interest. A cornerstone of this process will be ESA Datalabs\,\footnote{\url{https://datalabs.esa.int/}}, a new infrastructure built around the ESA science archives that provides unique archival data access capabilities, bringing the solution to the data rather than the other way around, playing a pivotal role in the development of the aforementioned system due to the huge volume of data that will be processed. After the discovery phase, we will follow a procedure similar to that discussed in Chapter \ref{chp:dtl_ucds} to characterise the low-resolution spectra of the identified ultracool dwarfs, creating a rich catalogue of spectroscopically characterised ultracool dwarfs that could be of great value to the astronomical community. The main strength of this methodology lies in its ability to eliminate biases that are present in other classical methodologies, in which objects are first selected on the basis of colours and then its ultracool nature is confirmed using spectra. Furthermore, we plan to test this methodology using different sets of atmosphere and evolutionary models, such as the ATMO 2020 \citep{atmo2020}, to study in detail the differences with our current setup.

As the data tsunami in observational astronomy continues to grow, artificial intelligence will become increasingly essential in processing, analysing, and interpreting the information about our cosmos. This thesis has demonstrated how machine and deep learning-driven methods, combined with the infrastructure of the Virtual Observatory, can empower the discovery and characterisation of M dwarfs and ultracool dwarfs. The data-driven techniques developed in this work pave the way for the automation of astrophysical analysis in the low-mass regime, enabling researchers to fully exploit the potential of the vast and growing datasets expected by upcoming astronomical surveys.

\chapter*{Data and Software Availability}\label{data_av}
\addcontentsline{toc}{chapter}{Data and Software Availability} 

During the development of this thesis, we have endeavoured to build several publicly accessible catalogues, codes and tools that can be exploited by the astronomical community. To help the researchers use our catalogues of ultracool dwarfs presented in Chapter \ref{chp:ucds_paper}, we provide an archive system that can be accessed  from  a webpage\,\footnote{\url{http://svocats.cab.inta-csic.es/jplus_ucds1; http://svocats.cab.inta-csic.es/jplus_ucds2}} or through a Virtual Observatory ConeSearch\,\footnote{e.g.\url{http://svocats.cab.inta-csic.es/jplus_ucds1/cs.php?RA=0.023&DEC=35.457&SR=0.1&VERB=2; http://svocats.cab.inta-csic.es/jplus_ucds2/cs.php?RA=238.569&DEC=52.742&SR=0.1&VERB=2}}. The archive system implements a  very simple search interface that allows queries by coordinates and radius as well as by other parameters of interest. The user can also select the maximum number of sources (with values from ten to unlimited). The result can be obtained as an HTLM table or downloaded as a VOTable or a CSV file. Detailed information on the output fields can be obtained placing the mouse over the question mark located close to the name of the column. The archive also implements the SAMP\,\footnote{\url{http://www.ivoa.net/documents/SAMP}} (Simple Application Messaging) Virtual Observatory protocol, which allows Virtual Observatory applications to communicate with each other in a seamless and transparent manner for the user. In this way, the results of a query can be easily transferred to other Virtual Observatory applications, such as, for instance, \texttt{TOPCAT}.

All the resources presented in Chapter \ref{chp:autoencoders_paper}, including the code developed to build the methodology described in Section \ref{acs_sec:methodology} and the code to reproduce the figures displayed in Section \ref{acs_sec:results} are publicly available at \texttt{GitHub}\,\footnote{\url{https://github.com/pedromasb/autoencoders-CARMENES}}. The catalogue of stellar atmospheric parameters for 286 CARMENES M dwarfs determined using our deep transfer learning methodology is available at VizieR\,\footnote{\url{https://cdsarc.cds.unistra.fr/viz-bin/cat/J/A+A/687/A205}}, and we also provide a data discovery interface that allows its interactive exploration\,\footnote{\url{https://cab.inta-csic.es/users/pmas/}}. Moreover, the files with the reduced spectra and processed TESS light curves used in Chapter \ref{chp:flares_paper}, and the code to reproduce the figures displayed in Section \ref{sec:results} are publicly available at \texttt{GitHub\,}\footnote{\url{https://github.com/pedromasb/flaring-MDwarfs}}.

On the other hand, we have also carried out several parallel projects dedicated to bringing new data analysis technology closer to the user. The author has created several tutorials, available at \texttt{GitHub}\,\footnote{\url{https://github.com/pedromasb/tutorials}}, that cover useful Python features, and developed a tool help the user create and share interactive visualizations without the need to write any code\,\footnote{\url{https://magicplotter.streamlit.app/}}. Moreover, the author has contributed to the organisation and development of sessions at the Centro de Astrobiología, aimed at sharing knowledge about Python tips and modules that are generally useful in Astrophysics data analysis\,\footnote{\url{https://github.com/PyCoffees/notebooks}}.

\chapter*{Acknowledgments}
\addcontentsline{toc}{chapter}{Acknowledgments} 

The author would like to thank Dra. Amelia Bayo and Dr. Chris Theissen for their insightful comments on the thesis, which have greatly enriched the value of the manuscript.

The work presented in this thesis has been possible thanks to the funding by INTA through grant PRE-OVE. This research has made use of the Spanish Virtual Observatory (\url{https://svo.cab.inta-csic.es}) project funded by MCIN/AEI/\\10.13039/501100011033/ through grant PID2020-112949GB-I00 and MDM-2017-0737 at Centro de Astrobiolog\'{i}a (CSIC-INTA), Unidad de Excelencia Mar\'{i}a de Maeztu. 

This research has made extensive use of the SIMBAD database \citep{Wenger00}, VizieR catalogue access tool \citep{och2000}, Aladin sky atlas \citep{aladin} provided by CDS, Strasbourg, France, and of the TOPCAT \citep{Taylor2005} and STILTS \citep{Taylor2006} tools. We also made use of VOSA \citep{vosa} and the SVO \textit{Carlos Rodrigo} Filter Profile Service \citep{fps}, developed under the Spanish Virtual Observatory project. We made extensive use of Python throughout the entire work, including the packages \texttt{pandas}\footnote{\url{https://github.com/pandas-dev/pandas}}, \texttt{seaborn} \citep{seaborn}, \texttt{numpy} \citep{numpy}, \texttt{matplotlib} \citep{matplotlib}, \texttt{scikit-learn} \citep{sklearn}, \texttt{tensorflow} \citep{tensorflow}, \texttt{umap-learn} \citep{McInnes2018}, \texttt{plotly}\footnote{\url{https://github.com/plotly/plotly.py}}, \texttt{lightkurve} \citep{lightkurve}, \texttt{scipy} \citep{scipy}, \texttt{astropy} \citep{astropy}, and \texttt{mocpy} \citep{mocpy}. We want to thank Jonathan Gagné and J. Lillo-Box for the development of the BANYAN $\Sigma$\footnote{\url{http://www.exoplanetes.umontreal.ca/banyan/}} and \texttt{tpfplotter}\footnote{\url{www.github.com/jlillo/tpfplotter}} tools, respectively. We also made use of the SPLAT tool \citep{splat}, a collaborative project of research students in the UCSD Cool Star Lab \footnote{\url{http://www.coolstarlab.org}}. 

Part of this work is based on observations made with the Gran Telescopio Canarias (GTC), installed at the Spanish Observatorio del Roque de los Muchachos of the Instituto de Astrofísica de Canarias, on the island of La Palma. We obtained data with the instrument OSIRIS, built by a Consortium led by the Instituto de Astrofísica de Canarias in collaboration with the Instituto de Astronomía of the Universidad Autónoma de México. OSIRIS was funded by GRANTECAN and the National Plan of Astronomy and Astrophysics of the Spanish Government. We collected spectra with the Nordic Optical Telescope, owned in collaboration by the University of Turku and Aarhus University, and operated jointly by Aarhus University, the University of Turku and the University of Oslo, representing Denmark, Finland and Norway, the University of Iceland and Stockholm University at the Observatorio del Roque de los Muchachos, La Palma, Spain, of the Instituto de Astrofisica de Canarias. The data presented here were obtained (in part) with ALFOSC, which is provided by the Instituto de Astrofisica de Andalucia (IAA) under a joint agreement with the University of Copenhagen and NOT. We thank NOT support astronomers David Jones and Tapio Pursimo for their help on site.

Funding for the J-PLUS project has been provided by the Governments of Spain and Arag\'on through the Fondo de Inversiones de Teruel; the Aragonese Government through the Research Groups E96, E103, E16\_17R, and E16\_20R; the Spanish Ministry of Science, Innovation and Universities (MCIU/AEI/FEDER, UE) with grants PGC2018-097585-B-C21 and PGC2018-097585-B-C22; the Spanish Ministry of Economy and Competitiveness (MINECO/FEDER, UE) under AYA2015-66211-C2-1-P, AYA2015-66211-C2-2, AYA2012-30789, and ICTS-2009-14; and European FEDER funding (FCDD10-4E-867, FCDD13-4E-2685). The Brazilian agencies FAPERJ and FAPESP as well as the National Observatory of Brazil have also contributed to this project. Part of the work is based on observations made with the JAST80 telescope at the Observatorio Astrofísico de Javalambre (OAJ) in Teruel, owned, managed and operated by the Centro de Estudios de Física del Cosmos de Aragón (CEFCA).We thank the OAJ Data Processing and Archiving Unit (UPAD) for reducing and calibrating the OAJ data used in this work. This work presents results from the European Space Agency (ESA) space mission \emph{Gaia}. \emph{Gaia} data are being processed by the \emph{Gaia} Data Processing and Analysis Consortium (DPAC). Funding for the DPAC is provided by national institutions, in particular the institutions participating in the Gaia MultiLateral Agreement (MLA). The \emph{Gaia} mission website is \url{https://www.cosmos.esa.int/gaia}. The \emph{Gaia} archive website is \url{https://archives.esac.esa.int/gaia}. We acknowledge the use of public TESS data from pipelines at the TESS Science Office and at the TESS Science Processing Operations Center. This paper includes data collected by the TESS mission that are publicly available from the Mikulski Archive for Space Telescopes (MAST)\footnote{\url{https://mast.stsci.edu/portal/Mashup/Clients/Mast/Portal.html}}. This work has made use of the Euclid Q1 data from the {\it Euclid} mission of the European Space Agency (ESA), 2025, \url{https://doi.org/10.57780/esa-2853f3b}.

We thank the anonymous referees for the comments that helped to improve the quality of the published papers. We thank the IRTF observers (D. Chih-Chun Hsu, N. Lodieu, C. Theissen, J.-Y. Zhang) for program 2022-A011 (PI A. Burgasser). The author wish to recognise and acknowledge the very significant cultural role and reverence that the summit of Maunakea has always had within the indigenous Hawaiian community. We are most fortunate to have the opportunity to conduct observations from this mountain.


\clearpage
\nocite{*}
\printbibliography

\clearpage

\begin{appendices}
\chapter{List of publications}
\label{app:pubs}

\section{First author}

\begin{itemize}

    \item[1.] \textbf{P. Mas-Buitrago}, J.-Y. Zhang, E. Solano, E. L. Martín. 
    
    \textcolor{gray}{\textit{``Ca~{\sc ii} and H$\alpha$ flaring M dwarfs detected with multi-filter photometry.''}}\\
    \href{https://ui.adsabs.harvard.edu/abs/2025A%26A...695A.182M/abstract}{Astronomy \& Astrophysics, 695, A182. March 2025.}

    \item[2.] \textbf{P. Mas-Buitrago}, A. González-Marcos, E. Solano, V. M. Passegger, et al. 
    
    \textcolor{gray}{\textit{``Using autoencoders and deep transfer learning to determine the stellar parameters of 286 CARMENES M dwarfs.''}}\\    \href{https://ui.adsabs.harvard.edu/link_gateway/2024A&A...687A.205M/doi:10.1051/0004-6361/202449865}{Astronomy \& Astrophysics, 687, A205. July 2024.}

    \item[3.] \textbf{P. Mas-Buitrago}, E. Solano, A. González-Marcos, C. Rodrigo, et al. 
    
    \textcolor{gray}{\textit{``J-PLUS: Discovery and characterisation of ultracool dwarfs using Virtual Observatory tools. II. Second data release and machine learning methodology.''}}\\    \href{https://ui.adsabs.harvard.edu/link_gateway/2022A&A...666A.147M/doi:10.1051/0004-6361/202243895}{Astronomy \& Astrophysics, 666, A147. October 2022.}

\end{itemize}

\section{Contributing author}

\begin{itemize}

    \item[1.] C. Dominguez-Tagle, M. {\v Z}erjal, N. Sedighi, E. L. Martín, et al. (including \textbf{P. Mas-Buitrago}).
    
    \textcolor{gray}{\textit{``Euclid Quick Data Release (Q1) - Euclid ultracool dwarfs. II: Spectroscopic search, classification and analysis''}}\\    \href{https://ui.adsabs.harvard.edu/abs/2025arXiv250322442D/abstract}{Submitted to Astrophysical Journal.}

    \item[2.] M.~{\v Z}erjal, C. Dominguez-Tagle, N. Sedighi, E. L. Martín, et al. (including \textbf{P. Mas-Buitrago}).
    
    \textcolor{gray}{\textit{``Euclid Quick Data Release (Q1) -- Euclid ultracool dwarfs. I: A photometric search''}}\\
    \href{https://ui.adsabs.harvard.edu/abs/2025arXiv250322497V/abstract}{Submitted to Astronomy \& Astrophysics.}

    \item[3.] E.~L. Martín, M. {\v Z}erjal, H. Bouy, D. Martin-Gonzalez, et al. (including \textbf{P. Mas-Buitrago}).
    
    \textcolor{gray}{\textit{``Euclid: Early Release Observations -- A glance at free-floating new-born planets in the sigma Orionis cluster.''}}\\
    \href{https://doi.org/10.1051/0004-6361/202450793}{Accepted for publication in Astronomy \& Astrophysics, as part of the special issue on ``Euclid on the sky''.}

    \item[2.] C. Rodrigo, P. Cruz, J. F. Aguilar, A. Aller, et al. (including \textbf{P. Mas-Buitrago}). 
    
    \textcolor{gray}{\textit{``Photometric segregation of dwarf and giant FGK stars using the SVO Filter Profile Service and photometric tools.''}}\\
    \href{https://doi.org/10.1051/0004-6361/202449998}{Astronomy \& Astrophysics, 689, A93 (September 2024).}

\end{itemize}

\section{Conference proceedings}

\begin{itemize}

    \item[1.] M. Galvez Ortiz, C. Rodrigo, P. Cruz, J. F. Aguilar, et al. (including \textbf{P. Mas-Buitrago}).
    
    \textcolor{gray}{\textit{``Photometric segregation between dwarfs and giants using AI and Spanish Virtual Observatory photometric tools''}}\\    
     \href{https://ui.adsabs.harvard.edu/abs/2024eas..conf.1392G/abstract}{European Astronomical Society Annual Meeting. July 2024.}

    \item[2.] E.~L. Martín, H. Bouy, D. Martin, M. $\check{Z}$erjal, et al. (including \textbf{P. Mas-Buitrago}).
    
    \textcolor{gray}{\textit{``Substellar science in the wake of the ESA Euclid space mission''}}\\    
     \href{https://articles.adsabs.harvard.edu/pdf/2023hsa..conf..275M}{Windows on the Universe 2023 - Rencontres du Vietnam. December 2023.}

    \item[3.] \textbf{P. Mas-Buitrago}, E. Solano, A. Aller, M. Cortés-Contreras, et al.
    
    \textcolor{gray}{\textit{``Remote Virtual Observatory schools.''}}\\    
     \href{https://articles.adsabs.harvard.edu/pdf/2023hsa..conf..500M}{Highlights of Spanish Astrophysics XI, Proceedings of the XV Scientific Meeting of the Spanish Astronomical Society. May 2023.}

    \item[4.] \textbf{P. Mas-Buitrago}, E. Solano, A. González-Marcos.
    
    \textcolor{gray}{\textit{``Identification of ultracool dwarfs in J-PLUS DR2 using Virtual Observatory tools and machine learning techniques''}}\\    
     \href{https://articles.adsabs.harvard.edu/pdf/2023hsa..conf..275M}{Highlights of Spanish Astrophysics XI, Proceedings of the XV Scientific Meeting of the Spanish Astronomical Society. May 2023.}

\end{itemize}
\chapter{Additional tables for Chapter \ref{chp:ucds_paper}}
\label{app:additional_ucds}

Table \ref{tab:binaries} lists the coordinates (J2000), parallaxes, proper motions, angular separations $\rho$, and projected physical separations $s$ of the six systems already tabulated as known binary systems by the Washington Double Star catalogue \citep[WDS;][]{WDS}

\begin{sidewaystable}
 \caption{Identified binary systems that are already tabulated by the WDS.}
 \label{tab:binaries}
 \centering          
 \begin{tabular}{l c c c c c c c}
  \hline\hline
  \noalign{\smallskip}
  
  Name\,$^{(a)}$ & Spectral & $\alpha$ & $\delta$ & $\varpi$ & $\mu_{\rm tot}$ & $\rho$ & $s$\\
    & type\,$^{(b)}$ & (deg) & (deg) & (mas) & (mas\,yr$^{-1}$) & (arcsec) & (au)\\
  
  \noalign{\smallskip}
  \hline
  \noalign{\smallskip}

  2MASS J01560053+0528562 & M6 & 29.00219 & 5.48230 & $11.66\pm0.09$ & $153.33\pm0.14$ & 7.3 & 621.4 \\
  2MASS J01560037+0528494 & M7.5 & 29.00194 & 5.47979 & $11.74\pm0.21$ & $154.39\pm0.32$ &  &  \\  

  \noalign{\smallskip}
  
  LSPM J0209+0732 & \ldots & 32.39985 & 7.54052 & $8.27\pm0.08$ & $166.00\pm0.11$ & 27.1 & 3881.2 \\
  2MASS J02093416+0732196 & \ldots & 32.39316 & 7.53862 & $6.99\pm1.00$ & $166.29\pm1.82$ &  &  \\ 

  \noalign{\smallskip}
  
  SLW J0851+4134A & M4.5 & 132.94640 & 41.57082 & $21.84\pm0.04$ & $114.65\pm0.05$ & 32.7 & 1503.1 \\
  2MASS J08514823+4134453 & M7\,V & 132.95016 & 41.57936 & $21.69\pm0.18$ & $114.10\pm0.24$ &  &  \\
  
  \noalign{\smallskip}
  
  Gaia EDR3 803883569892552320\,$^{(c)}$ & \ldots & 151.67670 & 41.58267 & $5.94\pm0.13$ & $52.53\pm0.15$ & 18.3 & 3110.5 \\
  SDSS J100640.86+413503.9 & M6\,V & 151.67006 & 41.58431 & $5.88\pm0.21$ & $52.66\pm0.23$ &  &  \\ 
  
  \noalign{\smallskip}
  
  SDSS J164331.99+634340.6 & \ldots & 250.88343 & 63.72784 & $13.55\pm0.12$ & $74.66\pm0.20$ & 5.8 & 415.1 \\
  Gaia EDR3 1631815065395291392\,$^{(c)}$ & \ldots & 250.88663 & 63.72773 & $13.71\pm0.12$ & $75.58\pm0.20$ &  &  \\

  \noalign{\smallskip}

  SLW J1840+4204A & M7.9 & 280.23178 & 42.06685 & $14.11\pm0.11$ & $59.67\pm0.19$ & 17.4 & 1217.5 \\
  SLW J1840+4204B & M8.1 & 280.23762 & 42.06474 & $14.59\pm0.39$ & $59.75\pm0.68$ &  &  \\ 
  
  \noalign{\smallskip}
  \hline
 \end{tabular}
  \tablefoot{$^{(a)}$ Primaries and secondaries are sorted as in the WDS. $^{(b)}$ When it is not available in Simbad, we show the spectral type given in the WDS. $^{(c)}$ The \textit{Gaia} source id is displayed, since the object is not reported in Simbad.}
\end{sidewaystable}
\chapter{Additional figures of Chapter \ref{chp:flares_paper}}
\label{app:app_flares}

\begin{figure}
        \centering
	\includegraphics[width=.48\columnwidth]{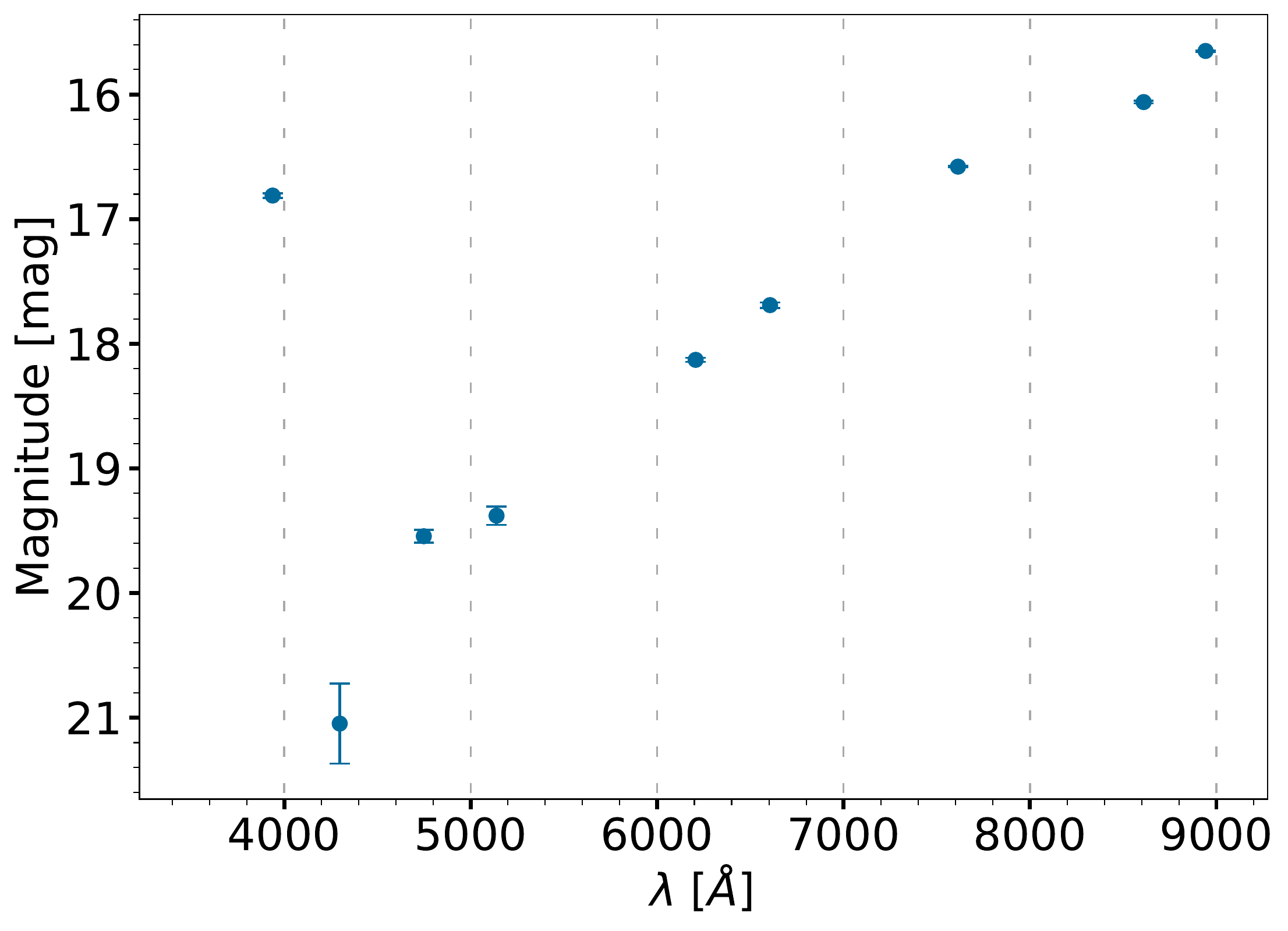}
        \includegraphics[width=.48\columnwidth]{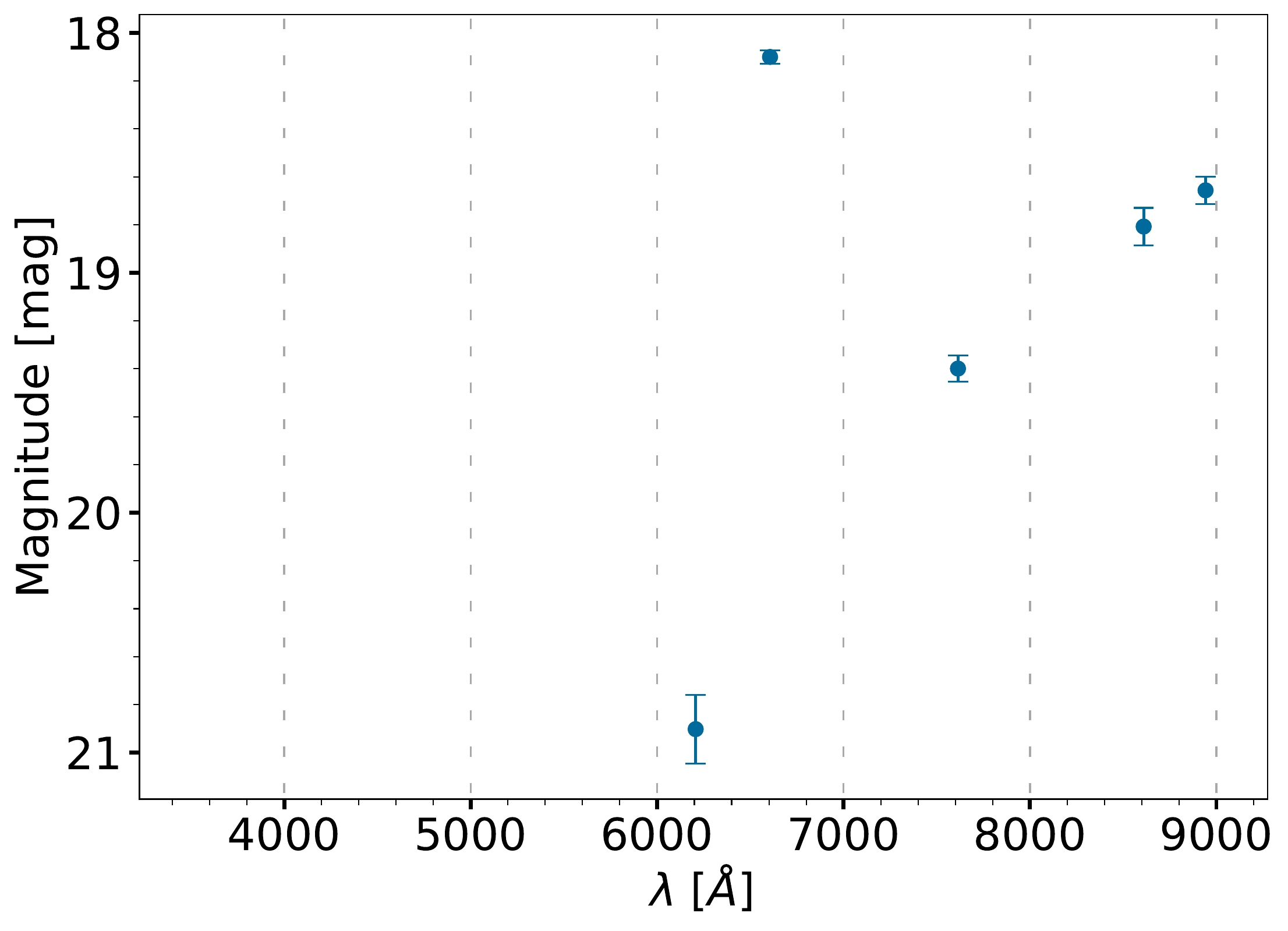}
    \caption{J-PLUS photometry for J-PLUS0114 and J-PLUS0226.}
    \label{fig:jplus_0114_0226}
\end{figure}

\begin{figure}
        \centering
	\includegraphics[width=.48\columnwidth]{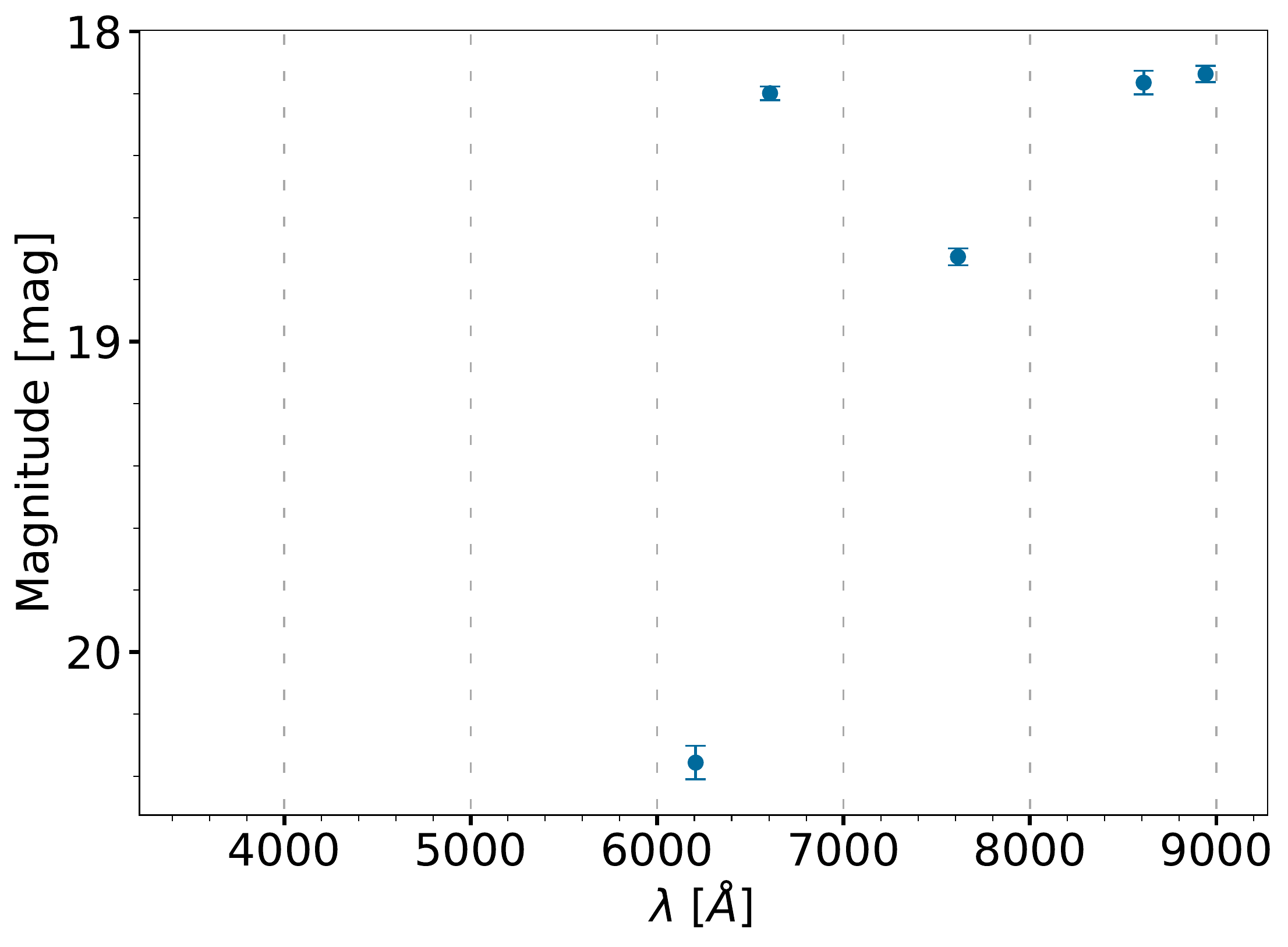}
        \includegraphics[width=.48\columnwidth]{Graphics/flares/sed_jplus0226.pdf}
    \caption{J-PLUS photometry for J-PLUS0708 and J-PLUS0744.}
    \label{fig:jplus_0708_0744}
\end{figure}

\begin{figure}
        \centering
	\includegraphics[width=.48\columnwidth]{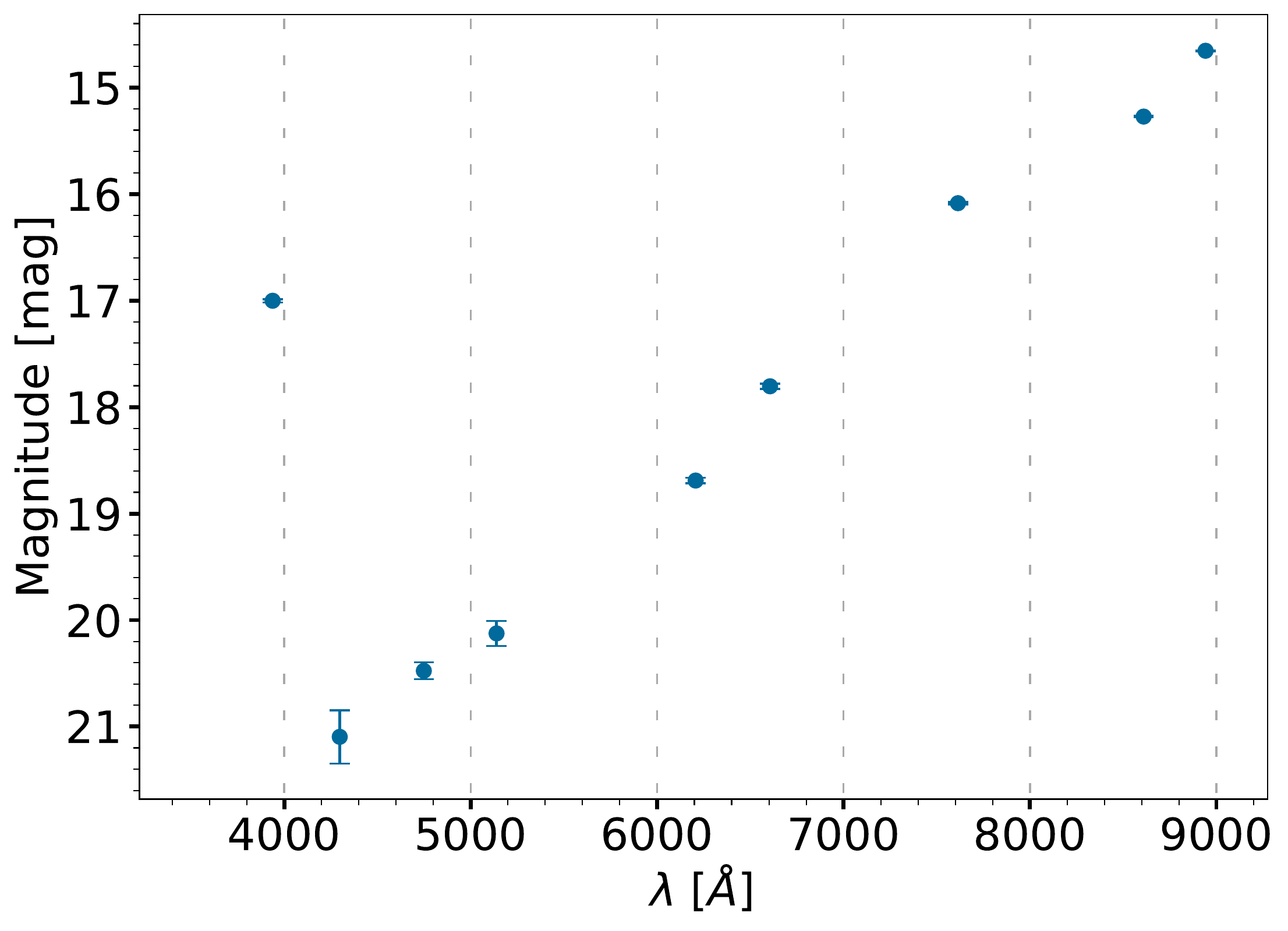}
        \includegraphics[width=.48\columnwidth]{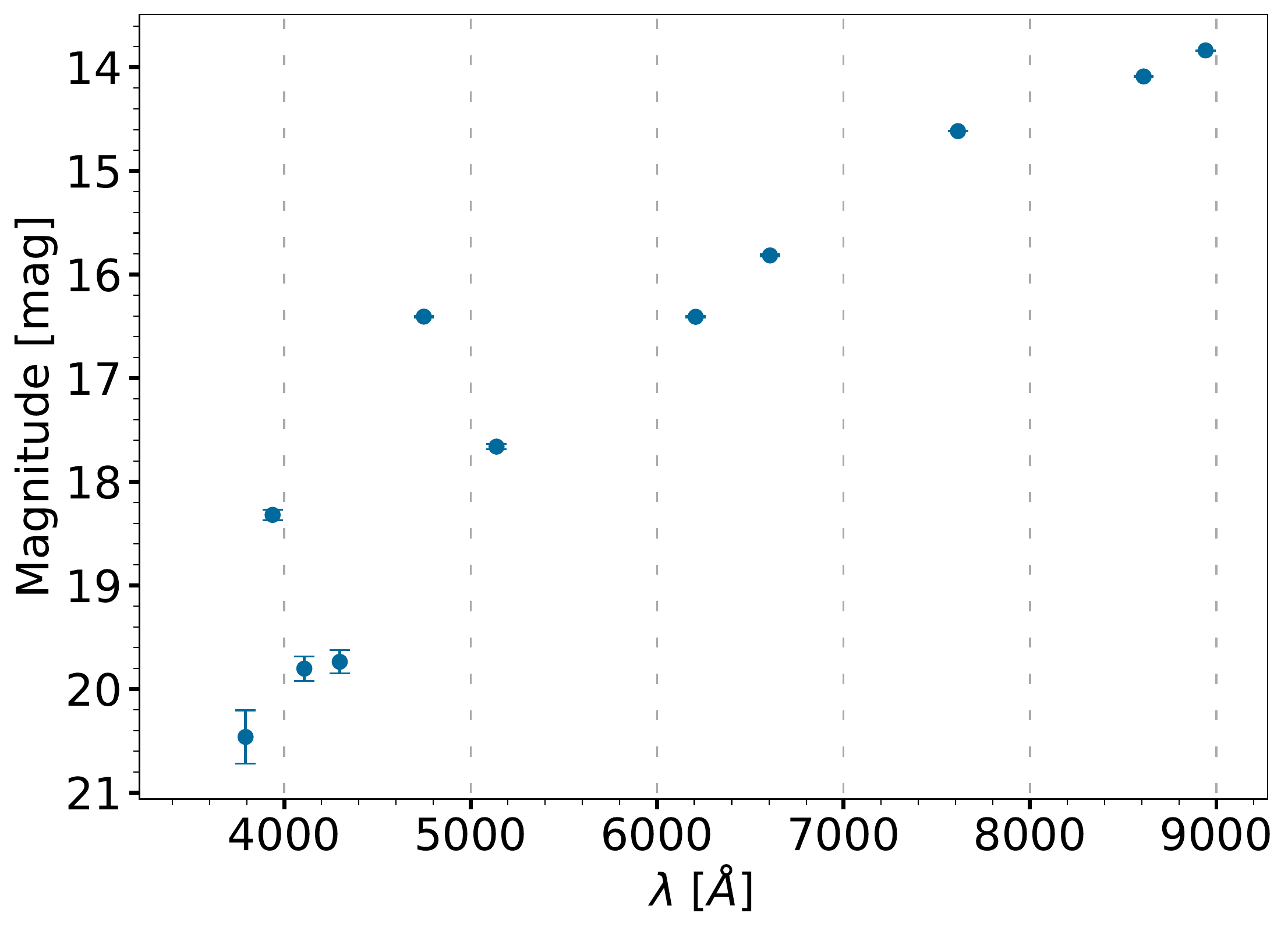}
    \caption{J-PLUS photometry for J-PLUS0807 and J-PLUS0903.}
    \label{fig:jplus_0807_0903}
\end{figure}

\begin{figure}
        \centering
	\includegraphics[width=.48\columnwidth]{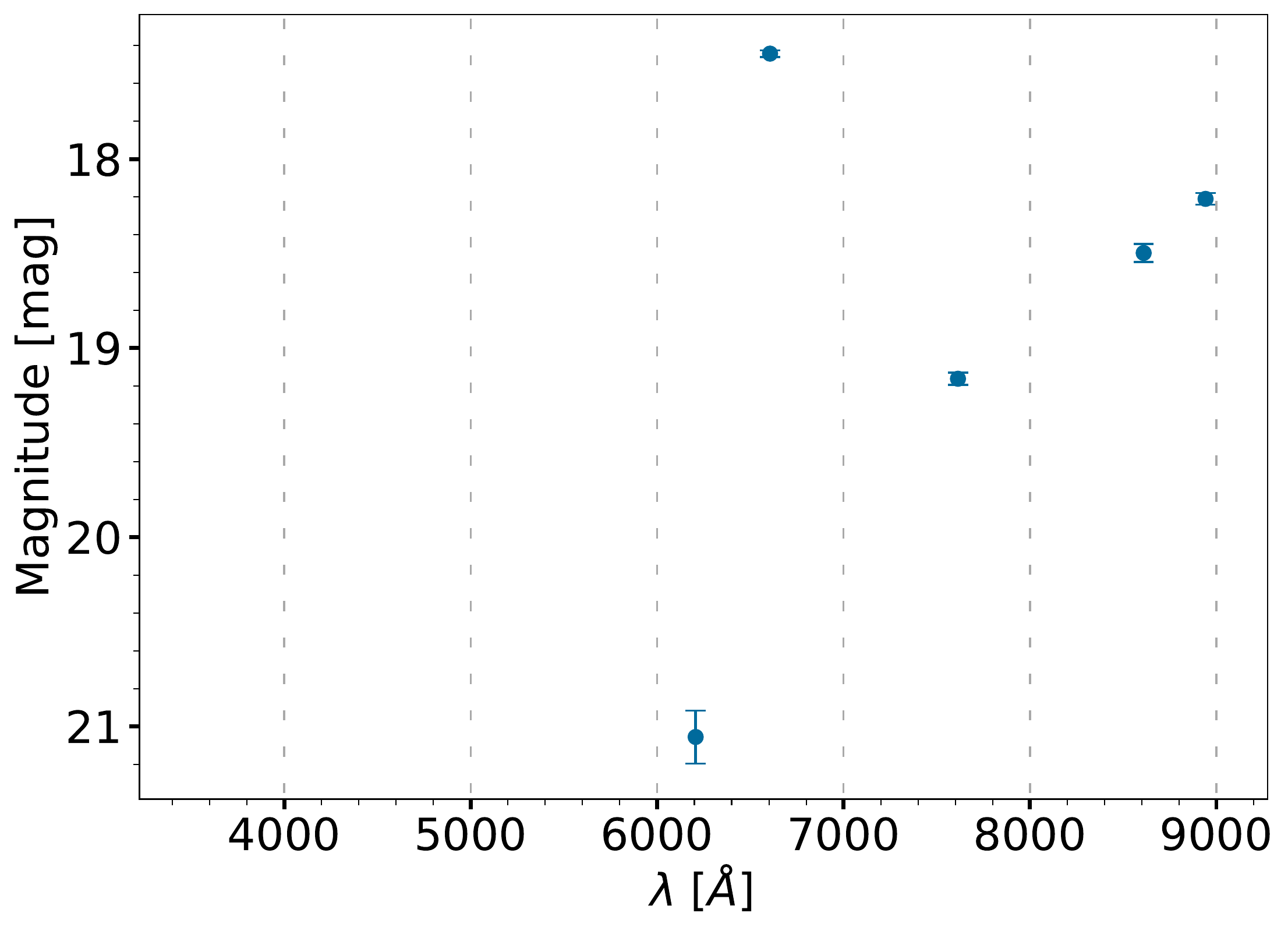}
    \caption{J-PLUS photometry for J-PLUS0914.}
    \label{fig:jplus_0914}
\end{figure}

In this appendix we provide the J-PLUS SEDs of each target star, as discussed in Section \ref{sec:reduced_sp}. Figures \ref{fig:jplus_0114_0226}, \ref{fig:jplus_0708_0744}, \ref{fig:jplus_0807_0903}, and \ref{fig:jplus_0914} show the SED of J-PLUS0114, J-PLUS0226, J-PLUS0708, J-PLUS0744, J-PLUS0807, J-PLUS0903, and J-PLUS0914, respectively.

\vspace*{-3em}

\chapter{Additional tables and figures of Chapter \ref{chp:autoencoders_paper}}
\label{app:additional_acs}

\section{Additional tables}\label{app:appb_tables}

Table \ref{tab:pars} is available in its entirety in electronic form at the CDS. This appendix only shows an extract of the table to facilitate its understanding. Table \ref{tab:cnn_arc} describes in detail the CNN architectures used for the estimation of each stellar parameter. Table \ref{tab:subd} details all the outliers identified with low-metallicity behaviour, along with the metallicity estimations found in the literature.

\begin{table}
\fontsize{11pt}{11pt}\selectfont
 \caption{Stellar atmospheric parameters, together with their uncertainties, determined with our methodology. Only the first five rows of the table are shown.}
 \label{tab:pars}
 \centering          
 \begin{tabular}{l c c c c c c}
  \hline\hline
  \noalign{\smallskip}
  
  Karmn & $\alpha\,^{(a)}$ & $\delta\,^{(a)}$ & $\textit{T}_{\rm eff}$ & log\,$\textit{g}$ & [Fe/H] & $\textit{v}\sin{i}$ \\

    & [J2016.0] & [J2016.0] & [K]  & [dex] & [dex] & [km\,s$^{-1}$] \\
  
  \noalign{\smallskip}
  \hline
  \noalign{\smallskip}
  
  J00051+457 & 00:05:12.22 & 03:03:08.6 & $3780_{-34}^{+41}$ & $4.70_{-0.04}^{+0.01}$ & $0.03_{-0.04}^{+0.06}$ & $3.19_{-0.16}^{+0.48}$ \\
  
  \noalign{\smallskip}
  
  J00067-075 &  00:06:42.32 & 23:29:48.8 & $3073_{-27}^{+18}$ & $5.10_{-0.09}^{+0.04}$ & $0.06_{-0.15}^{+0.08}$ & $3.02_{-0.66}^{+0.33}$ \\
  
  \noalign{\smallskip}

  J00162+198E &  00:16:16.96 & 01:19:26.6 & $3362_{-25}^{+34}$ & $4.90_{-0.06}^{+0.02}$ & $0.07_{-0.17}^{+0.02}$ & $2.13_{-0.22}^{+0.30}$ \\
  
  \noalign{\smallskip}

  J00183+440 &  00:18:27.17 & 02:56:05.9 & $3709_{-43}^{+15}$ & $4.80_{-0.07}^{+0.03}$ & $-0.33_{-0.17}^{+0.06}$ & $2.02_{-0.30}^{+0.20}$ \\
  
  \noalign{\smallskip}

  J00184+440 &  00:18:30.07 & 02:56:06.9 & $3251_{-13}^{+36}$ & $4.96_{-0.03}^{+0.05}$ & $-0.20_{-0.10}^{+0.09}$ & $2.82_{-0.24}^{+0.22}$ \\
  
  \noalign{\smallskip}
  \hline
 \end{tabular}
 \tablefoot{$^{(a)}$  From {\it Gaia} DR3.}
\end{table}

\begin{sidewaystable}
\fontsize{11pt}{11pt}\selectfont
 \caption{CNN architectures used for the estimation of $T_{\rm eff}$, log\,$\textit{g}$, [M/H], and $\textit{v}\sin{i}$.}
 \label{tab:cnn_arc}
 \centering
 \begin{tabular}{l c c c c c c c c c c c c}
 
  \hline\hline
  \noalign{\smallskip}

  Layer & \multicolumn{4}{c}{Output size} & \multicolumn{4}{c}{Number of filters} & \multicolumn{4}{c}{Number of parameters} \\

   & $\textit{T}_{\rm eff}$ & log\,$\textit{g}$ & [M/H] & $\textit{v}\sin{i}$ & $\textit{T}_{\rm eff}$ & log\,$\textit{g}$ & [M/H]  & $\textit{v}\sin{i}$ & $\textit{T}_{\rm eff}$ & log\,$\textit{g}$ & [M/H] & $\textit{v}\sin{i}$ \\
  
  \noalign{\smallskip}
  \hline
  \noalign{\smallskip}

  Conv1D & 32 & 32 & 32 & 32 & 64 & 16 & 32 & 64 & 192 & 48 & 96 & 192 \\
  Conv1D & 32 & 32 & 32 & 32 & 32 & 64 & 8 & 8 & 4\,128 & 2\,112 & 520 & 1\,032 \\
  Flatten & 1\,024 & 2\,048 & 256 & 256 & \ldots & \ldots & \ldots & \ldots & 0 & 0 & 0 & 0 \\
  Dense & 256 & 256 & 256 & 256 & \ldots & \ldots & \ldots & \ldots & 262\,400 & 524\,544 & 65\,792 & 65\,792 \\
  Dense & 128 & 128 & 128 & 128 & \ldots & \ldots & \ldots & \ldots & 32\,896 & 32\,896 & 32\,896 & 32\,896 \\
  Dense & 64 & 64 & 64 & 64 & \ldots & \ldots & \ldots & \ldots & 8\,256 & 8\,256 & 8\,256 & 8\,256 \\
  Dense & 1 & 1 & 1 & 1 & \ldots & \ldots & \ldots & \ldots & 65 & 65 & 65 & 65 \\
  
  \noalign{\smallskip}
  \hline
 \end{tabular}

\hfill \break

\caption{Low-metallicity stars identified in Fig. \ref{fig:par_diags}.}
\label{tab:subd}
\centering          
\begin{tabular}{l l c c c c c c}
\hline\hline
\noalign{\smallskip}

  Karmn & Name & [Fe/H]$_{\mathrm{AE}}^{\,(a)}$ & [Fe/H]$_{\mathrm{DTL}}^{\,(b)}$ & [Fe/H]$_{\mathrm{Mann15}}^{\,(c)}$ & [Fe/H]$_{\mathrm{corr, Mar21}}^{\,(d)}$ & [Fe/H]$_{\mathrm{Schw19}}^{\,(e)}$ & Pop.$^{\,(f)}$\\

    &  & [dex] & [dex]  & [dex] & [dex] & [dex] & \\
  
  \noalign{\smallskip}
  \hline
  \noalign{\smallskip}
  
  J00183+440 & GX And & $-0.33_{-0.17}^{+0.06}$ & $-0.26_{-\ldots}^{+\ldots}$ & $-0.30\pm0.08$ & $-0.52\pm0.11$ & $-0.25\pm0.16$ & D \\
  
  \noalign{\smallskip}
  
  J02123+035 & BD+02 348 & $-0.35_{-0.10}^{+0.12}$ & $-0.33_{-0.01}^{+0.01}$ & $-0.36\pm0.08$ & $-0.49\pm0.06$ & $-0.05\pm0.16$ & TD \\

  \noalign{\smallskip}
  
  J06371+175 & HD 260655 & $-0.41_{-0.13}^{+0.11}$ & $-0.37_{-0.02}^{+0.02}$ & $-0.34\pm0.08$ & $-0.43\pm0.04$ & $-0.42\pm0.16$ & TD-D \\

  \noalign{\smallskip}

  J11033+359 & Lalande 21185 & $-0.34_{-0.13}^{+0.08}$ & $-0.31_{-\ldots}^{+\ldots}$ & $-0.38\pm0.08$ & $-0.49\pm0.10$ & $-0.09\pm0.16$ & TD\\

  \noalign{\smallskip}
  
  J11054+435 & BD+44 2051A & $-0.40_{-0.18}^{+0.07}$ & $-0.35_{-\ldots}^{+\ldots}$ & $-0.37\pm0.08$ & $-0.56\pm0.09$ & $-0.3\pm0.16$ & TD-D \\

  \noalign{\smallskip}

  J12248-182\,$^{(g)}$ & Ross 695 & $-0.33_{-0.18}^{+0.06}$ & $-0.40_{-0.04}^{+0.02}$ & \ldots & $-0.60\pm0.09$ & $-0.17\pm0.16$ & TD \\

  \noalign{\smallskip}

  J13450+176 & BD+18 2776 & $-0.53_{-0.27}^{+0.09}$ & $-0.46_{-0.05}^{+0.06}$ & $-0.54\pm0.08$ & $-0.54\pm0.03$ & $-0.43\pm0.16$ & TD \\

  \noalign{\smallskip}

  J16254+543\,$^{(g)}$ & GJ 625 & $-0.33_{-0.15}^{+0.05}$ & $-0.32_{-0.03}^{+0.02}$ & $-0.35\pm0.08$ & $-0.28\pm0.07$ & $-0.26\pm0.16$ & YD\\

  \noalign{\smallskip}

  J17378+185 & BD+18 3421 & $-0.33_{-0.08}^{+0.11}$ & $-0.23_{-0.03}^{+0.01}$ & $-0.25\pm0.08$ & $-0.40\pm0.07$ & $-0.23\pm0.16$ & D \\

  \noalign{\smallskip}
  
  J19070+208\,$^{(g)}$ & Ross 730 & $-0.34_{-0.18}^{+0.05}$ & $-0.32_{-0.02}^{+0.01}$ & $-0.33\pm0.08$ & $-0.46\pm0.07$ & $-0.20\pm0.16$ & D \\

  \noalign{\smallskip}
  
  J19072+208\,$^{(g)}$ & HD 349726 & $-0.32_{-0.17}^{+0.05}$ & $-0.32_{-0.02}^{+0.02}$ & $-0.35\pm0.08$ & $-0.46\pm0.06$ & $-0.23\pm0.16$ & D \\

  \noalign{\smallskip}
  
  J23492+024 & BR Psc & $-0.43_{-0.12}^{+0.11}$ & $-0.40_{-0.03}^{+0.02}$ & $-0.45\pm0.08$ & $-0.55\pm0.08$ & $-0.13\pm0.16$ & TD \\

  \noalign{\smallskip}
  \hline
 \end{tabular}
 \tablefoot{As explained in \citet{pass20,passegger2022}, our [M/H] results directly translate into [Fe/H] values. $^{(a)}$ From this work. $^{(b)}$ From \citetalias{bello2023}. $^{(c)}$ From \citet{mann2015}. $^{(d)}$ From \citetalias{mar21}, corrected from the $\alpha$ enhancement. $^{(e)}$ From \citetalias{schw19}. $^{(f)}$ Galactic populations, including the thick disc (TD), the thick disc-thin disc transition (TD-D), the thin disc (D), and the young disc (YD), following Cortés-Contreras et al. in prep. $^{(g)}$ Reported to have a behaviour akin to subdwarfs in \citetalias{mar21} or \citetalias{schw19}. In particular, J19070+208 and J19072+208 are both components of the wide binary system LDS\,1017, and \citet{houdebine2008} already identified them as subdwarfs.}
\end{sidewaystable}

\section{Additional comparison with the literature} \label{app:appb_lit}

\begin{figure*}
    \centering
   	\includegraphics[width=0.68\linewidth]{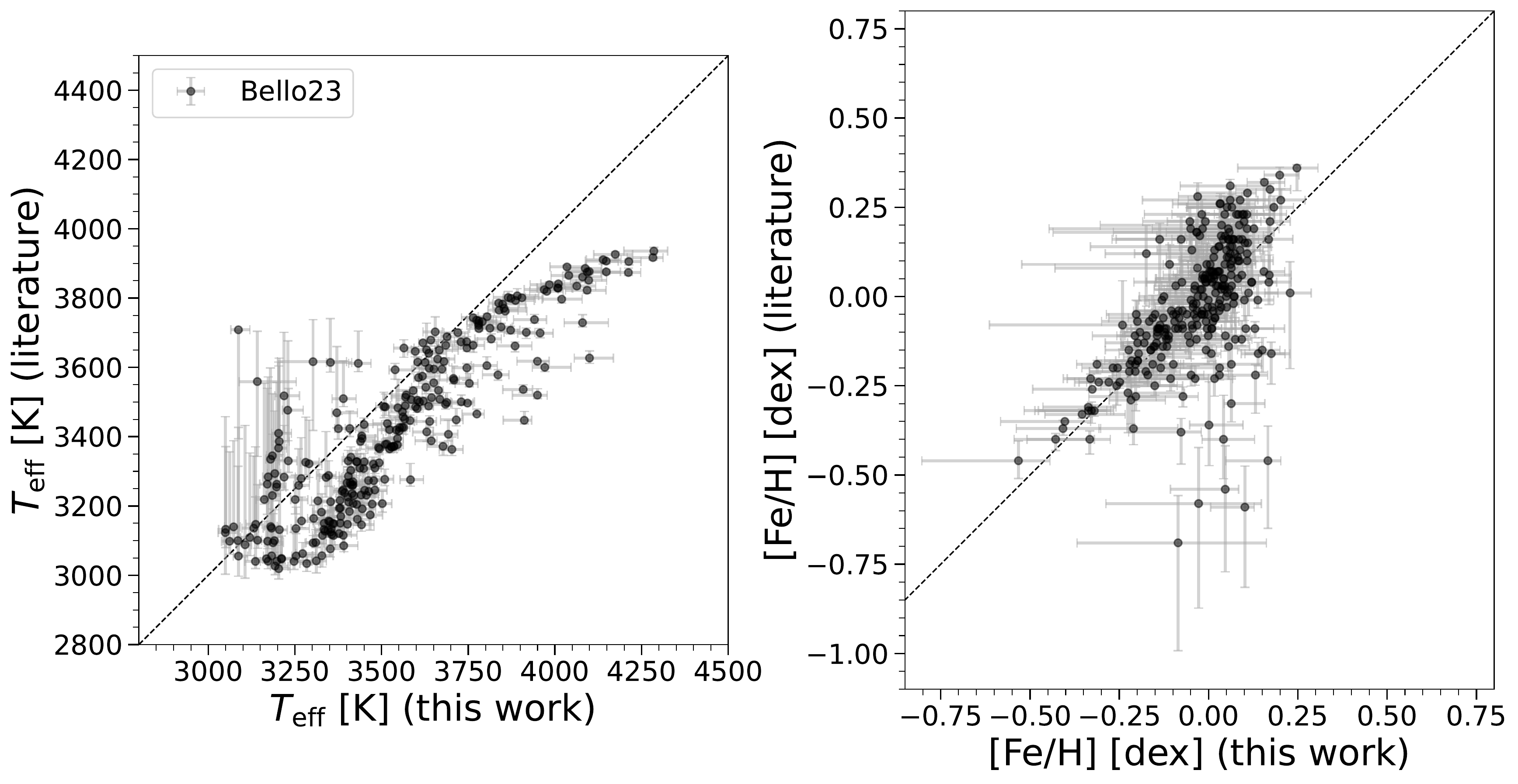}
    \caption{Comparison with \citetalias{bello2023}.}
    \label{fig:appendix_bello23}
\end{figure*}

\begin{figure*}
    \centering
    	\includegraphics[width=\linewidth]{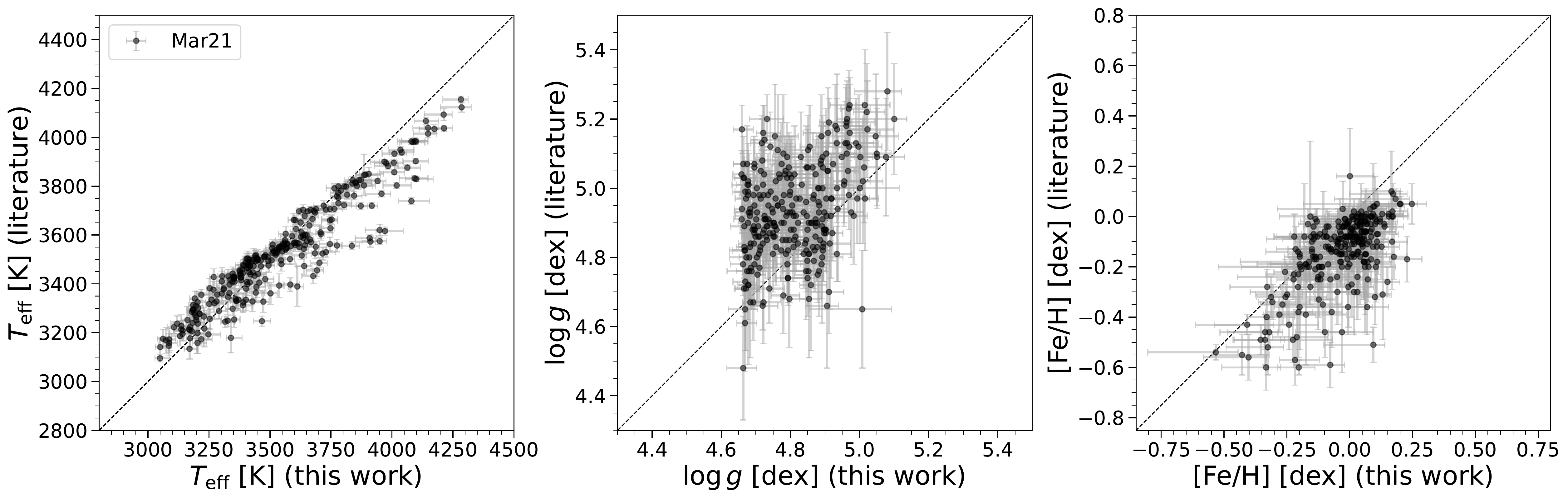}
    \caption{Comparison with \citetalias{mar21}.}
    \label{fig:appendix_mar21}
\end{figure*}

\begin{figure*}
    \centering
    	\includegraphics[width=\linewidth]{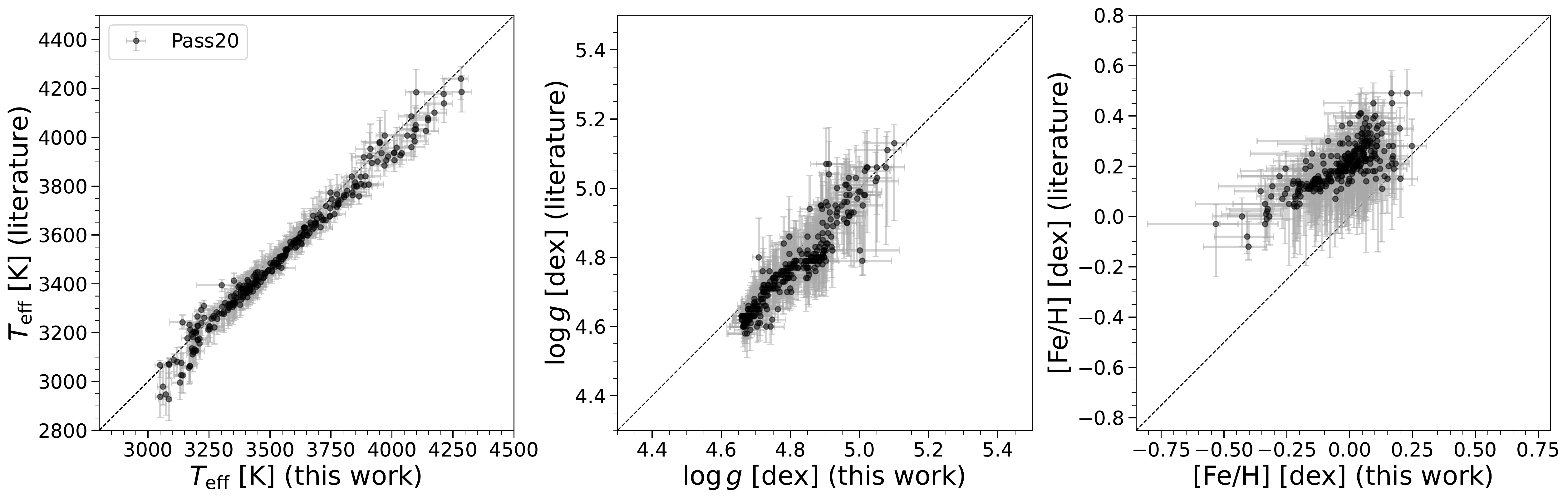}
    \caption{Comparison with \citetalias{pass20}.}
    \label{fig:appendix_pass20}
\end{figure*}

\begin{figure*}
    \centering
    	\includegraphics[width=\linewidth]{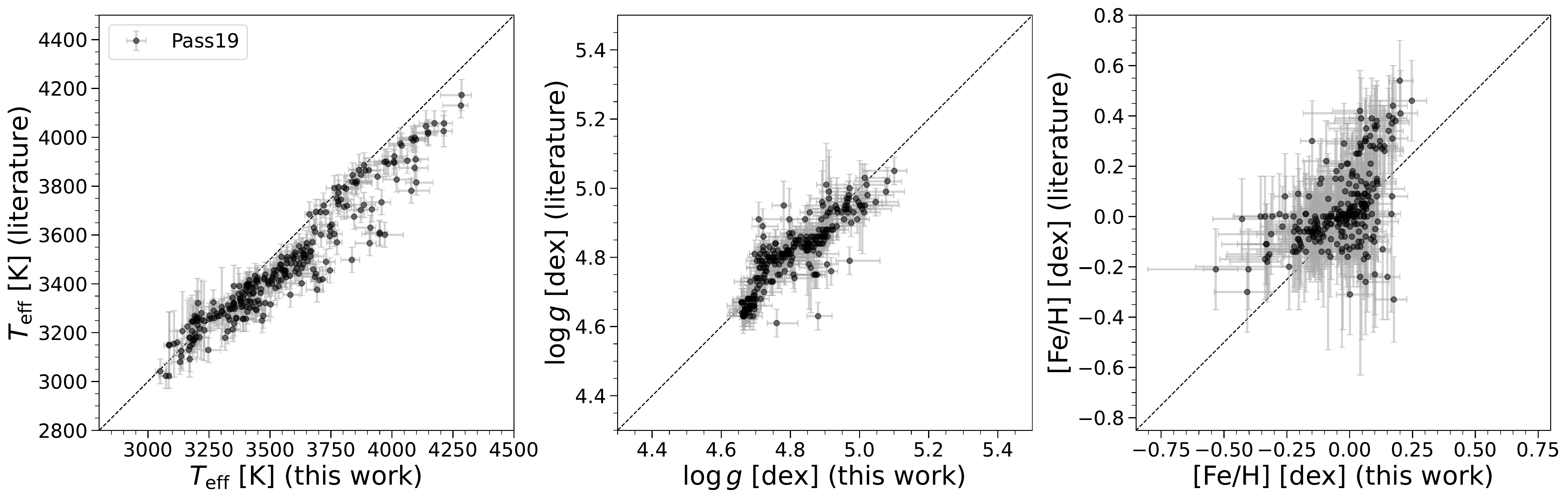}
    \caption{Comparison with \citetalias{pass2019}.}
    \label{fig:appendix_pass19}
\end{figure*}

\begin{figure*}
    \centering
    	\includegraphics[width=\linewidth]{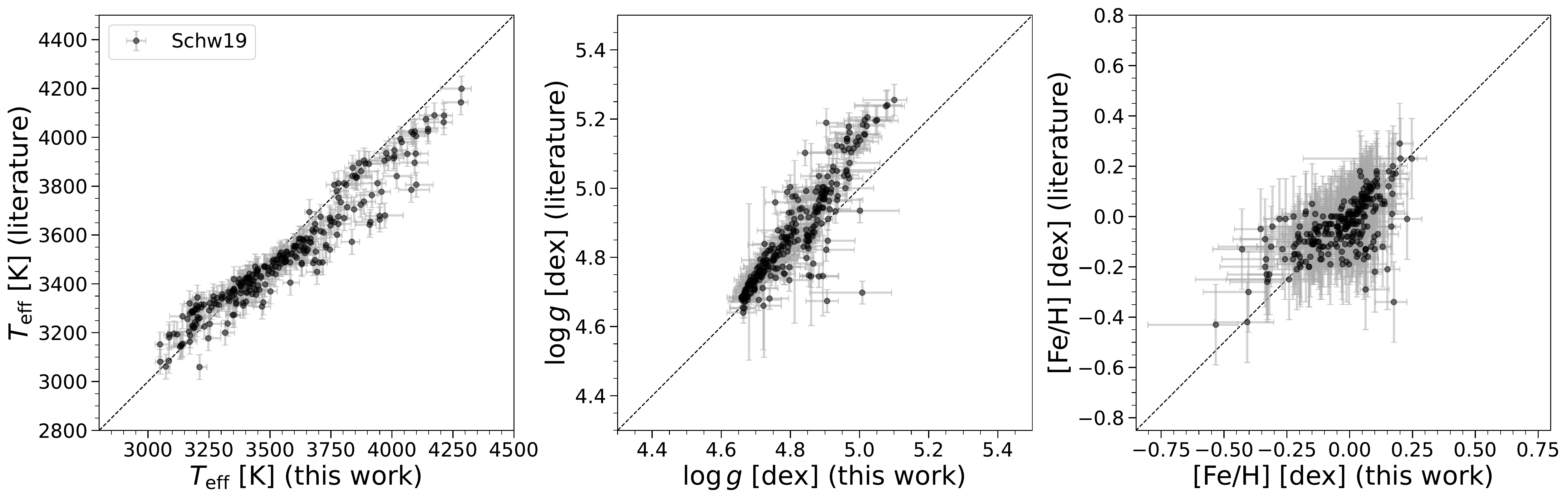}
    \caption{Comparison with \citetalias{schw19}.}
    \label{fig:appendix_schw19}
\end{figure*}

\begin{figure*}
    \centering
    	\includegraphics[width=\linewidth]{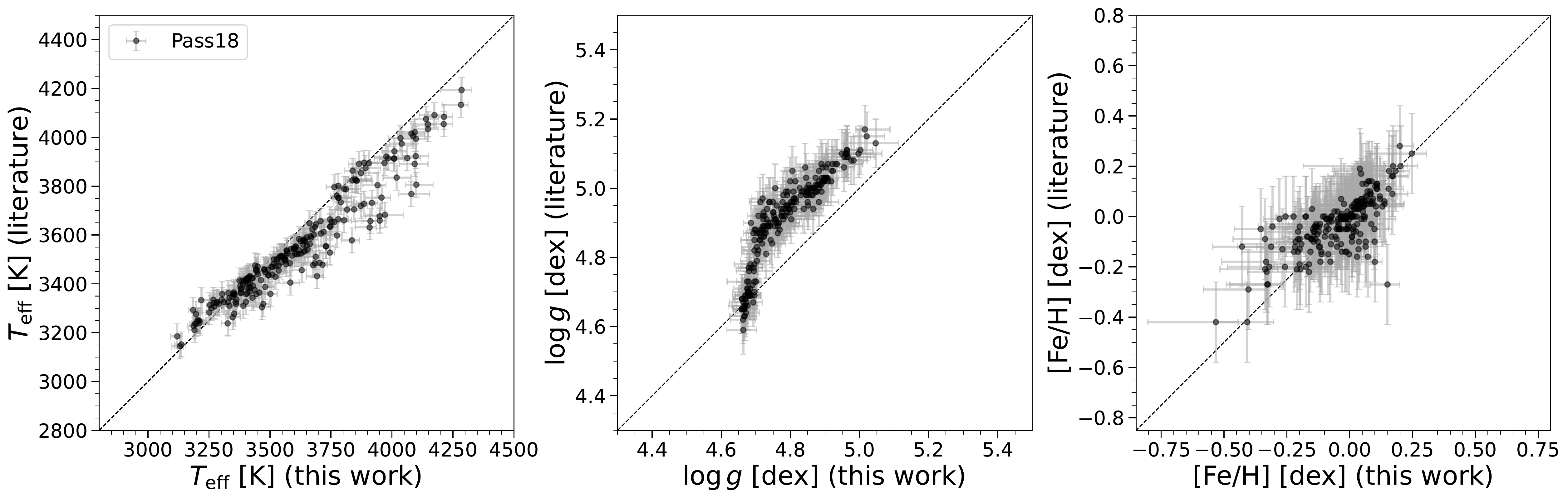}
    \caption{Comparison with \citet{pass18}.}
    \label{fig:appendix_pass18}
\end{figure*}

\begin{figure*}
    \centering
    	\includegraphics[width=0.68\linewidth]{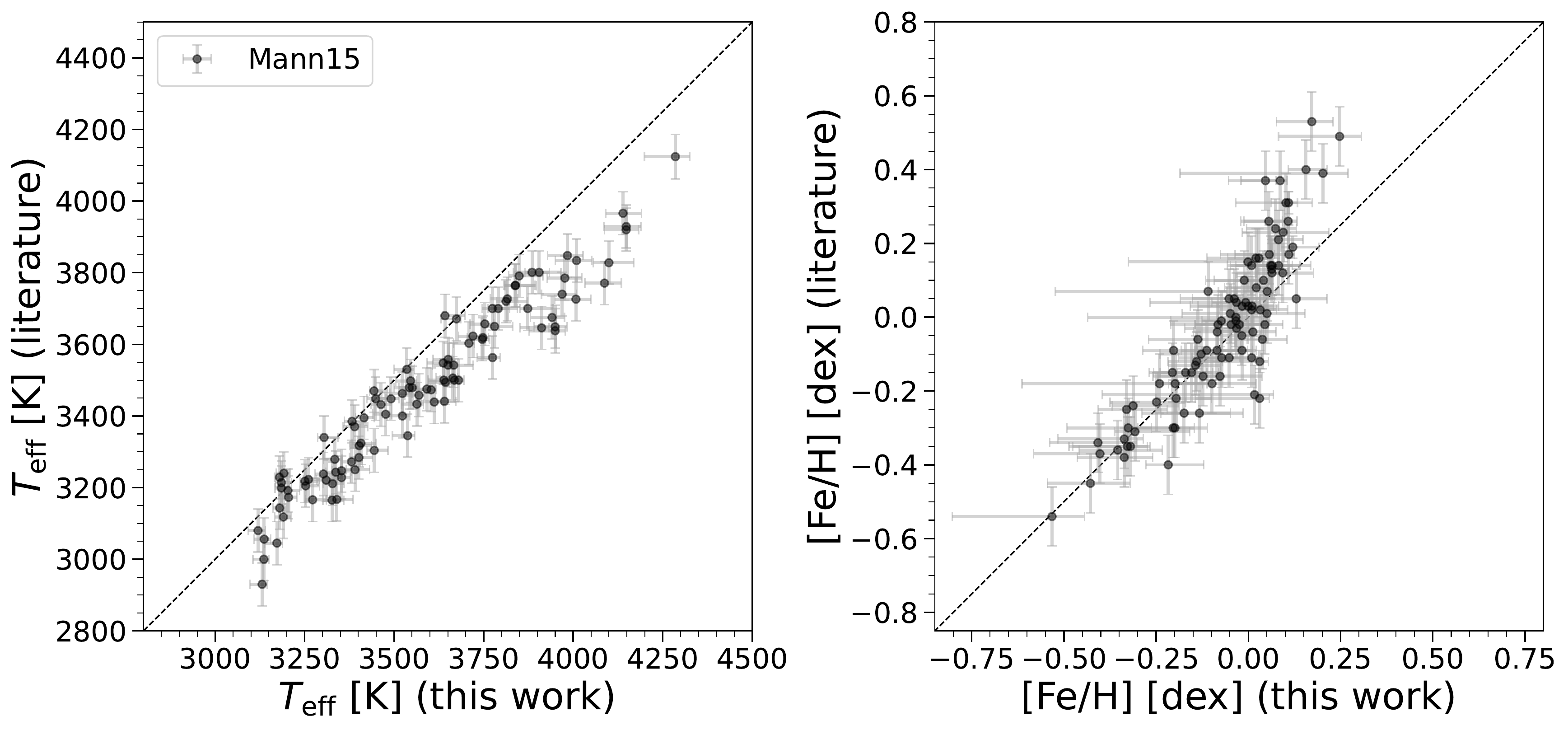}
    \caption{Comparison with \citet{mann2015}.}
    \label{fig:appendix_mann15}
\end{figure*}

\begin{figure*}
    \centering
    	\includegraphics[width=0.68\linewidth]{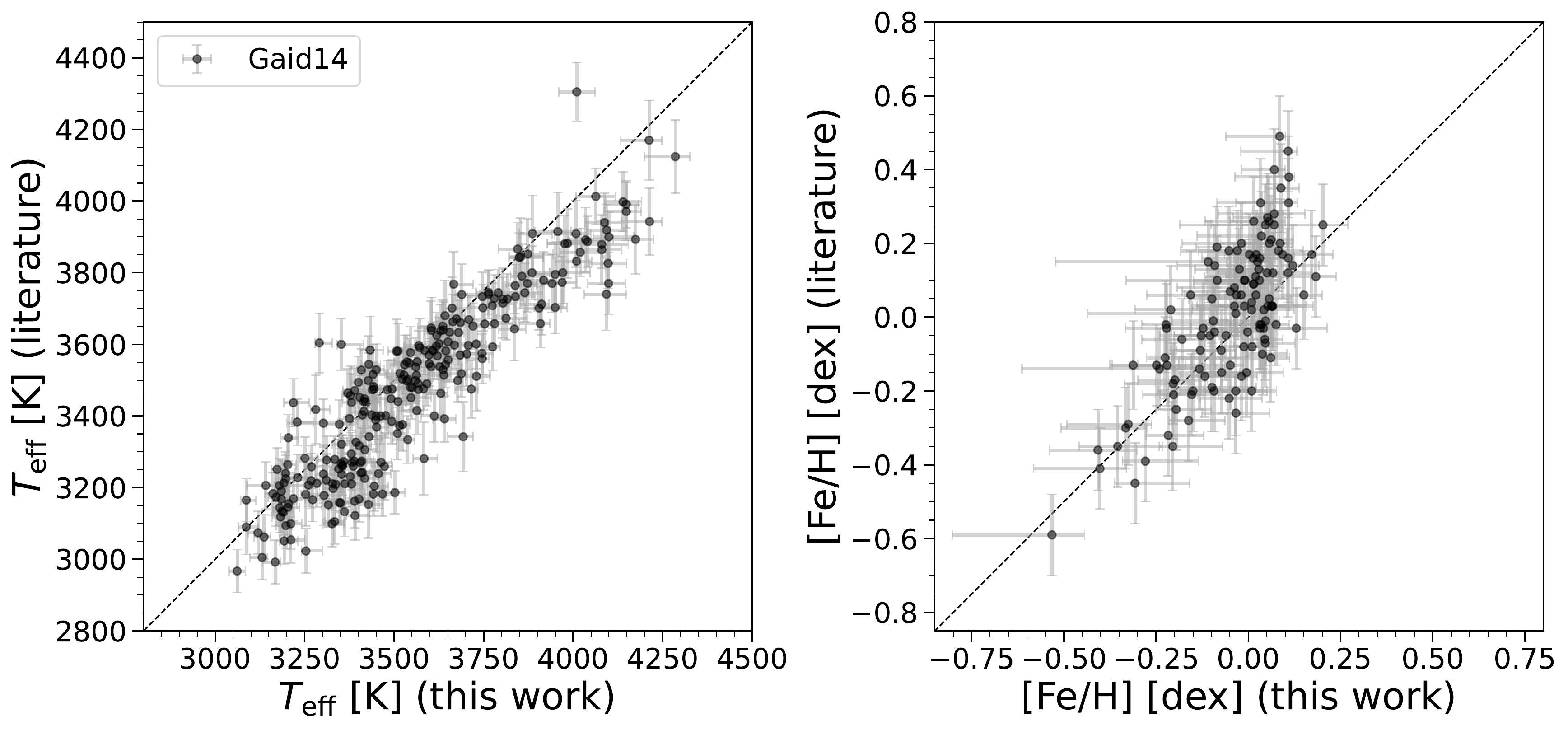}
    \caption{Comparison with \citet{gaid14}.}
    \label{fig:appendix_gaid14}
\end{figure*}

\begin{figure*}
    \centering
    	\includegraphics[width=0.68\linewidth]{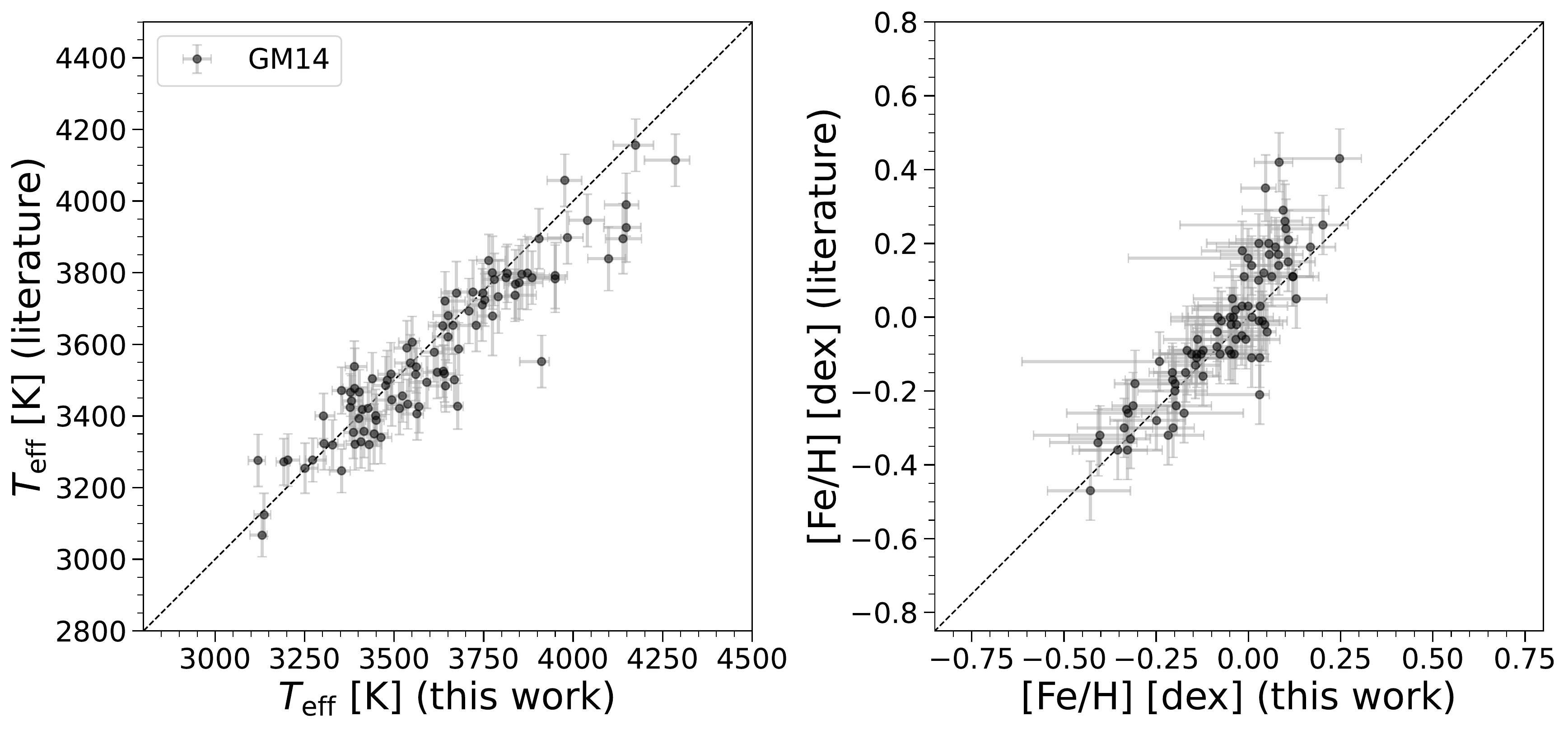}
    \caption{Comparison with \citet{GM14}.}
    \label{fig:appendix_gm14}
\end{figure*}

In this appendix, we provide an extensive comparison of this work with different results from the literature, as discussed in Section \ref{acs_sec:lit_comp}. Also, we repeat the comparison shown in Figs. \ref{fig:scatter_teff}, \ref{fig:scatter_logg} and \ref{fig:scatter_mh}, but including the error bars. Table \ref{tab:comparison_lit_app} replicates Table \ref{tab:comparison_lit} for the additional literature collections. Figures \ref{fig:appendix_bello23}, \ref{fig:appendix_mar21}, \ref{fig:appendix_pass20}, \ref{fig:appendix_pass19}, \ref{fig:appendix_schw19}, \ref{fig:appendix_pass18}, \ref{fig:appendix_mann15}, \ref{fig:appendix_gaid14}, and \ref{fig:appendix_gm14} show the comparison with \citetalias{bello2023}, \citetalias{mar21}, \citetalias{pass20}, \citetalias{pass2019}, \citetalias{schw19}, \citet{pass18}, \citet{mann2015}, \citet{gaid14}, and \citet{GM14}, respectively.

\begin{table}
 \caption{Comparison between our results and the additional literature collections. The structure is the same as in Table \ref{tab:comparison_lit}.}
 \label{tab:comparison_lit_app}
 \centering
 \begin{tabular}{l c c c c c c c c c}
 
  \hline\hline
  \noalign{\smallskip}

  Reference & \multicolumn{3}{c}{$\textit{T}_{\rm eff}$\,[K]} & \multicolumn{3}{c}{log\,$\textit{g}$\,[dex]} & \multicolumn{3}{c}{[Fe/H]\,[dex]}\\

    & $\overline{\Delta}$ & rmse & $r_{\rm p}$ & $\overline{\Delta}$ & rmse & $r_{\rm p}$ & $\overline{\Delta}$ & rmse & $r_{\rm p}$ \\
  
  \noalign{\smallskip}
  \hline
  \noalign{\smallskip}

  Pass18$^{\,(a)}$ & -59 & 98 & 0.96 & 0.12 & 0.14 & 0.89 & 0.01 & 0.09 & 0.73 \\    
  Mann15$^{\,(b)}$ & -109 & 136 & 0.96 & \ldots & \ldots & \ldots & 0.04 & 0.11 & 0.89 \\  
  Gaid14$^{\,(c)}$ & -69 & 151 & 0.87 & \ldots & \ldots & \ldots & 0.05 & 0.14 & 0.75 \\  
  GM14$^{\,(d)}$ & -42 & 102 & 0.93 & \ldots & \ldots & \ldots & 0.04 & 0.10 & 0.88 \\
  
  \noalign{\smallskip}
  \hline
 \end{tabular}
 \tablefoot{$^{(a)}$ From \citet{pass18}. $^{(b)}$ From \citet{mann2015}. $^{(c)}$ From \citet{gaid14}. $^{(d)}$ From \citet{GM14}.}
\end{table}

\chapter{Additional figures of Chapter \ref{chp:dtl_ucds}}
\label{app:dtl_ucds}

In this appendix we provide the effective temperatures determined for our sample of ultracool dwarfs with the developed deep transfer learning methodology discussed in Chapter \ref{chp:dtl_ucds}.

 \begin{longtable}{l c c c c c}
 \caption{Catalogue of determined effective temperatures for the sample of ultracool dwarfs discussed in Chapter \ref{chp:dtl_ucds}.}\\
  \hline\hline
  \noalign{\smallskip}
  
  Name$\,^{(a)}$ & $\alpha\,^{(a)}$ & $\delta\,^{(a)}$ & $\textit{T}_{\rm eff}$ & Lower error & Upper error\\

    & [deg] & [deg] & [K] & [K] & [K]\\
  
  \noalign{\smallskip}
  \hline
  \noalign{\smallskip}
  \endfirsthead

 \caption[]{Catalogue of determined effective temperatures for the sample of ultracool dwarfs discussed in Chapter \ref{chp:dtl_ucds} (continued).}\\
  \hline\hline
  \noalign{\smallskip}
  
  Name$\,^{(a)}$ & $\alpha\,^{(a)}$ & $\delta\,^{(a)}$ & $\textit{T}_{\rm eff}$ & Lower error & Upper error\\

    & [deg] & [deg] & [K] & [K] & [K]\\

  \noalign{\smallskip}
  \hline
  \noalign{\smallskip}
  \endhead
  
  \noalign{\smallskip}
  \hline
  \noalign{\smallskip}
  
  \multicolumn{3}{l@{}}{$^{(a)}$ From the UltracoolSheet catalogue.}\\
  \endfoot
  
  SDSS J000013.54+255418.6 & 0.0565 & 25.905 & 1104 & 99 & 71\\
  LP 584-4 & 0.5259 & 1.2600 & 2340 & 61 & 76\\
  2MASS J00054844-2157196 & 1.4519 & -21.9555 & 2352 & 42 & 129\\
  2MASSI J0006205-172051 & 1.5854 & -17.3475 & 1837 & 73 & 78\\
  2MASS J00070787-2458042 & 1.7829 & -24.9679 & 2355 & 147 & 209\\
  2MASS J00100009-2031122 & 2.5004 & -20.5201 & 2297 & 62 & 111\\
  2MASS J00132229-1143006 & 3.3430 & -11.7168 & 1158 & 56 & 139\\
  2MASSI J0013578-223520 & 3.4908 & -22.5890 & 1711 & 87 & 110\\
  2MASS J00145575-4844171 & 3.7324 & -48.7380 & 1982 & 124 & 92\\
  2MASSW J0015447+351603 & 3.9366 & 35.2674 & 1876 & 63 & 110\\
  SDSS J001637.62-103911.2 & 4.1568 & -10.6530 & 2293 & 60 & 125\\
  2MASS J00165953-4056541 & 4.2481 & -40.9483 & 1637 & 85 & 122\\
  SDSS J001911.65+003017.8 & 4.7986 & 0.5049 & 2210 & 120 & 127\\
  Koenigstuhl 1B & 5.2747 & -42.7454 & 2129 & 208 & 101\\
  SDSS J002209.31-011040.2 & 5.5388 & -1.1778 & 2458 & 58 & 101\\
  BRI 0021-0214 & 6.1026 & -1.9722 & 2160 & 94 & 94\\
  2MASS J00285545-1927165 & 7.2310 & -19.4546 & 2044 & 111 & 95\\
  2MASSW J0030438+313932 & 7.6827 & 31.6589 & 1866 & 93 & 103\\
  WISE J003110.04+574936.3 & 7.7892 & 57.8268 & 1662 & 195 & 105\\
  2MASS J00320509+0219017 & 8.0212 & 2.3172 & 2123 & 140 & 113\\
  2MASSI J0032431-223727 & 8.1796 & -22.6242 & 2180 & 43 & 154\\
  EROS-MP J0032-4405 & 8.2327 & -44.0849 & 1814 & 152 & 148\\
  SDSSp J003259.36+141036.6 & 8.2473 & 14.1769 & 1528 & 171 & 72\\
  2MASS J00332386-1521309 & 8.3495 & -15.3586 & 1634 & 96 & 79\\
  SDSS J003843.99+134339.5 & 9.6833 & 13.7276 & 2150 & 84 & 95\\
  HD 3651B & 9.8288 & 21.2547 & 682 & 49 & 169\\
  WISE J004024.88+090054.8 & 10.1039 & 9.0152 & 891 & 197 & 96\\
  SDSS J004154.54+134135.5 & 10.4773 & 13.6932 & 2173 & 71 & 112\\
  WISE J004542.56+361139.1 & 11.4276 & 36.1947 & 1013 & 99 & 55\\
  WISEP J004701.06+680352.1 & 11.7517 & 68.0651 & 1155 & 324 & 75\\
  WISEPC J004928.48+044100.1 & 12.3678 & 4.6826 & 1627 & 215 & 117\\
  WISE J004945.61+215120.0 & 12.4415 & 21.8556 & 809 & 96 & 76\\
  2MASS J00501994-3322402 & 12.5836 & -33.3775 & 1046 & 172 & 71\\
  SIPS J0050-1538 & 12.6017 & -15.6387 & 2264 & 82 & 88\\
  2MASSW J0051107-154417 & 12.7950 & -15.7380 & 1602 & 102 & 140\\
  SDSSp J005406.55-003101.8 & 13.5274 & -0.5173 & 2092 & 91 & 74\\
  SDSS J005705.55-084624.2 & 14.2731 & -8.7734 & 2309 & 86 & 63\\
  2MASSW J0058425-065123 & 14.6773 & -6.8567 & 2097 & 76 & 96\\
  LHS 132 & 15.7127 & -37.6288 & 2476 & 67 & 97\\
  2MASSI J0103320+193536 & 15.8837 & 19.5935 & 1684 & 207 & 50\\
  ULAS2MASS J0106+1518 & 16.6555 & 15.3153 & 2447 & 61 & 106\\
  2MASS J01165457-1357342 & 19.2274 & -13.9595 & 2427 & 100 & 118\\
  2MASSI J0117474-340325 & 19.4479 & -34.0572 & 1819 & 88 & 120\\
  2MASS J01194279+1122427 & 19.9283 & 11.3786 & 2148 & 55 & 106\\
  SSSPM J0124-4240 & 20.9961 & -42.6687 & 2320 & 39 & 93\\
  2MASSI J0125369-343505 & 21.4038 & -34.5847 & 1954 & 74 & 67\\
  CTI 012657.5+280202 & 21.9131 & 28.0982 & 2348 & 65 & 109\\
  WISE J013525.64+171503.4 & 23.8580 & 17.2514 & 923 & 26 & 126\\
  2MASSW J0135358+120522 & 23.8994 & 12.0894 & 1953 & 68 & 174\\
  SIMP J013656.5+093347.3 & 24.2357 & 9.5631 & 1262 & 34 & 115\\
  2MASSW J0141032+180450 & 25.2635 & 18.0806 & 1871 & 93 & 73\\
  2MASS J01443536-0716142 & 26.1475 & -7.2706 & 1609 & 112 & 114\\
  2MASS J01460119-4545263 & 26.5049 & -45.7573 & 2462 & 138 & 154\\
  2MASS J01472702+4731142 & 26.8625 & 47.5207 & 2054 & 108 & 117\\
  2MASSW J0147334+345311 & 26.8894 & 34.8864 & 2092 & 124 & 98\\
  2MASSW J0149090+295613 & 27.2875 & 29.9368 & 2035 & 70 & 109\\
  WISEPA J015010.86+382724.3 & 27.5423 & 38.4572 & 1433 & 152 & 94\\
  SDSS J015141.69+124429.6 & 27.9240 & 12.7412 & 1656 & 169 & 97\\
  SDSS J015354.23+140452.9 & 28.4759 & 14.0814 & 2452 & 109 & 160\\
  2MASS J02042212-3632308 & 31.0922 & -36.5419 & 2241 & 60 & 89\\
  2MASSW J0205034+125142 & 31.2645 & 12.8617 & 1631 & 217 & 76\\
  WISEPA J020625.26+264023.6 & 31.6048 & 26.6730 & 1327 & 197 & 134\\
  SDSS J020735.60+135556.3 & 31.8983 & 13.9323 & 1962 & 72 & 127\\
  2MASSW J0208183+254253 & 32.0766 & 25.7148 & 2029 & 90 & 100\\
  2MASSW J0208236+273740 & 32.0986 & 27.6277 & 1715 & 170 & 73\\
  2MASSI J0213288+444445 & 33.3700 & 44.7459 & 2382 & 102 & 235\\
  2MASSI J0218291-313322 & 34.6213 & -31.5564 & 1822 & 94 & 74\\
  2MASS J02192196+0506306 & 34.8415 & 5.1085 & 1946 & 88 & 75\\
  SSSPM J0219-1939 & 34.8669 & -19.6448 & 2141 & 112 & 65\\
  WISEPC J022322.39-293258.1 & 35.8406 & -29.5477 & 836 & 43 & 125\\
  HIP 11161B & 35.9029 & 52.6685 & 1885 & 26 & 229\\
  2MASS J02271036-1624479 & 36.7930 & -16.4132 & 1941 & 145 & 135\\
  2MASSW J0228110+253738 & 37.0460 & 25.6273 & 2071 & 83 & 126\\
  2MASS J02284243+1639329 & 37.1768 & 16.6592 & 2258 & 76 & 84\\
  WISE J023038.90-022554.0 & 37.6612 & -2.4319 & 1529 & 252 & 58\\
  DENIS J0230450-095305 & 37.6875 & -9.8848 & 2049 & 91 & 105\\
  WISE J023318.05+303030.5 & 38.3257 & 30.5086 & 1081 & 55 & 137\\
  SDSS J023547.56-084919.8 & 38.9482 & -8.8222 & 2073 & 98 & 91\\
  GJ 1048B & 38.9997 & -23.5223 & 2006 & 79 & 105\\
  2MASSI J0239424-173547 & 39.9269 & -17.5965 & 2367 & 102 & 95\\
  2MASSI J0241536-124106 & 40.4738 & -12.6853 & 1927 & 74 & 95\\
  2MASSW J0242435+160739 & 40.6815 & 16.1275 & 2049 & 110 & 105\\
  2MASSI J0243137-245329 & 40.8071 & -24.8917 & 939 & 146 & 75\\
  WISE J024512.62-345047.8 & 41.3029 & -34.8466 & 1039 & 208 & 22\\
  BR B0246-1703 & 42.1708 & -16.8561 & 2462 & 105 & 150\\
  TVLM 831-161058 & 42.8052 & 0.7934 & 2475 & 67 & 99\\
  2MASSI J0251148-035245 & 42.8125 & -3.8800 & 2267 & 173 & 167\\
  TVLM 832-10443 & 43.1096 & 0.9396 & 2792 & 416 & 213\\
  2MASS J02540582-1934523 & 43.5243 & -19.5812 & 2334 & 55 & 59\\
  PSO J043.5395+02.3995 & 43.5319 & 2.3989 & 941 & 143 & 93\\
  DENIS-P J025503.3-470049 & 43.7654 & -47.0143 & 2004 & 222 & 191\\
  2MASS J03001631+2130205 & 45.0680 & 21.5057 & 1816 & 65 & 165\\
  SSSPM J0306-3648 & 46.5483 & -36.7980 & 2430 & 45 & 98\\
  2MASSW J0309088-194938 & 47.2871 & -19.8275 & 1715 & 105 & 96\\
  2MASS J03101401-2756452 & 47.5583 & -27.9460 & 1646 & 115 & 68\\
  2MASS J03140344+1603056 & 48.5143 & 16.0515 & 2175 & 79 & 91\\
  2MASSI J0316451-284852 & 49.1881 & -28.8145 & 2080 & 142 & 79\\
  2MASS J03201720-1026124 & 50.0717 & -10.4368 & 2330 & 72 & 110\\
  LP 412-31 & 50.2488 & 18.9063 & 2345 & 76 & 131\\
  2MASS J03250136+2253039 & 51.2559 & 22.8843 & 1872 & 92 & 71\\
  WISE J032547.72+083118.2 & 51.4485 & 8.5218 & 869 & 102 & 53\\
  SDSS J032553.17+042540.1 & 51.4718 & 4.4279 & 960 & 186 & 47\\
  2MASSW J0326137+295015 & 51.5570 & 29.8376 & 1698 & 80 & 121\\
  2MASS J03264225-2102057 & 51.6761 & -21.0350 & 1432 & 204 & 89\\
  SDSSp J032817.38+003257.2 & 52.0725 & 0.5492 & 1723 & 93 & 66\\
  2MASSI J0328426+230205 & 52.1776 & 23.0348 & 1523 & 147 & 102\\
  SDSSp J033035.13-002534.5 & 52.6464 & -0.4264 & 1736 & 69 & 94\\
  2MASS J03320043-2317496 & 53.0019 & -23.2971 & 2401 & 48 & 83\\
  LEHPM 3396 & 53.5509 & -49.8922 & 2397 & 66 & 124\\
  2MASS J03354535+0658058 & 53.9390 & 6.9683 & 2452 & 62 & 79\\
  WISE J033651.90+282628.8 & 54.2160 & 28.4421 & 1116 & 75 & 52\\
  2MASSW J0337036-175807 & 54.2649 & -17.9688 & 1512 & 122 & 146\\
  LP 944-20 & 54.8969 & -35.4288 & 2234 & 54 & 123\\
  2MASP J0339527+245728 & 54.9700 & 24.9576 & 2424 & 89 & 96\\
  2MASS J03521086+0210479 & 58.0453 & 2.1800 & 2414 & 104 & 124\\
  SDSS J035308.54+103056.0 & 58.2853 & 10.5157 & 2143 & 139 & 102\\
  2MASS J03540135+2316339 & 58.5058 & 23.2761 & 2302 & 60 & 84\\
  2MASS J04012977-4050448 & 60.3741 & -40.8458 & 2309 & 145 & 63\\
  WISE J040137.21+284951.7 & 60.4066 & 28.8313 & 1834 & 88 & 55\\
  WISE J040418.01+412735.6 & 61.0753 & 41.461 & 1878 & 107 & 40\\
  2MASS J04070752+1546457 & 61.7814 & 15.7793 & 1628 & 135 & 119\\
  2MASS J04070885+1514565 & 61.7866 & 15.2491 & 1098 & 213 & 45\\
  2MASS J04081032+0742494 & 62.043 & 7.7137 & 2451 & 58 & 60\\
  2MASSI J0408290-145033 & 62.1211 & -14.8427 & 2014 & 96 & 93\\
  2MASSI J0409095+210439 & 62.2897 & 21.0775 & 1872 & 74 & 85\\
  2MASSI J0415195-093506 & 63.8324 & -9.5851 & 731 & 120 & 124\\
  2MASS J04174743-2129191 & 64.4477 & -21.4886 & 2528 & 101 & 158\\
  2MASS J04270723+0859027 & 66.7802 & 8.9841 & 2171 & 66 & 126\\
  2MASSI J0428510-225323 & 67.2124 & -22.8896 & 2061 & 85 & 150\\
  2MASS J04305157-0849007 & 67.7149 & -8.8169 & 2301 & 50 & 91\\
  2MASS J04362054-4218523 & 69.0856 & -42.3145 & 2173 & 85 & 97\\
  2MASS J04362788-4114465 & 69.1162 & -41.2462 & 2349 & 25 & 166\\
  2MASSI J0439010-235308 & 69.7542 & -23.8857 & 1685 & 164 & 97\\
  2MASSI J0443058-320209 & 70.7743 & -32.0358 & 1840 & 210 & 30\\
  2MASS J04441479+0543573 & 71.0616 & 5.7326 & 2571 & 151 & 151\\
  2MASSI J0445538-304820 & 71.4746 & -30.8057 & 2234 & 182 & 71\\
  WISEPA J044853.29-193548.5 & 72.2194 & -19.5987 & 1334 & 103 & 140\\
  2MASSI J0451009-340214 & 72.7539 & -34.0375 & 2192 & 129 & 94\\
  2MASSI J0453264-175154 & 73.3603 & -17.8651 & 1874 & 58 & 93\\
  WISE J045746.08-020719.2 & 74.4418 & -2.1217 & 1352 & 69 & 80\\
  WISEPA J050003.05-122343.2 & 75.0146 & -12.397 & 947 & 56 & 260\\
  2MASS J05002100+0330501 & 75.0875 & 3.5139 & 1710 & 85 & 122\\
  2MASS J05012406-0010452 & 75.3504 & -0.1793 & 1649 & 94 & 92\\
  2MASSI J0502134+144236 & 75.556 & 14.7102 & 2290 & 41 & 77\\
  LSR J0510+2713 & 77.5837 & 27.2339 & 2450 & 55 & 102\\
  2MASSI J0512063-294954 & 78.0265 & -29.8316 & 1339 & 191 & 73\\
  2MASS J05160945-0445499 & 79.0392 & -4.7640 & 1185 & 95 & 64\\
  2MASS J05170548-4154413 & 79.2728 & -41.9115 & 2423 & 120 & 322\\
  2MASS J05173766-3349027 & 79.4071 & -33.8175 & 2297 & 65 & 74\\
  WISE J052126.29+102528.4 & 80.3571 & 10.4267 & 786 & 147 & 46\\
  2MASSI J0523382-140302 & 80.9093 & -14.0506 & 2009 & 110 & 80\\
  2MASS J05264348-4455455 & 81.6812 & -44.9293 & 2111 & 117 & 154\\
  2MASS J05301261+6253254 & 82.5525 & 62.8904 & 2092 & 126 & 87\\
  2MASS J05345844-1511439 & 83.7435 & -15.1955 & 2309 & 51 & 110\\
  HIP 26653B & 84.9564 & 52.8999 & 1937 & 74 & 102\\
  SDSSp J053951.99-005902.0 & 84.9667 & -0.9837 & 1786 & 130 & 61\\
  2MASS J05441150-2433018 & 86.0480 & -24.5507 & 2580 & 157 & 210\\
  WISE J054601.19-095947.5 & 86.5050 & -9.9965 & 1215 & 118 & 72\\
  WISEA J055007.94+161051.9 & 87.5323 & 16.1819 & 1845 & 90 & 72\\
  2MASS J05591914-1404488 & 89.8299 & -14.0803 & 1191 & 111 & 70\\
  2MASS J06020638+4043588 & 90.5265 & 40.7330 & 995 & 107 & 103\\
  2MASS J06022216+6336391 & 90.5924 & 63.6108 & 1904 & 86 & 93\\
  LSR J0602+3910 & 90.6269 & 39.1829 & 2202 & 102 & 112\\
  2MASS J06050196-2342270 & 91.2582 & -23.7075 & 2338 & 71 & 110\\
  WISEP J060738.65+242953.4 & 91.9126 & 24.4991 & 1490 & 111 & 157\\
  WISEA J060742.13+455037.0 & 91.9251 & 45.8453 & 1885 & 43 & 163\\
  SIPS J0614-2019 & 93.5499 & -20.3218 & 2051 & 109 & 66\\
  DENIS-P J0615493-010041 & 93.9557 & -1.0116 & 2227 & 138 & 85\\
  2MASS J06195260-2903592 & 94.9692 & -29.0664 & 2120 & 208 & 155\\
  WISE J062442.37+662625.6 & 96.174 & 66.442 & 2098 & 107 & 58\\
  2MASS J06244595-4521548 & 96.1914 & -45.3652 & 1539 & 228 & 137\\
  SDSS J062621.22+002934.2 & 96.5884 & 0.4928 & 2153 & 96 & 73\\
  WISEPA J062720.07-111428.8 & 96.8337 & -11.2401 & 1051 & 156 & 47\\
  WISE J062905.13+241804.9 & 97.2715 & 24.3022 & 1564 & 127 & 95\\
  2MASS J06411840-4322329 & 100.3268 & -43.3757 & 2028 & 76 & 99\\
  DENIS-P J0652197-253450 & 103.0823 & -25.5807 & 2353 & 55 & 92\\
  PSO J103.0927+41.4601 & 103.0927 & 41.4601 & 1414 & 103 & 104\\
  2MASSI J0652307+471034 & 103.128 & 47.1764 & 1631 & 155 & 83\\
  WISEPA J065609.60+420531.0 & 104.039 & 42.0916 & 1571 & 87 & 162\\
  ESO 207-61 & 106.972 & -49.014 & 2099 & 58 & 167\\
  WISEA J071552.38-114532.9 & 108.9661 & -11.758 & 1833 & 81 & 196\\
  DENIS-P J0716478-063037 & 109.1996 & -6.5103 & 2074 & 79 & 99\\
  2MASSW J0717163+570543 & 109.3178 & 57.0953 & 1653 & 124 & 59\\
  UGPS J072227.51-054031.2 & 110.6164 & -5.6761 & 624 & 29 & 169\\
  2MASS J07231462+5727081 & 110.8109 & 57.4522 & 2158 & 106 & 151\\
  2MASSI J0727182+171001 & 111.8265 & 17.1665 & 925 & 135 & 38\\
  2MASS J07290002-3954043 & 112.2501 & -39.9008 & 877 & 126 & 118\\
  SDSS J073519.59+410850.4 & 113.8318 & 41.1474 & 2409 & 110 & 84\\
  SDSS J074149.15+235127.5 & 115.4549 & 23.8578 & 966 & 128 & 112\\
  2MASS J07415784+0531568 & 115.4910 & 5.5325 & 2112 & 93 & 71\\
  SDSS J074201.41+205520.5 & 115.5050 & 20.9221 & 965 & 142 & 53\\
  SDSS J074756.31+394732.9 & 116.9847 & 39.7924 & 2322 & 67 & 201\\
  SDSS J075054.74+445418.7 & 117.7282 & 44.9056 & 2262 & 94 & 86\\
  DENIS-P J0751164-253043 & 117.8183 & -25.5120 & 2148 & 105 & 93\\
  2MASSI J0753321+291711 & 118.3840 & 29.2866 & 1835 & 57 & 80\\
  2MASSI J0755480+221218 & 118.9496 & 22.2048 & 967 & 70 & 123\\
  HIP 38939B & 119.5057 & -25.6499 & 1242 & 95 & 56\\
  SDSS J075840.33+324723.4 & 119.6681 & 32.7900 & 1283 & 51 & 123\\
  2MASS J08041429+0330474 & 121.0595 & 3.5132 & 2297 & 81 & 66\\
  WISE J080700.23+413026.8 & 121.7511 & 41.5084 & 1615 & 209 & 157\\
  SDSS J080959.01+443422.2 & 122.4948 & 44.5719 & 1415 & 231 & 79\\
  SDSS J081110.35+185527.9 & 122.7932 & 18.9245 & 2247 & 114 & 115\\
  DENIS-P J0812316-244442 & 123.1322 & -24.7451 & 2285 & 117 & 136\\
  SDSS J081757.49+182405.0 & 124.4895 & 18.4014 & 2307 & 132 & 116\\
  SDSS J081812.28+331048.2 & 124.5513 & 33.1800 & 2054 & 134 & 133\\
  2MASS J08194602+1658539 & 124.9417 & 16.9816 & 2468 & 72 & 78\\
  WISEPA J081958.05-033529.0 & 124.9924 & -3.5908 & 1229 & 26 & 111\\
  2MASSW J0820299+450031 & 125.1249 & 45.0087 & 1375 & 75 & 220\\
  WISEPA J082131.63+144319.3 & 125.3821 & 14.7228 & 1218 & 122 & 133\\
  2MASS J08230838+6125208 & 125.7847 & 61.4224 & 1892 & 64 & 103\\
  2MASS J08234818+2428577 & 125.9507 & 24.4827 & 1811 & 73 & 103\\
  2MASSI J0825196+211552 & 126.3317 & 21.2643 & 1354 & 217 & 116\\
  SDSS J082642.65+193922.0 & 126.6776 & 19.6562 & 2279 & 102 & 172\\
  SSSPM J0829-1309 & 127.1424 & -13.1555 & 1984 & 75 & 112\\
  2MASSW J0829066+145622 & 127.2776 & 14.9395 & 1891 & 53 & 115\\
  SDSSp J083008.12+482847.4 & 127.534 & 48.4799 & 1537 & 184 & 144\\
  LHS 2021 & 127.6357 & 9.7876 & 2402 & 90 & 108\\
  SDSS J083048.80+012831.1 & 127.7034 & 1.4754 & 1012 & 44 & 108\\
  2MASSW J0832045-012835 & 128.0188 & -1.4767 & 2011 & 52 & 119\\
  WISE J083450.79+642526.8 & 128.7124 & 64.4248 & 2428 & 57 & 92\\
  2MASS J08352366+1029318 & 128.8486 & 10.4922 & 2463 & 82 & 89\\
  2MASSI J0835425-081923 & 128.9272 & -8.3232 & 1673 & 106 & 88\\
  SDSS J083545.33+222430.9 & 128.939 & 22.4086 & 2269 & 63 & 95\\
  2MASS J08355829+0548308 & 128.9929 & 5.8086 & 1963 & 95 & 107\\
  SDSS J083621.98+494931.5 & 129.0917 & 49.8255 & 2165 & 105 & 89\\
  SDSS J083646.35+052642.6 & 129.1932 & 5.4452 & 2086 & 55 & 111\\
  SDSSp J083717.22-000018.3 & 129.3215 & -0.0051 & 1687 & 207 & 45\\
  2MASS J08391608+1253543 & 129.817 & 12.8984 & 2417 & 57 & 102\\
  SDSS J084106.85+603506.3 & 130.2785 & 60.5852 & 2108 & 111 & 99\\
  SDSS J084307.95+314129.2 & 130.7831 & 31.6915 & 1839 & 37 & 159\\
  SDSS J084333.28+102443.5 & 130.8885 & 10.4131 & 2003 & 129 & 68\\
  SDSS J084457.38+120825.4 & 131.2391 & 12.1405 & 2198 & 109 & 86\\
  2MASSI J0847287-153237 & 131.8698 & -15.5437 & 2174 & 179 & 99\\
  SDSS J085234.90+472035.0 & 133.1454 & 47.3431 & 1547 & 157 & 80\\
  SDSSp J085758.45+570851.4 & 134.4936 & 57.1476 & 1244 & 93 & 214\\
  SDSS J085834.42+325627.7 & 134.643 & 32.9407 & 1486 & 193 & 109\\
  SDSS J085836.98+271050.8 & 134.6539 & 27.1811 & 2077 & 72 & 121\\
  2MASSI J0859254-194926 & 134.856 & -19.824 & 1532 & 149 & 166\\
  2MASS J08593854+6341355 & 134.9106 & 63.6932 & 2448 & 85 & 155\\
  2MASS J08594029+1145325 & 134.9179 & 11.759 & 2374 & 100 & 138\\
  2MASS J09054654+5623117 & 136.4439 & 56.3866 & 1652 & 172 & 119\\
  2MASSI J0908380+503208 & 137.1584 & 50.5356 & 1814 & 257 & 49\\
  SDSS J090948.13+194043.9 & 137.4509 & 19.6786 & 2605 & 320 & 108\\
  DENIS-P J090957.1-065806 & 137.4895 & -6.9718 & 2091 & 54 & 126\\
  2MASS J09161504+2139512 & 139.0625 & 21.6642 & 2245 & 61 & 76\\
  WISEA J091657.18-112104.7 & 139.2379 & -11.3501 & 2529 & 257 & 186\\
  2MASSW J0918382+213406 & 139.6592 & 21.5682 & 1827 & 81 & 73\\
  2MASS J09211410-2104446 & 140.3088 & -21.079 & 2116 & 131 & 98\\
  SDSS J092308.70+234013.7 & 140.7859 & 23.671 & 2202 & 138 & 84\\
  2MASSW J0928397-160312 & 142.1654 & -16.0535 & 1861 & 113 & 46\\
  SDSS J093237.47+672514.5 & 143.1562 & 67.4209 & 2164 & 101 & 182\\
  2MASS J09352803-2934596 & 143.8668 & -29.5832 & 2206 & 105 & 73\\
  2MASSI J0937347+293142 & 144.3952 & 29.5282 & 819 & 77 & 129\\
  2MASS J09384022-2748184 & 144.6676 & -27.8051 & 2431 & 55 & 85\\
  SDSS J093858.88+044343.9 & 144.7453 & 4.7288 & 2747 & 408 & 106\\
  2MASS J09393548-2448279 & 144.8979 & -24.8077 & 843 & 165 & 129\\
  SDSS J094047.88+294653.0 & 145.1996 & 29.7815 & 2105 & 95 & 162\\
  SDSS J094134.92+100942.0 & 145.3955 & 10.1619 & 2388 & 108 & 116\\
  2MASSW J0944027+313132 & 146.0116 & 31.5258 & 1954 & 121 & 95\\
  2MASS J09474477+0224327 & 146.9366 & 2.4091 & 2370 & 49 & 90\\
  2MASS J09490860-1545485 & 147.2859 & -15.7635 & 1464 & 50 & 68\\
  LHS 2195 & 147.3426 & 8.1125 & 2470 & 72 & 128\\
  WISEPC J095259.29+195507.3 & 148.2473 & 19.919 & 1008 & 42 & 159\\
  2MASS J09532126-1014205 & 148.3386 & -10.2391 & 2171 & 106 & 149\\
  2MASS J10031918-0105079 & 150.8298 & -1.0856 & 2434 & 47 & 91\\
  2MASSW J1004392-333518 & 151.1638 & -33.5886 & 1818 & 79 & 62\\
  SDSS J100711.74+193056.2 & 151.7994 & 19.5155 & 1534 & 267 & 90\\
  WISE J100926.40+354137.5 & 152.3598 & 35.6947 & 2306 & 74 & 87\\
  2MASSI J1010148-040649 & 152.5615 & -4.1139 & 1666 & 168 & 71\\
  2MASS J10163470+2751497 & 154.1445 & 27.8636 & 2297 & 73 & 118\\
  SDSS J101742.51+431057.9 & 154.4271 & 43.1828 & 2125 & 45 & 97\\
  2MASSW J1018588-290953 & 154.7449 & -29.1649 & 2023 & 90 & 112\\
  WISEPA J101905.63+652954.2 & 154.7737 & 65.4979 & 831 & 77 & 108\\
  DENIS J1019245-270717 & 154.8518 & -27.1214 & 2374 & 54 & 110\\
  2MASS J10213232-2044069 & 155.3846 & -20.7353 & 2541 & 110 & 155\\
  SDSS J102204.88+020047.5 & 155.5204 & 2.0133 & 2415 & 71 & 104\\
  HD 89744B & 155.562 & 41.2407 & 2171 & 99 & 64\\
  2MASS J10224821+5825453 & 155.7009 & 58.4293 & 2030 & 66 & 123\\
  WISEA J102304.04+155616.4 & 155.7672 & 15.9397 & 2329 & 66 & 113\\
  SDSS J102552.43+321234.0 & 156.4681 & 32.2097 & 1568 & 222 & 149\\
  2MASSI J1029216+162652 & 157.3404 & 16.4478 & 1820 & 106 & 42\\
  ULAS J102940.52+093514.6 & 157.4201 & 9.5877 & 802 & 120 & 138\\
  2MASS J10315064+3349595 & 157.9609 & 33.8332 & 1750 & 66 & 172\\
  2MASS J10321706+0501032 & 158.0711 & 5.0175 & 2444 & 67 & 110\\
  SDSS J103309.11+121626.0 & 158.2878 & 12.274 & 2378 & 98 & 188\\
  SDSS J103405.67+035016.3 & 158.5235 & 3.8379 & 2120 & 119 & 82\\
  2MASSW J1035245+250745 & 158.8522 & 25.1291 & 2148 & 135 & 74\\
  WISE J103907.73-160002.9 & 159.783 & -16.0003 & 965 & 149 & 122\\
  2MASS J10430758+2225236 & 160.7815 & 22.4234 & 1314 & 205 & 83\\
  SDSS J104335.08+121314.1 & 160.8962 & 12.2208 & 1556 & 190 & 153\\
  SDSS J104409.43+042937.6 & 161.0393 & 4.4938 & 1610 & 218 & 88\\
  2MASSI J1045240-014957 & 161.3499 & -1.8327 & 2115 & 71 & 118\\
  2MASS J10461875+4441149 & 161.5783 & 44.6874 & 1833 & 101 & 87\\
  DENIS-P J104731.1-181558 & 161.8794 & -18.266 & 2235 & 148 & 64\\
  2MASSI J1047538+212423 & 161.9735 & 21.4063 & 769 & 98 & 100\\
  SDSS J104842.84+011158.5 & 162.1784 & 1.1995 & 2125 & 124 & 81\\
  SDSS J105151.25+131116.3 & 162.9636 & 13.1879 & 2092 & 140 & 119\\
  WISE J105257.95-194250.2 & 163.2405 & -19.7132 & 847 & 45 & 167\\
  2MASS J10554733+0808427 & 163.9473 & 8.1453 & 2432 & 63 & 88\\
  DENIS-P J1058.7-1548 & 164.6993 & -15.8048 & 1865 & 73 & 79\\
  2MASS J11000965+4957470 & 165.0402 & 49.963 & 1787 & 69 & 102\\
  2MASSI J1104012+195921 & 166.0053 & 19.9894 & 1723 & 59 & 114\\
  2MASS J11145133-2618235 & 168.7131 & -26.3066 & 1181 & 117 & 160\\
  2MASSI J1117369+360936 & 169.4039 & 36.16 & 2113 & 116 & 89\\
  2MASS J11220826-3512363 & 170.5344 & -35.2102 & 1496 & 49 & 142\\
  2MASSW J1122362-391605 & 170.651 & -39.2682 & 1985 & 153 & 140\\
  WISEPC J112254.73+255021.5 & 170.7323 & 25.8403 & 1536 & 74 & 201\\
  2MASS J11240487+3808054 & 171.0204 & 38.1348 & 2403 & 57 & 102\\
  WISE J112438.12-042149.7 & 171.1602 & -4.3638 & 874 & 197 & 70\\
  2MASS J11263991-5003550 & 171.6658 & -50.0652 & 1937 & 168 & 46\\
  SDSS J112647.03+581632.2 & 171.6959 & 58.2757 & 2100 & 145 & 80\\
  2MASS J11414406-2232156 & 175.4335 & -22.5375 & 2354 & 52 & 110\\
  SDSS J114912.31-015300.6 & 177.3013 & -1.8835 & 2474 & 83 & 144\\
  2MASS J11533966+5032092 & 178.4153 & 50.5359 & 2006 & 68 & 121\\
  2MASS J11544223-3400390 & 178.6759 & -34.0109 & 2045 & 50 & 96\\
  2MASSW J1155395-372735 & 178.9147 & -37.4599 & 2099 & 56 & 69\\
  LP 851-346 & 178.9285 & -22.4163 & 2506 & 110 & 147\\
  SDSS J115553.86+055957.5 & 178.9747 & 5.9994 & 1634 & 198 & 67\\
  DENIS-P J1157480-484442 & 179.4504 & -48.7452 & 1971 & 133 & 109\\
  DENIS-P J1159+0057 & 179.9104 & 0.9574 & 2219 & 93 & 83\\
  SDSS J115940.72+540938.6 & 179.9198 & 54.1607 & 1996 & 103 & 130\\
  SDSSp J120358.19+001550.3 & 180.9922 & 0.2639 & 1846 & 93 & 81\\
  2MASSI J1204303+321259 & 181.1265 & 32.2165 & 2192 & 119 & 120\\
  SDSS J120602.51+281328.7 & 181.5103 & 28.2247 & 1255 & 67 & 71\\
  SDSS J120610.49+624257.2 & 181.5438 & 62.716 & 1933 & 128 & 106\\
  DENIS J1206501-393725 & 181.7088 & -39.6239 & 2076 & 109 & 93\\
  2MASS J12070374-3151298 & 181.7655 & -31.8583 & 1820 & 143 & 72\\
  2MASS J12073804-3909050 & 181.9085 & -39.1514 & 2093 & 160 & 90\\
  2MASS J12123389+0206280 & 183.1412 & 2.1078 & 2039 & 68 & 152\\
  2MASSI J1213033-043243 & 183.2639 & -4.5455 & 1791 & 71 & 87\\
  2MASSI J1217110-031113 & 184.2958 & -3.1869 & 937 & 139 & 124\\
  SDSS J121951.45+312849.4 & 184.9647 & 31.4804 & 1521 & 187 & 76\\
  2MASS J12212770+0257198 & 185.3655 & 2.9555 & 2109 & 44 & 194\\
  WISE J122152.28-313600.8 & 185.4657 & -31.6014 & 1100 & 198 & 41\\
  BRI B1222-1222 & 186.2174 & -12.6433 & 2191 & 73 & 153\\
  WISE J122558.86-101345.0 & 186.4959 & -10.2283 & 900 & 187 & 52\\
  2MASS J12312141+4959234 & 187.8392 & 49.9898 & 1982 & 63 & 108\\
  2MASS J12314753+0847331 & 187.9482 & 8.7924 & 900 & 167 & 37\\
  2MASS J12321827-0951502 & 188.0762 & -9.864 & 2302 & 92 & 109\\
  2MASS J12373919+6526148 & 189.4129 & 65.4373 & 760 & 123 & 92\\
  2MASSW J1246467+402715 & 191.6949 & 40.4542 & 1718 & 174 & 39\\
  SDSS J124908.66+415728.6 & 192.2864 & 41.958 & 2273 & 59 & 81\\
  WISE J125448.52-072828.4 & 193.7018 & -7.4742 & 1118 & 95 & 147\\
  SDSSp J125453.90-012247.4 & 193.7246 & -1.3798 & 1348 & 44 & 96\\
  2MASS J12565688+0146163 & 194.237 & 1.7712 & 1844 & 96 & 68\\
  WISE J125715.90+400854.2 & 194.315 & 40.148 & 1046 & 202 & 78\\
  2MASSW J1300425+191235 & 195.1771 & 19.2096 & 2179 & 82 & 81\\
  2MASS J13015465-1510223 & 195.4778 & -15.1729 & 2171 & 93 & 131\\
  2MASSI J1305410+204639 & 196.4211 & 20.7776 & 1677 & 181 & 103\\
  2MASS J13061727+3820296 & 196.572 & 38.3416 & 2133 & 59 & 99\\
  WISEPC J132004.16+603426.2 & 200.0212 & 60.5743 & 910 & 95 & 86\\
  2MASS J13204427+0409045 & 200.1845 & 4.1513 & 1935 & 116 & 93\\
  DENIS-P J1323-1806 & 200.8999 & -18.1105 & 2185 & 106 & 105\\
  2MASS J13243553+6358281 & 201.1479 & 63.9744 & 1376 & 118 & 130\\
  2MASSW J1326201-272937 & 201.5836 & -27.4936 & 1367 & 229 & 97\\
  SDSSp J132629.82-003831.5 & 201.6242 & -0.6421 & 1505 & 240 & 120\\
  SDSS J132715.21+075937.5 & 201.8136 & 7.9938 & 2285 & 163 & 93\\
  2MASSW J1328550+211449 & 202.2293 & 21.2467 & 1604 & 115 & 105\\
  2MASS J13313310+3407583 & 202.888 & 34.1328 & 2230 & 111 & 138\\
  SDSS J133148.92-011651.4 & 202.9538 & -1.2809 & 1809 & 207 & 49\\
  SDSS J133312.79+150956.6 & 203.3034 & 15.1658 & 2293 & 60 & 124\\
  SDSS J133345.36-021600.2 & 203.4391 & -2.2667 & 2124 & 90 & 122\\
  2MASS J13364062+3743230 & 204.1691 & 37.723 & 2202 & 107 & 102\\
  2MASSW J1343167+394508 & 205.8194 & 39.7525 & 1731 & 135 & 72\\
  SDSSp J134646.45-003150.4 & 206.6934 & -0.5307 & 1037 & 163 & 54\\
  LHS 2803B & 207.0113 & -13.7356 & 941 & 97 & 59\\
  SDSS J135852.68+374711.9 & 209.7197 & 37.7869 & 1077 & 148 & 114\\
  2MASS J13595510-4034582 & 209.9796 & -40.5829 & 2025 & 73 & 142\\
  SDSS J140023.12+433822.3 & 210.0966 & 43.6394 & 1587 & 215 & 133\\
  WISE J140035.40-385013.5 & 210.1475 & -38.8363 & 1474 & 79 & 93\\
  2MASS J14022235+0648479 & 210.5932 & 6.8133 & 2255 & 48 & 90\\
  2MASS J14044495+4634297 & 211.1873 & 46.575 & 2214 & 65 & 70\\
  SDSS J140601.47+524931.0 & 211.5063 & 52.8252 & 2397 & 116 & 296\\
  2MASS J14075361+1241099 & 211.9734 & 12.6861 & 1735 & 105 & 71\\
  2MASS J14090310-3357565 & 212.263 & -33.9657 & 2060 & 136 & 98\\
  2MASSW J1411175+393636 & 212.8224 & 39.6101 & 2102 & 68 & 78\\
  2MASS J14122268+2354108 & 213.0945 & 23.903 & 2488 & 83 & 100\\
  2MASSW J1412244+163312 & 213.1021 & 16.5533 & 2203 & 86 & 64\\
  2MASS J14182962-3538060 & 214.6235 & -35.635 & 2004 & 116 & 67\\
  SDSS J142058.30+213156.6 & 215.2434 & 21.5321 & 2190 & 107 & 94\\
  2MASSW J1421314+182740 & 215.3809 & 18.4613 & 2207 & 73 & 126\\
  SDSS J142257.15+082752.1 & 215.7383 & 8.465 & 2048 & 84 & 120\\
  GD 165B & 216.1629 & 9.2863 & 1822 & 86 & 90\\
  DENIS-P J142527.97-365023.4 & 216.3666 & -36.8398 & 1493 & 221 & 73\\
  2MASS J14283132+5923354 & 217.1304 & 59.3932 & 1856 & 84 & 74\\
  LHS 2924 & 217.1801 & 33.1776 & 2373 & 87 & 102\\
  SDSS J143242.10+345142.7 & 218.1757 & 34.8619 & 2135 & 109 & 101\\
  2MASSI J1438082+640836 & 219.5345 & 64.1434 & 2402 & 106 & 80\\
  2MASSW J1438549-130910 & 219.7292 & -13.1529 & 1764 & 94 & 69\\
  2MASSW J1439284+192915 & 219.8673 & 19.4878 & 2184 & 96 & 98\\
  2MASS J14403186-1303263 & 220.1329 & -13.0573 & 2135 & 156 & 51\\
  SDSSp J144600.60+002452.0 & 221.5025 & 0.4144 & 1735 & 94 & 88\\
  2MASSW J1448256+103159 & 222.1068 & 10.5331 & 1582 & 226 & 82\\
  ULAS2MASS J1452+1114 & 223.0076 & 11.2498 & 2120 & 139 & 144\\
  SDSS J145255.58+272324.4 & 223.2318 & 27.3904 & 2340 & 137 & 108\\
  WISEA J145408.03+005325.7 & 223.5343 & 0.8903 & 2116 & 62 & 171\\
  2MASSI J1456014-274735 & 224.0058 & -27.7935 & 2173 & 61 & 128\\
  LHS 3003 & 224.1594 & -28.1635 & 2528 & 71 & 115\\
  Gliese 570D & 224.3129 & -21.364 & 769 & 108 & 129\\
  WISEPC J145715.03+581510.2 & 224.3153 & 58.2531 & 1101 & 58 & 213\\
  2MASS J14582453+2839580 & 224.6022 & 28.6661 & 2404 & 65 & 75\\
  TVLM 513-46546 & 225.2841 & 22.8339 & 2319 & 58 & 76\\
  2MASS J15031961+2525196 & 225.8317 & 25.4222 & 938 & 157 & 58\\
  2MASSW J1506544+132106 & 226.7264 & 13.3517 & 1893 & 78 & 80\\
  TVLM 868-110639 & 227.5701 & -2.6856 & 2242 & 50 & 93\\
  SDSS J151240.67+340350.1 & 228.1693 & 34.0639 & 1913 & 99 & 140\\
  SDSS J151506.11+443648.3 & 228.7753 & 44.6134 & 1469 & 188 & 81\\
  SDSS J152039.82+354619.8 & 230.1655 & 35.7725 & 1588 & 181 & 133\\
  SDSS J152103.24+013142.7 & 230.2635 & 1.5285 & 1248 & 43 & 81\\
  2MASS J15230657-2347526 & 230.7773 & -23.798 & 2237 & 83 & 95\\
  Gl 584C & 230.8443 & 30.2489 & 1573 & 253 & 103\\
  2MASP J1524248+292535 & 231.1032 & 29.4254 & 2443 & 45 & 91\\
  2MASSI J1526140+204341 & 231.5584 & 20.7279 & 1796 & 208 & 62\\
  SDSS J153453.33+121949.2 & 233.722 & 12.3304 & 1643 & 105 & 123\\
  DENIS-P J153941.96-052042.4 & 234.9247 & -5.3452 & 1834 & 109 & 68\\
  SDSS J154009.36+374230.3 & 235.0392 & 37.7088 & 1554 & 220 & 145\\
  2MASS J15461461+4932114 & 236.5611 & 49.5362 & 1259 & 52 & 63\\
  2MASSI J1546271-332511 & 236.6134 & -33.4198 & 953 & 110 & 122\\
  SDSS J154849.02+172235.4 & 237.2046 & 17.3766 & 1775 & 85 & 165\\
  2MASS J15485834-1636018 & 237.243 & -16.6006 & 2376 & 86 & 78\\
  SDSS J155120.86+432930.3 & 237.837 & 43.4918 & 1885 & 109 & 67\\
  2MASSW J1552591+294849 & 238.2461 & 29.8135 & 2072 & 41 & 123\\
  2MASSW J1555157-095605 & 238.8157 & -9.935 & 2137 & 132 & 117\\
  SDSS J155644.35+172308.9 & 239.1848 & 17.3858 & 2031 & 52 & 133\\
  WISE J155755.29+591425.3 & 239.4821 & 59.2398 & 2372 & 75 & 78\\
  2MASS J16150413+1340079 & 243.7681 & 13.6688 & 1032 & 113 & 73\\
  2MASS J16154255+4953211 & 243.9275 & 49.8893 & 1042 & 41 & 357\\
  2MASSW J1615441+355900 & 243.934 & 35.9832 & 1781 & 47 & 98\\
  2MASS J16184503-1321297 & 244.6876 & -13.3583 & 2285 & 110 & 73\\
  SDSS J161928.31+005011.9 & 244.868 & 0.8366 & 2108 & 193 & 75\\
  GJ 618.1B & 245.109 & -4.2755 & 1836 & 53 & 76\\
  WISEPA J162208.94-095934.6 & 245.5371 & -9.993 & 1095 & 169 & 54\\
  SDSSp J162414.37+002915.6 & 246.0597 & 0.4877 & 979 & 220 & 62\\
  SDSS J162603.03+211313.0 & 246.5126 & 21.2204 & 1912 & 127 & 62\\
  WISEPA J162725.64+325525.5 & 246.8572 & 32.9245 & 848 & 138 & 64\\
  PSO J247.3273+03.5932 & 247.3267 & 3.5936 & 1384 & 61 & 78\\
  SDSS J163030.53+434404.0 & 247.6273 & 43.7343 & 1568 & 222 & 106\\
  2MASS J16304139+0938446 & 247.6725 & 9.6457 & 2040 & 72 & 97\\
  WISE J163236.47+032927.3 & 248.1518 & 3.4908 & 1078 & 179 & 69\\
  SDSS J163256.13+350507.2 & 248.2338 & 35.0854 & 2137 & 77 & 74\\
  SDSS J163359.23-064056.5 & 248.4972 & -6.6821 & 1785 & 125 & 63\\
  2MASS J16351919+4223053 & 248.83 & 42.3848 & 2393 & 50 & 94\\
  WISE J163645.56-074325.1 & 249.1902 & -7.7234 & 1401 & 63 & 137\\
  2MASS J16452207+3004071 & 251.3419 & 30.0686 & 1844 & 98 & 107\\
  2MASSW J1645221-131951 & 251.3421 & -13.3312 & 2235 & 115 & 135\\
  WISEPA J164715.59+563208.2 & 251.8158 & 56.535 & 1421 & 188 & 115\\
  2MASS J16490419+0444571 & 252.2675 & 4.7492 & 2338 & 59 & 83\\
  WISEPA J165311.05+444423.9 & 253.2963 & 44.7408 & 791 & 92 & 162\\
  SDSS J165329.69+623136.5 & 253.3737 & 62.5268 & 2115 & 131 & 93\\
  SDSS J165450.79+374714.6 & 253.7116 & 37.7874 & 2138 & 109 & 88\\
  2MASS J16573454+1054233 & 254.394 & 10.9065 & 1986 & 97 & 103\\
  WISE J165842.56+510335.0 & 254.6786 & 51.0605 & 1779 & 150 & 124\\
  SDSS J165850.26+182000.6 & 254.7096 & 18.3334 & 2191 & 93 & 105\\
  SDSS J165950.91+351508.0 & 254.9621 & 35.2523 & 2239 & 94 & 104\\
  SDSS J170316.71+190636.0 & 255.8197 & 19.11 & 2138 & 83 & 176\\
  DENIS-P J170548.38-051645.7 & 256.4514 & -5.2795 & 2134 & 114 & 93\\
  2MASS J17065487-1314396 & 256.7286 & -13.2444 & 1936 & 118 & 106\\
  2MASSI J1707333+430130 & 256.8889 & 43.0251 & 2222 & 54 & 95\\
  SDSS J171049.35+332325.2 & 257.7056 & 33.3903 & 2296 & 51 & 110\\
  2MASS J17111353+2326333 & 257.8063 & 23.4426 & 2091 & 63 & 125\\
  2MASS J17114559+4028578 & 257.9401 & 40.4827 & 1882 & 152 & 72\\
  SDSS J171714.10+652622.2 & 259.3086 & 65.4395 & 1511 & 217 & 82\\
  2MASSI J1721039+334415 & 260.2651 & 33.738 & 1953 & 80 & 95\\
  WISE J172134.46+111739.4 & 260.3939 & 11.2939 & 1142 & 162 & 108\\
  SDSS J172244.32+632946.8 & 260.6847 & 63.4963 & 2097 & 99 & 65\\
  SDSS J172543.84+532534.9 & 261.4327 & 53.4264 & 2433 & 62 & 108\\
  VVV BD001 & 261.6693 & -27.6333 & 2101 & 128 & 76\\
  WISEPA J172844.93+571643.6 & 262.1869 & 57.2782 & 1123 & 69 & 147\\
  2MASS J17312974+2721233 & 262.8739 & 27.3565 & 2296 & 83 & 93\\
  2MASS J17320014+2656228 & 263.0006 & 26.9397 & 1973 & 69 & 138\\
  WISE J173332.50+314458.3 & 263.3865 & 31.7493 & 1752 & 114 & 78\\
  DENIS-P J1733423-165449 & 263.4261 & -16.9139 & 2095 & 161 & 100\\
  2MASS J17343053-1151388 & 263.6272 & -11.8608 & 2302 & 66 & 109\\
  WISE J174102.78-464225.5 & 265.2615 & -46.7059 & 1097 & 271 & 133\\
  WISE J174113.12+132711.9 & 265.3049 & 13.4539 & 1176 & 24 & 155\\
  2MASSW J1743415+212707 & 265.9229 & 21.452 & 1750 & 75 & 101\\
  DENIS-P J1745346-164053 & 266.3944 & -16.6817 & 2185 & 95 & 77\\
  2MASS J17461199+5034036 & 266.5499 & 50.5676 & 1759 & 105 & 60\\
  SDSS J175024.01+422237.8 & 267.5993 & 42.3771 & 1632 & 100 & 60\\
  2MASS J17502484-0016151 & 267.6034 & -0.2708 & 1857 & 214 & 31\\
  SDSSp J175032.96+175903.9 & 267.6372 & 17.9844 & 1252 & 61 & 105\\
  2MASS J17545447+1649196 & 268.726 & 16.8221 & 1041 & 162 & 35\\
  SDSS J175805.46+463311.9 & 269.5227 & 46.5531 & 1030 & 225 & 70\\
  2MASS J18000116-1559235 & 270.0049 & -15.99 & 1903 & 204 & 69\\
  2MASSI J1807159+501531 & 271.8164 & 50.2588 & 1986 & 75 & 131\\
  WISE J180901.07+383805.4 & 272.2566 & 38.6363 & 1067 & 135 & 169\\
  2MASS J18212815+1414010 & 275.3673 & 14.2336 & 1692 & 186 & 99\\
  2MASS J18283572-4849046 & 277.1488 & -48.8179 & 1259 & 105 & 70\\
  2MASSW J1841086+311727 & 280.2859 & 31.2912 & 1721 & 151 & 64\\
  WISE J185101.83+593508.6 & 282.7574 & 59.5845 & 1761 & 84 & 110\\
  WISEPA J185215.78+353716.3 & 283.0649 & 35.6221 & 944 & 153 & 45\\
  2MASS J19010601+4718136 & 285.2752 & 47.304 & 1080 & 103 & 86\\
  WISEPA J190624.75+450808.2 & 286.6031 & 45.1362 & 981 & 157 & 67\\
  WISEP J190648.47+401106.8 & 286.7003 & 40.1857 & 1964 & 63 & 153\\
  DENIS-P J1909081-193748 & 287.2842 & -19.63 & 1961 & 84 & 86\\
  vB 10 & 289.2401 & 5.1504 & 2431 & 64 & 87\\
  WISE J191915.54+304558.4 & 289.8134 & 30.7651 & 1849 & 83 & 136\\
  WISE J192841.35+235604.9 & 292.1729 & 23.9339 & 914 & 168 & 33\\
  2MASS J19285196-4356256 & 292.2166 & -43.9404 & 1657 & 80 & 114\\
  WISE J195113.62-331116.7 & 297.8064 & -33.1869 & 2090 & 117 & 100\\
  2MASS J19561542-1754252 & 299.0643 & -17.907 & 2209 & 55 & 101\\
  2MASS J20025073-0521524 & 300.7113 & -5.3646 & 1227 & 238 & 82\\
  WISE J200804.71-083428.5 & 302.0188 & -8.5742 & 907 & 104 & 91\\
  DENIS J2013108-124244 & 303.2951 & -12.7126 & 2042 & 111 & 201\\
  WISE J203042.79+074934.7 & 307.6764 & 7.8266 & 1496 & 138 & 31\\
  2MASS J20343769+0827009 & 308.657 & 8.4503 & 2064 & 86 & 75\\
  2MASS J20360316+1051295 & 309.0132 & 10.8582 & 1889 & 72 & 92\\
  2MASS J20414283-3506442 & 310.4285 & -35.1123 & 1830 & 74 & 113\\
  WISE J204356.42+622048.9 & 310.9834 & 62.3455 & 1390 & 79 & 58\\
  SDSS J204749.61-071818.3 & 311.9567 & -7.305 & 1478 & 157 & 118\\
  2MASS J20491972-1944324 & 312.3323 & -19.7424 & 2453 & 78 & 99\\
  SSSPM J2052-4759 & 313.1171 & -47.979 & 2400 & 119 & 169\\
  2MASSI J2057540-025230 & 314.4754 & -2.8751 & 1995 & 61 & 100\\
  SDSS J205755.92-005006.7 & 314.483 & -0.8352 & 2310 & 52 & 103\\
  2MASSI J2107316-030733 & 316.882 & -3.126 & 2180 & 72 & 66\\
  2MASS J21075409-4544064 & 316.9754 & -45.7351 & 2139 & 124 & 72\\
  HB88 M18 & 319.6324 & -45.0979 & 2406 & 49 & 116\\
  SDSS J212413.89+010000.3 & 321.0579 & 0.9999 & 1219 & 84 & 45\\
  2MASS J21263403-3143224 & 321.6418 & -31.7229 & 2389 & 69 & 125\\
  HB88 M19 & 321.8589 & -42.2551 & 2376 & 86 & 125\\
  HB88 M20 & 322.536 & -44.7745 & 2406 & 52 & 116\\
  2MASSW J2130446-084520 & 322.686 & -8.7557 & 2146 & 118 & 72\\
  SDSS J213240.36+102949.4 & 323.1681 & 10.497 & 1823 & 90 & 70\\
  2MASS J21324898-1452544 & 323.2041 & -14.8818 & 1295 & 92 & 88\\
  2MASS J21371044+1450475 & 324.2935 & 14.8466 & 2060 & 97 & 74\\
  2MASS J21373742+0808463 & 324.4058 & 8.1462 & 1677 & 129 & 131\\
  DENIS J2139136-352950 & 324.8069 & -35.4974 & 2132 & 69 & 100\\
  2MASS J21392676+0220226 & 324.8615 & 2.3396 & 1357 & 109 & 69\\
  2MASS J21403907+3655563 & 325.1629 & 36.9323 & 2208 & 96 & 156\\
  2MASS J21420580-3101162 & 325.5243 & -31.0212 & 1736 & 113 & 82\\
  2MASS J21481628+4003593 & 327.0679 & 40.0665 & 1253 & 268 & 68\\
  2MASS J21495655+0603349 & 327.4857 & 6.0597 & 2357 & 58 & 73\\
  2MASS J21512543-2441000 & 327.8562 & -24.6833 & 1567 & 189 & 73\\
  2MASS J21513839-4853542 & 327.9101 & -48.8984 & 1318 & 104 & 160\\
  2MASS J21542494-1023022 & 328.6038 & -10.3841 & 1275 & 38 & 109\\
  2MASS J21543318+5942187 & 328.6382 & 59.7051 & 1087 & 43 & 137\\
  2MASS J21580457-1550098 & 329.519 & -15.8361 & 1844 & 77 & 79\\
  DENIS J220002.0-303832B & 330.0085 & -30.6423 & 2381 & 125 & 198\\
  2MASS J22092183-2711329 & 332.341 & -27.1925 & 1465 & 55 & 86\\
  2MASS J22114470+6856262 & 332.9361 & 68.9406 & 1955 & 96 & 73\\
  WISEPC J221354.69+091139.4 & 333.4779 & 9.1943 & 856 & 151 & 37\\
  WISE J222219.93+302601.4 & 335.5829 & 30.433 & 1575 & 121 & 177\\
  WISEPC J222623.05+044003.9 & 336.597 & 4.6688 & 789 & 106 & 109\\
  2MASS J22282889-4310262 & 337.1205 & -43.174 & 1258 & 134 & 119\\
  WISEPC J223729.53-061434.2 & 339.373 & -6.2432 & 1415 & 102 & 147\\
  WISEPC J223937.55+161716.2 & 339.9052 & 16.287 & 1310 & 29 & 117\\
  2MASS J22425317+2542573 & 340.7217 & 25.7159 & 1730 & 72 & 83\\
  2MASSI J2254188+312349 & 343.5788 & 31.3972 & 1236 & 42 & 115\\
  2MASSI J2254519-284025 & 343.7164 & -28.6737 & 2189 & 87 & 97\\
  SDSSp J225529.09-003433.4 & 343.8712 & -0.576 & 2280 & 89 & 135\\
  WISEPC J225540.74-311841.8 & 343.9189 & -31.3112 & 782 & 135 & 197\\
  WISE J230133.32+021635.0 & 345.3891 & 2.2768 & 970 & 162 & 46\\
  DENIS J2308113-272200 & 347.0475 & -27.3668 & 2089 & 100 & 68\\
  SSSPM J2310-1759 & 347.577 & -17.986 & 2168 & 106 & 75\\
  2MASS J23174712-4838501 & 349.4463 & -48.6473 & 1582 & 215 & 39\\
  2MASS J23185497-1301106 & 349.729 & -13.0197 & 1092 & 147 & 165\\
  WISEPC J231939.13-184404.3 & 349.9128 & -18.7349 & 1375 & 115 & 185\\
  2MASS J2320292+412341 & 350.122 & 41.3949 & 2209 & 63 & 105\\
  2MASS J23211254-1326282 & 350.3024 & -13.4412 & 2054 & 97 & 87\\
  2MASS J23224684-3133231 & 350.6951 & -31.5565 & 2112 & 121 & 107\\
  2MASS J23231347-0244360 & 350.8062 & -2.7433 & 2504 & 73 & 158\\
  WISEPC J232728.75-273056.5 & 351.8689 & -27.5159 & 1464 & 143 & 115\\
  2MASS J23302258-0347189 & 352.5942 & -3.7886 & 2153 & 100 & 103\\
  2MASS J23312378-4718274 & 352.8491 & -47.3076 & 1335 & 88 & 125\\
  SDSS J233129.35+155222.5 & 352.8723 & 15.873 & 2187 & 55 & 111\\
  SDSS J233358.42+005012.1 & 353.4934 & 0.8366 & 2363 & 94 & 171\\
  SDSS J233526.42+081721.3 & 353.86 & 8.2893 & 2239 & 56 & 126\\
  2MASSI J2339101+135230 & 354.7926 & 13.8749 & 1095 & 102 & 54\\
  WISEPC J234026.62-074507.2 & 355.1105 & -7.7516 & 992 & 177 & 41\\
  ULAS J234228.97+085620.1 & 355.62 & 8.939 & 907 & 168 & 85\\
  2MASS J23440624-0733282 & 356.026 & -7.5579 & 1756 & 82 & 102\\
  SIMP J23444256+0909020 & 356.1773 & 9.1506 & 2065 & 81 & 189\\
  2MASS J23453903+0055137 & 356.4126 & 0.9205 & 2372 & 67 & 87\\
  APMPM J2347-3154 & 356.7281 & -31.8983 & 2331 & 82 & 163\\
  WISEPC J234841.10-102844.4 & 357.1694 & -10.4794 & 967 & 148 & 81\\
  2MASS J23512200+3010540 & 357.8417 & 30.1816 & 1642 & 124 & 100\\
  2MASS J23520507-1100435 & 358.0211 & -11.0122 & 2463 & 55 & 102\\
  2MASS J23535946-0833311 & 358.4977 & -8.5588 & 2450 & 82 & 116\\
  LHS 4039C & 358.5387 & -33.2741 & 2276 & 49 & 72\\
  DENIS J2354599-185221 & 358.7497 & -18.8727 & 2085 & 108 & 107\\
  SSSPM J2356-3426 & 359.0451 & -34.4346 & 2431 & 65 & 100\\
  2MASSI J2356547-155310 & 359.2282 & -15.8868 & 1237 & 129 & 37\\
  WISE J235716.49+122741.8 & 359.3187 & 12.4629 & 973 & 237 & 73\\
  SSSPM J2400-2008 & 359.9902 & -20.1279 & 2385 & 74 & 118

 \label{tab:ucds_teffs}
 \end{longtable}

\end{appendices}

\listoffigures
\listoftables
\clearpage

\clearpage
\thispagestyle{empty}
\null
\newpage

\clearpage
\thispagestyle{empty}
\vspace*{\fill}
\begin{center}
    \includegraphics[width=0.5\textwidth]{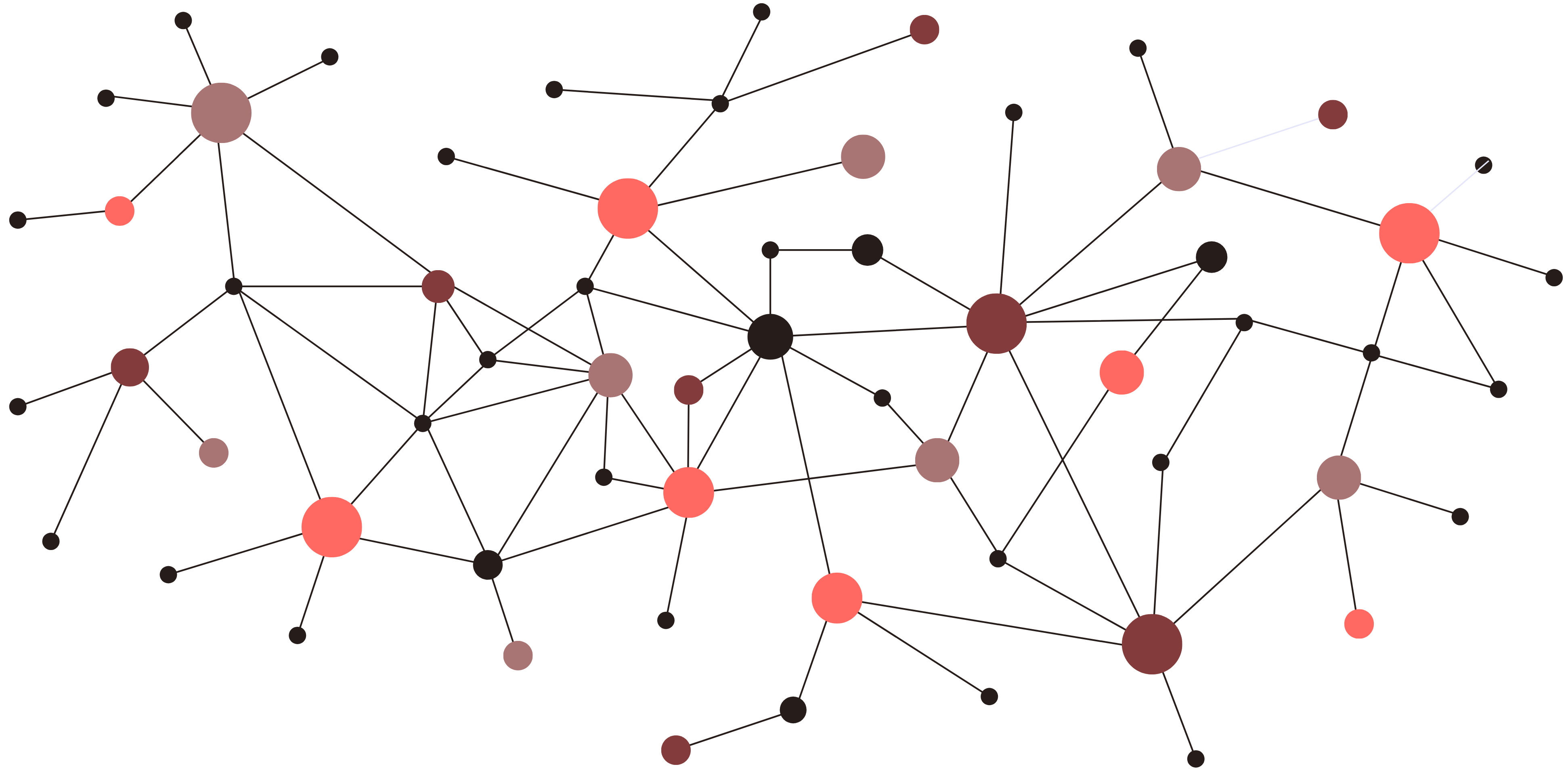}
\end{center}
\vspace*{.5cm} 
\clearpage

\includepdf[pages=2]{Thesis_Cover.pdf}

\end{document}